\documentclass{report}
\usepackage{graphicx} % Required for inserting images
\usepackage[a4paper, total={6in, 9in}]{geometry}
\usepackage[hidelinks]{hyperref}
\usepackage{braket}
\usepackage{slashed}
\usepackage{tabstackengine}
\stackMath
\usepackage[mathscr]{euscript}
\usepackage[dvipsnames]{xcolor}
\usepackage{amsmath} 
\usepackage{amssymb}
\usepackage{tocloft} % Package to customize the table of contents
\usepackage[numbers,sort&compress]{natbib}
\usepackage{cancel}
\usepackage{pdfpages}
\usepackage{rotating}

\usepackage{tikz} % Package for drawing graphics

\newlength{\dotspacing}
\usepackage{pgfplots}
\usepackage{pgfplotstable}
\usepackage{booktabs}

\pgfplotsset{compat=1.18}

\hypersetup{colorlinks,linkcolor = {teal}, citecolor={teal},urlcolor={teal},pdfborder = {0 0 0}, linktocpage=true}

\usepackage[]{appendix}

\newcommand{\mypapertitle}[1]{\def\papertitle{#1}}

\newcommand{\mypaperdate}[1]{\def\paperdate{#1}}
\newcommand{\mypaperabstract}[1]{\def\paperabstract{#1}}

\newcommand{\makepapertitle}{%
    { % Open group to limit the scope of \centering
    \centering
    {\LARGE \papertitle \par} % Title
    \vskip 1em

    \vskip 0.5em
    {\large \paperdate \par} % Date
    \vskip 2em
    \textbf{Abstract} % Abstract title
    \vskip 0.5em
    \begin{quote} % Abstract text
        \paperabstract
    \end{quote}
    \vskip 2em
    \par
    } % Close group to end centering
}

\newcommand{\Higgs}{\phi}
\newcommand{\lep}{L}
\newcommand{\Lnum}{\mathcal{L}}
\newcommand{\iflav}{\ell}
\newcommand{\dmsol}{\Delta m_{12}}
\newcommand{\dmatm}{\Delta m_{23}}
\newcommand{\MDI}{M_{\rm DI}}
\newcommand{\mnu}{\hat{m}_\nu}

\newcommand{\Mpl}{{M}_{\rm Pl}}
\newcommand{\Msol}{{M}_\odot}
\newcommand{\vEW}{v_{\rm EW}}
\newcommand{\etasphal}{\eta_{\rm sphal}}
\newcommand{\fiducial}{\mathcal{C}}
\newcommand{\GeV}{\mathrm{GeV}}
\newcommand{\eV}{\mathrm{eV}}
\newcommand{\g}{\mathrm{g}}
\newcommand{\Lagr}{\mathcal{L}}
\newcommand{\drm}{\mathrm{d}}

\newcommand{\Tsphal}{T_{\rm sphal}}
\newcommand{\Tewpt}{T_{\rm EWPT}}
\newcommand{\TEWPT}{T_{\rm EWPT}}
\newcommand{\TBBN}{T_{\rm BBN}}
\newcommand{\Tflav}{T_{\rm flav}}

\newcommand{\YB}{Y_B}
\newcommand{\YBm}{\tilde{Y}_B}
\newcommand{\YBmax}{Y_B^{\rm max}}
\newcommand{\YBL}{Y_{B-L}}
\newcommand{\YBobs}{Y_B^{\rm obs}}

\newcommand{\varrhopbh}{\varrho_{\rm PBH}}
\newcommand{\varrhorad}{\varrho_{\rm rad}}
\newcommand{\rhopbh}{\rho_{\rm PBH}}
\newcommand{\rhorad}{\rho_{\rm rad}}

\newcommand{\entropy}{\mathcal{S}} 

\newcommand{\MN}{M_N}
\newcommand{\MNi}{M_{N_i}}
\newcommand{\MNo}{M_{N_1}}
\newcommand{\MNt}{M_{N_2}}
\newcommand{\MNth}{M_{N_3}}
\newcommand{\ENi}{E_{N_i}}
\newcommand{\Mdegen}{M}
\newcommand{\Nc}{\bar{N}^c}
\newcommand{\UPMNS}{U_{\rm PMNS}}

\newcommand{\GNt}{\Gamma_{N_2}}
\newcommand{\GNi}{\Gamma_{N_i}}
\newcommand{\GNiT}{\Gamma_{N_i}^T}
\newcommand{\GNiID}{\Gamma_{N_i}^{ID}}
\newcommand{\GNTo}{\Gamma_{N_1}^T}
\newcommand{\Tevap}{T_{\rm evap}}
\newcommand{\alphaevap}{\alpha_{\rm evap}}
\newcommand{\MPBH}{M_{\rm PBH}}
\newcommand{\MPBHini}{M_{\rm PBH}^{\rm ini}}
\newcommand{\Tform}{T_{\rm form }}
\newcommand{\alphaform}{\alpha_{\rm form }}
\newcommand{\Sr}{\mathcal{S}_{\beta}/\mathcal{S}_{\beta = 0}}

\newcommand{\NNo}{\mathcal{N}_{N_1}}

\newcommand{\NNi}{\mathcal{N}_{N_i}}

\newcommand{\NBLl}{\mathcal{N}_{B-\Lnum_\iflav}}
\newcommand{\NNeqo}{\mathcal{N}_{N_1}^{\rm eq}}
\newcommand{\NNeq}{\mathcal{N}_{N_i}^{\rm eq}}
\newcommand{\Nleq}{\mathcal{N}_{\ell}^{\rm eq}}
\newcommand{\NPBH}{\mathcal{N}_{\rm PBH}}
\newcommand{\equaref}[1]{Eq.\,\ref{#1}}
\newcommand{\chapref}[1]{Chapter\,\ref{#1}}
\newcommand{\secref}[1]{Sec.\,\ref{#1}}
\newcommand{\partref}[1]{Part\,\ref{#1}}
\newcommand{\figref}[1]{Fig.\,\ref{#1}}

\newcommand{\rcrit}{r_{\rm core}}
\newcommand{\rdec}{r_{\rm dec}}

\newcommand{\rHS}{r_{\rm HS}}
\newcommand{\rsphal}{r_{\rm sphal}}
\newcommand{\rcore}{r_{\rm core}}
\newcommand{\rFO}{r_{\rm FO}}

\newcommand{\Tcore}{T_{\rm core}}
\newcommand{\Tplasma}{\tilde{T}}
\newcommand{\TBH}{T_{\rm BH}}
\newcommand{\TFO}{T_{\rm FO}}
\newcommand{\Tmax}{T_{\rm max}}

\newcommand{\mDM}{m_{\rm DM}}
\newcommand{\mZp}{m_{\rm Z^\prime}}
\newcommand{\DM}{\psi}

\newcommand{\mHiggs}{m_{\Higgs}}
\newcommand{\mHiggsT}{M_{\Higgs}}
\newcommand{\mLepT}{M_{\lep}}

\newcommand{\tini}{t_{\rm ini}}

\begin{document}

\title{\huge Thesis}
\author{\LARGE Jacob Gunn\vspace{10mm} \\ 
\LARGE Università degli Studi di Napoli Federico II}
\date{\today}

\includepdf[pages=1]{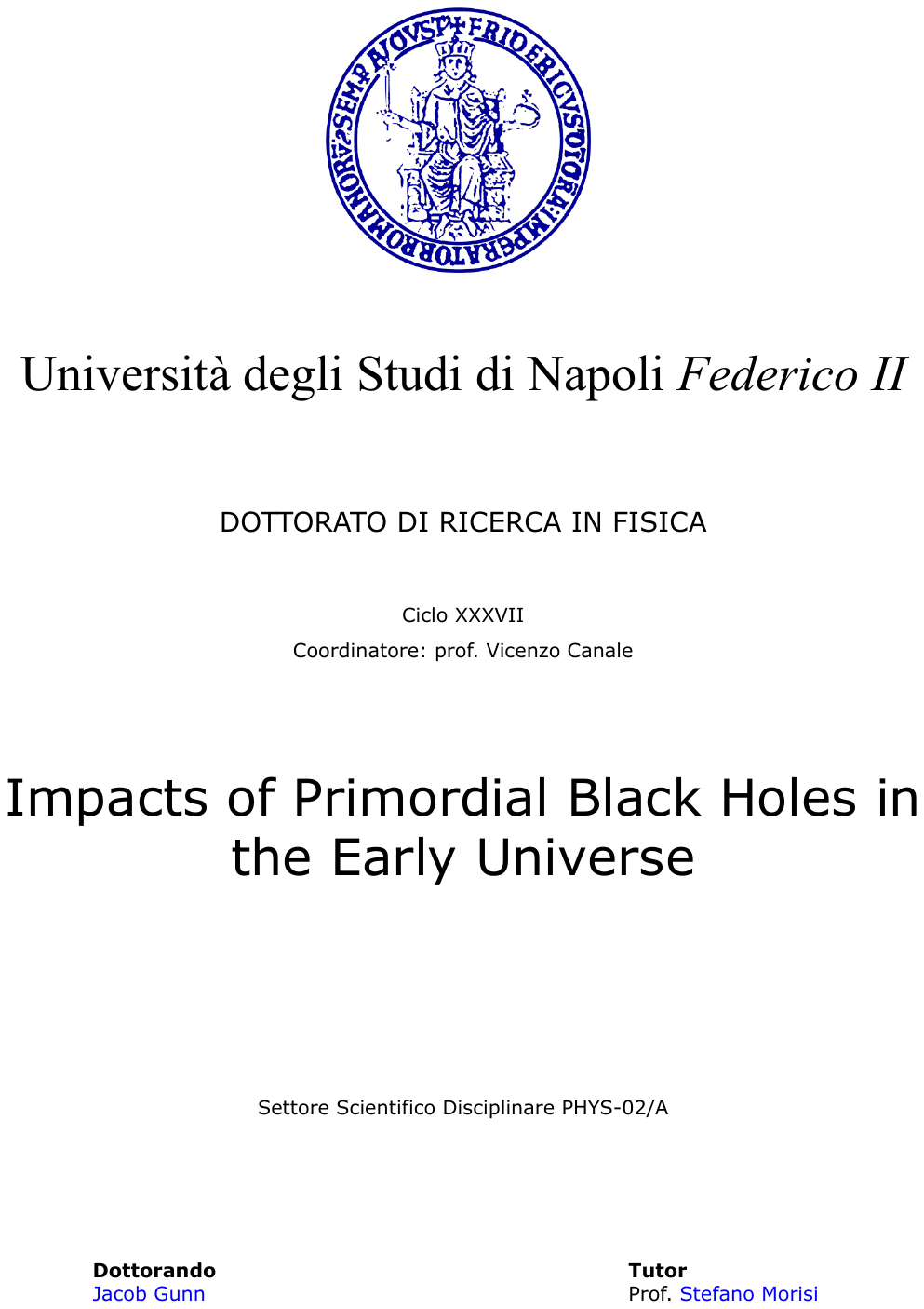}

\newpage
\begin{abstract}
    I dedicate this thesis to my late father, Phillip John Gunn, who's tenacity and integrity in life guide my aspirations, and who's memory inspires me to live without compromise. The work composing this thesis would not have been possible without my mentors, family and colleagues. I am grateful to Professor Stefano Morisi for his mentorship as my PhD supervisor, which shaped the research presented in this thesis and informed my professional and personal growth over the last 3 years. During my time in Naples I have been humbled by the support of many of my colleagues especially Dr Marco Chianese, Professor Pietro Santorelli, Dr Ninetta Saviano and Guido Celentano. I also owe much thanks to Professor Jessica Turner, who has selflessly and relentlessly supported me academically and personally since we met. To my amazing fiancée Saranya Joseph, thank you for everything you do for me. If not for your patience and empathy, it would have been impossible to finish this journey. To my mother Samantha Gunn, your guidance, support and love opened all of the doors through which I have stepped so far in life, this journey would never have begun without your influence. I still aspire to your strength, resilience and compassion. Finally to my grandmother Gillian Gunn, I appreciate that our conversations sparked many of my questions about the universe, and I have always been moved by the genuine curiosity you have shown in my work.
\end{abstract}

\tableofcontents 

\newpage

\section*{Introduction}

For over two decades the unexplained origin of the neutrino mass has stood as a continual provocation to look beyond the Standard Model (SM) of particle physics. Despite its remarkable success at describing a wide range of particle physics phenomena, the SM offers no explanation for the $\sim 0.1 \eV$ scale masses of the neutrinos - particles which are emphatically massless in the SM framework. No cosmological argument is necessary to see that particle physics' concordance model falls exactly 3 fermions short of explaining the masses of all known particles. 26 years on from the discovery of oscillations in atmospheric neutrinos at Super Kamiokande \cite{Super-Kamiokande:1998kpq}, the interesting question to ask is no longer whether or not the SM must be extended, but how?\\

Promoting the SM to a cosmological theory, underwritten by the geometry of Einstein's general relativity, one discovers yet more deficiencies. There exists no viable mechanism in the SM by which an asymmetry between baryons and anti-baryons could come about and persist until today. Annihilation amongst ancient baryons would render the late-time universe devoid of ordinary matter, unless some physics beyond the SM (BSM) drove a dynamical generation of the relic asymmetry. Perhaps most strikingly, the particle nature of Dark Matter (DM) which according to Planck constitutes $84.3\%$ of the matter in the universe \cite{Planck:2018vyg}, is completely unknown. Since the SM contains no suitable candidates, the existence of DM - corroborated by an overwhelming array of gravitational probes - demands extension of the SM.\\

Beyond the requirement for DM production and baryogenesis, very little is known about the earliest epochs of the universe. Big Bang Nucleosynthesis (BBN) confers some of the tightest constraints, being highly sensitive to conditions at $\TBBN \sim \mathrm{MeV}$, when the universe was around $1\rm s$ old. At earlier times, the history of our universe is cloaked in shadow. A period of exponential expansion, known as \textit{inflation} and driven by the \textit{inflaton}, is widely expected to precede the hot big bang. Canonically, inflaton decays and populates cosmological populations of particles at the so-called reheating temperature, though this temperature remains a mystery. Besides necessarily occurring before BBN, inflation could reasonably be expected to reheat the universe at any temperature up to and including the Grand Unified Theory (GUT) scale $\Lambda_{\rm GUT} \sim 10^{16} \GeV$. Furthermore since the nature of DM is unknown, it is also unknown when or how DM was produced in the early universe. DM production must have occurred prior to recombination in order to generate the baryon acoustic oscillations observed in the Cosmic Microwave Background (CMB). However recombination occurs long after BBN, leaving the first seconds of the universe unconstrained by DM production. Similarly to inflation, BBN constrains baryogenesis to occur at $T > \TBBN$ so that the correct abundances of light elements are synthesised. Clearly baryogenesis must also occur post-reheating by the inflaton, but given the huge freedom in the reheating temperature, baryogenesis remains free to occur almost anywhere in the cosmological chronology. Painting a precise picture of the primordial universe will be essential to resolve some of the most profound mysteries in modern physics, and progress here will rely on novel theoretical and experimental methods in the coming years.\\

Particularly appealing are theoretical frameworks in which multiple aspects of this project can be addressed simultaneously. For example, the Type-1 seesaw mechanism \cite{Yanagida:1980xy,Yanagida:1979as} is a simple extension of the SM which would explain the origin of the neutrino mass and provides a mechanism of baryogenesis via leptogenesis. In some circumstances the seesaw mechanism could even produce the relic abundance of DM. The Type-1 seesaw mechanism is sensitive to the masses of the SM (active) neutrinos and the masses of the sterile neutrinos it introduces (as well as other parameters). Throughout this work, the gauge singlet fermions introduced by the Type-1 seesaw mechanism are referred to interchangeably as sterile neutrinos as Right Handed Neutrinos (RHNs). Sensitivity projections for near-future experiments suggest that $\sim \GeV$ scale sterile neutrinos could soon be probed, while the active neutrino mass is subject to constraints at the $\sim \eV$ level. However many variations of the seesaw model feature sterile neutrinos with masses much larger than the $\GeV$ scale, up to the Planck scale. While the active neutrino mass must be within an order of magnitude or so of current constraints, sterile neutrinos at the Planck scale would surely remain beyond the reach of experiments for the foreseeable future. For successful leptogenesis, low scale $\sim \GeV$ sterile neutrinos require very small couplings to the SM. While detection possibilities do present in the margins of viable parameter space, we would be exceptionally lucky to discover such light and weakly coupled sterile neutrinos. Moreover a discovery of $\sim \GeV$ sterile neutrinos would not on its own provide enough information to claim leptogenesis occurred at all. As is the case with DM production and inflation, the Type-1 seesaw mechanism has proved incredibly difficult to constrain experimentally. Perhaps it will be necessary to approach these problems from an entirely different perspective.\\

One of the most fascinating ways this may be possible is through the lens of Primordial Black Holes (PBHs), black holes which may have formed in the early universe (as opposed to astrophysical black holes which are formed by collapsing stars). The present day cosmological abundance of PBHs is tightly constrained by various probes such as the non-observation of extragalactic gamma-rays and gravitational microlensing events. However PBHs in the ultralight mass range, ranging from a few grams to a few billion grams, would have evaporated completely before the onset of BBN and therefore cannot be constrained by direct observations. Ultralight PBHs form and die in the early cosmos, and could drastically alter the evolution of the universe. The violent final phase of their lifecycle produces abundant high energy particles, including BSM states, such that cosmologically significant populations inevitably have strong impacts on particle physics processes in the early universe. These impacts have been studied in the literature since PBHs were proposed by Hawking in the 1970s (see \secref{PBHs} for relevant references), but the field has experienced something of a renaissance in recent years following the detection of Gravitational Waves (GWs) by LIGO in 2016 \cite{LIGOScientific:2016aoc}. PBHs leave GW signatures associated with their formation and evaporation. It is especially interesting that for ultralight PBHs both of these signals can overlap with the projected sensitivity regions of upcoming GW experiments \cite{Bhaumik:2023wmw}. The tantalising prospect of detecting the gravitational echos of long dead black holes brings with it the possibility to draw conclusions on the unresolved issues of the neutrino mass, baryogenesis and DM. If the impacts of PBHs in the early universe are well understood, information from GW experiments could illuminate the darkest era of cosmological history.\\

By exploring these impacts, this thesis seeks to demonstrate how PBHs shed light on some of the most difficult to constrain physics of the primordial epoch. Particular attention is paid to the interplay of PBHs with leptogenesis, and the phenomenology of the locally heated regions around PBHs, called \textit{hot spots}. Anticipating future information on ultralight PBHs from GW experiments, one of the key aims of this thesis is to show how this information could be leveraged to rule out leptogenesis scenarios far beyond what is possible with particle detection experiments. This would help narrow the search for sterile neutrinos and focus theoretical efforts. Furthermore the work presented in this thesis pioneers the study of Hawking radiation in hot-spots. Efforts to understand the role that Hawking radiation plays in cosmology have historically missed the impact of hot spots. By treating the propagation of Hawking radiation in the extreme temperature gradients of the hot spot, this thesis sets out to elucidate some of the rich phenomenology occurring in hot spots; absorption of DM and leptogenesis sustained after sphalerons freeze-out.\\

The structure of this work is as follows. The thesis as a whole is structured in two main parts. Part 1 is concerned broadly with the particle physics occurring at the earliest times in the universe, intending to establish key cosmological and thermodynamic results necessary for approaching the open questions of the neutrino mass origin, baryogenesis and DM. Part 2 covers the physics of PBHs and Hawking radiation, addressing the most important known results and current state-of-the-art. This is followed by the original research and key results of the thesis concerning the impacts of PBHs in the early universe.\\

In Part 1, \chapref{SM} looks briefly at the origin of the neutrino mass, the nature of DM and the baryon asymmetry of the universe in more detail, motivating the search for new physics. \secref{SM/vMass} starts from the SM Electroweak Lagrangian to discuss why neutrinos are massless in the SM and then overviews the existing evidence for their masses, \secref{SM/DM} reviews the evidence for DM and clues as to its nature before \secref{SM/Baryogenesis} introduces the problem of baryogenesis in detail. \chapref{Early Universe} introduces the general relativistic framework necessary for studying the early universe and establishes key results concerning cosmological particle processes. \secref{Early Universe/Cosmology} derives the Friedmann equations and crucial thermodynamic results from Einstein's gravitational field equations, \secref{Early Universe/Abundance} sets out the framework for calculating cosmological abundances and introduces the Boltzmann equations formalism, then \secref{Early Universe/Out-of-Equilibrium} delves into the physics of out-of-equilibrium processes in the early universe before \secref{Early Universe/TFT} summarises some important results concerning finite temperature effects. \chapref{vMass} explores in detail the Type-1 seesaw mechanism as an explanation for the neutrino mass. \secref{vMass/Seesaw} shows how the seesaw mechanism generates neutrino masses starting from the seesaw Lagrangian, then \secref{vMass/Casas-Ibarra} demonstrates a convenient parameterisation for the seesaw mechanism, then \secref{vMass/Mixing} discusses relevant aspects of active-sterile neutrino mixing. \chapref{Leptogenesis} extends considerations to the mechanism of baryogenesis via leptogenesis in the Type-1 seesaw framework. \secref{Leptogenesis/Sphalerons} introduces sphalerons starting from the SU(2) Lagrangian and covers the sphaleron rate and freeze-out, \secref{Leptogenesis/CP} is concerned with the origin of CP violation in leptogenesis and its behaviour, while \secref{Leptogenesis/Out-of-Equilibrium} covers the departure from equilibrium, also deriving the sterile neutrino decay rates and thermal averaging. \secref{Leptogenesis/Thermal} applies finite temperature considerations to leptogenesis before \secref{Leptogenesis/BEs} details the Boltzmann equations for leptogenesis. Then, \secref{Leptogenesis/Resonant} tackles the enhancement of CP asymmetry in the special case of resonant leptogenesis, and the origin of the required small mass splitting, after which \secref{Leptogenesis/Experiment} overviews the experimental status of leptogenesis.

In Part 2, \chapref{PBHs} reviews the historical context for PBHs and elucidates the physics necessary for studying their role in cosmology. \secref{PBHs/Formation} derives the formation temperature of PBHs, discusses formation by inflationary perturbation collapse and the parameter space of PBHs. \secref{PBHs/Hawking} revisits Hawking's famous result for the black hole surface temperature, and gives expressions for the rates of particle production by black holes. \secref{PBHs/Cosmology} reviews the cosmological evolution of a universe with a PBH population, \secref{PBHs/Hot Spots} then gives a detailed account of the state-of-the-art understanding of hot spots around PBHs. The key results of this thesis begin in \chapref{PBHLepto}, investigating the interplay between PBHs and leptogenesis and drawing mutual exclusion limits between their parameter spaces. \secref{PBHLepto/High Scale} applies the considerations of Part 1 and Part 2 to the model of high scale leptogenesis, demonstrating how PBHs can constrain leptogenesis. The PBH-induced modification of the sphaleron freeze-out temperature is calculated for the first time and future neutrino-mass results are anticipated. \secref{PBHLepto/Low Scale} then considers the impact of PBHs on the low scale resonant leptogenesis model, constraints on PBHs are derived depending on the sterile neutrino mass such that future detections of PBHs or sterile neutrinos are inextricably linked. Finally, \secref{Hot Spots} tackles the phenomenology of Hawking radiation in PBH hot spots. \secref{Hot Spots/Evolution} looks for the first time at the evolution of a hot spot in an expanding universe, then \secref{Hot Spots/Hawking} sets out a formalism for treating the propagation of Hawking radiation in a hot spot. \secref{Hot Spots/Lepto} applies the hot spot considerations to the case of leptogenesis, deriving regions which support successful leptogenesis after sphaleron freeze-out, before \secref{Hot Spots/DM} considers DM production revealing that hot spots efficiently absorb DM and drastically alter constraints one would draw from DM production on PBHs. \chapref{Conclusions} concludes the thesis and discusses the outlook. Appendix \ref{A} provides a detailed derivation of the thermal averaging factor for a $2\leftrightarrow 2$ cross section where the incoming particles have different temperatures and different masses, as is the case in Hawking radiation scattering. Appendix \ref{B} provides analytical expressions for the scattering cross sections of DM used in \secref{Hot Spots/DM}. Appendix \ref{C} considers the impact of hot spots on the constraints derived in \chapref{PBHLepto}, justifying \textit{a posteriori} the treatment of homogeneous heating.

\part*{Part 1}

\chapter{\textbf{Going Beyond the Standard Model}}\label{SM}

\section{Neutrino Mass}\label{SM/vMass}
Writing the SM Electroweak Lagrangian, one must include every term compatible with the $SU(2)_L \times U(1)_Y$ gauge symmetry
\begin{equation}\label{SM/Neutrino Mass/EWLagr}
    \Lagr_{\rm EW} = \Lagr_g + \Lagr_k + \Lagr_\Higgs + \Lagr_y
\end{equation}
where $\Lagr_g$ describes the interactions of the $W,B$ vector bosons, $\Lagr_k$ contains the kinetic terms for the SM fermions, $\Lagr_\phi$ is the Higgs part of the Lagrangian describing its quartic and gauge interactions, and $\Lagr_y$ is the Yukawa interaction coupling the Higgs to the SM fermions
\begin{equation}\label{SM/Neutrino Mass/Lagry}
    \Lagr_y \supset -h_{\ell\,l} \bar{\lep}_\ell e_R^l \Higgs + h.c.
\end{equation}
where $\ell,l$ are flavour indices, $h$ is the Yukawa matrix coupling the Higgs doublet $\Higgs$ to the right handed charged lepton fields $e_R^\iflav$ and left handed lepton doublets
\begin{equation}
    \lep_\iflav = \begin{pmatrix}
        \nu_\iflav \\
        e_\iflav 
    \end{pmatrix}
\end{equation}
where $\nu$ are the neutrinos. Mass terms like $\bar{f}_Rf_L + \bar{f}_L f_R$ where $f$ is some fermion are forbidden by the gauge symmetry, the EW Lagrangian contains no mass terms for fermions before symmetry breaking and so at high energies the SM fermions are formally massless. Via the Higgs mechanism \cite{Higgs:1964ia,Higgs:1964pj,Higgs:1966ev}, the SM fermions including the leptons gain masses when the Higgs acquires a vacuum expectation value and terms like \equaref{SM/Neutrino Mass/Lagry} generate masses between the left and right handed fields. In the SM, neutrinos have only left handed components. Clearly then in the SM neutrinos are massless fields.\\

If some mechanism beyond the SM were to generate masses for the neutrinos, it would be possible for neutrinos to oscillate between flavours. This could occur because the neutrino weak interaction eigenstates are not identical to the mass eigenstates
\begin{equation}
    \nu_\iflav = \sum_i \UPMNS^{\iflav i}\nu_i
\end{equation}
where $\iflav,i$ index the flavour and mass eigenstates respectively. $\UPMNS$ is the Pontecorvo-Maki-Nakagawa-Sakata matrix \cite{Pontecorvo:1967fh,pontecorvo1957mesonium,pontecorvo1957inverse,Maki:1962mu} describing the mixing between flavour and mass eigenstates and can be parameterised in terms of three mixing angles. For simplicity, consider the mixing between electron and muon type neutrinos, so that there is a single relevant mixing angle $\theta$. Electron neutrinos produced as their weak interaction eigenstate may oscillate $\nu_e \to \nu_\mu$ with probability
\begin{equation}
    P(\nu_e \to \nu_\mu) = \sin^2(2\theta)\sin^2\left( \frac{\Delta m_{21}^2 L}{4E}\right) \end{equation}
where $L$ is the distance propagated, $E$ is the neutrino energy and $\Delta m_{21}^2 = m_2^2 - m_1^2$ is the square mass splitting between the two states. Clearly if the neutrinos are massive and not exactly degenerate, there is a non-zero probability of oscillations occurring. That such oscillations have been observed experimentally offers perhaps the best evidence that the SM has to be extended, at least to explain neutrino masses.

For example the disappearance of $\nu_\mu$ due to oscillations was observed in atmospheric neutrinos at Super-Kamiokande \cite{Super-Kamiokande:1998kpq,Super-Kamiokande:1998zvz} and subsequently in reactor neutrinos at the K2K \cite{K2K:2006yov}, MINOS \cite{PhysRevLett.97.191801} and T2K experiments \cite{T2K:2012qoq}. $\nu_e$ disappearance was observed by the KamLAND experiment \cite{KamLAND:2002uet}.
$\nu_e-\nu_\mu$ oscillations have been directly observed by the NOvA and T2K experiments \cite{NOvA:2019cyt,T2K:2019bcf}, while evidence for appearance of $\nu_\tau$ due to oscillations was found by Super-Kamiokande \cite{Super-Kamiokande:2012xtd}. Currently the best fit values to the available oscillation data are \cite{Esteban:2020cvm}
\begin{eqnarray}\label{SM/vMass/splittings}
    \Delta m_{21}^2 = (7.42\pm 0.2)\times 10^{-5}\mathrm{eV}^2\\
    \Delta m_{31}^2 = (2.517\pm 0.027)\times 10^{-3}\mathrm{eV}^2.
\end{eqnarray}
In this discussion, and in the remainder of this work, normal ordering of the neutrino mass hierarchy is assumed, such that $m_1 < m_2 < m_3$. Therefore the notation $m_l = m_1$, $m_m = m_2$, $m_h = m_3$ will sometimes be used to make explicit the heaviest, median and lightest neutrinos. In practice it is not known whether the neutrino mass hierarchy is normal or inverted hierarchy (IH) ($m_3 < m_1 < m_2$), nor is the overall scale of the neutrino mass matrix known. More information from experiments is needed to answer these crucial questions, but it is nonetheless obvious that the origin of the neutrino mass must come from beyond the SM.

\newpage

\section{Dark Matter}\label{SM/DM}
According to the Planck collaboration, $84.3\%$ of the matter content of the universe is non-baryonic \cite{Planck:2018vyg}. Of this, leptons make up a vanishingly small amount and the precise particle nature of the vast majority of the matter in the universe is unknown, Dark Matter (DM). The existence of DM has repeatedly been confirmed at a wide variety of experiments ranging from cosmological to astrophysical probes. Notably it is well known that the rotation curves of stars are sustained at much larger distances from the galactic centre than expected from the visible matter content, implying a large DM component. For a review see \cite{Battaner:2000ef}. The evidence provided by the Planck collaboration that almost all of the matter content of the universe is non-baryonic combines data from baryon acoustic oscillations, CMB polarisation and the CMB power spectrum. The so-called Bullet Cluster (in fact two colliding clusters) provides intriguing, direct evidence for Dark Matter because X-ray observations show that the baryonic, visible matter is dragged relative to the centre of masses of the clusters, which gravitational lensing shows have passed through one another apparently unhindered \cite{Clowe:2003tk}.\\

All of these probes of DM rely on its gravitational impacts; on the rotations of starts in galaxies, the dynamics of colliding clusters and its role in the cosmological expansion. There currently exists no evidence for DM using visible sector probes, whatever the particle nature of DM it must be so weakly coupled to the SM as to have evaded every detection effort made thus far. Therefore, clues as to the particle nature of DM must also come from gravitational probes \cite{Buckley:2017ijx}. In $\Lambda$CDM DM is modeled as cold and completely dark, and while simulations have been generally very successful at reproducing large scale structure in this paradigm, multiple challenges to the model persist at small scales, for reviews see \cite{Bullock:2017xww,Efstathiou:2024dvn}. For example, many authors interpret the tension between $\Lambda$CDM N-body simulations of galaxy formation and the morphologies observed in Dwarf galaxies as as suggesting that DM is composed of ultralight scalars \cite{Hui:2016ltb} which form solitonic cores in halos and exhibit fascinating wave-like behaviors on astrophysical and cosmological scales. A natural and suitable candidate with these properties is the hypothetical QCD axion \cite{Peccei:1977hh}, which was introduced by Peccei and Quinn to explain the strong CP problem. For a review of axions in cosmology see \cite{Marsh:2015xka}.\\

On particle physics grounds, it has been suggested that RHNs, appearing in extensions of the SM seeking to explain neutrino mass (see \chapref{vMass}), could additionally play the role of Dark Matter \cite{Asaka:2005an,Canetti:2012kh}. The well known WIMP miracle, whereby weakly-interacting-massive-particles (WIMPs) with weak scale cross sections and masses produce the correct relic abundance of DM, has failed to return any direct detection thus far with limits on cross sections pushing many orders of magnitude below the weak scale, however efforts are ongoing. For a recent review see \cite{Roszkowski:2017nbc}. Along with those mentioned, a plethora of BSM candidates for DM remain possible, but there are no suitable candidates in the SM. Neutrinos are weakly coupled but as DM candidates they are too warm, failing to produce structure in the universe \cite{Tremaine:1979we}. Astrophysical Black Holes (BHs) are dark and massive, but were not present in the early universe to drive structure formation. Interestingly, Primordial Black Holes (PBHs) would have formed in the early universe and may still be a viable candidate for DM, requiring no explicit particle extension of the SM although most formation mechanisms rely on the physics of the inflaton or other exotic fields to generate the large overdensities required to form them. See \chapref{PBHs} for more details. \\

The existence of DM then, and the lack of viable particle candidates in the SM, provides an additional imperative to extend the SM. 
\newpage

\section{Baryon Asymmetry}\label{SM/Baryogenesis}
Considering the conditions from which the complex present day cosmos must have evolved, it is tempting to presume that the universe was initially symmetric with respect to baryon number
\begin{equation}
    B(t = 0) = n_B(t=0) - n_{\bar{B}}(t = 0) = 0
\end{equation}
where $n_B$ and $n_{\bar{B}}$ are the number densities of baryons and anti-baryons respectively. Immediately though a problem presents itself, baryons and anti-baryons annihilate when they meet yet the present universe is obviously composed entirely of baryons, there are almost no anti-baryons left in the universe. One may therefore imagine that the relic density of baryons comes from an initial asymmetry $B(t = 0) \neq 0$ which survived until the present day. Observations are sensitive to the ratio of the baryon asymmetry today to the number density of photons, and is measured by Planck to be \cite{Planck:2018vyg}
\begin{equation}
\eta_B=(n_B-n_{\bar{B}})/n_\gamma = (6.21\pm 0.16)\times 10^{-10}\,.
\end{equation}
The measured value is extremely unnatural from a technical standpoint as an initial condition\footnote{Technically natural means in this context the property of a theory to have order unity parameters}. Taking the perspective of eternal inflation, if $B(t=0)$ can be treated as a parameter which the inflationary multiverse randomly distributes across patches exiting inflation, such a tiny value would appear extremely infrequently and observers ought to be surprised to find themselves in such a universe. Anthropic selection effects might assuage this conclusion if it turned out that $\eta_B$ must have its observed value or close to it in order for structure and intelligent life to form. Interestingly, Bernard Carr and Martin Rees explored this idea in 1977, finding no particular requirement for the baryon-photon-ratio other than an lower limit beyond which stars and galaxies would not form, of order $\sim 3\times 10^{-10}$ \cite{Carr:1979sg}. In fact this lower limit is close to the modern day observed value today, a coincidence for which no explanation is forthcoming. \\

Dynamical generation of the observed baryon asymmetry, or baryogenesis, is therefore appealing. An initially symmetric (or almost symmetric) baryonic sector could evolve into the asymmetric one observed today through the out-of-equilibrium dynamics of some $B$-violating process. Sakharov wrote down the three crucial conditions for successful baryogenesis in 1967 \cite{Sakharov:1967dj}; CP violation is necessary so that the mechanism can preferentially create baryons over antibaryons, $B$-violation ensures that an overall baryon number can be generated, and departure from equilibrium prevents produced asymmetry from being washed out completely by inverse processes. Remarkably all of these conditions can be satisfied in the SM, field configurations known as sphalerons violate $B$ \& $CP$ and would depart from equilibrium around the time of the Electroweak Phase Transition (EWPT), for details on sphalerons see \secref{Leptogenesis/Sphalerons}. However baryogenesis in the SM, known as electroweak baryogenesis \cite{Kuzmin:1985mm,Shaposhnikov:1987tw}, is not viable following the measurement of the Higgs mass by the ATLAS collaboration, $\mHiggs \approx 125\GeV$ \cite{ATLAS:2012yve}. The Higgs mass was also measured by the CMS collaboration \cite{CMS:2012qbp}, for a recent measurement see \cite{CMS:2024eka}. If the Higgs were significantly lighter, $\mHiggs \approx 75\GeV$, the EWPT could have been a first order phase transition, with the sphaleron rate in the bubbles of broken phase sufficiently suppressed to conserve some relic asymmetry. However with the Higgs mass now known, the EWPT would have been a second order phase transition and the mechanism is known not to work without extending the SM. Therefore it is widely accepted that any realistic mechanism of baryogenesis requires some extension of the SM. If nature were completely described by the SM then a initially symmetric universe would be completely devoid of baryons and therefore readers of this thesis. The very fact that you are reading these words is evidence that the SM must be extended.\\

\chapref{Leptogenesis} focuses on one of the most interesting and popular baryogenesis mechanisms, leptogenesis. Indeed, the model on which leptogenesis is based, the type-1 seesaw model, can also explain the origin on neutrino masses, this is explored in \chapref{vMass}. For these purposes, the parameterisation 
\begin{equation}
    \YB \equiv \frac{n_B - n_{\bar{B}}}{s}
\end{equation}
is used, where $s$ is the entropy density of the universe. The observed value of this quantity is straightforwardly related to $\eta_B$ and is
\begin{equation}
    \YBobs = 8.7\times 10^{-11}\,.
\end{equation}
\newpage

\chapter{The Early Universe}\label{Early Universe}
This chapter aims to introduce notation and establish the key physics necessary for studying particle processes in the early universe. First, \secref{Early Universe/Cosmology} explores the general relativistic foundations for studying the evolution of the universe. \secref{Early Universe/Abundance} motivates the Boltzmann equations formalism for tracking the evolution of cosmological number densities and establishes some key results and notation. Then, \secref{Early Universe/Out-of-Equilibrium} is concerned with the out-of-equilibrium dynamics of particle processes in an expanding universe. Finally \secref{Early Universe/TFT} reviews some important results of thermal field theory for early universe cosmology.

\section{Cosmology}\label{Early Universe/Cosmology}
The dynamics of the expansion of the universe can be described by the evolution of the scale factor, $a$, with respect to the cosmic time $t$. A common choice is to set $a = 1$ today, but any choice of scale factor normalisation is equally valid. In 1915 Albert Einstein wrote down his famous field equations of gravitation \cite{Einstein:1916vd,Einstein:1915ca}
\begin{equation}\label{Early Universe/Cosmology/Einstein}
    R_{\mu\nu} -\frac{1}{2}R = 8\pi G T_{\mu\nu}
\end{equation}
where $R_{\mu\nu}$ is the Ricci tensor, $g^{\mu \nu}$ is the metric tensor, $R = g^{\mu \nu}R_{\mu\nu}$ is the scalar curvature, $T_{\mu \nu}$ is the energy-momentum tensor and $G$ is Newton's gravitational constant $G =\Mpl^{-2} = 6.7\times10^{-39}\GeV^{-2}$. The Ricci tensor is given in terms of the Christoffel symbols as
\begin{equation}
    R_{\mu\nu} = \partial_\lambda \Gamma^\lambda_{\mu\nu} + \Gamma^\lambda_{\mu\nu}\Gamma^\sigma_{\mu\sigma} - (\partial_\mu \Gamma^\lambda_{\nu\lambda} + \Gamma^\lambda_{\mu\sigma}\Gamma^\sigma_{\lambda\nu})
\end{equation}
while the Christoffel symbols are calculated as
\begin{equation}
    \Gamma^\mu_{\nu\lambda} = \frac{1}{2}g^{\mu\sigma}(\partial_\nu g_{\lambda\sigma}+ \partial_\lambda g_{\nu\sigma} - \partial_\sigma g_{\nu\lambda})\,.
\end{equation}
Clearly one must choose a metric with which to work. The cosmological principle asserts that the universe should be homogeneous and isotropic on sufficiently large scales. These properties are captured by the Friedmann-Lemaître-Robertson-Walker (FLRW) metric 
\begin{eqnarray}
    g_{00} = 1 \\
    g_{ij} = -a^2(t)\gamma_{ij}(x)
\end{eqnarray}
where $\gamma_{ij}$ is the metric of a 3-sphere, 3-hyperboloid or 3-plane depending on the radius of curvature. Therefore, the nonvanishing components of the Christoffel symbols are
\begin{eqnarray}\label{Early Universe/Cosmology/Christoffel}
    \Gamma^i_{0j} = \frac{\dot{a}}{a}\delta^i_j \nonumber\\
    \Gamma^0_{ij} = a\dot{a}\gamma_{ij} \\
    \Gamma^i_{jk}  = ^{(3)}\Gamma^i_{jk} \nonumber 
\end{eqnarray}
where $^{(3)}\Gamma^i_{jk}$ are the Christoffel symbols with metric $\gamma_{ij}$. Using these non-vanishing components, the $00$ component of the Ricci tensor can be calculated as
\begin{eqnarray}
    R_{00} = -3\frac{\Ddot{a}}{a}
\end{eqnarray}
while the spatial components are given by
\begin{equation}
    R_{ij} = (\Ddot{a}a + 2\dot{a}^2 + 2\mathcal{R})\gamma_{ij}
\end{equation}
where $\mathcal{R}$ parameterises the curvature of the universe. $\mathcal{R} = 0$ gives the 3-plane metric, used throughout this thesis since the most recent available data suggests that our universe is nearly flat \cite{Planck:2018vyg}. The above results for the Ricci tensor components allow to calculate the scalar curvature as 
\begin{eqnarray}
    R &=& g^{00}R_{00} + g^{ij}R_{ij}\\
    &=& -6\left(\frac{\Ddot{a}}{a} + \frac{\dot{a}^2}{a^2} \right)\,.
\end{eqnarray}
Modelling the cosmological fluid as homogeneous with energy density $\rho(t)$ and pressure $p(t)$, the 00 and spatial components of the energy-momentum tensor are related by
\begin{equation}\label{Early Universe/Cosmology/T00}
    T_{00} = \rho(t)\\
    T_{ij} = -g_{ij}p(t)
\end{equation}
such that the $00$ component of the Einstein equation, \equaref{Early Universe/Cosmology/Einstein}, is
\begin{eqnarray}\label{Early Universe/Cosmology/Friedmann}
    -3\frac{\Ddot{a}}{a} +3\left(\frac{\Ddot{a}}{a} + \frac{\dot{a}^2}{a^2} \right) = 8\pi G\rho(t) \nonumber\\
    \left(\frac{\dot{a}^2}{a^2}\right) = \frac{8\pi G}{3}\rho(t)
\end{eqnarray}
which is the flat-universe Friedmann equation, relating the time evolution of the scale factor $a(t)$, to that of the cosmological fluid energy density $\rho(t)$. In general relativity, the conservation of energy and momentum is ensured via the vanishing of the covariant derivative of the energy-momentum tensor
\begin{equation}
    \nabla_\mu T^{\mu\nu} =  \partial^\mu T^{\mu \nu} + \Gamma^\mu_{\mu \sigma} T^{\sigma \nu} + \Gamma^\nu_{\mu \sigma}T^{\mu \sigma}= 0 
\end{equation}
which when combined with the expressions \equaref{Early Universe/Cosmology/T00} and \equaref{Early Universe/Cosmology/Christoffel} yields the constraint
\begin{equation}\label{Early Universe/Cosmology/Econs}
    \dot{\rho} + \frac{\dot{a}}{a}\left(\rho + p\right)\,.
\end{equation}
The system of equations needed to track the thermodynamic evolution of the expanding universe is closed by an equation of state for the cosmological fluid $p = p(\rho)$. For \partref{Part2} the relevant era of the universe is before the onset of BBN at high temperatures where in the standard $\Lambda$CDM picture the energy density of the universe would be dominated by radiation, for which the equation of state is
\begin{equation}
    \rhorad = 3p
\end{equation}
therefore \equaref{Early Universe/Cosmology/Econs} gives
\begin{equation}
    \rhorad \propto a^{-4}
\end{equation}
whereas the energy density of non-relativistic matter evolves as $\rho_{\rm matter} \propto a^{-3}$. Relativistic matter is diluted by an additional power of the scale factor due to cosmological redshift. \partref{Part2} will also be concerned with the case where significant populations of black holes exist in the early universe. The component of the energy density composed of black holes scales like non-relativistic matter. \\

Defining the Hubble rate of expansion of the universe as 
\begin{equation}
    H \equiv \frac{\dot{a}}{a}
\end{equation}
the Friedmann equation, \equaref{Early Universe/Cosmology/Friedmann}, for a universe with cosmologically significant radiation-like and matter-like components leads to
\begin{equation}\label{Early Universe/Cosmology/H}
    H^2 = \frac{8\pi G}{3}\left(\frac{\varrhorad}{a^4} + \frac{\varrho_{\rm matter}}{a^3}\right)
\end{equation}
where the comoving energy densities are defined by $\varrhorad \equiv \rhorad a^4$ and $\varrho_{\rm matter} \equiv \rho_{\rm matter} a^3$.\\

When the universe is dominated by radiation, taking $\rho \approx \rhorad$ the Friedmann equation has the solution
\begin{equation}\label{Early Universe/Cosmology/at}
    a(t) \propto t^{\frac{1}{2}}
\end{equation}
so the radiation dominated universe expands at a decreasing rate as cosmic time progresses. The cosmological particle horizon, defined as the distance light could have travelled during the age of the universe is given by
\begin{equation}\label{Early Universe/Cosmology/dHint}
    d_H(t) = a(t)\int^t_0 \frac{\drm t^\prime}{a(t^\prime)}
\end{equation}
such that during an initial radiation dominated era, the particle horizon is simply given in natural units as 
\begin{equation}\label{Early Universe/Cosmology/dHanalytical}
    d_H(t) = 2t = \frac{1}{H(t)}
\end{equation}
which is finite and grows linearly with the cosmic time. \\

Furthermore assuming that during this initial radiation dominated era particle species are in thermal equilibrium, the energy density of radiation in terms of its temperature is given by the Stefan-Boltzmann law 
\begin{equation}\label{Early Universe/Cosmology/rhorad}
    \rhorad(T) = \frac{\pi^2 g_*}{30}T^4 
\end{equation}
so that the temperature falls linearly with the inverse of the scale factor
\begin{equation}\label{Early Universe/Cosmology/Ta}
    T \propto a^{-1}\,.
\end{equation}
Finally, the entropy density of a radiation dominated fluid is given by
\begin{equation}
    s = \frac{2\pi^2 g_*}{45}T^3
\end{equation}
which clearly scales as $s \propto a^{-3}$.
With these considerations, it is possible to, either analytically in the limiting cases such as $\rho \approx \rhorad$ or numerically more broadly, solve for the evolution of the energy densities, temperature, entropy density and cosmic time in terms of the scale factor $a$ in the early universe epochs of interest for this thesis.
\newpage

\section{Cosmological Abundances}\label{Early Universe/Abundance}
In addition to the macroscopic thermodynamic quantities studied in the previous section, physics in the early universe depends sensitively on the dynamics of individual species. Cosmological particle populations are distributed in momentum space according to the cosmic temperature $T$, a good approximation is often
\begin{equation}\label{Early Universe/Abundances/MB}
    f(k) = \frac{1}{(2\pi)^3}\frac{1}{e^{E(k)/T} \pm 1}
\end{equation}
where $E(k)$ is the particle energy as a function of its three-momentum, given by $E(k)^2 = m_X^2 + k^2$ for a particle with mass $m_X$ and the $\pm 1$ distinguishes Fermi-Dirac distributed fermions ($+$) from Bose-Einstein distributed bosons ($-$). Chemical potentials are assumed to vanish throughout this section, $\mu_X = 0$. In the limit of high temperature and low particle density the above distributions both approach the Maxwell-Boltzmann distribution $f(k) \propto \mathrm{exp}(-E(k)/T)$. Integrating $f(k)$ over the momentum gives the equilibrium number density of a species in thermal equilibrium
\begin{equation}\label{Early Universe/Abundance/nint}
n_X^{\rm eq} = g_X\int f(k) \drm^3 k
\end{equation}
where $g_X$ is the number of degrees of freedom of species $X$. Assuming that $X$ is a boson, in the ultra-relativistic limit $T \gg m_X$ the equilibrium number density is
\begin{equation}\label{Early Universe/Abundance/neqrel}
    n_X^{\rm eq} \underset{T \gg m_X}{\longrightarrow} \frac{g_X\zeta(3)}{\pi^2}T^3
\end{equation}
where $\zeta(3) \approx 1.2$. If $X$ were a fermion, there would be an additional factor of $3/4$. In the non-relativistic limit $m_X \gg T$ the species $X$ is Boltzmann suppressed due to the fact that only those particles in the high energy tails of the distribution have sufficient energy to produce $X$ while the $X$-annihilating channels remain unsuppressed. The non-relativistic limit of \equaref{Early Universe/Abundance/nint} is
\begin{equation}
    n_X^{\rm eq} \underset{m_X \gg T}{\longrightarrow} g_X\left(\frac{m_XT}{2\pi}\right)^{\frac{3}{2}}e^{-m_X/T}\,.
\end{equation}

The evolution of the number density of a particle species not necessarily in thermal equilibrium can be tracked by solving the associated Boltzmann Equation (BE),
\begin{equation}\label{Early Universe/Abundance/dnx}
    \frac{\drm n_X}{\drm t} - 3Hn_X = \mathcal{C}(n_X)
\end{equation}
where $\mathcal{C}$ is the collisional operator integrated with respect to the momentum, giving the net number density of $X$ created by interactions with other particles per unit time. Following Gondolo and Gelmini \cite{Gondolo:1990dk} and assuming that the number density of $X$ is modified only by a unspecified number of $2\leftrightarrow2$ processes 
\begin{equation}
    \mathcal{C}(n_X) = -\sum_{\sigma(XY \to Z)} \langle\sigma v_{\rm Mol}\rangle(n_Xn_Y - n_X^{\rm eq}n_Y^{\rm eq})
\end{equation}
where the sum runs over all possible diagrams in which $X$ scatters against $Y$ into any two-particle final state $Z$, and $v_{\rm Mol}$ is the Moller velocity. Also allowing $X$ to have a decay mode $X \to a,b$, where $a,b$ are assumed to be in thermal equilibrium, and comparing to \equaref{Early Universe/Abundance/dnx} leads to the general form of the Boltzmann equation
\begin{eqnarray}\label{Early Universe/Abundance/BE}
    \frac{\drm n_X}{\drm t} - 3Hn_X &=& -\sum_{\sigma(XY \to Z)} \langle\sigma v_{\rm Mol}\rangle(n_Xn_Y - n_X^{\rm eq}n_Y^{\rm eq}) - \Gamma_X n_X
+ \Gamma_X^{ID} n_a^{\rm eq}\\
    \frac{\drm n_X}{\drm t} - 3Hn_X &=& -\left(\Gamma_X + \sum_{\sigma(XY \to Z)} \langle\sigma v_{\rm Mol}\rangle n_Y^{\rm eq} \right)(n_X
- n_X^{\rm eq})\
\end{eqnarray}
where $n_a^{\rm eq}$ is the (equilibrium) number density of the daughter particle of $X$ and $\Gamma_X,\Gamma_X^{ID}$ are the forward and inverse decay widths for the mode $X \to a,b$. In the second line it was assumed that the species off which $X$ scatters are always in thermal equilibrium (this approximation will turn out to always be sufficient of the purposes of this thesis) and the relationship
\begin{equation}\label{Early Universe/Abundance/ID}
\Gamma_X^{ID} = \frac{n_X^{\rm eq}}{n_a^{\rm eq}}\Gamma_X
\end{equation}
was imposed. \\

\equaref{Early Universe/Abundance/BE} gives the general form of the BE tracking the evolution of the number density of the unstable species $X$. That the cross sections are enclosed in $\langle \rangle$ indicates a thermal average, and the decay width $\Gamma_X$ should also be thermally averaged to reflect the distribution of particle energies. For the cross sections, it is often sufficient to assume that both initial state as well as both final state particles all have the same mass, and the same temperature. This calculation yields a simple result and was done in \cite{Gondolo:1990dk}. In \secref{Hot Spots} the case where the initial state particles have different temperatures will be of interest, and the result for the thermal averaging in this case can be found in \cite{Cheek:2022mmy}. The case where the initial state particles have different masses and different temperatures will be relevant in \chapref{Hot Spots} and this calculation is given in detail in Appendix \ref{C}.\\

It will prove convenient for calculations to rewrite the above BE in terms of the logarithmic scale factor $\alpha = \rm{log}_{10}(a)$ and the comoving number densities $\mathcal{N}_X \equiv n_Xa^3$. Doing so gives
\begin{eqnarray}\label{Early Universe/Abundance/BEalpha}
    \frac{\drm \mathcal{N}_X}{\drm \alpha} &=& -\frac{\ln(10)}{H}\left(\Gamma_X + \sum_{\sigma(XY \to Z)} \langle\sigma v_{\rm Mol}\rangle n_Y^{\rm eq} \right)(\mathcal{N}_X
- \mathcal{N}_X^{\rm eq})\
\end{eqnarray}

As well as the number density of a species, in the context of leptogenesis (see \chapref{Leptogenesis}) it will be necessary to track the evolution of the asymmetry between the number density of leptons, and that of anti-leptons. An overall asymmetry can only survive if there is a departure from equilibrium, otherwise the inverse modes of the same processes which generate a net lepton number would wash it out. 
\newpage 

\section{Out-of-Equilibrium Dynamics}\label{Early Universe/Out-of-Equilibrium}
Consider a generic unstable particle, $X$, which decays and is produced by the forward and inverse modes of $X \to a,b$. The decay (forward) mode has temperature dependent width $\Gamma_X(T)$ while the inverse mode has the width $\Gamma_X^{ID}(T)$ which is related to $\Gamma_X(T)$ by \equaref{Early Universe/Abundance/ID}. $X$ has relativistic equilibrium abundance at high temperatures $T \gg m_X$ for $m_X$ the mass of $X$. The forward and inverse widths diverge around $T \sim m_X$ because the inverse decay becomes Boltzmann suppressed. When $\Gamma_X^{ID}(T) < H(T)$ the particle is said to decay out-of-equilibrium, the inverse decay is no longer efficient compared to Hubble. At some smaller temperature $\Gamma_X(\TFO) = H(\TFO)$, such that after this moment the lifetime of $X$ becomes longer than the age of the universe. From the perspective of cosmological expansion, whatever the comoving abundance of $X$ was at $T = \TFO$ will persist in perpetuity. Broadly, this mechanism by which a relic abundance of $X$ may survive is known as \textit{freeze-out}.\\

It could also be the case that $X$ initially has vanishing abundance in the universe, in this case the number density of $X$ is populated by the inverse decay mode. At some temperature $T_{\rm FI}$ before $X$ has reached equilibrium abundance, $\Gamma_X^{ID}(T_{\rm FI}) < H(T_{\rm FI})$ so that the (comoving) number density of $X$ stops evolving as the inverse decay mode producing it falls out of equilibrium. This mechanism is known as \textit{freeze-in}. The simple pictures of \textit{freeze-out} and \textit{freeze-in} painted here would be complicated by any scattering diagrams which produce and destroy $X$.\\

\begin{figure}[h!]
    \centering    \includegraphics[width=0.9\linewidth]{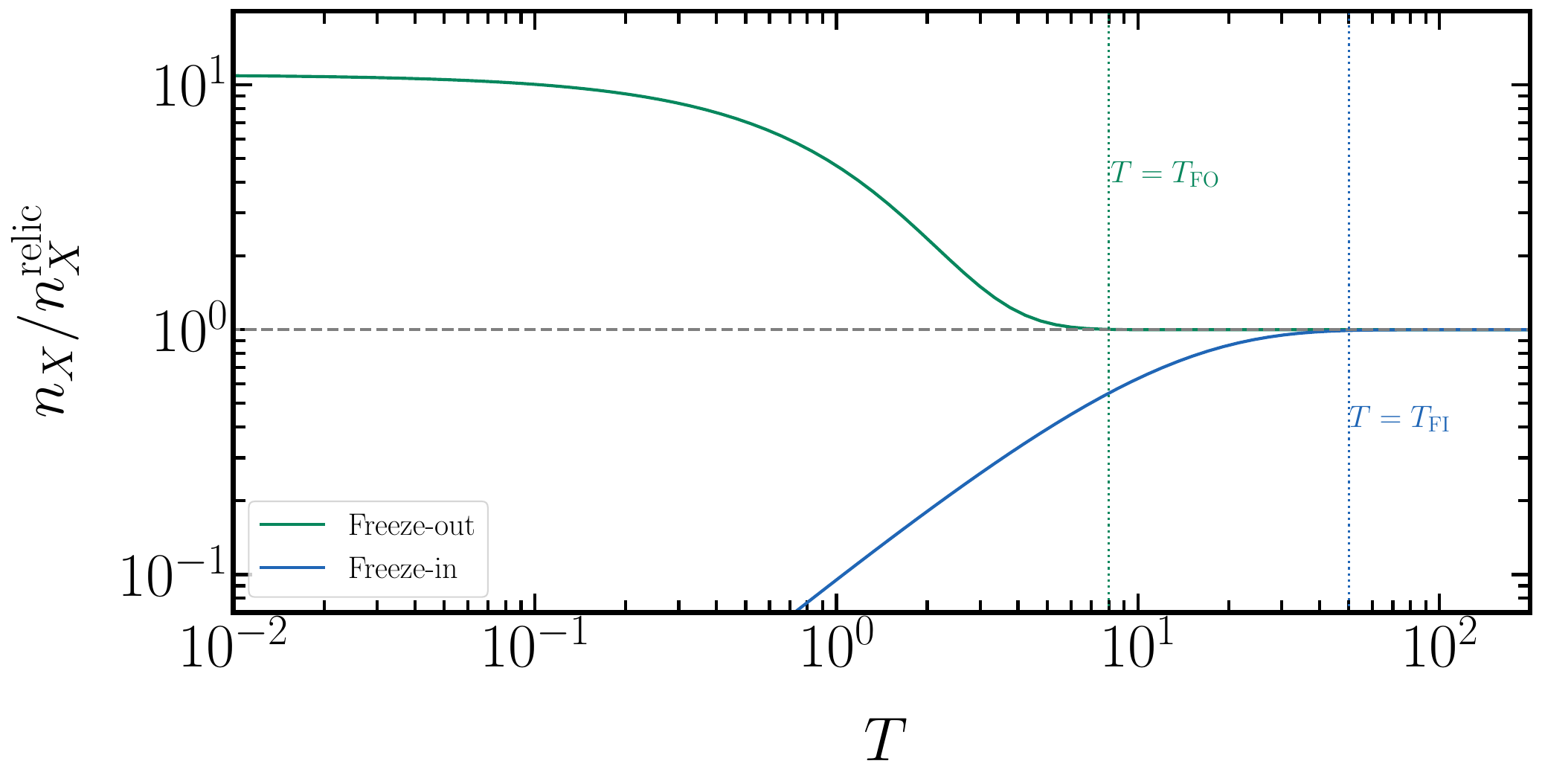}
    \caption{Schematic depiction of the \textit{freeze-in} and \textit{freeze-out} processes. The green line shows the number density of the species $X$ relative to the relic density as $X$ undergoes \textit{freeze-out}. The navy blue line shows \textit{freeze-in}. The green and blue vertical dotted lines indicate respectively the \textit{freeze-out} and \textit{freeze-in} temperatures, $\TFO,T_{\rm FI}$. The curves shown are schematic only, and do not precisely represent any particle processes.}
\end{figure}

Either via \textit{freeze-in} or \textit{freeze-out}, the departure of the processes populating $X$ from equilibrium can lead to a relic comoving abundance of $X$ which remains constant as the universe evolves. It is also possible that in an analogous way, an asymmetry between $X$ and $\bar{X}$ could survive until the present. This will be particularly important in the context of baryogenesis via leptogenesis (see \chapref{Leptogenesis}), where an asymmetry between baryons and anti-baryons is produced in the early universe and survives because the processes responsible \textit{freeze-out} around the time of the EWPT. In order to predict what the surviving baryon asymmetry will be, it is necessary to track the evolution of the asymmetry between leptons and anti-leptons. Anticipating notation from thermal leptogenesis, the rate of change in the asymmetry is given by the difference between the rate of lepton production and the rate of anti-lepton production
\begin{equation}
\frac{d\mathcal{N}_{\Delta \Lnum}^\ell}{dt} = \frac{d\mathcal{N}_{\ell}}{dt} - \frac{d\mathcal{N}_{\bar{\ell}}}{dt} \end{equation}
where $\mathcal{N}_{\ell,\bar{\ell}}$ is the number density of leptons/antileptons of flavour $\ell$. For the purposes of the current discussion it is sufficient to presume that the particle $X$ decays CP asymmetrically into leptons and antileptons so that 
\begin{equation}
\frac{d\mathcal{N}_{\Delta \Lnum}^\ell}{dt} = (\Gamma_{\ell X}\mathcal{N}_{X} - \Gamma^{ID}_{\ell X}\mathcal{N}_\ell) - (\bar{\Gamma}_{\ell X}\mathcal{N}_{X} - \bar{\Gamma}^{ID}_{\ell i}\mathcal{N}_{\bar{\ell}}) \end{equation}
where $\Gamma_{\ell X}$ is the rate of decay into lepton flavour $\ell$, and $\bar{\Gamma}_{\ell X}$ is the CP conjugate rate. The inverse decay width, $\Gamma^{ID}_{\ell X}$ is related to the decay width by \equaref{Early Universe/Abundance/ID}. Making use of this relation, and decomposing 
\begin{eqnarray}
    \mathcal{N}_\ell &=& \Nleq + \frac{1}{2}\mathcal{N}_{\Delta \Lnum}^\ell \\
    \mathcal{N}_{\bar{\ell}} &=& \mathcal{N}^{eq}_{\bar{\ell}} - \frac{1}{2}\mathcal{N}_{\Delta \Lnum}^\ell
\end{eqnarray}
one obtains
\begin{equation}\label{Early Universe/Equilibrium/DeltaLBE}
\frac{d\mathcal{N}_{\Delta \Lnum}^{\ell}}{dt} =\epsilon_{\ell \ell}\Gamma_{X}(\mathcal{N}_{X}-\mathcal{N}^{\rm eq}_{X}) -   \frac{\mathcal{N}^{eq}_{X}}{2\Nleq}\mathcal{N}_{\Delta \Lnum}^{\ell} P_{\ell X}
\end{equation}
where the flavour projector of $X$ onto $\ell$ is defined as
\begin{equation}\label{Early Universe/Equilibrium/Pflav}
    \Gamma_{\ell X} + \bar{\Gamma}_{\ell X} = \Gamma_{X}P_{\ell X}
\end{equation}
for $\Gamma_{X}$ the total decay width of $X$, and the CP asymmetry parameter is defined as 
\begin{equation}
    \epsilon_{\ell \ell} = \frac{\Gamma_{\ell X} - \bar{\Gamma}_{\ell X}}{\Gamma_{X}}\,,
\end{equation}
see also \equaref{Leptogenesis/CP/epsilon}. The BE \equaref{Early Universe/Equilibrium/DeltaLBE} accounts for the asymmetry generation and washout due to the decays and inverse decays of $X$. Terms proportional to $-\mathcal{N}_{\Delta \Lnum}^\ell$ wash out the asymmetry. One should also account for $2\to2$ diagrams involving $X$, this is done in detail for the case $X = N_i$ in \secref{Leptogenesis/BEs}.

\newpage

\section{Finite Temperature Effects}\label{Early Universe/TFT}

At the temperatures of interest for thermal leptogenesis, $T \geq \Tsphal$, the Higgs potential is unbroken except in the narrow temperature range $\Tsphal \leq T \leq \Tewpt$. Therefore all the SM fermions and the Higgs are formally massless for almost all of the relevant temperature range. Consider a fermion with momentum $k$. Interactions with the finite temperature plasma, on timescales much shorter than that of the expansion of the universe, can be resummed such that they introduce corrections to the self-energy $\Pi$ and shift the pole of the propagator \cite{Weldon:1982bn}
\begin{eqnarray}
    \frac{\slashed{k}}{k^2 + i\epsilon} \to \frac{\slashed{k}}{k^2 + \Pi(k,T) + i\epsilon}
\end{eqnarray}
where $\slashed{k} = \gamma^\mu k_\mu$. Similar arguments apply to the Higgs and the gauge bosons. In shifting the propagator pole, fast interactions with the thermal bath generate temperature dependent thermal masses $M^2 \sim \lambda T^2$ where $\lambda$ is a coefficient depending on the couplings.\\

In the range $T \lesssim \Tewpt$ all of the SM particles (except neutrinos, and photons) gain a mass depending on the dynamics of the EWPT. It is beyond the scope of this thesis to discuss the details of the EWPT. For the Higgs \cite{Kapusta:2006pm}
\begin{eqnarray}\label{Early Universe/TFT/MHiggs}
    \mHiggsT(T)^2 = m_\Higgs^2 v^2(T)\Theta(\Tewpt - T) + \delta_{M_\Higgs}\Theta(T - \Tewpt)
\end{eqnarray}
where $v^2(T) = (1-T^2/\Tewpt^2)$, $m_\Higgs = 125\GeV$ is the zero-temperature Higgs mass, and
\begin{eqnarray}
	\delta_{M_\Higgs} = \left(\frac{\lambda_\Higgs}{2} +\frac{3g_2^2 + g_1^2}{16}  + \frac{y_t^2}{4} \right)\left(T^2-\Tewpt^2\right) 
\end{eqnarray}
where $\lambda_\Higgs$ is the Higgs quartic coupling, $g_2$ and $g_1$ are the $SU(2)_L$ and $U(1)_Y$ couplings respectively, and $y_t$ is the top quark Yukawa coupling. \\

\begin{figure}[h!]
    \centering
    \includegraphics[width=0.8\linewidth]{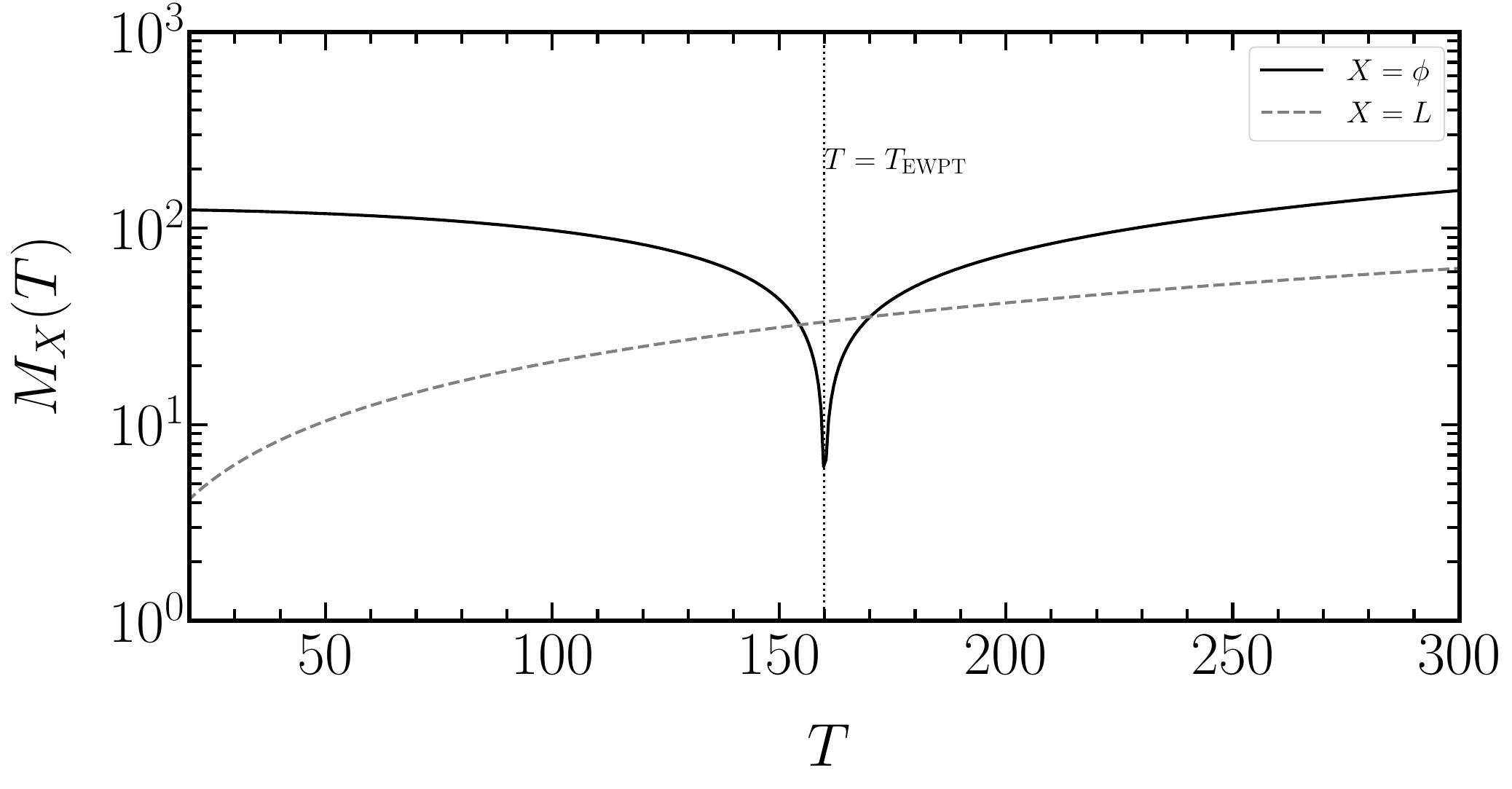}
    \caption{The thermally corrected masses of the SM leptons and Higgs are shown with respect to the temperature, in the vicinity of the EWPT. The black solid line depicts the variation of the Higgs mass with the temperature while the grey dashed line depicts the lepton mass. The vertical dotted line indicates the EWPT temperature. In the temperature range shown, the lepton bare mass is always much smaller than the thermal component. In contrast, the Higgs bare mass is quickly reached as $T$ drops below $\TEWPT$ while the Higgs mass is purely thermal for $T\gg \TEWPT$.}
    \label{fig:enter-label}
\end{figure}

The SM charged lepton masses evolve with temperature according to \cite{Kapusta:2006pm}
\begin{eqnarray}\label{Early Universe/TFT/MLep}
\mLepT(T)^2 = m_\lep^2v^2(T)\Theta(\Tewpt - T) + \delta_{M_\lep}
\end{eqnarray}
where
\begin{equation}
\delta_{M_\lep} = (3g_2^2 + g_1^{2})\frac{T^2}{32}
\end{equation}
and $m_L$ are the bare lepton masses.

\subsection*{Landau-Pomeranchuk-Migdal suppression}
Another important finite-temperature effect is the suppression of Bremsstrahlung radiation via the Landau-Pomeranchuk-Migdal (LPM) effect \cite{Landau:1953gr,Landau:1953um,Migdal:1956tc}. A high energy particle with momentum $p$ for which $p \gg T$ in a medium of temperature $T$ undergoes small angle elastic scatterings enhanced by t-channel gauge boson exchange, while large angle scatterings are suppressed by the large $p$. 
The dominant mechanism by which the high energy parent loses its energy is through splitting into (nearly collinear) daughter particles \cite{Harigaya:2014waa}. However quantum interference between the  daughter particles is destructive, and inelastic scatterings cannot occur until the overlap between the daughter particles is lost. \\
\begin{figure}[h!]
    \centering
    \includegraphics[width=0.7\linewidth]{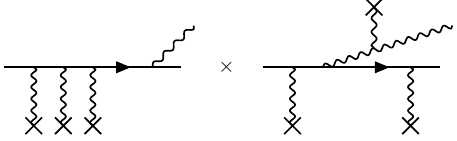}
    \caption{Feynman diagrams representing interference between a high energy parent particle, and its soft daughter particle. Crosses represent particles originating in the thermal plasma. Interference of this type leads to the LPM effect.}
    \label{Cosmology/TFT/LPMFeynman}
\end{figure}

Feynman diagrams corresponding to the interference between parent and daughter particle, responsible for the LPM effect, are shown in \figref{Cosmology/TFT/LPMFeynman}. Multiple soft scatterings with the thermal plasma, represented by the crosses in \figref{Cosmology/TFT/LPMFeynman}, destructively interfere and suppress the rate of splitting for the high energy parent.
Therefore for high energy particles their rate of splitting in a medium can be estimated by the LPM rate, given by \cite{He:2022wwy}
\begin{equation}\label{Early Universe/TFT/GLPM}
 \Gamma_{\rm LPM}(k,T) \sim A^2 \frac{T^{\frac{3}{2}}}{\sqrt{|\vec{k}|}}\,,
\end{equation}
where $\vec{k}$ is the three-momentum of the daughter particle, and \( A = g^2/(4\pi) \) is the relevant gauge coupling constant. A parent particle with initial energy \( E \), loses its energy at a rate given by
\begin{equation}\label{Early Universe/TFT/dELPM}
\frac{1}{E}\frac{d E}{d t} \sim \frac{1}{E} \int^{E/2} \Gamma_{\rm{LPM}}(k,T) \, d k \propto A^2 T \sqrt{\frac{T}{E}} \,,
\end{equation}
such that the energy loss rate is dominated by large $k$ but the splitting rate is largest for small $k$. This effect will turn out to be crucial for understanding the formation of temperature gradients around black holes in the early universe, see \secref{PBHs/Hot Spots} and \chapref{Hot Spots}.

\newpage

\chapter{Origin of the Neutrino Mass}\label{vMass}
\section*{Introduction}
The outstanding problem of the origin of the neutrino mass is one of the most important open questions in theoretical physics. As discussed in \secref{SM/vMass}, no explanation exists within the framework of the SM because neutrinos only have a left handed component in the SM. This section explores one of the most theoretically appealing and historically popular mechanisms by which the neutrino mass might arise, the type-1 seesaw mechanism. This section is organised as follows. \secref{vMass/Seesaw} introduces the seesaw mechanism and demonstrates how the neutrino mass is generated. Then \secref{vMass/Casas-Ibarra} discusses the particularly convenient Casas-Ibarra parameterisation before \secref{vMass/Mixing} explores the sterile-active neutrino mixing.

\section{The Type-1 Seesaw Mechanism}\label{vMass/Seesaw}
The SM is extended by $n$ gauge singlet fermions, called Right Handed Neutrinos (RHNs), $N_i\, , \, i = 1...n$, which are coupled to the SM leptons through a Yukawa coupling with the Higgs. Since they are gauge singlets, they are also often known as sterile neutrinos. The Lagrangian of the model reads
\begin{equation}\label{vMass/Seesaw/Lagrseesaw}
    \Lagr_{\rm seesaw} = i\Nc \slashed{\partial} N - \frac{1}{2}\Nc \MN N - \bar{L} Y\tilde{\phi} N + {\rm h.c.}
\end{equation}
where $\MN$ is the $n\times n$ Majorana mass matrix of the RHNs, $\tilde{\phi} = i \sigma_2\phi^*$ with $\Higgs$ the SM Higgs doublet, and $Y$ is the Yukawa matrix couplings the SM lepton doublets $L_\iflav = (e_\iflav,\nu_\iflav)^T$ to the Higgs and RHNs. RHNs obey the Majorana condition 
\begin{equation}
    \Nc \equiv C(\bar{N})^T = N\,,
\end{equation}
where $C$ is the charge conjugation matrix. As such, the gauge singlet Majorana mass can be included in the Lagrangian as a bare mass term.
One can always choose the basis in which the matrix $\MN$ is diagonal without loss of generality, $\MN = \mathrm{diag}(M_{N_1}...M_{N_n})$. Therefore throughout this work $\MN$ is understood to be always diagonal.\\

The Higgs doublet breaks the gauge group $SU(2)\times U_Y(1)$ down to the electroweak group $U(1)$, generating masses for the fermions at $\Tewpt = 159.5 \pm 1.5$GeV \cite{DOnofrio:2014rug} so that for $T \ll \Tewpt$ the Higgs can be replaced by its vacuum expectation value $\vEW = 174$GeV. Then the relevant part of the Lagrangian becomes
\begin{equation}\label{vMass/Seesaw/Lagr0}
      -\frac{1}{2}\Nc \MN N  - \bar{\nu} m_D N + {\rm h.c.}
\end{equation} which can be recast (see \cite{Bilenky:1987ty} for a detailed calculation) as
\begin{equation}\label{vMass/Seesaw/LagrMatrix}
    \Lagr_{\rm seesaw} \supset \frac{1}{2}\begin{pmatrix}
        \bar{\nu} & \Nc 
    \end{pmatrix}
    \begin{pmatrix}
        0 & m_D^T\\
        m_D & \MN
    \end{pmatrix}
    \begin{pmatrix}
        \nu^c \\ N
    \end{pmatrix} = \frac{1}{2}\begin{pmatrix}
        \bar{\nu} & \Nc 
    \end{pmatrix}
    M_\nu
    \begin{pmatrix}
        \nu^c \\
        N
    \end{pmatrix}
\end{equation}
where the Dirac mass matrix is given by $m_D \equiv \vEW Y$. In the limit $m_D \ll \MN$ one can integrate out the fields $N$, which leads to the well known seesaw relation \cite{Yanagida:1980xy,MohapatraRabindraSenjanovi}
\begin{equation}\label{vMass/Seesaw/seesaw}
     m_\nu \approx - m_D^T \MN^{-1}m_D = - \vEW^2 Y^T\MN^{-1}Y\,,
\end{equation}
giving the neutrino mass matrix $m_\nu$ approximately in terms of the RHN mass matrix $\MN$ and the Yukawa matrix $Y$. This relation is called the seesaw relation because of the inverse relationship between the active and sterile neutrino masses, heavier RHNs result in smaller neutrino masses for constant $Y$. For $\MN$ at the GUT scale, $Y$ must be $\sim \mathcal{O}(10^{-1})$ to give $\sim 0.1\eV$ neutrino masses. The overall scale and texture of $\MN$ are not determined \textit{a priori}. If $\MN \sim \GeV$ as in resonant leptogenesis (see \secref{Leptogenesis/Resonant}) then the Yukawa matrix elements must be very small, $|Y| \sim 10^{-7}$. In general, $Y$ is fixed by the observed neutrino mass splittings as well the known mixing parameters, given the mass matrices $\MN$ and $m_\nu$.\\

\newpage

\section{The Casas-Ibarra Parameterisation}\label{vMass/Casas-Ibarra}
Since the Yukawa matrix $Y$ must be fixed as to reproduce the observed neutrino mass splittings \equaref{SM/vMass/splittings} and mixing data, it is convenient to parameterise $Y$ in terms of the known low-energy parameters \cite{Casas:2001sr}. Starting from the basis in which the charged lepton Yukawas and gauge interactions are diagonal in flavour space, the active neutrino mass matrix $m_\nu$ is diagonalised by the Pontecorvo-Maki-Nakagawa-Sakata (PMNS) matrix \cite{Pontecorvo:1967fh,pontecorvo1957mesonium,pontecorvo1957inverse,Maki:1962mu} which rotates between the neutrino mass and flavour eigenstates 
\begin{equation}
    \begin{pmatrix}\nu_e\\
             \nu_\mu\\ 
             \nu_\tau \end{pmatrix} = \UPMNS \begin{pmatrix}\nu_1\\
             \nu_2\\ 
             \nu_3 \end{pmatrix}\,.
\end{equation}
If the SM neutrinos are Dirac neutrinos, $\UPMNS$ can be expressed in analogy to the CKM matrix for quarks as a $3\times 3$ complex matrix parameterised in terms of a single physical (Dirac) phase, $\delta$, and three complex mixing angles $\theta_{12},\theta_{13},\theta_{23}$
\begin{equation}
    \UPMNS = \begin{pmatrix}
        c_{12}c_{23} & s_{12}c_{13} & s_{13}e^{-i\delta}\\
        -s_{12}c_{23}-c_{12}s_{12}s_{23}e^{i\delta} & c_{12}c_{23}-s_{12}s_{13}s_{23}e^{i\delta} &  c_{13}s_{23}\\
        s_{12}s_{23}-c_{12}s_{12}c_{23}e^{i\delta} & -c_{12}s_{23}-s_{12}s_{13}c_{23}e^{i\delta} & c_{13}c_{23}
    \end{pmatrix}
\end{equation}
where $c_{ij} \equiv \cos \theta_{ij}$,$s_{ij} \equiv \sin \theta_{ij}$. In 1980 Bilenky, Hosek and Petcov pointed out that in the case of Majorana neutrinos, there are two additional phases which cannot be reabsorbed, the Majorana phases $\eta_1\,,\,\eta_2$ \cite{Bilenky:1980cx}. \\

$\UPMNS$ diagonalises the active neutrino mass matrix
\begin{equation}
    \hat{m}_\nu = \UPMNS^T m_\nu \UPMNS.
\end{equation}
 such that by replacing $m_\nu$ according to the seesaw relation \equaref{vMass/Seesaw/seesaw}, one obtains
\begin{equation}
    \hat{m}_\nu = \vEW^2 \UPMNS^T Y^T\MN^{-1}Y \UPMNS\,.
\end{equation}
Now multiplying on both the left and right by $\sqrt{\hat{m}_\nu^{-1}}$ 
\begin{eqnarray}
    I_3 &=& \vEW^2 \sqrt{\hat{m}_\nu^{-1}}\UPMNS^T Y^T\MN^{-1}Y\UPMNS\sqrt{\hat{m}_\nu^{-1}}\\
    &= & \vEW^2 (\sqrt{\MN^{-1}}Y\UPMNS\sqrt{\hat{m}_\nu^{-1}})^T(\sqrt{\MN^{-1}}Y \UPMNS\sqrt{\hat{m}_\nu^{-1}})
\end{eqnarray}
where $I_3$ is the $3\times 3$ identity matrix. This equation is solved by an arbitrary complex and orthogonal matrix
\begin{equation}\label{vMass/Casas-Ibarra/R}
    R \equiv \vEW \sqrt{\MN^{-1}}Y \UPMNS\sqrt{\hat{m}_\nu^{-1}}
\end{equation}
which allows us to express the Yukawa matrix in terms of the physical neutrino masses and mixing parameters as 
\begin{equation}\label{vMass/Casas-Ibarra/Y}
    Y = v_{\rm EW}^{-1} \sqrt{\MN} R \,\sqrt{\hat{m}_\nu} \, \UPMNS^\dagger.
\end{equation}
Therefore the Yukawa matrix $Y$ depends on the parameters of the PMNS matrix, the $n$ masses of the RHNs, the active neutrino masses and the three complex mixing angles which describe the rotational matrix $R$
\begin{equation}
    R = R_{12}(\theta_{12})R_{23}(\theta_{23})R_{13}(\theta_{13})
\end{equation}
where in many cases $R$ can be adequately described in terms of a single mixing angle $\theta = x + iy$ where $x,y$ are real parameters. The overall scale of $\hat{m}_\nu$ is not known but two mass splittings are well measured so that a single parameter, the mass of the heaviest neutrino $m_h$ (or equivalently the lightest $m_l$), determines $\hat{m}_\nu$ 
\begin{eqnarray}
    \hat{m}_\nu  &=& \mathrm{diag}(m_l,m_m,m_h)\\
    m_m &=& \sqrt{-\dmatm^2 + m_h^2}\\
    m_l &=& \sqrt{-\dmsol^2 + m_m^2}\,.
\end{eqnarray}
The Casas-Ibarra parameterisation is convenient because it facilitates calculation of the Yukawa matrix elements based on the desired scale and texture of the RHN mass matrix and fixed such that the observed neutrino mass splittings are recovered. 
\newpage

\section{Active-Sterile Neutrino Mixing}\label{vMass/Mixing}
The seesaw Lagrangian after symmetry breaking \equaref{vMass/Seesaw/Lagr0} can be rewritten as
\begin{equation}\label{vMass/Mixing/LagrMatrix}
    \Lagr_{\rm seesaw} \supset \frac{1}{2}\begin{pmatrix}
        \bar{\nu} & \Nc 
    \end{pmatrix}
    \begin{pmatrix}
        0 & M_D^T\\
        M_D & \MN
    \end{pmatrix}
    \begin{pmatrix}
        \nu^c \\ N
    \end{pmatrix} = \frac{1}{2}\begin{pmatrix}
        \bar{\nu} & \Nc 
    \end{pmatrix}
    M_\nu
    \begin{pmatrix}
        \nu^c \\
        N
    \end{pmatrix}
\end{equation}
where the overall neutrino mass matrix $M_\nu$ is complex and symmetric and therefore can be diagonalised by a $(3+n)\times(3+n)$ unitary matrix
\begin{equation}
    \mathrm{diag}(m_1,m_2,m_3,\MNo...M_{N_n}) = U^T M_\nu U\,.
\end{equation}
Therefore, the mass eigenstates can be expressed as a linear superposition of the interaction basis eigenstates
\begin{equation}
    \vec{\nu}_{\rm mass} = U^\dagger\begin{pmatrix}
       \nu^c\\
       N 
    \end{pmatrix} + U^*\begin{pmatrix}
        \nu \\
        N^c 
        
    \end{pmatrix}
\end{equation}
where $\vec{\nu}_{\rm mass} = (\nu_1,\nu_2,\nu_3,N_1...N_n)$ is a $(3 + n)$ component vector containing the neutrino mass eigenstates. This fact leads to mixing between the active and sterile neutrinos, opening the door to potentially detecting RHNs experimentally.
Experimental efforts to detect RHNs are sensitive to the mixing of the RHNs with each active neutrino flavour. For instance at SHiP \cite{SHiP:2018xqw,SHiP:2021nfo} RHNs are produced by a proton beam incident upon a fixed target which then decay in the detector into a SM final state. In this case, and given some RHN mass spectrum, the production of lepton flavour $\iflav$ is a function of \cite{Chianese:2018agp} 
\begin{equation}\label{vMass/Mixing/Uflav}
    U_\iflav^2 = \sum_{i=1}^n|U_{\iflav (3 + i)}|^2
\end{equation}
quantifying the the mixing of all $n$ RHNs into flavour $\iflav$, while the total mixing of the SM neutrinos to RHN $N_i$ is given by
\begin{equation}
    U_i^2 = \sum_\iflav U_{\iflav (3+i)}^2\,.
\end{equation}
The total mixing between the active and sterile neutrinos is given by
\begin{equation}\label{vMass/Mixing/U2}
    U^2 = \sum_{\iflav}U_{\iflav}^2 = \sum_{i}U_{i}^2
\end{equation}
which along with the relevant RHN masses often forms the parameter space of low scale leptogenesis models, see \secref{Leptogenesis/Resonant}. Current and future generation experiments are not sensitive to $\MN \gtrsim \GeV$ so $U^2$ is not typically used for high scale models of leptogenesis where $\MN \gg \GeV$.\\

It is also not known whether the SM neutrinos are Dirac neutrinos or Majorana neutrinos. Setting $\MN=0$ in \equaref{vMass/Mixing/LagrMatrix} leads to a vanishing Majorana mass term such that when the overall mass matrix $M_\nu$ is diagonalised, the eigenvalues are simply $\pm m_D$. In this case, neutrinos are purely Dirac neutrinos. In the opposite case where $\MN \gg M_D$ (or in the pseudo-Dirac case $M_D \gg M_N > 0$)  neutrinos are Majorana-type neutrinos allowing processes where lepton number is violated by two units. The most important of these is the ``neutrinoless double-$\beta$ decay" (denoted $(\beta\beta)_{0\nu}$) which could occur in some even-even nuclei \cite{PETCOV1982245} 
\begin{equation}
    (A,Z) \to (A,Z+2) + 2e^-
\end{equation}
at a rate proportional to the effective Majorana mass of the neutrinos,
\begin{equation}
    \langle m_\nu \rangle = \sum_i U_{ei}^2\hat{m}_\nu^i\,,
\end{equation}
which is acutely sensitive to the neutrino mass hierarchy, as well as the Majorana CP phases $\eta_{1,2}$ \cite{Bilenky:2001rz}. $(\beta\beta)_{0\nu}$ has not yet been observed but if it is, it would confirm the Majorana nature of neutrinos as lead us closer to a complete picture of the origin of the neutrino mass.

\newpage

\chapter{Baryogenesis via Leptogenesis}\label{Leptogenesis}
That baryons dominate the ordinary matter of the universe at the expense of anti-baryons could be seen as so self evident as to be almost Cartesian, \textit{cogito, ergo sum} \footnote{Of course, the specific size of the observed asymmetry is not self-evident. However an observer who knows about annihilation may deduce \textit{cogito, ergo BAU}.}. And yet the cosmological mechanism by which it was dictated so remains unknown. What is clear is that in the early universe, baryogenesis should have completed before the onset of BBN, yielding approximately one baryon per billion photons. Furthermore, as discussed in \secref{SM/Baryogenesis}, any would-be mechanism of baryogenesis must satisfy Sakharov's eponymous conditions \cite{Sakharov:1967dj} of CP violation, baryon number ($B$) violation and departure from equilibrium. Following the measurement of the Higgs mass by the ATLAS collaboration \cite{ATLAS:2012yve}, there are no remaining possibilities for the BAU to have been generated entirely within the SM. At least some extension of the SM will ultimately be required to understand the generation of the BAU.\\

Leptogenesis, based on the type-I seesaw mechanism introduced in \secref{vMass/Seesaw}, explains the observed BAU by generating a net leptonic asymmetry through the out-of-equilibrium dynamics of at least two gauge singlet fermions, RHNs, $N_i$ where $i$ is a generational index. This leptonic asymmetry is then translated into the baryonic sector by the $B+L$ violating EW sphaleron processes. First proposed by Fukugita and Yanagida \cite{Fukugita:1986hr}, leptogenesis is one of the most attractive models of baryogenesis because the seesaw mechanism on which it is based may simultaneously explain the generation of the active neutrino masses. In its most minimal realisation, leptogenesis requires extending the SM only by the inclusion of two RHNs which generate the active neutrino masses via the seesaw mechanism and the BAU via their CP violating decay (or production) in the early universe. More complicated theories based on the type-I seesaw mechanism, such as the so-called neutrino-Minimal-Standard-Model ($\nu$MSM) \cite{Asaka:2005an,Canetti:2012kh}, even cast RHNs as DM candidates, killing all three birds (neutrino mass, baryogenesis, DM) with one stone.\\

Despite requiring relatively minimal extension of the SM to achieve successful baryogenesis, the parameter space of leptogenesis is high-dimensional and difficult to probe experimentally. Establishing a complete theory of leptogenesis would require precise measurements of at least two RHN masses, the active neutrino mass scale, at least one complex angle in the Yukawa matrix, as well as the Dirac and Majorana phases of the PMNS matrix. This project is a long way from completion, and presents a significant obstacle on the road to understanding leptogenesis.\\

Naturally one may wonder if there are other angles from which to approach the goal, and in \secref{PBHLepto} and \secref{Hot Spots} it will be shown that PBHs offer a tantalising opportunity to do just that. First it is necessary to set out in this section the thermal theory of type-I leptogenesis in detail and introduce the necessary tools to predict $\YB$. The structure of this section is as follows. In \secref{Leptogenesis/Sphalerons}, \secref{Leptogenesis/CP}, and \secref{Leptogenesis/Out-of-Equilibrium} it is described how leptogenesis satisfies each of Sakharov's conditions. The following section \secref{Leptogenesis/Thermal} delineates the importance of thermal effects on leptogenesis before \secref{Leptogenesis/BEs} introduces the Boltzmann equations used to calculate the resulting baryon asymmetry. Then \secref{Leptogenesis/Resonant} considers the experimentally accessible low scale resonant leptogenesis model. Finally, the experimental status of leptogenesis is analysed in \secref{Leptogenesis/Experiment}.

\newpage

\section{Sphalerons}\label{Leptogenesis/Sphalerons}
The first of Sakharov's conditions for successful baryogenesis is the requirement of baryon number, $B$, violation. Baryon number violation occurs in the SM at high temperatures through the EW sphalerons. Consider the Lagrangian for the SU(2) gauge interactions of fermions 
\begin{equation}
\Lagr_{SU(2)} = \bar{\psi}^k_L\gamma^\mu(\partial_\mu - \frac{ig_2}{2}\sigma W_\mu)\psi_L^k
\end{equation}
where $\psi^k = \{q_k^C,\lep_{k}\}$ are the SM fermions ie the quark and lepton doublets with $C$ a colour index and $k$ a generational index, $\sigma$ are the Pauli matrices and $W_\mu$ are the gauge fields. Note that the Lagrangian contains no term which directly produces fermions. The Lagrangian has 12 global symmetries and an associated vector current 
\begin{equation}
    j_\mu^{kl} = \bar{\psi}^k_L\gamma^\mu \psi_L^l
\end{equation}
and axial vector current
\begin{equation}
    J_\mu^{kl} = \bar{\psi}^k_L\gamma^\mu \gamma^5\psi_L^l
\end{equation}
for each. The vector current and the traceless part of the axial vector current are conserved at tree level, but the current $J_\mu^{kk}$ is anomalous at loop level \cite{tHooft:1976rip}. At one-loop one finds
\begin{equation}
    \partial^\mu J^{kk}_\mu = \frac{F_{\mu \nu} \Tilde{F^{\mu\nu}}}{32 \pi^2}
\end{equation}
where $F_{\mu \nu}$ is the SU(2) field strength tensor. The space-time integral of this anomaly from $t=0$ to some later time is the change in the topological Chern-Simons quantum number
\begin{equation}
    \Delta N_{\rm cs} = \frac{1}{32 \pi^2}\int^t_0 \drm t \int \drm^3x \, F_{\mu \nu} \Tilde{F^{\mu\nu}}\,.
\end{equation}
That $\Delta N_{\rm cs}$ can be a non-zero integer implies that there exist field configurations which tunnel at zero temperature between minima of the SU(2) vacuum, labelled by $N_{\rm cs}$ \cite{Belavin:1975fg}. These tunneling configurations source all 9 quarks and 3 leptons by inducing a change in $N_{\rm cs}$, with $\Delta B = \Delta \Lnum = 3$ such that the combination $B-\Lnum$ is conserved. $B + \Lnum$ is not conserved. However, the rate of such tunnelling is so exponentially suppressed as to never occur in practise.

At finite temperature however, thermal fluctuations have non-zero probability to overcome the potential barrier between minima. Such configurations are known as sphalerons \cite{PhysRevD.36.581,Kuzmin:1985mm,PhysRevD.30.2212} and their rate of occurrence can be fast even relative to the Hubble rate at high temperatures. Therefore at high temperatures in the SM, sphalerons efficiently source both baryons and leptons, violating $B$ (and $\Lnum$) but leaving $B - \Lnum$ unchanged. The first condition for baryogenesis is provided already in the SM.

Computation of the sphaleron rate in the SM has long been pursued via non-perturbative lattice simulations \cite{DOnofrio:2012phz,Moore:1998zk,Bodeker:1998hm} and was known already in the symmetric phase (ie $T > \Tewpt$) in 1999 \cite{Moore:1999fs}. In order to compute the sphaleron freeze-out temperature $\Tsphal$, one requires the sphaleron rate across the EWPT which became possible following the precise measurement of the Higgs mass. $\Tsphal$ is defined by
\begin{equation}
    \Gamma_{\rm sphal}(\Tsphal) / \Tsphal^3 = \alpha_{\rm sphal}\, H(\Tsphal)
\end{equation}
where $\Gamma_{\rm sphal}$ is the rate of the sphaleron processes and $\alpha_{\rm sphal} \approx 0.1015$. In the SM, $\Tsphal = (131.7 \pm 2.3)\GeV$ \cite{DOnofrio:2014rug}. \\

Therefore, in the early universe the EW sphalerons are efficient compared to Hubble for $10^{12} \lesssim \frac{T}{\GeV} \lesssim 131$ \cite{Bento:2003jv}. It is natural to assume that the universe was initially symmetric in the baryonic and leptonic sectors, $B(\tini) = \Lnum(\tini) = 0$. Since sphalerons conserve $B-\Lnum$ the universe remains symmetric until leptogenesis generates a net lepton number at time $t$. Then immediately before the action of sphalerons, $B(t)-\Lnum(t) = -\Lnum(t)$. Sphalerons then source both quarks and leptons such that while $B-L$ is conserved, $B+L$ is violated and the universe gains a net baryon number. The conversion of lepton asymmetry into baryon asymmetry
follows the differential equation \cite{Eijima:2017cxr}
\begin{equation}\label{Leptogenesis/Sphalerons/dNB}
\frac{{\rm d}\mathcal{N}_B}{{\rm d}\alpha} =   -\frac{\ln(10)}{H}\Gamma_{\rm sphal}(T)( \mathcal{N}_B + \etasphal(T)\mathcal{N}_{\Delta \Lnum} )\,,
\end{equation}
where $\mathcal{N}_{\Delta \Lnum}  = \sum_{\ell} \mathcal{N}_{\Delta \Lnum_\ell}$ is the total lepton asymmetry and the sphaleron efficiency factor is given by \cite{Harvey:1990qw}
\begin{eqnarray}\label{Leptogenesis/Sphalerons/etasphal}
    \etasphal(T) = 4\frac{77 + 27(v(T)/T)^2}{869 + 333(v(T)/T)^2}\,.
\end{eqnarray}
The sphaleron rate found in \cite{DOnofrio:2014rug} is well approximated by \cite{Eijima:2017cxr}
\begin{equation}
\Gamma_{\rm sphal} = 9 \frac{869 + 333(v(T)/T)^2}{792 + 306(v(T)/T)^2} \frac{\Gamma_{\rm CS}(T)}{T^3} \,,
\end{equation}
where $\Gamma_{\rm CS}$ is the Chern-Simons diffusion rate
\begin{equation}
    \Gamma_{\rm CS}  = \begin{cases}
        T^4 \, {\rm exp}(-14.7 + 0.83\,T) & T < \Tewpt \\
        18 \,T^4\, a^5_W & T > \Tewpt
    \end{cases}\,,
\end{equation}
with $a_W \approx 0.0073$ being the fine structure constant. Knowing the evolution of the lepton asymmetry produced during leptogenesis, $\mathcal{N}_{\Delta \Lnum}(\alpha)$, then allows one to solve for the exact BAU by integrating \equaref{Leptogenesis/Sphalerons/dNB} between $\alpha_{\rm ini}$ and $\alpha_{\rm sphal}$ which are respectively the logarithmic scale factors at the start of leptogenesis and at the sphaleron freezeout ie $T(\alpha_{\rm sphal}) = \Tsphal$.\\

If the leptonic asymmetry has stopped evolving long before $\alpha_{\rm sphal}$ then one can approximate the freezeout of sphalerons as occurring instantaneously, with the final baryon asymmetry being given by $\chi(\Tsphal)\mathcal{N}_{\Delta \Lnum}$. Since in the SM $\Tsphal < \Tewpt$, $\chi(\Tsphal) = 12/39$ \cite{Harvey:1990qw}. This will be the case for example in high scale models of leptogenesis (see \secref{PBHLepto/High Scale}) whereas if the evolution of the lepton asymmetry is fast close to the sphaleron freeze-out, the late-time evolution is not fully translated into the baryonic sector due to the suppression of $\Gamma_{\rm sphal}$ for $T\leq \Tewpt$. In this case, as will be the case for low scale leptogenesis scenarios (see \secref{Leptogenesis/Resonant}), the exact BAU must be found by integrating \equaref{Leptogenesis/Sphalerons/dNB}.

\newpage

\section{CP violation}\label{Leptogenesis/CP}
The second of Sakharov's conditions for baryogenesis is the violation of both the C and CP symmetries. C is already maximally violated by weak interactions. In leptogenesis, CP violation occurs in the asymmetric decays (and inverse decays) of RHNs into SM leptons and antileptons \footnote{Depending on the temperature of the thermal bath, CP violation also occurs in the Higgs decay into RHNs and SM leptons, see \secref{Leptogenesis/Out-of-Equilibrium} and \secref{Leptogenesis/Thermal}.}. The amount of CP violation in the decay of RHN $N_i$ into lepton flavour $\iflav$ is given by
\begin{equation}\label{Leptogenesis/CP/epsilon}
\epsilon_{\iflav\iflav}^i =\frac{\Gamma(N_i\to \lep_\iflav \, \Higgs)-\Gamma(N_i\to \bar{\lep}_\iflav\, \bar{\Higgs})}{\Gamma(N_i\to \lep_\iflav \,\phi)+\Gamma(N_i\to \bar{\lep}_\iflav\, \bar{\Higgs})}\,,
\end{equation}
where $\Gamma$ indicates the width of the decay mode. The numerator of this expression gives the difference between the partial decay rate of $N_i$ and its CP conjugate process whereas the denominator is equal to the total decay rate of $N_i$. The inverse decay process, $ \lep_\iflav \, \Higgs \to N_i$ carries with it $\epsilon$ of the same magnitude as the decay and of opposite sign. \\

At tree level, the decay diagram of $N_i$ is\\

\begin{figure}[h!]
    \centering
    \includegraphics[width=0.3\linewidth]{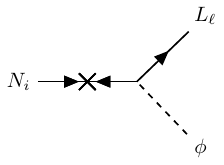}
    \caption{Tree level decay of $N_i$ in to SM leptons and Higgs. The cross indicates a Majorana mass insertion. \label{Leptogenesis/CP/Nitree}}
\end{figure}
while the CP violation arises at one-loop level from interference with loop diagrams containing $N_j$ for $j \neq i$. No CP asymmetry would occur in the decay of a lone RHN, at least one other state is required. To see why this is the case, consider that the partial rate $\Gamma(N_i\to \lep_\iflav \, \Higgs)$ is proportional to the matrix element squared by Fermi's golden rule 
\begin{equation}\label{Leptogenesis/CP/Fermi}
    \Gamma(N_i\to \lep_\iflav \, \Higgs)  = \mathcal{C}|c_0A_0 + c_1A_1|^2
\end{equation}
for some constant $\mathcal{C}$. $c_{0,1}$ are the tree level and one-loop coupling constants respectively whereas $A_{0,1}$ are the tree level and one-loop amplitudes. From \figref{Leptogenesis/CP/Nitree} it is clear the tree level coupling constant is given by $c_0 = Y_{i\iflav}^*$, while the amplitude is $A_0 = |A_0|e^{i\Phi}$ where the phase $\Phi$ is inherited from the imaginary part of the coupling $Y_{i\iflav}^*$. The one-loop level diagrams which interfere with the tree level decay are shown in \figref{Leptogenesis/CP/Niloops},
\begin{figure}[h]
    \centering
    \includegraphics[width=0.9\linewidth]{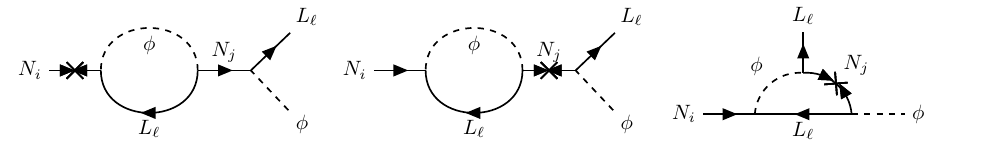}
    \caption{The loop diagrams which interfere with the tree level decay of $N_i$ providing CP violation. Crosses indicate Majorana mass insertions, while the flavour of the internal lepton lines are summed over. In order to achieve nonvanishing CP asymmetry, $i\neq j$. \label{Leptogenesis/CP/Niloops}}
\end{figure}
since the loop diagrams have couplings with two RHNs $N_i$ and $N_j$, the loop amplitude has two different phases and so can be written $A_1 = |A_1|e^{i(\Phi + \zeta)}$.

Crucially, the loop amplitude $A_1$ has an imaginary part if and only if the internal Higgs and lepton lines can go on-shell. If $i = j$ then $A_1 = |A_1|e^{i\Phi}$ and the phases would cancel in \equaref{Leptogenesis/CP/Fermi}. Therefore, the difference
\begin{equation}
    \Gamma(N_i\to \lep_\iflav \, \Higgs)-\Gamma(N_i\to \bar{\lep}_\iflav\, \bar{\Higgs}) = 2\mathrm{Im}\{\Gamma(N_i\to \lep_\iflav \, \Higgs)\}
\end{equation}
would vanish. It follows that successful leptogenesis requires at least two RHNs $N_{i,j}$. \\

The precise form of $\epsilon_{\iflav \iflav}^i$ depends on the texture of the RHN mass matrix $\MN$, the overall scale of the active neutrino mass matrix $\mnu$, and may receive important thermal corrections (see \secref{Leptogenesis/Thermal}). Given a hierarchical $\MN$, ie $\MNo \ll M_{N_j}$ for all $j$, the propagators of the heavy $N_{j}$ can be replaced with effective 4-fermion contact interactions in the diagrams shown in \figref{Leptogenesis/CP/Niloops} for $i = 1$. In this case, careful summation of spins would give \cite{Davidson:2008bu,Hambye:2016sby}
\begin{equation} \label{Leptogenesis/CP/epsilon1}
    \epsilon^1_{\iflav \iflav} = -\sum_{j}\frac{3\MNo}{16\pi M_{N_j}}\frac{\mathrm{Im}\{ (Y^\dagger_{1\iflav}Y_{\iflav j})^2\}}{|Y^\dagger Y|_{11}}
\end{equation}
where the sum is over all other states $N_j$ which exist. One can show that the total CP asymmetry (summed over all final state lepton flavours) is bounded from above \cite{Davidson:2002qv}
\begin{equation}
    |\epsilon^1| \leq \frac{3\MNo}{16\pi\vEW^2}(m_3 - m_1)
\end{equation}
which translates into a lower bound on the mass of $N_1$
\begin{equation}
    \MNo \geq \MDI \approx 10^9\GeV
\end{equation}
known as the Davidson-Ibarra (DI) limit. In fact, this bound is only strictly defined in the unflavoured regime and for an infinite hierarchy of RHN masses. An exhaustive search of the relevant parameter space done in \cite{Moffat:2018wke} found that hierarchical RHNs can in fact support successful leptogenesis in some regions of parameter space for $\MN \sim 10^6\GeV$. In general, the DI limit can be evaded when the RHN masses approach degeneracy or when the complex angles in the $R$ matrix have large imaginary parts. It was shown in \cite{Hambye:2003rt} that for $j = 2,3$ and in the simple limit $m_1 = m_2 = 0$ that
\begin{equation}
    |\epsilon^1| \leq \max\left(\frac{\MNo^3}{\MNt \MNth^2},\frac{3\MNo}{16\pi\vEW^2}(m_3 - m_1)\right)
\end{equation}
and that the DI limit breaks down for $\MNth/\MNt = \MNt/\MNo \leq 100$.

When RHNs have small mass splittings, the CP asymmetry can be resonantly enhanced to much larger values. The resonant enhancement of the CP asymmetry due to mixing between nearly-states $N_1$ and $N_2$ is, for $\MNo \leq \MNt$ and a not too degenerate spectrum $\Gamma_{N_{1,2}} \ll |\MNt - \MNo|$, proportional to \cite{Klaric:2021cpi,Flanz:1996fb,Liu:1993tg}
\begin{equation}
    \epsilon^1_{\rm resonant} \propto \frac{\MNo \MNt}{\MNt^2 - \MNo^2}
\end{equation}
which clearly diverges in the exactly degenerate limit. This is an artifact of the fact that unstable RHNs cannot in reality be S-matrix asymptotic states, to ameliorate this the resonant CP asymmetry should have a regulator of order the decay rate \cite{DeSimone:2007gkc,Klaric:2021cpi,Garny:2009qn}. The resonant enhancement of the CP asymmetry is greatest when the mass splitting between the degenerate states is close to the decay width, and is expected to vanish with zero mass splitting, the lepton number conserving limit.

\newpage

\section{Departure from Equilibrium}\label{Leptogenesis/Out-of-Equilibrium}
Sakharov's final condition for successful baryogenesis is the departure from equilibrium. In leptogenesis, this occurs due to the production or the decay of the population of RHNs out of equilibrium. The initial conditions of the universe, that is to say the conditions at the moment $t = t_{\rm ini}$ when leptogenesis begins, are important in this regard and two distinct cases can be defined
\begin{eqnarray}
    \NNi(t=\tini) = \NNeq\\
    \NNi(t=\tini) = 0 
\end{eqnarray}
which are referred to as the Thermal Initial Abundance (TIA) and Vanishing Initial Abundance (VIA) respectively. In the TIA case the RHNs already have an equilibrium abundance when leptogenesis begins, their decay is the only necessary departure from equilibrium.
In the VIA case no cosmologically significant population of RHNs exists prior to the onset of leptogenesis. Both the thermal production and subsequent decay of the RHN population occur out of equilibrium. 

\subsection*{Thermal Initial Abundance}
A cosmologically significant population of RHNs may exist prior leptogenesis most obviously due to inflaton decay populating the RHNs following inflation. It has been shown in several studies that abundant production of RHNs during reheating is possible and a generic feature of minimal and next-to-minimal inflationary scenarios \cite{Co:2022bgh,PhysRevD.105.075005,PhysRevD.109.063543}.

A population of RHNs with thermal abundance satisfies Sakharov's second condition by decaying out-of-equilibrium. The tree level decay rate of $N_i$ in its rest frame can be expressed via the optical theorem as 
\begin{equation}\label{Leptogenesis/Out-of-Equilibrium/Optical}
 \GNi = \frac{\mathrm{Im}\{\mathcal{M}(N_i\to N_i)\}}{\MNi} 
\end{equation}
where the imaginary part of the matrix element for the $N_i$ propagator can be found using the Cutkosky rules. First, writing the one-loop amplitude for the RHN self energy diagram 

\begin{figure}[h!]
    \centering
    \includegraphics[width=0.5\linewidth]{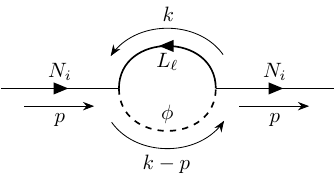}
    \caption{The self energy diagram for $N_i$ at one-loop. The lepton and Higgs propagators can be on-shell, introducing a discontinuity in the diagram and therefore an imaginary part of the matrix element. }
    \label{Leptogenesis/Out-of-Equilibrium/NiSE}
\end{figure}
yields 
\begin{equation}\label{Leptogenesis-Out-of-Equilbrium/MNiSE}
\mathcal{M} = -|Y^\dagger Y|_{ii}^2 \sum_{\rm spins}u_N(p)\int \frac{d^4k}{(2\pi)^4} \frac{\slashed{k} + \mLepT}{k^2 -\mLepT^2 + i\varepsilon} \frac{1}{(p-k)^2 - \mHiggsT^2 + i\varepsilon}\bar{u}_N(p)
\end{equation}
$\mHiggsT$ and $\mLepT$ are the Higgs and lepton masses including thermal corrections. At the temperatures of interest for thermal leptogenesis the Higgs and lepton are formally massless but receive thermal masses, see \secref{Early Universe/TFT}. Note that the flavour of the internal lepton has been summed over. $u_N(p)$ is the two component spinor for the RHNs. The internal Higgs and lepton lines are then cut and their propagators replaced according to \cite{PhysRevLett.4.624}
\begin{equation}
\frac{i}{p^2 -m^2 + i\varepsilon} \to 2\pi \delta(p^2-m^2)\theta(p^0)
\end{equation}
where the Heaviside function selects only the positive energy states, giving the discontinuity in the self energy diagram \figref{Leptogenesis/Out-of-Equilibrium/NiSE}. The discontinuity is simply related to the imaginary part of the matrix element by $\mathrm{Disc}(i\mathcal{M}) = -2\mathrm{Im}\{\mathcal{M}_{\rm loop}\}$, so that
\begin{equation}
\mathrm{Im}\{\mathcal{M}_{\rm loop}\} = -\frac{|Y^\dagger Y|_{ii}^2}{8\pi^2} \sum_{\rm spins}u_N(p)\int d^4k(\slashed{k} + \mLepT) \delta(k^2 - \mLepT^2)\theta(k^0) \delta((p-k)^2 - \mHiggsT^2)\theta((p-k)^0)\bar{u}_N(p)
\end{equation}
which when integrated can be substituted back into \equaref{Leptogenesis/Out-of-Equilibrium/Optical}, leading to the result
\begin{equation}\label{Leptogenesis/Out-of-Equilibrium/GNifull}
\GNi = \frac{(Y^\dagger Y)_{ii}\MNi}{8\pi} \lambda^{\frac{1}{2}}(1,a_\Higgs,a_\lep) \times (1-a_\Higgs + a_\lep)\,\Theta (1-a_\Higgs - a_\lep)
\end{equation}
where $\lambda(a,b,c) = (a-b-c)^2 - 4bc$, $a_X = (m_X/M)^2$, and the Heaviside functions reflect the kinematic restrictions.
If the Higgs and lepton masses can be considered small compared to $\MNi$, the result simplifies to
\begin{equation}\label{Leptogenesis/Out-of-Equilibrium/GNi}
    \GNi = \frac{|Y^\dagger Y|_{ii}^2\MNi}{8\pi}\,.
\end{equation}

For a cosmological RHN population, the rest frame decay rates \equaref{Leptogenesis/Out-of-Equilibrium/GNi}, \equaref{Leptogenesis/Out-of-Equilibrium/GNifull} should be corrected by an average inverse time dilation factor to reflect the distribution of particle energies. The RHNs are assumed to be Maxwell-Boltzmann distributed. Then \label{Leptogenesis/Out-of-Equilbrium/TAint}
\begin{equation} 
 \GNiT  \equiv \frac{\int
 \frac{\MNi}{\ENi} f\, \drm^3p }{\int f \drm^3p }\GNi
\end{equation}
where $f = e^{-\ENi/T}$. The volume element can be recast as
\begin{equation}
    d^3p = \sqrt{\ENi^2 - \MNi^2}\ENi \drm \ENi d\Omega
\end{equation}
where $d\Omega = \sin \theta \hspace{1mm}d\theta \hspace{1mm}d\phi$ is the solid angle element, so that the numerator becomes
\begin{equation} 
\int_{\MNi}^\infty \MNi e^{-\ENi/T} \sqrt{\ENi^2 - \MNi^2} d\ENi \drm \Omega\,.
\end{equation}
Integrating in the solid angle immediately gives a factor $4\pi$. Changing variables to
\begin{eqnarray}
    x_i \equiv \frac{\ENi}{\MNi}
\end{eqnarray}
allows the numerator to be expressed as
\begin{eqnarray}
4\pi \MNi^3 \int_1^\infty  e^{-x_i\MNi/T} \sqrt{x^2 - 1} \hspace{1mm}\drm x_i 
\end{eqnarray}
such that by comparison to the integral form of the modified Bessel functions \cite{abramowitz+stegun}
\begin{equation}\label{Leptogenesis/Out-of-equilibrium/Bessel}
 K_n(z) = \frac{\sqrt{\pi}}{(n-1/2)!}\left(\frac{z}{2}\right)^n \int^\infty_1 e^{-zx}(x^2 - 1)^{n-1/2}dx
\end{equation}
the numerator simplifies to
\begin{equation}
    4\pi K_1\left(\frac{\MNi}{T}\right) \MNi^2 T\,.
\end{equation}
The denominator leads to the known result (see \cite{Davidson:2008bu} Eqn. 13.2)
\begin{equation}
    \int f(p) d^3p = 4\pi K_2\left(\frac{\MNi}{T}\right)\MNi^2 T
\end{equation}
such that the thermally averaged rate is given by
\begin{equation}\label{Leptogenesis/Out-of-equilibrium/GNiT}
    \GNiT = \frac{K_1(z)}{K_2(z)}\GNi
\end{equation}
where $z = \MNi/T$, and $K_n$ is the modified Bessel function of order $n$. It is important to note that the temperature $T$ is the temperature of the RHN population, not necessarily the temperature of the universe. This is relevant for example when the RHN population is produced by Hawking radiation, as will be the case in \secref{Hot Spots/Lepto}.\\

The inverse decay mode $\Higgs \, \lep_\ell \to N_i$ proceeds with rate \cite{DiBari:2012fz}
\begin{equation}
    \GNiID = \frac{\NNeq(z)}{\Nleq(z)}\GNiT
\end{equation}
where the SM leptons are always relativistic at the temperatures relevant. Therefore for $z \ll 1$ the rate is identical to the decay rate , while for $z > 1$ the average energy of Higgs and lepton particles in the thermal bath is insufficient to produce $N_i$ and so $\GNiID$ is suppressed. The inverse decay mode does not produce any lepton asymmetry, indeed any process without leptons in the final state would not. However, as mentioned in \secref{Leptogenesis/CP}, the inverse decay is CP asymmetric with the same magnitude and opposite sign as the decay. Therefore, the inverse decay depletes any existing asymmetry between leptons and antileptons. Processes which act only to reduce existing asymmetry are referred to as washout processes.\\

For all $T$ for which
\begin{equation}
    \GNiID < H(T) 
\end{equation}
where $H$ is the Hubble rate, given by \equaref{Early Universe/Cosmology/H}, the inverse decay rate is inefficient compared to Hubble and is effectively out-of-equilibrium. The condition
\begin{equation}\label{Leptogenesis/Out-of-equilibrium/washout}
    \GNiT > H(T = \MNi) 
\end{equation}
defines the parameter space referred to as strong washout, while the opposite case is referred to as weak washout. In the TIA scenario, the thermal population of RHNs will decay out-of-equilibrium after $\GNiT > H(T)$, and produce a relic lepton asymmetry which is converted to baryon asymmetry by sphalerons independently of whether washout is weak or strong. In the VIA case however, the strength of washout is crucial.

\subsection*{Vanishing Initial Abundance}\label{Leptogenesis/Out-of-equilibrium/VIA}
When RHNs have a VIA, there initially exists no population of RHNs which can decay and produce leptonic asymmetry. In the strong washout regime $N_i$ reaches equilibrium abundance by $z\sim 1$, produced via inverse decays and depending on the temperature, Higgs decay $\Higgs \to N_i\,\lep$. The crossing symmetry for the Higgs decay gives
\begin{equation}
    \mathcal{M}(\Higgs \to N_i\,\bar{\lep}) = \mathcal{M}(\Higgs N_i \to \bar{\lep}) = \mathcal{M}( N_i \to \bar{\lep}\Higgs^\dagger)
\end{equation}
such that by \equaref{Leptogenesis/CP/Fermi} and \equaref{Leptogenesis/CP/epsilon} the CP asymmetry in the Higgs decay is equal in magnitude and opposite in sign to that in the $N_i$ decay. The same is true of the inverse decay mode. Therefore, the production of the $N_i$ population brings with it an initial ``anti-asymmetry" in the leptons. In the weak washout regime, $N_i$ decays are slow compared to Hubble at $z\sim 1$, therefore the RHN population decays at a later time when washout processes are weaker. This leads to an exact cancellation of the asymmetry from decays against the initial ``anti-asymmetry". In general, an additional mechanism beyond thermal production is required for VIA leptogenesis to produce sufficient baryon asymmetry in the weak washout case.\\\\

In the strong washout regime $\NNi \sim \NNeq$ as the temperature of the universe approaches $\MNi$, and identically to the case of TIA, the inverse decay rate becomes kinematically suppressed and the population decays. The asymmetry produced during the $N_i$ decay again cancels against any ``anti-asymmetry" left over from the production. If there were no washout processes, the asymmetry would cancel exactly and there would be no relic asymmetry. However since washout is strong, erasure of the ``anti-asymmetry" means that exact cancellation does not occur and leptonic asymmetry can survive once the washout processes go out of equilibrium. It is also important to note that since in the strong-washout regime $\NNi$ will reach thermal abundance before the population decays, the final asymmetry has no memory of the initial condition. \\

In order for the initial abundance of RHNs to vanish, assuming a period of inflation prior to leptogenesis, the inflaton must not decay into $N_i$. This could occur either because the reheating temperature of the universe is lower than $\MNi$, or because the coupling between the inflaton and the RHN is suppressed. When RHNs are very heavy, $\MNi \sim M_{\rm GUT}$, VIA would be expected for most reheating temperatures, whereas lighter RHNs would be expected to have a TIA.\\

It was shown in 1998 by Akhmedov, Rubakov and Smirnov \cite{Akhmedov:1998qx} that since RHNs produced in the early universe oscillate between their interaction and mass bases, mixing with the active neutrinos unevenly distributes the total lepton number between flavours. Therefore if at least one RHN, initially with VIA, fails to reach equilibrium abundance before sphaleron freeze-out, the oscillations of RHNs provide an additional source of baryon asymmetry not accounted for in the picture of thermal leptogenesis painted above. In fact, as shown by Klaric, Shaposhnikov, and Timiryasov, the leptogenesis via neutrino oscillations mechanism is described by the same set of quantum kinetic equations as thermal resonant leptogenesis (see \secref{Leptogenesis/Resonant}) in a different region of parameter space \cite{Klaric:2020phc,Klaric:2021cpi}. Indeed, the Boltzmann equations formalism detailed in \secref{Leptogenesis/BEs} appears as the limit in which RHNs are heavy and hierarchical, or have TIA. The contribution of neutrino oscillations to $\YB$ may be strong when RHNs are degenerate and have VIA. Thus, in the remainder of this work the models of leptogenesis considered do not receive contributions to the BAU from $N_i$ oscillations because the model in \secref{PBHLepto/High Scale} features heavy and hierarchical RHNs and that in \secref{PBHLepto/Low Scale} assumes TIA.

\newpage

\section{Thermal Effects}\label{Leptogenesis/Thermal}
Leptogenesis does not occur in a vacuum, amid the extreme temperature and density of the primordial universe it is influenced by a host of spectator processes which do not directly produce lepton asymmetry. Interactions taking place on timescales much slower than the expansion rate of the universe can be safely neglected while as established in \secref{Early Universe/TFT}, those which are very fast can be resummed and treated in terms of thermal masses and corrections to couplings. In particular since leptogenesis takes place almost exclusively at $T \geq \Tewpt$, the Higgs and SM leptons are formally massless. However they are also assumed to be in thermal equilibrium with the plasma and so gain thermal masses. The Higgs and lepton masses including thermal corrections are given by \equaref{Early Universe/TFT/MHiggs} and \equaref{Early Universe/TFT/MLep} respectively. \\

The decay rate \equaref{Leptogenesis/Out-of-Equilibrium/GNifull} accounts for the Higgs and lepton thermal masses and so is a function of the background plasma temperature. Since the thermal masses of the Higgs and leptons grow with the temperature, at high temperatures $T \gg \MNi$, then $\MNi < \mHiggsT + \mLepT$ and the decay of $N_i$ is kinematically suppressed. At very high temperatures, the thermal mass of the Higgs can be large enough that the decay $\Higgs \to N_i + \lep$ is possible and occurs at a rate given by \cite{Hambye:2016sby}
\begin{equation}\label{Leptogenesis/Thermal/GHiggs}
    \Gamma_{\Higgs \to N_i \lep} = \frac{(Y^\dagger Y)_{ii} \mHiggsT} {8\pi a_\Higgs^2}\lambda^{\frac{1}{2}}(a_\Higgs,1,a_\lep)\times (a_\Higgs - a_\lep-1)\,\Theta(a_\Higgs - a_\lep-1)\, .
\end{equation}
It follows that only one of the decays $N_i \to \Higgs + \lep$ and $\Higgs \to N_i + \lep$ is kinematically unsuppressed at any particular temperature. At low temperatures, the RHN decay is dominant while at high temperatures the Higgs decay dominates. At intermediate temperatures, neither decay mode is kinematically unsuppressed. It has been shown that corrections due to the scattering of nearly collinear soft gauge bosons enhance the low temperature decay rate of the HIggs by an order unity factor, and bridge the phase space gap between the high and low temperature regimes \cite{Anisimov:2010gy}, allowing the continued production of $N_i$ in the intermediate regime.\\

The Higgs decay also contributes to the lepton asymmetry with the same magnitude and sign of CP asymmetry as the RHN decay. Therefore when formulating the Boltzmann equations to track the evolution of the $N_i$ and $\Delta \Lnum$ densities it will be useful to define
\begin{equation}
    \Gamma_{NH} = \GNiT + \Gamma_{\Higgs \to N_i \lep}^T
\end{equation}
which accounts for the rate of $N_i$ and $\Higgs$ decays, the superscript $T$ indicates a thermal average as in \equaref{Leptogenesis/Out-of-Equilbrium/TAint}. Using the definitions \equaref{Leptogenesis/Thermal/GHiggs} and \equaref{Leptogenesis/Out-of-Equilibrium/GNifull} does not account for the effects of soft gauge scatterings with the thermal bath. \\

The CP violating parameter defined by \equaref{Leptogenesis/CP/epsilon} also receives thermal corrections. In addition to the effects of the Higgs and lepton thermal masses, the CP asymmetry parameter is also affected at high temperatures by interaction of the intermediate Higgs and leptons (see \figref{Leptogenesis/CP/Niloops}) with the background \cite{Frossard:2012pc,Davidson_2008}. This results in thermal corrections to the RHN self energy quantified by \cite{Granelli_2021,Frossard:2012pc,Hambye:2016sby}
\begin{equation}\label{Leptogenesis/Thermal/gamma}
    \langle \gamma \rangle = \frac{k_\mu L_k(p)^\mu}{k_\mu p^\mu}
\end{equation}
where $k_\mu,p_\mu$ are the lepton and RHN momenta respectively as in \figref{Leptogenesis/Out-of-Equilibrium/NiSE}. The absorptive function $L_k(p)$ is given by
\begin{equation}
    L_k(p)^\mu = 16\pi \int \hspace{1mm} \drm \Pi^\phi\hspace{1mm} \drm \Pi^L (2\pi)^4 \delta(p-k-(p-k)) \slashed{k} (1-f_\phi(k-p) -f_L(k))
\end{equation}
where $\Pi^\phi(\Pi^L)$ and $f_\phi(f_L)$ are the Higgs (lepton) self energies and distribution functions respectively. Analytical expressions for $L_k$ can be found for various kinematic regimes in Appendix D of \cite{Frossard:2012pc}. These expressions do not account for the scattering of soft gauge bosons \cite{Anisimov:2010gy} and so the function $\langle \gamma \rangle$ vanishes in the intermediate temperature regime in which both RHN and Higgs decays are kinematically forbidden. As done in \cite{Granelli_2021}, one could interpolate between the high and low temperature regimes of $\langle \gamma \rangle$ in order to approximate the effects of multiple soft scatterings. In this work, $\langle \gamma \rangle$ is computed using the non-interpolated form reported in \cite{Granelli_2021}.

In the case of resonant leptogenesis, which is discussed in detail in \secref{Leptogenesis/Resonant}, the difference in thermal masses between the degenerate states is important. For the states $N_i$ and $N_j$ with mass splitting $\Delta \Mdegen_{ij} = \MNi - M_{N_j}$, the thermal correction to the mass splitting is given by
\begin{equation}\label{Leptogenesis/Thermal/dMT}
    \Delta \Mdegen^T_{ij} = \frac{\pi }{z^2} \sqrt{ \left(\Gamma_{jj}-\Gamma_{ii}\right)^2 + 4|\Gamma_{ij}|^2}
\end{equation}
where 
\begin{equation}
    \Gamma_{ij} = \frac{|Y^\dagger Y|_{ij}\Mdegen}{8 \pi}\,.
\end{equation}\\
with $\Mdegen$ the common mass of the degenerate states.

\newpage

\section{Boltzmann Equations}\label{Leptogenesis/BEs}
Calculating the final BAU can be achieved by solving the coupled set of Boltzmann equations governing the evolution of the number density of $N_i$ and of lepton asymmetry $\Delta \Lnum$, see \secref{Early Universe/Abundance}. \equaref{Early Universe/Abundance/BEalpha} gives the Boltzmann equation for the comoving number density of an unstable particle, specialising to the case of $N_i$ it becomes
\begin{equation}\label{Leptogenesis/BEs/dNNi}
    \frac{\drm \NNi}{\drm \alpha} = -\frac{\ln(10)}{H}\left(\Gamma_{NH} + \sum_{\sigma(N_i a \to ij)} \langle\sigma v_{\rm Mol}\rangle n_a^{\rm eq} \right)(\NNi
- \NNeq)\
\end{equation}
where the sum is over all $2\to 2$ diagrams with external $N_i$, $N_i\,a\leftrightarrow ij$, and $\Gamma_{NH}$ is the sum of the $N_i$ and Higgs decay rates (see \secref{Leptogenesis/Thermal}). The above equation is coupled to the Boltzmann equation governing the evolution of the number density of lepton asymmetry. The number density of leptonic asymmetry due to a general species is given by \equaref{Early Universe/Equilibrium/DeltaLBE}. The lepton asymmetry number density is defined as
\begin{equation}
   \mathcal{N}_{\Delta \Lnum_\iflav} \equiv \mathcal{N}_{\lep_\iflav} - \mathcal{N}_{\bar{\lep_\iflav}}
\end{equation}
while the baryon asymmetry number density is given by
\begin{equation}
   \mathcal{N}_{\Delta B} \equiv \mathcal{N}_{B} - \mathcal{N}_{\bar{B}}
\end{equation}
both of which are expected to vanish at very early times. For leptogenesis, it will be useful to track the evolution of the flavoured $B-\Lnum$ asymmetry densities since they are preserved by sphalerons
\begin{equation}\label{Leptogenesis/BEs/B-L}
    \frac{\drm \NBLl}{\drm \alpha} =  \frac{1}{3}\frac{\drm \mathcal{N}_{\Delta B}}{\drm \alpha} - \frac{\drm \mathcal{N}_{\Delta \Lnum_\iflav}}{\alpha}\, .
\end{equation}
The baryon asymmetry evolves only via the sphaleron processes in leptogenesis, which source it at the same rate as $\Delta \Lnum$. Therefore in the subtracted $B-\Lnum_\ell$ Boltzmann equations \equaref{Leptogenesis/BEs/B-L} the sphaleron terms cancel leaving simply the negative equation for $\mathcal{N}_{\Delta \Lnum_\ell}$ with no sphaleron term
\begin{equation}\label{Leptogenesis/BEs/dNBL}
\frac{\drm \NBLl^i}{\drm \alpha} = \frac{\ln(10)}{H} \left[ \epsilon_{\iflav \iflav}(\NNi-\NNeq)\Gamma_{NH} - P_{\iflav i}\frac{\NBLl}{\Nleq}\left(\frac{\GNiT}{2}\NNeq + 2\gamma_{S_t} + \gamma_{S_s}\frac{\NNi}{\NNeq}\right) \right] \end{equation}

where the superscript $i$ indicates that the Boltzmann equation tracks the change in $\NBLl$ due to the RHN $N_i$, $P_{\iflav i}$ is the projector of $N_i$ onto flavour $\iflav$ as defined in \equaref{Early Universe/Equilibrium/Pflav}, the sum over $\sigma$ now counts all diagrams with external leptons and $\mathcal{N}_a$ is the number density of the particles off which the leptons scatter. The Boltzmann equations \equaref{Leptogenesis/BEs/dNNi}, \equaref{Leptogenesis/BEs/dNBL} account for all processes involving RHNs and the SM leptons up to order $Y^4$ in the Yukawa couplings. \\

The Boltzmann equation \equaref{Leptogenesis/BEs/dNBL} describes the evolution of the asymmetry in lepton flavour $\iflav$. At the Lagrangian level the SM leptons are distinguished by their Yukawa couplings $h_{ij}$, which mediate elastic scattering interactions in the early universe, see \equaref{SM/Neutrino Mass/Lagry}. These processes produce neither lepton asymmetry nor RHNs and are generally spectator processes. However as shown in \secref{Early Universe/TFT}, when fast compared to the Hubble rate they ought to generate thermal masses for the SM leptons. The charged lepton Yukawas vary greatly in size and so one would expect similarly disparate thermal masses. Estimating the charged lepton Yukawa interaction rate as 
\begin{equation}\label{Leptogenesis/BEs/Tflav}
    \Gamma_\iflav \approx 5\times 10^{-3} h_{\iflav \iflav}^2 T
\end{equation}
where $h_{\iflav\,\iflav}$ is the diagonal element of the charged lepton Yukawa matrix of \equaref{SM/Neutrino Mass/Lagry}, it turns out that for $T \gtrsim 10^{12}$\,GeV none of the charged lepton Yukawa interactions are fast compared to the Hubble rate \cite{Davidson:2008bu}. At extremely high temperatures therefore, the total BAU could be calculating by summing \equaref{Leptogenesis/CP/epsilon} over the flavour index $\epsilon = \sum_\iflav \epsilon_{\iflav \iflav}$, leptogenesis is blind to lepton flavour and the total asymmetry evolves as if in a single flavour. It was demonstrated in \cite{Granelli:2021fyc} that flavour effects may persist at higher scales in the low-energy CP violation scenario. For the purposes of this thesis, at temperatures lower than $T \gtrsim 10^{12}$, the flavoured Boltzmann equations \equaref{Leptogenesis/BEs/dNBL} are employed to account for flavour effects. 

$\Gamma_{NH}$ denotes the rate of $1\leftrightarrow 2$ decays and inverse decays of $N_i$ and the Higgs
\begin{eqnarray}\label{Leptogenesis/BEs/12}
    N_i &\leftrightarrow& \lep \Higgs^\dagger\\
    \Higgs &\leftrightarrow& N_i \lep
\end{eqnarray}
Depending on the temperature of the thermal bath, one or both decay pathways may be kinematically suppressed and both cannot occur simultaneously, see \secref{Leptogenesis/Thermal}. These processes are responsible for the generation of lepton asymmetry.\\

$N_i$ also undergoes $2 \leftrightarrow 2$ scatterings with gauge bosons and quarks
\begin{eqnarray}\label{Leptogenesis/BEs/Nscatter}
    N_i \,\lep &\leftrightarrow& Q_3 \,Q_3\\
    N_i \,Q_3 &\leftrightarrow& \lep \,Q_3\\
    N_i \,\lep &\leftrightarrow& \Higgs \,A_\mu\\
    N_i \,A_\mu &\leftrightarrow& \Higgs\, \lep \\
    N_i \,\Higgs &\leftrightarrow& \lep \,A_\mu\\
\end{eqnarray}
where $A_\mu$ are the $SU(2)\times U(1)$ gauge bosons \cite{Giudice:2003jh} and $Q_3$ is the top quark (scattering involving the lighter quarks is suppressed by the quark Yukawas). These processes produce (destroy) $N_i$ in the forward (backward) time direction and washout asymmetry since there are external leptons.
As well as appearing in all of the processes listed in \equaref{Leptogenesis/BEs/Nscatter}, the SM leptons also take part in $2 \leftrightarrow 2$ scatterings with $\Delta \Lnum = 2$ mediated by $N_i$
\begin{eqnarray}\label{Leptogenesis/BEs/Lscatter}
    \lep\, \Higgs & \leftrightarrow&\bar{\lep} \,\Higgs^\dagger\\
   \lep \,\bar{\lep}& \leftrightarrow& \Higgs \,\Higgs^\dagger
\end{eqnarray}
which should be considered in addition when summing over cross sections in \equaref{Leptogenesis/BEs/dNBL}. Care must be taken when calculating the rate for the $\Delta \Lnum = 2$ processes to subtract the contribution from on-shell intermediate $N_i$ states which are already counted in the rate $\Gamma_{NH}$.\\

In \equaref{Leptogenesis/BEs/dNNi} the $s-$channel quark scattering diagram is grouped with the gauge-boson scattering diagram for which the $U(1)$ contribution is the $s-$channel diagram
\begin{eqnarray}
    S_s &=& (N_i \lep \leftrightarrow Q_3 Q_3) + (N_i \lep \leftrightarrow \Higgs A_\mu) \,,
\end{eqnarray}
corresponding to the Feynman diagrams below\\
\begin{figure}[h!]
    \centering
    \includegraphics[width=0.65\linewidth]{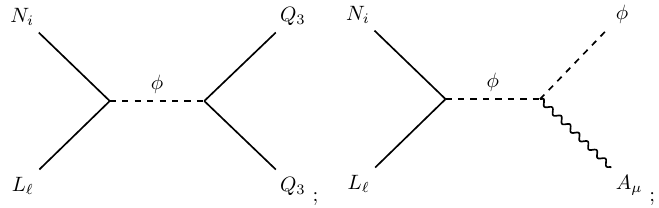}
    \caption{Feynman diagrams corresponding to the rates denoted by $S_s$ are shown. The $SU(2)$ ($t-$channel) contribution to the process $N_i L \leftrightarrow \phi A_\mu$ must also be included.}
    \label{Leptogenesis/Feynmans}
\end{figure}

such that $S_t$ counts the $t-$channel quark scattering diagram $N_i Q_3 \leftrightarrow L Q_3$ as well as the remaining diagrams involving gauge bosons for which the $s-$channel diagram is the $SU(2)$ contribution. \newpage 

The Feynman diagrams composing $S_t$ are shown below.\\

\begin{figure}[h!]
    \centering
    \includegraphics[width=0.9\linewidth]{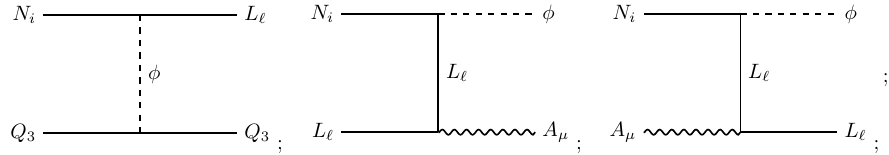}
    \caption{Feynman diagrams corresponding to the rates denoted by $S_t$ are shown. The $U(1)$ contributions to the central and rightmost processes must also be included.}
    \label{Leptogenesis/Feynmans}
\end{figure}

The reaction density $\gamma$ is given in thermal equilibrium by \cite{Giudice:2003jh}
\begin{equation}
    \gamma^{\rm eq}(Na\rightarrow ij) = \frac{1}{8 \pi} \int \frac{\rm{d}^4 p}{(2\pi)^4} \hat{\sigma}(Na \leftrightarrow ij) e^{E/T}
\end{equation}
where the reduced cross section is given by $\hat{\sigma}(s) \equiv 2s \lambda (1,\MNi/s,M_a/s)\sigma(s)$, with $s$ the usual Mandelstam variable.
Expressions for cross sections corresponding to the processes listed in \equaref{Leptogenesis/BEs/Nscatter} as well as the subtracted cross sections for \equaref{Leptogenesis/BEs/Lscatter} can be found in \cite{Giudice:2003jh,Pilaftsis:2003gt}. The expressions given in \cite{Giudice:2003jh} for the gauge boson scattering cross sections account for the thermal masses of $A_\mu,\lep$ where they regulate IR divergences but neglect their thermal motion with respect to the plasma. The authors state that at $T \gg \MNi$ this approach neglects terms suppressed by higher powers of the couplings. Those found in \cite{Pilaftsis:2003gt} regulate IR divergences by cutting the phase space of integration according to a universal regulator which varies between the Higgs and lepton thermal masses. Since the expression for the $N_i \,\Higgs \leftrightarrow \lep \,A_\mu$ cross section in \cite{Giudice:2003jh} is negative, the expressions used throughout the rest of this work are those found in \cite{Pilaftsis:2003gt}.

\newpage

\section{Resonant Leptogenesis}\label{Leptogenesis/Resonant}
As mentioned in \secref{Leptogenesis/CP}, the DI limit for successful leptogenesis $\MNi \geq \MDI$ can be evaded in a number of ways. In particular, if two RHN states $N_{i,j}$ are quasi degenerate with a small mass splitting
\begin{equation}
    \Delta \Mdegen = \MNi - M_{N_j} 
\end{equation}
the CP asymmetry parameter $\epsilon$, and in particular the self energy contribution (see the leftmost two diagrams in \figref{Leptogenesis/CP/Niloops}) can be resonantly enhanced by many orders of magnitude \cite{Covi:1996fm}. Neglecting the contribution of the vertex CP violation, the amount of CP violation in the decay of $N_i$ summed over flavour is given by \cite{Pilaftsis:1997jf}
\begin{equation}\label{Leptogenesis/Resonant/epsi}
    \epsilon^i_{SE} = \frac{\rm{Im}\{(Y^\dagger Y)_{ij}\}^2}{(Y^\dagger Y)_{ii}(Y^\dagger Y)_{jj}}\frac{(\MNi^2-M_{N_j}^2)\MNi\Gamma_{N_j}}{(\MNi^2-M_{N_j}^2)^2 + \MNi^2\Gamma_{N_j}^2}
\end{equation}
where the regulating term in the denominator, being absent in finite order perturbation theory, is present due to the physical expectation that the divergence in the asymmetry should be regulated by the finite width of the RHNs \cite{Pilaftsis:1997jf,Hambye:2016sby}. Very generally, the resonant enhancement is strongest when $\Delta \Mdegen_{ij} \sim \frac{\Gamma_{N_j}}{2}$. Considering $\Delta \Mdegen_{ij}$ as a free parameter, leptogenesis can be successful with RHNs as light as $\MNi \sim \mathcal{O}(\rm GeV)$ \cite{Granelli_2021,Klaric:2020phc}. Leptogenesis scenarios in which the resonantly enhanced decay of two quasi-degenerate RHNs are responsible for producing the BAU are referred to as resonant leptogenesis. This is distinguished as a particular regime in the low scale leptogenesis parameter space in which all 3 RHNs reach equilibrium before sphaleron freeze-out, precluding any contribution to $\YB$ from the quantum kinetic evolution of sterile neutrino oscillations \cite{Akhmedov:1998qx,Klaric:2021cpi}. Alternatively the resonant leptogenesis case arises necessarily when degenerate RHNs have TIA (see \secref{Leptogenesis/Out-of-Equilibrium}).

The correct CP asymmetry parameter for the resonant decay of $N_i$, degenerate with $N_j$, should include the self-energy contribution given by \equaref{Leptogenesis/Resonant/epsi}, the vertex contribution which is not resonantly enhanced, and thermal corrections (see \secref{Leptogenesis/Thermal}). In resonant leptogenesis not only the Higgs and lepton thermal masses are important but also the finite temperature contribution to the zero-temperature mass splitting $\Delta \Mdegen_{ij}^T$. Furthermore, the important dynamics of resonant leptogenesis occur at $T \ll 10^{12} \GeV$ so flavour effects must also be accounted for (see \equaref{Leptogenesis/BEs/Tflav}). The flavoured, thermally corrected CP asymmetry parameter is given by ~\cite{Granelli:2020ysj} (see also Refs.~\cite{BhupalDev:2014pfm,BhupalDev:2014oar, Hambye:2016sby})
\begin{equation}
\label{Leptogenesis/Resonant/epsi}
\epsilon_{\iflav\iflav}^i = \sum_{j,\,j\neq i}\frac{2\, \mathrm{sgn}(\MNi-M_{N_j})\,I_{ij,\iflav\iflav}\,\Delta M_{ij}\gamma(z)}{4\frac{\GNt}{\Gamma_{jj}^2}(\Delta M_{ij} + \Delta M_{ij}^T)^2 + \frac{\Gamma_{jj}^2}{\GNt}\gamma^2(z)} \,,
\end{equation}
where $z = \MNi/T$, and the thermal mass splitting $\Delta \Mdegen_{ij}^T$ is given by \equaref{Leptogenesis/Thermal/dMT}. The function $I_{ij,\iflav\iflav}$ describes the Yukawa dependence of $\epsilon^i_{\ell \ell}$ including both the self energy and vertex contributions, and is given by
\begin{equation}
\label{Leptogenesis/Resonant/I}
I_{ij,\iflav\iflav} =  \frac{\mathrm{Im}\{Y^\dagger_{i\iflav}Y_{\iflav\,j}(Y^\dagger Y)_{ij} \} + \frac{\MNi}{M_{N_j}}\mathrm{Im}(Y^\dagger_{i\iflav}Y_{\iflav\,j}(Y^\dagger Y)_{ji} ) }{(Y^\dagger Y)_{ii}(Y^\dagger Y)_{jj}} \,,
\end{equation}
and the function $\gamma(z)$ describes the temperature dependence of the RHN self-energy cut, given by \equaref{Leptogenesis/Thermal/gamma}.

The requirement to reproduce the neutrino masses no larger than the existing upper limit from KATRIN \cite{KATRIN:2021uub} $m_h \lesssim 1\rm{eV}$ places tight constraints on the Yukawa couplings in low scale leptogenesis. Taking $\Mdegen \sim \GeV$, the seesaw formula \equaref{vMass/Seesaw/seesaw} gives 
\begin{equation}
    |Y^\dagger Y| \lesssim 10^{-12}
\end{equation}
and one expects small mixing angles $U^2$ (see \equaref{vMass/Mixing/U2}). However when there is a small mass splitting between a pair of RHNs $N_{i,j}$ and a third decoupled RHN, larger mixing angles can be compatible with the observed neutrino masses \cite{Drewes:2016gmt}. In the approximate two-neutrino limit, the imaginary part of the single relevant angle in the Casas-Ibarra $R$ matrix defined by \equaref{vMass/Casas-Ibarra/R} can be arbitrarily large without affecting the size of the active neutrino masses. Therefore in principle RHNs as light as the $\GeV$ scale which mix strongly with the SM are allowed. In practice however, larger mixing angles also mean stronger washout and the viable thermal resonant leptogenesis parameter space does not extend to arbitrarily large $U^2$, see for example Fig.\,9 in \cite{Klaric:2021cpi} and \figref{PBHLepto/LowScale/standard}. \\

Light RHNs with potentially strong mixing with the SM offer the tantalising possibility of being directly detected. Multiple near future experiments project sensitivity curves which can probe the viable parameter space including the SHiP \cite{SHiP:2018xqw}, SHADOWS \cite{Baldini:2021hfw}, MATHUSLA \cite{MATHUSLA:2022sze} and ANUBIS \cite{Hirsch:2020klk} experiments. However only the $\Mdegen\sim\GeV$ regime of resonant leptogenesis is coincident with these experimental sensitivities. Nonetheless, much attention has been devoted to models of leptogenesis with light RHNs \cite{Pilaftsis:1997jf,Asaka:2005an,Canetti:2012kh,Granelli_2021,Klaric:2020phc}. \\

While it is appealing from an experimental point of view that resonant leptogenesis may be probed by experiments, some may view it as theoretically problematic that successful thermal resonant leptogenesis at $\MNi \sim \mathcal{O}(\rm GeV)$ requires such a tiny mass splitting $\Delta \Mdegen_{ij} \lesssim 10^{-12}\, \GeV$. If this mass splitting originates at the Lagrangian level it can be viewed as a free parameter of the theory and its smallness could be considered unnatural. Alternatively, the mass splitting could arise from radiative corrections.\\

Suppose that the states $N_{i,j}$ are exactly degenerate at some high scale $\Lambda$ and are the only states to exist. Radiative corrections will induce slightly different corrections to the masses $\MNi$ and $M_{N_j}$ at low energies and result in a non-zero mass splitting. The Renormalisation Group Equation (RGE) for the Majorana mass matrix is given by
\begin{equation}
    (4\pi)^2\left(\frac{\drm \MN}{\drm t}\right)_{ij} = (Y^\dagger Y)_{ij}M_j + (Y^\dagger Y)_{ji}M_i
\end{equation}
here $t \equiv \ln{\mu_/\mu_0}$, where $\mu$ is the energy scale and $\mu_0$ is some reference scale. Since at high energies $\MNi = M_{N_j} = \Mdegen$ it follows that
\begin{equation}
    (4\pi)^2\left(\frac{\drm \MN}{\drm t}\right)_{ij} = \Mdegen((Y^\dagger Y)_{ij} + (Y^\dagger Y)_{ji})
\end{equation}
which can be integrated directly. Approximating that the running of the common mass $\Mdegen$ is insignificant due to the small Yukawa couplings, and using the fact that $Y^\dagger Y$ is a RGE invariant, integrating from the scale $\mu_0$ to $\Lambda$ gives the result
\begin{equation}\label{Leptogenesis/Resonant/deltaMRG}
    (4\pi)^2((\MN)_{ij}(\Lambda) - (\MN)_{ij}(\mu_0)) = \Mdegen\ln{\frac{\Lambda}{\mu_0}} ((Y^\dagger Y)_{ij} + (Y^\dagger Y)_{ji})
\end{equation}
which for vanishing mass splitting at $\mu = \Lambda$ gives the low energy mass splitting
\begin{equation}\label{Leptogenesis/Resonant/dMrad}
    \Delta \Mdegen_{ij}(\mu_0) = - \frac{\Mdegen}{8\pi^2}\ln{\frac{\Lambda}{\mu_0}} ((Y^\dagger Y)_{ii} - (Y^\dagger Y)_{jj})\,.
\end{equation}
Taking $\Lambda = \Mpl$ and $\mu_0 = \Tewpt$, the radiatively induced mass splitting is approximately $\Delta \Mdegen_{ij}/\Mdegen \sim 10^{-16}$. This simple scenario however is not viable for resonant leptogenesis. As shown in \cite{Dev:2015wpa}, when $\MN(\Lambda) = \Mdegen \times I_2$, then since \equaref{Leptogenesis/Resonant/deltaMRG} is proportional to $\rm{Re}\{Y^\dagger Y\}$ and necessarily diagonal, then $\rm{Re}\{Y^\dagger Y\}$ is also diagonal. Since the CP asymmetry parameter is proportional to the off diagonal elements of $\rm{Re}\{Y^\dagger Y\}$ at $\mathcal{O}(Y^4)$ (see \equaref{Leptogenesis/Resonant/epsi}), it vanishes. There must necessarily be an additional source of flavour breaking in the RHN mass matrix for resonant leptogenesis at this order in the Yukawas. The requirement of tiny mass splittings for thermal resonant leptogenesis could be viewed as fine tuning.

\newpage

\section{Experimental Status of Leptogenesis}\label{Leptogenesis/Experiment}
In its most general realization, type-1 leptogenesis depends on 18 different parameters. These are the 3 RHN masses in $\MN$, the three active neutrino masses, the Dirac and Majorana CP violating parameters in $\UPMNS$ along with its three mixing angles, and the real and imaginary parts of the 3 angles in the Casas-Ibarra $R$ matrix. Of these, the three angles in $\UPMNS$ have been measured experimentally and the $3\sigma$ ranges are \cite{Esteban:2020cvm}
\begin{eqnarray}
    \theta_{12} = (31.31 \to 35.74)^\circ\\
    \theta_{23} = (39.7 \to 51.0)^\circ\\
    \theta_{13} = (8.23 \to 8.91)^\circ
\end{eqnarray}
while the Dirac CP violating phase $\delta$ is constrained to lie in the range $144^\circ \to 350^\circ$ by the T2K experiment \cite{T2K:2019bcf}. The Majorana CP violating phases (of which, only the difference $\eta_{2} - \eta_{1}$ is physical) are not currently subject to any constraints, but might be measurable in the case of IH if $(\beta\beta)_{0\nu}$ is observed with sufficient precision \cite{Simkovic:2012hq} or if quantum decoherence in neutrino oscillations exists it may be possible to infer the value of the difference from the DUNE measurement of $\delta$ \cite{Carrasco-Martinez:2020mlg}. As explained in \secref{SM/vMass}, the overall scale of the active neutrino mass matrix is not known but the two mass splittings $\dmatm^2$ and $\dmsol^2$ are well measured, having the central values given by \equaref{SM/vMass/splittings}, 
so that the lightest possible $m_h$, corresponds to the case where the lightest of the neutrinos is massless, $m_l = 0$. The most stringent laboratory constraint on $m_h$ is that given by the KATRIN experiment \cite{KATRIN:2021uub}
\begin{equation}
    m_h \leq 0.64 \rm{eV}
\end{equation}
while cosmological bounds on the neutrino mass are more model dependent. If cosmological neutrinos decay when non-relativistic, the 2018 Planck data \cite{Planck:2018vyg} excludes $m_h \geq 0.42 \rm{eV}$ \cite{FrancoAbellan:2021hdb}.\\

The 9 remaining parameters, the three RHN masses and 3 complex $R$ matrix angles, are completely unconstrained experimentally. Normally, models of leptogenesis consider the case where at least one of the RHNs is decoupled in such a way as to not affect leptogenesis. For example, in high-scale models with a hierarchical $\MN$, where $\MNo \ll \MNt \ll \MNth$, the dynamics of leptogenesis depend only on $\MNo$. In resonant leptogenesis, the two degenerate states $N_i$ and $N_j$ are responsible for all of the asymmetry generation while the third state is often considered to be so light as to never depart from equilibrium while the EW sphalerons are active, or so heavy as to only generate asymmetry at high temperatures which is efficiently washed out. If the three masses of $\MN$ and the 3 complex $R$ matrix angles are taken completely general, then it is highly complicated to make any predictions for leptogenesis.\\

A common approximation is to consider that since $\dmatm \gg \dmsol$, from the perspective of $\nu_3$, $m_l \approx m_m$ and therefore one may consider that the $R$ matrix is effectively described by a single mixing angle, $R = R_{13}(\theta_{13})$ (or equivalently, $R=R_{23}(\theta_{23})$). Then the real and imaginary parts of the angle, $\theta = x + iy$ play crucial roles. In this case, the dependence of the CP asymmetry parameter $\epsilon$ on the real part $x$ is periodic and so it can be fixed to maximize the predicted value of $\YB$. This leaves the scale $m_h$, the imaginary part of the $R$ matrix angle $y$, the relevant 1 or 2 masses in $\MN$ and the Majorana phases unspecified. The parameter space to which direct detection experiments are sensitive is the total mixing of the RHNs to each lepton flavour $|U_\iflav|^2$ (see \equaref{vMass/Mixing/Uflav}) and the RHN masses. Many experiments such as MATHUSLA are sensitive to mixing between RHNs and all 3 flavours of SM leptons \cite{MATHUSLA:2022sze}. Constraints are then projected onto the three $|U_\iflav|^2$ vs $\MNi$ planes assuming the RHNs mix exclusively to each flavour.
\begin{figure}[h]
    \centering
    \includegraphics[width=0.8\linewidth]{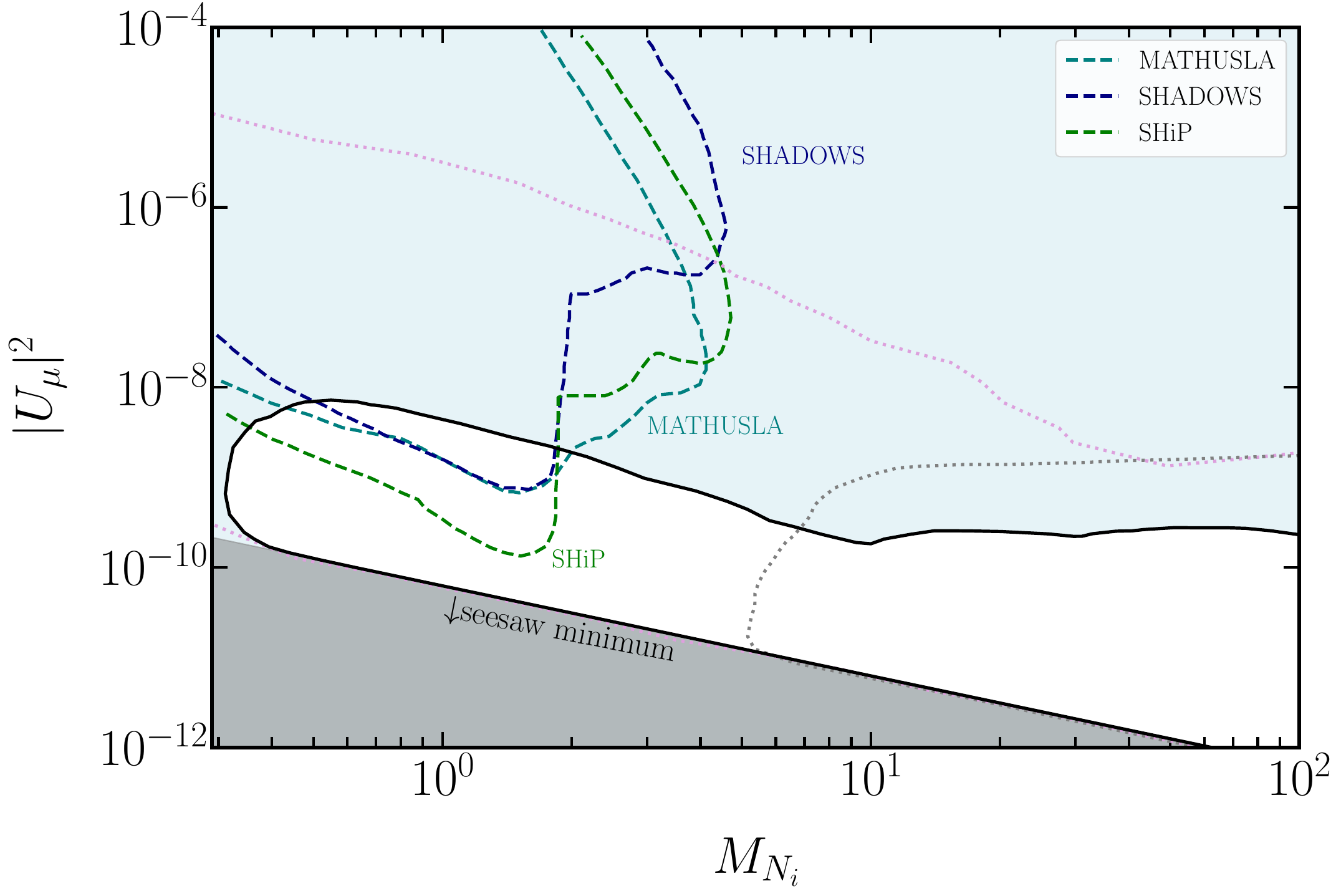}
    \caption{The projected sensitivity curves of the SHiP \cite{SHiP:2018xqw}, SHADOWS \cite{Baldini:2021hfw} and MATHUSLA \cite{MATHUSLA:2022sze} assuming RHNs couple exclusively to muons. The white unshaded region is the viable resonant leptogenesis parameter space assuming no contribution to $\YB$ from neutrino oscillations and VIA, according to \cite{Granelli_2021}. The grey shaded region is theoretically excluded by the seesaw mechanism. In the blue shaded region, resonant leptogenesis with VIA does not produce enough asymmetry for any combination of free parameters. The purple and grey dotted lines show respectively the parameter space for successful leptogenesis, and leptogenesis with TIA according to \cite{Klaric:2021cpi}.  }
    \label{Leptogenesis/Experiment/mu}
\end{figure}
In \figref{Leptogenesis/Experiment/mu} the projected sensitivity curves of the SHiP \cite{SHiP:2018xqw}, SHADOWS \cite{Baldini:2021hfw} and MATHUSLA \cite{MATHUSLA:2022sze} experiments are shown with dashed lines and the viable parameter space for resonant leptogenesis with VIA as calculated in \cite{Granelli_2021} is shown as the white region. The viable parameter space for low scale leptogenesis according to \cite{Klaric:2021cpi} is bounded by the pink dotted contour, while the grey dotted contour is the same assuming TIA. Specifically the purple and grey contours are calculated using the quantum kinetic equations formalism, with the purple contour accounting for the contribution due to neutrino oscillations, whereas the grey dotted line bounds the region which is calculated with TIA and therefore receives no contribution from oscillations. \\

The experimental sensitivity curves shown in \figref{Leptogenesis/Experiment/mu} overlap significantly with the low scale leptogenesis parameter space when neutrino oscillations contribute, and only overlap the resonant leptogenesis parameter spaces when VIA is assumed. In \figref{PBHLepto/LowScale/standard} the same sensitivity curves are compared to the resonant leptogenesis parameter space with TIA as calculated in this work, where the very edge of the low mass end of the viable parameter space overlaps the sensitivity curves. The prospects that experiments might observe RHNs directly then depends on leptogenesis occurring at very low scales and likely with VIA. It was also shown that in this case upcoming charged lepton flavour violation experiments could probe the leptogenesis parameter space \cite{Granelli:2022eru}. Larger values of the mixing $U^2$ may be compatible with low scale leptogenesis if CP violation comes soley from the $U$ matrix \cite{Granelli:2022eru}. Furthermore all of the leptogenesis parameter space at these low scales requires some degree of mass degeneracy, therefore if one of these experiments were to see something in the future it would be necessary to also resolve the two mass degenerate states in order to say anything about leptogenesis.

It is clear that for high scale models where $\MNi \geq \MDI$, experiments are extremely far from being sensitive to such heavy RHNs. It will be necessary to look for alternative ways to constrain such high scale models.
\newpage

\part*{Part 2}\label{Part2}
\newpage

\chapter{Primordial Black Holes}\label{PBHs}
\section*{Introduction}
Primordial Black Holes (PBHs) are black holes which form in the early universe, long before stars form and collapse into astrophysical black holes. Unconstrained by the Chandrasekhar limit, PBHs could form with masses spanning over 50 orders of magnitude. Recently interest in PBHs has re-emerged following the first detection of Gravitational Waves (GWs) by LIGO \cite{LIGOScientific:2016aoc}, because PBHs would produce specific GW signals associated with their formation and evaporation \cite{Papanikolaou:2020qtd,Domenech:2020ssp,Bhaumik:2022pil,Bhaumik:2022zdd,Ghoshal:2023sfa}. Formation mechanisms include inflationary overdensity collapse, and bubble collisions during phase transitions. The vast range of possible formation (and therefore evaporation) scales of PBHs opens up almost all early universe phenomena to their influence. DM  \cite{Lennon:2017tqq,Morrison:2018xla,Hooper:2019gtx,Auffinger:2020afu,Gondolo:2020uqv,Bernal:2020bjf,Bernal:2020ili,Bernal:2020kse,Baldes:2020nuv,Masina:2020xhk,Masina:2021zpu,Sandick:2021gew,Bernal:2021bbv,Bernal:2021yyb,Cheek:2021odj,Cheek:2021cfe,Barman:2021ost,Bernal:2022oha,Cheek:2022mmy,Chen:2023lnj,Chen:2023tzd,Kim:2023ixo,Barman:2024iht,Barman:2024slw,Haque:2023awl,RiajulHaque:2023cqe} and Dark Radiation production \cite{Hooper:2019gtx,Lunardini:2019zob,Masina:2020xhk,Masina:2021zpu,Hooper:2020evu,Arbey:2021ysg,Cheek:2022dbx,Eby:2024mhd}, the stability of the Standard Model (SM) Higgs potential~\cite{Burda:2016mou,Burda:2015isa,Hamaide:2023ayu} and baryogenesis ~\cite{Barrow:1990he,Majumdar:1995yr,Upadhyay:1999vk,Dolgov:2000ht,Bugaev:2001xr,Baumann:2007yr,Hooper:2020otu,Gehrman:2022imk} are all subject to modification by PBHs. PBHs have also been shown to heat their surroundings, producing hot spots with extreme temperature gradients in the primordial plasma \cite{Das:2021wei,He:2022wwy,He:2024wvt}. In \chapref{PBHLepto} it will be demonstrated that the profound impact of PBHs in the early universe allows GW information on PBHs to be translated into constraints on leptogenesis, while in \chapref{Hot Spots} the impact of hot spots on out-of-equilibrium dynamics is explored for the first time and the remarkable capacity of hot spots to absorb DM and sustain leptogenesis after the freeze out of sphalerons is uncovered. \\

This chapter aims to establish the physics necessary for quantifying the impacts of PBHs on early universe physics. The structure of this chapter is as follows. In \secref{PBHs/Formation} the formation of PBHs is briefly reviewed, discussing formation by inflationary perturbation collapse and the PBH parameter space. \secref{PBHs/Hawking} focuses on the evaporation of PBHs via Hawking radiation, following which \secref{PBHs/Cosmology} studies the cosmological evolution of a population of PBHs. \secref{PBHs/Hot Spots} discusses the formation of hot spots around PBHs before finally \secref{PBHs/Constraints} reviews the status of existing constraints on PBHs.

\newpage

\section{PBH Formation}\label{PBHs/Formation}
Common to all mechanisms of PBH formation is the necessity to generate hugely overdense regions in the plasma. A primordial black hole has a radius given by \cite{Schwarzschild:1916uq}
\begin{equation}
    r_s = 2 G \MPBH
\end{equation}
where $G = 6.7\times 10^{-39}\GeV^{-2}$ is Newton's gravitational constant and $\MPBH$ is the (instantaneous) mass of the PBH. In order to form a solar mass PBH then, a density of spike of magnitude
\begin{equation}
    \rho \sim 7\times 10^{16} \g \rm{cm}^{-3}
\end{equation}
would have to form in the early universe. To put this value in context it is necessary to know the average density in the universe at the moment at which the PBH forms. The initial mass of a PBH, $\MPBHini$, is related to the particle horizon mass by
\begin{equation}\label{PBHs/Formation/MPBHini}
    \MPBHini = \gamma_{\rm BH}M_H(\Tform)
\end{equation}
where $M_H$ is the total mass contained within the particle horizon at the PBH formation temperature $T = \Tform$, and $\gamma_{\rm BH}$ describes the gravitational collapse efficiency. $\gamma_{\rm BH} = 0.2$ is often taken \cite{Carr:1975qj,Carr:2016drx}, its value depends on the details of gravitational collapse and in particular on which value is considered for the threshold overdensity for gravitational collapse $\Delta_{\rm th}$ \cite{Green:2004wb}. Effective values greater than unity could occur if PBHs accrete through the inflow of fluid during and following collapse \cite{1979ApJ...232..670B}.

In terms of the cosmic time $t$ in the radiation-dominated era, the particle horizon size is given by \equaref{Early Universe/Cosmology/dHanalytical}, therefore since $\rho \approx \rhorad$ using \equaref{Early Universe/Cosmology/H} gives
\begin{equation}\label{PBHs/Formation/eqm}
    \MPBHini = \frac{\gamma_{\rm BH}}{8G}t_{\rm form}
\end{equation}
where $t_{\rm form}$ is the age of the universe when the PBHs form. Relating the cosmic time to the temperature via
\begin{equation}
    t = \sqrt{ \frac{90}{32\pi^3 G g_*}}T^{-2}
\end{equation}
allows \equaref{PBHs/Formation/eqm} to be solved for the temperature of the universe when a PBH of mass $\MPBHini$ would form
\begin{eqnarray}\label{PBHs/Formation/Tform}
    \Tform(\MPBHini) = \left(\frac{45}{1024\pi^3 G^3 g_*}\right)^{\frac{1}{4}}\left(\frac{\gamma_{\rm BH}}{\MPBHini}\right)^{\frac{1}{2}} \,.
\end{eqnarray}
Here $g_*$ is a function of the cosmic temperature, so this equation can be solved numerically for the average density of the plasma when $T = \Tform(\Msol)$
\begin{equation}
\rhorad(\Tform(\Msol)) \approx 6\times 10^{13}\g \rm{cm}^{-3}
\end{equation}
such that the density contrast of the overdensity required to form a solar mass PBH is about 3 orders of magnitude compared to the average density. The probability of such a large density contrast forming via thermal fluctuations will be, assuming that fluctuations can be described by a Gaussian distribution, suppressed by the negative exponential of the squared contrast $P \propto \rm{exp}(-(\Delta \rho)^2/\sigma_\Delta^2)$ where $\sigma$ is the variance. Clearly, the suppression of the probability by the density contrast can be ameliorated by having a large variance of perturbations. For a relativistic fluid $\sigma_\Delta \propto T$, such that the probability of PBH forming overdensities occurring due to thermal fluctuations is always highly suppressed. If PBHs formed in the early universe, there must have been some additional mechanism producing the large density contrasts which seed them.\\

The variance of Gaussian distributed density perturbations of magnitude $|\Delta \rho|$ on the length scale $R$ can be expressed as
\begin{equation}
    \sigma_\Delta^2 = \int^\infty_0 W^2(kR) \mathcal{P}_\Delta (k)\frac{\drm k}{k}
\end{equation}
where $\mathcal{P}_\Delta = k^3/2\pi^2\langle|\Delta \rho|^2\rangle$ is the power spectrum of perturbations, and $W(k)$ is a window function with $k$ the wavenumber. If the power spectrum is large for some range in $k$, the resulting variance of the perturbations may be large enough for PBH forming perturbations to occur in the tails. On large scales (small $k$), the magnitude of perturbations is constrained by the CMB to be around $\mathcal{P}(k_{\rm CMB}) \sim 2\times 10^{-9}$ \cite{Planck:2019nip}, but the small scale power spectrum is almost entirely unconstrained and there could have been significant power in small scale perturbations in the early universe.

\subsection*{Inflationary perturbation collapse}
It is widely believed that the radiation dominated universe was preceded by a period of cosmic inflation, first proposed by Guth \cite{Guth:1980zm} and Starobinsky \cite{Starobinsky:1980te}, for a review of inflationary cosmology see \cite{Linde:2007fr}. The canonical model of single field inflation features a scalar particle called the \textit{inflaton} in a simple quadratic potential 
\begin{equation}
    V(\varphi) = \frac{1}{2}m_\varphi^2 \varphi^2
\end{equation}
where $m_\varphi$ is the mass of the inflaton $\varphi$. In this simple scenario often known as slow-roll inflation, the primordial power spectrum is given by \cite{Sasaki:2018dmp}
\begin{equation}
    \mathcal{P}_\Delta(k) = \left(\frac{H^2}{\dot{\varphi}}\right)_{aH = k}
\end{equation}
where $\dot{\varphi}$ is the time derivative of the inflaton and the condition $aH = k$ specifies the the moment during inflation at which a perturbation of wavenumber $k$ exits the Hubble horizon. Single field slow-roll inflation of this type typically predicts a flat power spectrum which fits the large scale CMB data well, but falls drastically short of producing enough power at small scales to form PBHs \cite{Chongchitnan:2006wx}. In general, the formation of cosmologically significant populations of PBHs is only possible when the inflaton potential departs significantly from the slow-roll regime, requiring features in the potential or multi-field models. In general, the abundance of PBHs formed is given by \cite{Sasaki:2018dmp}
\begin{equation}\label{PBHs/Formation/beta}
    \beta \equiv \frac{\rho_{\rm PBH}}{\rho_{\rm rad}}\Bigg|_{T = \Tform} = \gamma_{\rm BH}\int^1_{\Delta_{\rm th}} P(\Delta) \drm \Delta
\end{equation}
where $P(\Delta)$ gives the probability that a perturbation exceeds the threshold for collapse $\Delta_{\rm th}$. Therefore if the probability function $P(\Delta)$ can be found then the abundance of PBHs can be calculated. PBHs are often assumed to form from Gaussian distributed perturbations, so since PBHs form in the tails of the distribution, PBH formation is very sensitive to non-Gaussianity \cite{Gow:2022jfb,Passaglia:2018ixg}. For the remainder of this work it will be assumed that PBHs form with an initially Gaussian distribution. It has also been implicitly assumed that the PBHs form during a radiation-dominated era and are therefore Schwarzchild BHs, whereas Kerr PBHs may form during a matter dominated phase with significant angular momentum \cite{Harada:2017fjm}.\\

\subsection*{The parameter space of PBHs}
For Schwarzschild PBHs the population is completely described by the initial mass and abundance $\{\MPBHini,\beta\}$. Often, the initial abundance of PBHs is rescaled as 
\begin{equation}
    \beta^\prime \equiv \gamma_{\rm BH}^{\frac{1}{2}}\beta
\end{equation}
when projecting constraints in the PBH parameter space. Throughout the remainder of this work, both $\beta$ and $\beta^\prime$ will be used with the latter primarily employed to enable comparison to existing constraints in the literature. \\

A population in which all PBHs form at the same time and have the same initial mass is known as a monochromatic distribution. This is typical of PBHs which form from the collisions of bubbles during first order phase transitions since the PBH-forming perturbations occur at a time determined essentially only by the symmetry breaking scale \cite{Jedamzik:1999am,Liu:2021svg}. Alternatively PBHs may form with an extended mass distribution, this occurs for example if an inflationary model predicts significant power across a range of scales. Then the PBHs form across a range in time and with a range of masses.\\

In either case, the physics of the PBHs can be effectively studied by considering monochromatic mass distributions and varying the initial mass. This is the approach adopted in this work, extended mass distributions are not explicitly considered. Instead, a population of PBHs with initial mass $\MPBHini$ and abundance $\beta$ is assumed to form in the early universe and the parameter space $\{\MPBHini,\beta\}$ is scanned over. Any point in this parameter space which is constrained for monochromatic mass distributions would be expected to be constrained in the same manner for an extended distribution which includes that point.

PBHs with greater initial masses form later, see \equaref{PBHs/Formation/MPBHini}. However the temperatures at which they form, given by \equaref{PBHs/Formation/Tform}, are comfortably in the radiation-dominated era for $\MPBHini \leq 10^{15}\Msol$. Later in \chapref{PBHLepto} and \chapref{Hot Spots} the impact of PBHs on early universe phenomena such as leptogenesis (see \chapref{Leptogenesis}) and DM production will be of interest. Therefore the lightest PBHs will be of particular interest since they form the earliest. For $\MPBHini \leq 10^{17}\g$ PBHs form in the symmetric Higgs phase which is the relevant period for leptogenesis. In fact, only the very lightest PBHs in the ultralight mass range $\MPBHini \lesssim 10^9\g$ will be relevant because PBHs quickly evaporate in the early universe.

\newpage

\section{Hawking Radiation and Evaporation}\label{PBHs/Hawking}
In 1974 Hawking showed that black holes emit radiation at the same rate one would expect of a body with temperature $1/(8\pi M_{\rm BH})$ where $M_{\rm BH}$ is the black hole mass \cite{Hawking:1974rv}. Beckenstein had previously suggested that some multiple of the BH surface gravity $\kappa = 1/4M_{\rm BH}$ should be viewed as the BH temperature \cite{Bekenstein:1973ur}, an idea which was initially seen only as an interesting analogy by Hawking and his collaborators. Calculating the ratio between the expected number of particles emitted by a PBH to those that would be absorbed from an incident wavepacket, Hawking found exactly $(\rm{exp}(2\pi  \omega/\kappa)-1)^{-1}$ for scalars with frequency $\omega$, and for fermions $(\rm{exp}(2\pi  \omega/\kappa)+1)^{-1}$. Remarkably, these are simply the ratios between emission and absorption cross sections for a body with temperature $\kappa/2\pi$. This lead Hawking to the interpretation that BHs are in fact continuous emitters, radiating particles at the Hawking temperature
\begin{equation}\label{PBHs/Evaporation/TBH}
    \TBH \equiv \frac{1}{8\pi M_{\rm BH} G}.
\end{equation}
Conservation of energy dictates that this emission reduces the mass of the BH, in turn increasing the Hawking temperature. The production of particles by a BH by can be thought of as occurring due to the quantum tunnelling of particles inside the BH through the event horizon, which acts as a gravitational potential barrier of size $\kappa$. An alternative heuristic is that of particle pair creation near the event horizon, with the negative energy state crossing the event horizon when the wavelength is order the BH radius, and the positive energy state escaping as thermal radiation. An uncharged, Schwarzschild BH with temperature $\TBH$ produces a particle $X$ with mass $m_X$, spin $s$, internal degrees of freedom $g_X$ and energy $E$ at the rate \cite{Hawking:1975vcx,MacGibbon:1990zk}
\begin{equation}\label{PBHs/Evaporation/dn}
\frac{\drm^2 n_X}{\drm t \drm E} = \frac{\Gamma_X^s(M_{\rm BH},E) g_X}{2\pi}\left(\rm{exp}\left(\frac{2\pi E}{\TBH}\right) -(-1)^{2s}\right)^{-1}
\end{equation}
where $\Gamma_X^s$ is the spin-dependent absorption probability for $X$, and $n_X$ is the physical number density of $X$ produced by the PBH. The absorption probability for $X$ can be interpreted as the fractional gain (or loss) in energy for a wave scattering off a BH and analytical formulae in the case of small frequency waves were derived by Starobinsky and Churilov in 1973 \cite{Starobinskil:1974nkd}, later generalised by Teukolsky and Press who developed numerical methods to calculate it \cite{TeukolskyPress}. At high energies $GM_{\rm BH}E \gg 1$, $\Gamma_X^s$ approaches the geometric optics limit
\begin{equation}
    \Gamma_X^s \to 27G^2M_{\rm BH}^2E^2
\end{equation}
while at low energies the result can be expressed analytically for a massless particle as 
\begin{equation}
    \Gamma_X^s = \vartheta^sG^2M_{\rm BH}^2E^2
\end{equation}
where 
\begin{equation}
    \vartheta^s = \begin{cases}
        16 & s = 0\\
        2 & s = \frac{1}{2}\\
        64 & s = 1\\
        256 & s = 2
    \end{cases}
\end{equation}
comes from averaging all BH orientations with respect to the spin $s$ \cite{MacGibbon:1991tj}. The general form, for intermediate energies and massive particles, however must be calculated numerically and this was done consistently for example in \cite{Cheek:2021cfe,Masina:2021zpu}. By conservation of energy, the mass loss rate for a Schwarzschild BH is given by
\begin{equation}\label{PBHs/Evaporation/dM}
    \frac{\drm M_{\rm BH}}{dt} = \sum_X \int^\infty_0  \frac{\drm^2  ] n_X}{\drm t \drm E} E \drm E
\end{equation}
summed over all possible particle species $X$. Note that gravitational production of particles will produce all particle species in nature, even those BSM. As the BH loses mass, its rate of mass loss and the resulting flux of Hawking radiation increases. This results in a finite lifetime for a BH of order $10^{71}s$ for a solar mass BH \cite{Hawking:1974rv}, much longer than the lifetime of the universe. Therefore the present mass loss rate of astrophysical black holes (which have the lower limit $1.4\Msol \lesssim M_{\rm BH}$) is extremely small, far too small to be detected or significantly impact the BH surroundings.\\

However PBHs can have much smaller masses and would therefore evaporate much more quickly.
Near the end of the PBH lifetime, the Hawking temperature rapidly increases during the explosive final stages of the evaporation. Whether or not a black hole may radiate away more than $\sim 1/2$ of its initial mass is subject to debate in the literature \cite{Almheiri:2020cfm,Buoninfante:2021ijy,Juarez-Aubry:2023kvl,Burman:2023kko,Vachaspati:2006ki,Dvali:2020wft,Dvali:2024hsb}. Since it is not currently known if nor how \equaref{PBHs/Evaporation/dn} should be replaced due to the memory-burden effect, in this work it is assumed that the semiclassical evaporation proposed by Hawking continues until PBHs leave $\sim \Mpl$ relics.  In this limit, a PBH with initial mass $\MPBHini \approx 4.3\times10^9\g$ would have a lifetime of just $1s$. Such light PBHs would form and evaporate away completely before the onset of BBN. Only those PBHs for which $\MPBHini \geq 10^{15}\g$ would remain in the present day universe. The initial mass of a population of PBHs, $\MPBHini$, determines their lifetime and therefore their cosmological implications of their existence.

\newpage

\section{PBHs in Cosmology}\label{PBHs/Cosmology}
Cosmological populations of PBHs form when the universe has temperature $\Tform$ dictated by their initial mass, given by \equaref{PBHs/Formation/Tform}, and with surface temperature given by \equaref{PBHs/Evaporation/TBH}. At the moment of formation it is always the case that $\Tform > \TBH$ and so the evaporation of PBHs does not begin coincidentally with their formation. For $t > t_{\rm form}$, the radiation dominated universe continues its decelerating expansion according to equation \equaref{Early Universe/Cosmology/at}, so that the temperature of the universe cools with the inverse of the scale factor, see \equaref{Early Universe/Cosmology/Ta}. As the universe expands and cools, the dominant radiation component redshifts as $\rhorad \propto a^{-4}$ while the PBH energy density is only diluted as $\rhopbh \propto a^{-3}$. Therefore the relative fraction of the energy budget of the universe composed of PBHs grows as 
\begin{equation}\label{PBHs/Cosmology/frac}
    \frac{\rhopbh}{\rhorad} = \beta \frac{a}{a_{\rm form}}
\end{equation}
where the relative abundance of PBHs at formation $\beta$ is defined in \equaref{PBHs/Formation/beta}, and $a_{\rm form}$ is the scale factor at formation. This fraction can become $\gg 1$ if the initial abundance of PBHs is large enough, then PBHs become the dominant component of the universe and instantiate an early matter(PBH)-dominated era. As mentioned in \secref{PBHs/Formation}, PBHs which form in a matter dominated era may form with large angular momentum. In an extended PBH mass distribution scenario, a period of PBH domination could be brought about by the lighter PBHs which in turn causes the heavier, later forming PBHs to gain angular momentum. This work assumes a monochromatic mass distribution and Schwarzschild PBHs.
The Hubble rate for a universe containing a significant population of PBHs is given by
\begin{equation}\label{PBHs/Cosmology/H}
    H^2 = \frac{8\pi}{3\Mpl^2}\left(\frac{\varrhopbh}{a^3}+\frac{\varrhorad}{a^4} \right)\,\,,
\end{equation}
where $\varrhopbh \equiv a^3\rhopbh$ is the comoving PBH energy density. Since a PBH population with $\beta \ll 1$ may still eventually dominate the universe, when this occurs the Hubble rate will be enhanced relative to what it would have been if no PBHs formed. The enhancement of the Hubble rate could have many interesting consequences, for example in \secref{PBHLepto/High Scale} it will be shown that the freeze-out of the EW sphaleron processes occurs at higher temperatures in a PBH dominated era.\\

As well as large enough $\beta$, to eventually come to dominate the universe, the PBH population must also be heavy enough as to not evaporate entirely before this occurs. Indeed, the relation \equaref{PBHs/Cosmology/frac} assumes negligible loss of mass from the PBHs. This is accurate in the initial stages of the PBH lifetime, the relative growth of the PBH energy density due to the expansion of the universe far outstrips the rate of evaporation. However, as the PBH loses its mass the rate of evaporation increases sharply. Eventually the process of Hawking evaporation will radiate the PBH energy density in the form of high energy particles and quickly restore radiation domination. The evolution of the PBH and radiation energy densities follow the set of coupled differential equations~\cite{Lunardini:2019zob, Perez-Gonzalez:2020vnz, Bernal:2022pue}
\begin{eqnarray}\label{PBHs/Cosmology/Friedmann}
\frac{\drm\varrhorad}{\drm\alpha} &=& - f_{\rm Dark}^{-1} 10^{\alpha} \frac{\drm\ln \MPBH}{\drm\alpha}  \varrhopbh\,,\\
    \frac{\drm\varrhopbh}{\drm\alpha} &=& \frac{\drm\ln \MPBH}{\drm\alpha}\varrhopbh\,
\end{eqnarray}
where $f_{\rm Dark}$ is the fraction of Hawking radiation composed of dark sector particles which do not contribute to the radiation energy density, $\alpha \equiv \mathrm{log}_{10}(a)$ and 
\begin{equation}
\frac{\drm\ln \MPBH}{\drm\alpha} =  \frac{\ln(10)}{H\MPBH}\frac{\drm\MPBH}{\drm t} 
\end{equation}
with $\drm \MPBH / \drm t$ given by \equaref{PBHs/Evaporation/dM}.
\newpage
  
\begin{figure}[h]
    \centering
    \includegraphics[width=0.8\linewidth]{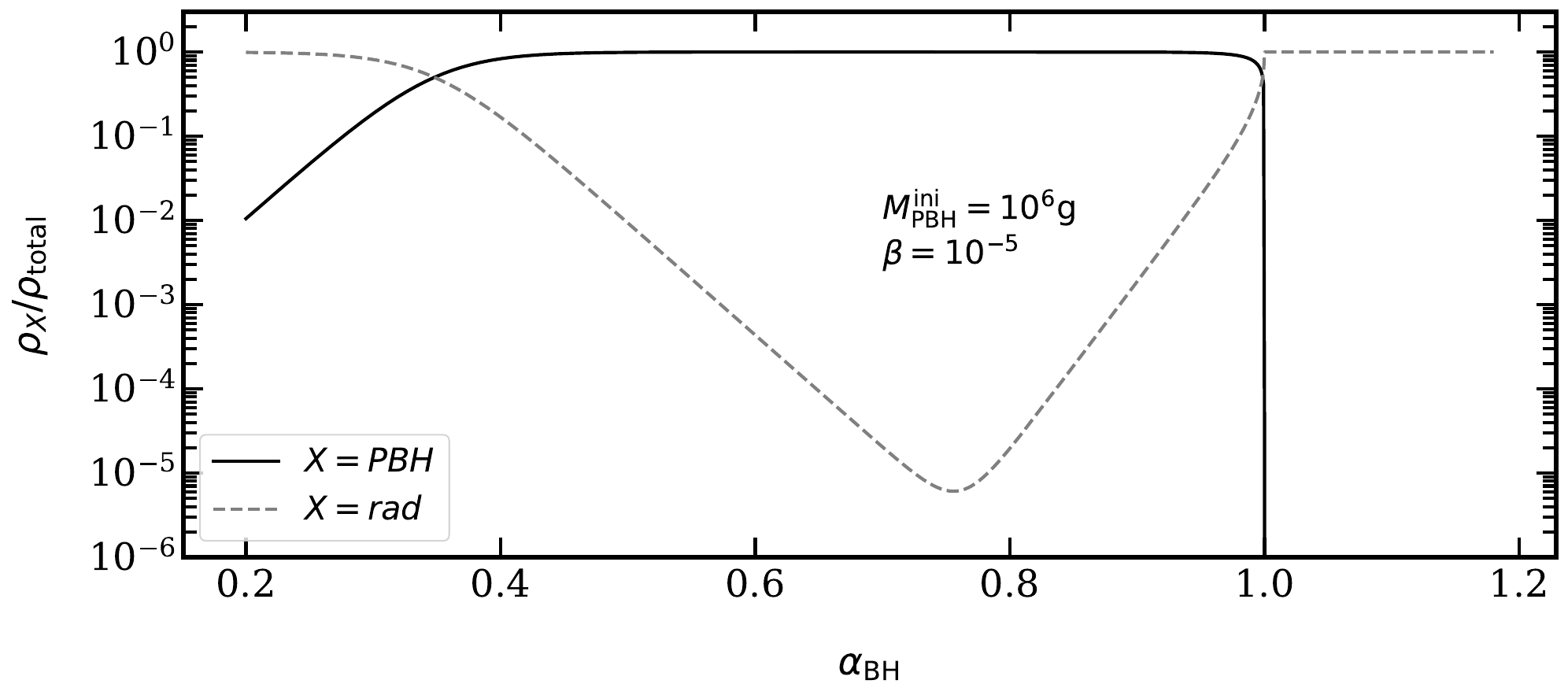}
    \includegraphics[width=0.8\linewidth]{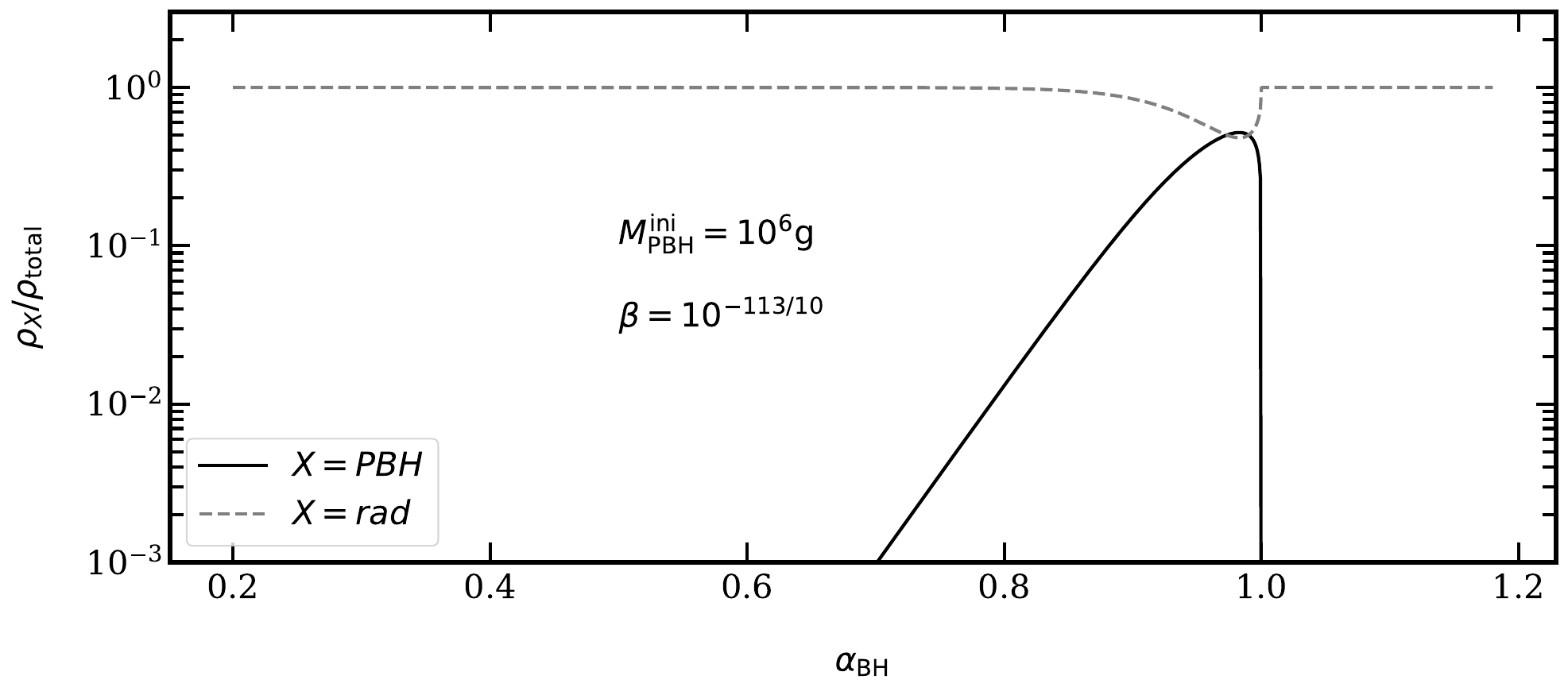}
    \caption{The evolution of the relative fraction of the energy budget of the universe composed of PBHs (black solid line) and radiation (grey dashed line) is shown as a function of $\alpha_{\rm BH} \equiv \alpha/\alpha_{\rm evap}$. In both panels $\MPBHini = 10^6\g$, in the upper panel $\beta = 10^{-5}$ while in the lower panel $\beta = 10^{-11.3}$.}
    \label{PBHs/Cosmology/evol}
\end{figure}

The evolution of the PBH and radiation energy densities is shown in \figref{PBHs/Cosmology/evol} in terms of $\alpha_{\rm BH} \equiv \alpha/\alpha_{\rm evap}$, where $\alpha_{\rm evap}$ is the logarithmic scale factor when the PBHs evaporate completely. In both cases the initial energy density made up of PBHs is very small compared to the radiation density in the universe. As the universe expands, the relative fraction of PBHs grows as \equaref{PBHs/Cosmology/frac} until eventually  $\rhopbh > \rhorad$. In the case of $\beta = 10^{-5}$, the upper panel, this occurs quite quickly and PBHs dominate the energy budget of the universe for most of the PBH lifetime. In the lower panel, $\beta$ is smaller and $\rhopbh > \rhorad$ only occurs fleetingly at the very end of the PBH lifetime. Lighter, or less abundant PBHs would never achieve PBH-radiation equality.\\

In general, heavier and more abundant PBHs tend to bring about a PBH dominated era. The evolution of a PBH dominated universe departs from the standard radiation dominated one, so one must carefully account for the impact on particle processes. This is captured by the non-standard evolution of the Hubble rate, given by \equaref{PBHs/Cosmology/H}. Furthermore when the period of PBH domination ends following evaporation, the PBHs produce huge numbers of high energy photons and other particles, such that the energy density in radiation and therefore the entropy $s$ of the universe, given by
\begin{equation}
    s = \frac{4}{3}\left(\frac{\pi^2g_*}{30}\right)^{\frac{1}{4}}\rhorad^{\frac{3}{4}}\,
\end{equation}
are both greatly enhanced compared to the case $\beta = 0$. As will be demonstrated in \secref{PBHLepto}, this is crucially important for baryogenesis scenarios since the observable $\YB$ is inversely proportional to the entropy of the universe. The evolution of the entropy density is given by
\begin{equation}\label{PBHs/Cosmology/S}
    \frac{\drm\entropy}{\drm \alpha} = -\frac{f_{\rm Dark}^{-1}}{T(\alpha)} \frac{{\rm d}\ln \MPBH}{{\rm d}\alpha} \varrhopbh
\end{equation}
where $\entropy \equiv sa^3$ is the comoving entropy density of the universe.\\

\begin{figure}[h]
    \centering
    \includegraphics[width=0.8\linewidth]{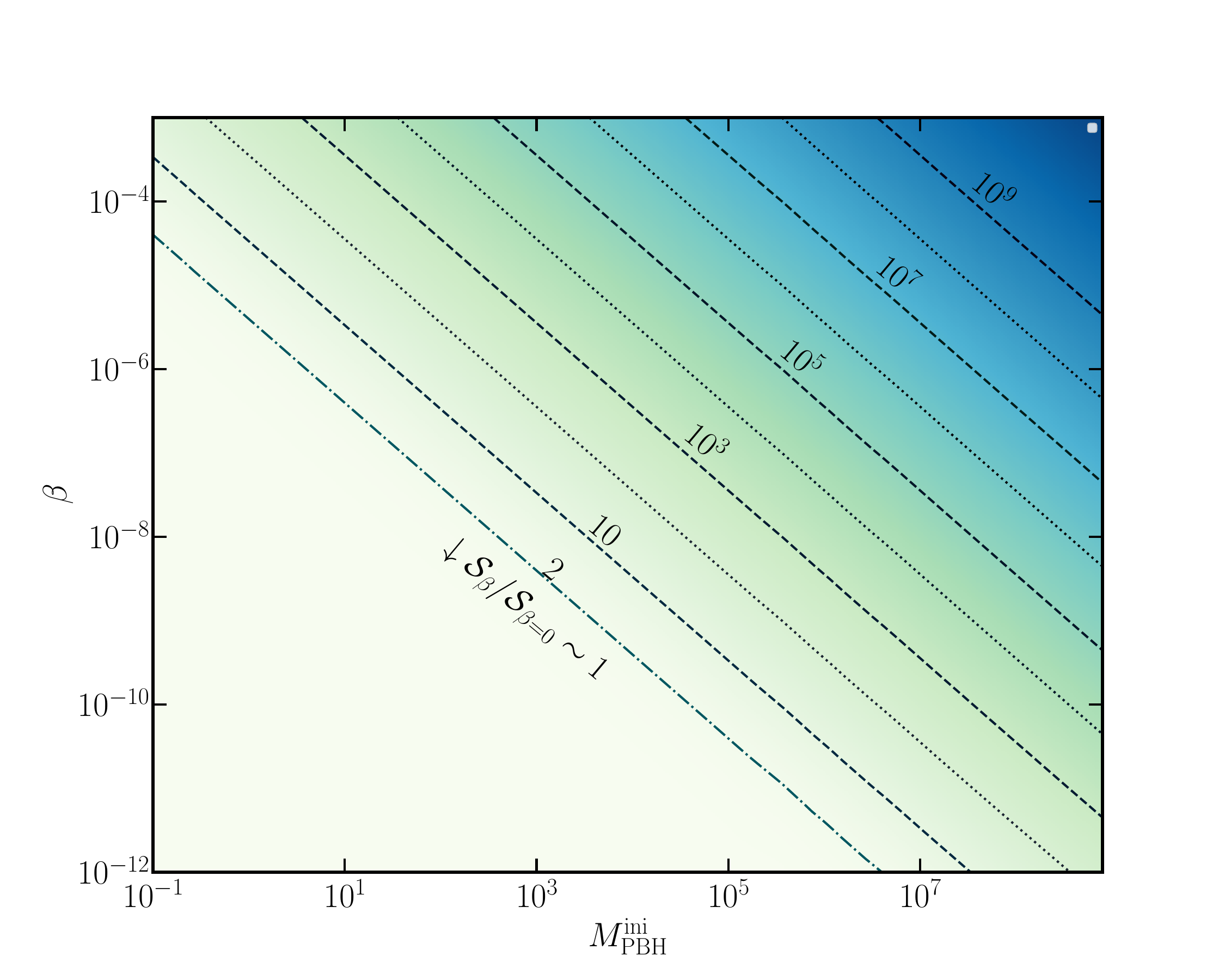}
    \caption{Contours of the ratio $\mathcal{S}_\beta/\mathcal{S}_{\beta = 0}$ in PBH parameter space, representing the relative increase in comoving entropy density due to the evaporation of a population of PBHs.}
    \label{PBHs/Cosmology/entropy}
\end{figure}
The injection of entropy by PBH evaporation is depicted in \figref{PBHs/Cosmology/entropy} by contours of the ratio $\mathcal{S}_\beta/\mathcal{S}_{\beta = 0}$, evaluated immediately following the cessation of PBH evaporation. The ratio therefore quantifies the relative increase in entropy due to the evaporation of a population of PBHs with initial abundance $\beta$ compared to the case where $\beta = 0$. Entropy injection is most significant when PBHs dominate the universe, but non-dominant PBHs still produce entropy. Indeed, any $\beta \neq 0$ gives $\mathcal{S}_\beta/\mathcal{S}_{\beta = 0}\neq 1$. In the radiation dominated universe, the entropy density is usually expressed in terms of the radiation temperature as
\begin{equation}
    s = \frac{2\pi^2g_*}{45}T^3
\end{equation}
and therefore the evaporation of cosmological PBH populations has generally been considered (see for example the treatments in \cite{Bernal:2022pue,Perez-Gonzalez:2020vnz,Barman:2024slw}) to reheat the universe according to 
\begin{equation}\label{PBHs/Cosmology/T}
    \frac{\drm T}{\drm\alpha} \simeq -T \left[\ln(10) - \frac14 f_{\rm Dark}^{-1} 10^\alpha \frac{\drm \ln \MPBH}{\drm\alpha}\frac{\varrhopbh}{\varrhorad}\right]
\end{equation} 
which describes an increase in the average plasma temperature, ie the temperature of the universe is considered to evolve homogeneously. However recent studies have shown that PBHs heat their surroundings locally, producing potentially extreme temperature gradients in the early universe.
\newpage

\section{Hot spots}\label{PBHs/Hot Spots}
In this section, the formation and evolution of hot spots around PBHs is explored. Historically PBHs have been treated as affecting the universe homogeneously, heating the plasma everywhere by an average amount. This picture neglects the interaction of Hawking radiation with the surrounding plasma. In fact, PBHs do not evaporate in a vacuum! Especially for PBHs in the ultralight mass regime ($\MPBHini \leq 10^9\g$), the early universe into which they radiate is extremely hot and dense. With the obvious exception of dark sector particles, Hawking radiation is bound to eventually interact with its surroundings, and since an evaporating PBH always satisfies $\TBH \geq \Tplasma$, the boosted radiation deposits energy into its surroundings and heats the region around the PBH. Here $\Tplasma$ is the temperature of the background universe unaffected by the PBH. The existence of a non-trivial temperature profile $T(r)$ where $r$ is the radial distance from the Schwarzschild radius, around a PBH was studied in 2021 by Das and Hook \cite{Das:2021wei}, and then a detailed calculation of the hot spot profile was made by He, Kohri, Mukaida and Yamada first using analytical methods ~\cite{He:2022wwy}, later confirming these results with numerical simulations \cite{He:2024wvt}.\\

%%%%%%%%%%%%%%%%%%%%%%%%%%%%%%%%%%%%%%%%%%%%%%%%%%%%%%%%%%%%%%%
\subsubsection{Thermalisation and diffusion} \label{sec:LPMsection}
%%%%%%%%%%%%%%%%%%%%%%%%%%%%%%%%%%%%%%%%%%%%%%%%%%%%%%%%%%%%%%%
\noindent Once a PBH begins to evaporate and the
Hawking radiation travels radially outwards, the typical momenta of these emitted particles is \( \langle \vec{p} \rangle \sim \TBH > \Tplasma \). The high energy Hawking particle will eventually deposit its energy into the surrounding plasma via scattering. One can show that the dominant mechanism by which this occurs is by nearly collinear splittings of the Hawking radiation into soft daughter particles \cite{Harigaya:2014waa}, suppressed by the LPM effect, see \secref{Early Universe/TFT}. The LPM suppressed rate of inelastic scatterings in a medium of temperature $T$ is $\Gamma_{\rm LPM}$ given by \equaref{Early Universe/TFT/GLPM}. The resulting rate of energy loss is given by \equaref{Early Universe/TFT/dELPM}, $\propto E^{-1/2}$ where $E$ is the energy of the Hawking radiation, such that with each successive scattering and associated decrease in $E$, the Hawking radiation thermalises more quickly. Therefore, the radius at which the radiated particle deposits its energy into the plasma can be approximated by
\begin{equation}
    \rcore(\TBH,T) \equiv \Gamma_{\rm LPM}^{-1}(\TBH,T)\,.
\end{equation}

Energy deposited by Hawking radiation at \( r = r_{\rm core} \) diffuses both outwards and inwards back towards the PBH, through a random succession of large angle elastic scatterings. Modelling the diffusion processes as a random walk, the average time taken to walk the distance $\rcore$ back to the PBH is given by
\begin{equation}
    t_d(\TBH,T) = \frac{\rcore(\TBH,T)^2}{t_{\rm el}(T)}
\end{equation}
where $t_{\rm el} \sim (A^2 T)^{-1}$ is the timescale of the diffusive scattering processes between particles in the plasma. If $t_d(\TBH,T) \ll t_{\rm ev}(\TBH)$ where $t_{\rm ev}$ is the timescale of PBH evaporation, then diffusion is efficient in the region $r \leq \rcore$ compared to the evaporation of the PBH and so the temperature profile in this region should be flat, being efficiently smoothed over by diffusion. In fact, the condition $t_d(\TBH,T) \ll t_{\rm ev}(\TBH)$ is equivalent to \cite{He:2022wwy}
\begin{equation}
    \MPBH \gg M_* \sim 0.8\g
\end{equation}
such that very light PBHs or those which have radiated away all but $M_*$ of their mass are evaporating so quickly that diffusion is no longer able to smooth the region $r\leq \rcore$. For $\MPBHini \gg M_*$, the PBH radiates away almost all of its mass before this occurs and so the final short moments of evaporation are unlikely to be phenomenologically relevant. The remainder of this chapter is concerned with the case $\MPBH \gg M_*$.\\

While diffusion in the core ($r\leq \rcore$) is efficient until $\MPBH \leq M_*$, for any instantaneous PBH mass, diffusion can only keep up with the associated evaporation rate out to some large radius \cite{He:2022wwy}
\begin{equation}\label{PBHs/Hot Spots/rdec}
 \rdec \approx 4.56\times 10^{14} \GeV^{-1}\left[\frac{A}{0.1}\right]^{-8/5}\left[\frac{g_*}{106.75}\right]^{1/5}\left[\frac{g_{H*}}{106.75}\right]^{-4/5} \left[\frac{T_{\rm BH}}{10^4\rm GeV}\right]^{-11/5}\,,
\end{equation}
beyond which diffusion cannot keep the universe in thermal contact with the PBH. Therefore, the longest length scale over which diffusion can smooth during the whole lifetime of the PBH defines $\rdec^{\rm ini}\equiv \rdec(\MPBH^{\rm ini})$. As the PBH evaporates faster and faster, $\rdec$ shrinks. Later in the PBH life it can be the case that large radii $\rdec \leq r \leq \rdec^{\rm ini}$, were once in thermal contact with the PBH, but freeze out when the PBH evaporation rate exceeds the diffusion rate out to $r$. In the above, as in \secref{Early Universe/TFT}, $A \sim 0.1$ is taken as a fiducial value for the SM gauge couplings.

\subsubsection{Hot Spot Profile}\label{PBHs/Hot Spots/profile}
Considering that in time $t_d$ a PBH radiates away energy $E \approx t_d \times \drm \MPBH / \drm t$, the temperature of a homogeneous sphere of radius $\rcore$ and energy $E$ is given by
\begin{equation}\label{PBHs/Hot Spots/Tcore}
 \Tcore(\alpha) = 2 \times 10^{-4} \left( \frac{A}{0.1} \right)^{8/3} \left( \frac{g_{H*}} { g_*} \right)^{2/3} \TBH(\MPBH(\alpha))\,,
\end{equation}
where $g_*$ is the total relativistic degrees of freedom and $g_{H*}$ is the number of degrees of freedom of the Hawking radiation. $\Tcore$ is the temperature to which Hawking radiation would heat the region $r\leq \rcore$ surrounding the PBH. Observe that the core temperature of the would-be hot spot is always approximately four orders of magnitude below the Hawking temperature. This result can be used to express the core radius as a function of the PBH temperature
\begin{equation}
 \rcore \approx 6\times 10^7 \left(\frac{A}{0.1}\right)^{-6}\left(\frac{g_*}{g_{H*}}\right)\TBH^{-1}\,.
\end{equation}
so that it is clear that as the core gets hotter, it shrinks in size also. At very late times when $\MPBH \to M_*$, the length scales $\rcore$ and $\rdec$ merge by definition and the hot spot core reaches its maximum temperature
\begin{equation}\label{PBHs/Hot Spots/Tmax}
    \Tmax = 2\times 10^9\GeV \left(\frac{A}{0.1}\right)^{19/3}\left(\frac{g_*}{106.75}\right)^{-4/3}\left(\frac{g_{H_*}}{108}\right)^{5/3}
\end{equation}
which is remarkably independent of the initial PBH mass. Outside the hot spot core, but at small enough radii that diffusion keeps the local plasma temperature $T(r)$ in thermal contact with the PBH, $\rcore(\MPBH) \leq r \leq \rdec(\MPBH)$, conservation of energy leads to the temperature decreasing as $T(r) \propto r^{-\frac{1}{3}}$. For radii $\rdec < r < \rdec^{\rm ini}$ the plasma once in thermal contact with the PBH freezes out into an envelope with temperature given by
\begin{equation}
     T_{\rm envelope} = 0.4\,\mathrm{MeV}\left(\frac{\alpha}{0.1}\right)^{6/5}\left(\frac{g_*}{106.75}\right)^{-2/5}\left(\frac{g_{H*}}{106.75}\right)^{3/5}\left(\frac{T_{\rm BH}^{\rm ini}}{10^4 \rm GeV}\right)^{7/5}
\end{equation}
and the temperature of the envelope falls with $r^{-7/11}$ until the background temperature of the Universe $\Tplasma$ is reached. Finally, the temperature profile around a PBH following hot spot formation can be expressed as \cite{He:2022wwy}
\begin{equation}\label{eq:tprof}
T(r) = 
\begin{cases}
 \Tcore & r_{\rm BH} < r < \rcrit \\
 \mathrm{max}\left[\Tplasma, \Tcore \left(\frac{r}{\rcrit} \right)^{-1/3}\right] & \rcrit < r < \rdec\\
 \mathrm{max}\left[\Tplasma,T_{\rm envelope} \left(\frac{\rdec^{\rm ini}}{r}\right)^{7/11}\right] & \rdec < r < \rdec^{\rm ini} \\
 \Tplasma & r > \rdec^{\rm ini}
\end{cases}
\end{equation} 
where $\rcore, \Tcore$ and $\rdec$ are all implicitly time ($\alpha$) dependent through their dependence on $\TBH$ which increases as the PBH evaporates, and $\Tplasma$ is explicitly a function of $\alpha$ being the solution to the Friedmann equations in the background Universe. This hot spot profile was established by \cite{He:2022wwy} apart from the possibility for $\Tplasma$ to exceed any of the relevant temperatures. This is imposed because the work in \cite{He:2022wwy,He:2024wvt} makes no attempt to consider the evolution of the background universe around the PBHs. Accounting for the Hubble expansion is crucial for understanding the evolution of hot spots in the universe, and this exercise will be done in \chapref{Hot Spots}.

\newpage

\section{Constraints on PBHs}\label{PBHs/Constraints}

Since PBHs can take a wide range of initial masses spanning over 50 orders of magnitude, the experimental methods necessary to detect them vary just as widely. The heaviest PBHs, if present in intergalactic regions, would induce peculiar velocities and are constrained against the observed CMB dipole \cite{Carr:2021bzv}. PBHs in the $\sim \Msol$ mass range are constrained by their propensity to cause perturbations of wide binaries and star clusters, X-ray binary accretion, and supernova microlensing \cite{Carr:2021bzv}. In the so-called asteroid mass range $10^{17}\rm{g} \lesssim \MPBHini \lesssim 10^{23}$g, PBHs are sufficiently unconstrained as to potentially constitute all of the relic DM density, see \cite{Gorton:2024cdm} for a recent perspective. It should be noted that all of these constraints are model dependent to some degree and subject to change as our understanding of PBH physics improves.\\

The primary focus of this thesis is the ultralight mass range $\MPBHini \lesssim 10^9$g. PBHs this light would evaporate completely before the onset of BBN, making them impossible to detect directly today. However LIGO's detection of the first GW signal in 2016 \cite{LIGOScientific:2016aoc} opened up a new window into the early universe, PBHs are expected to be associated with GW signatures which in principle may be observable at current and future generation GW experiments. Firstly, the significant scalar perturbations necessary for PBH formation (see \secref{PBHs/Formation}) induce tensor perturbations at second order leading to a GW signal \cite{Bugaev:2009zh,Cai:2018dig}. Similarly if PBHs dominate the universe, the density fluctuation field formed of the PBHs themselves induces GWs at second order. The associated GW energy density could be so large as to exceed the background energy density, leading to a backreaction problem which would occur for $\MPBHini \sim 10^7$g with $\beta^\prime \gtrsim 10^{-8}$ \cite{Papanikolaou:2020qtd,Papanikolaou:2022chm}. For smaller $\beta^\prime$ the signal is within the sensitivity of the upcoming Laser Interferometer Space Antenna (LISA) \cite{LISA:2017pwj,Papanikolaou:2022chm}. Lastly the evaporation of large populations of PBHs causes a sudden transition from matter domination to radiation domination, which induces a stochastic GW signal associated with evaporation \cite{Inomata:2019ivs}. So ultralight PBHs may produce at least 3 separate GW signatures, arising from their formation, their interactions as a diffuse gas of PBHs, and their evaporation. This is a compelling scenario for detection prospects if these signatures can all be probed by experiments such as LISA.\\

Interestingly it has been shown recently that the GW signal produced by the evaporation of a population of PBHs with $\MPBHini = 10^8$g and $\rm{log}_{10}(\beta) = -9.75$ could explain the stochastic GW background observed at the NANOGrav \cite{NANOGrav:2023hde} experiment \cite{Bhaumik:2023wmw}. Furthermore this same signal would be observable by the Nancy Roman telescope \cite{Wang:2020pmf} and the signature associated with the formation of the PBH population would overlap with the Einstein Telescope's projected sensitivity \cite{Punturo:2010zz}. The fact that independent GW signals coming from ultralight PBH formation and evaporation will both be separately observable at upcoming experiments sets the stage favorably for detection prospects in the short to medium term.\\

It is therefore important to anticipate the quite plausible ultralight PBH discovery scenario. In that scenario, understanding the implications for early universe physics would facilitate advancement in multiple directions. PBHs strongly affect particle processes, such as DM production and baryogenesis, occurring before BBN by vociferously producing high energy particles as Hawking radiation. Since PBHs can be probed with GWs they offer a unique window into these crucial cosmological phenomena. Progress in the understanding of Hawking radiation in the early universe will therefore extend its reach into the most difficult to probe epochs of the universe. 
\newpage

\chapter{The Impacts of PBHs in the Early Universe}
This chapter seeks to demonstrate how the rich interplay between PBHs and crucial cosmological and particle processes in the earliest times of our universe shines light on physics which may be otherwise unobservable. By carefully considering the impact of Hawking radiation on the particle physics occurring before $\TBBN$, and consistently accounting for the production of entropy and the possibility of early PBH domination, mutual exclusion limits are discovered between the parameter spaces of leptogenesis and PBHs. Future experimental evidence for PBHs (RHNs) is therefore shown to also constrain leptogenesis (PBHs). \\

Then, the impact of PBH hot spots on particle physics is uncovered, revealing the crucial role played by hot spots in calculating cosmological observables. First developing a formalism with which to treat the propagation of Hawking radiation through hot spots, the interplay between leptogenesis/WIMP DM and PBHs is revisited and profound phenomenological consequences demonstrated. In the case of leptogenesis, hot spots support EW sphaleron processes even after freeze-out in the background enabling the production of the BAU from PBHs previously though too heavy and in testable regions of neutrino parameter space too strongly coupled for successful thermal leptogenesis. WIMP DM produced by PBHs is then shown to be effectively absorbed by hot spots, reducing its contribution to the DM relic density and significantly altering the constraints one should draw from PBHs which overclose the universe.\

\section{PBHs Constrain Leptogenesis}\label{PBHLepto}
Cosmologically contemporaneous, the eras of leptogenesis and ultra-light PBHs are extremely difficult to probe experimentally. As established in \secref{Leptogenesis/Experiment}, realistic leptogenesis models reside in a high-dimensional space and prefer heavy, weakly coupled RHNs which are certain to evade detection in all but the smallest margins of parameter space. Ultralight PBHs would no longer even exist in the universe and detection prospects rely on measuring the GW signals remnant after their evaporation, see \secref{PBHs/Constraints}. Their coincidence however produces fascinating interplay in the early universe, PBHs may either enhance or inhibit leptogenesis depending on the specific characteristics of each. \\

The very lightest PBHs $\MPBHini \lesssim 10^5$g evaporate completely before the temperature of the background universe cools to $\Tsphal$. Therefore, when RHNs produced as Hawking radiation decay, resulting leptonic asymmetry is readily transferred to the baryonic sector by sphalerons. Ref.~\cite{Bernal:2022pue} demonstrated that very light PBHs ($\mathcal{O}(1)$ g) could expand the viable parameter space for high-scale leptogenesis up to the GUT scale. In this scenario very heavy RHNs are produced when the washout processes are kinematically suppressed and sphalerons are active, leading to an enhancement of $\YB$. $\YB$ is also affected by PBHs when RHNs are of intermediate scale, $10^6-10^9$GeV, although this only partially ameliorates the dilution by PBH produced photons. In fact, an early period of PBH domination was shown to be incompatible with intermediate scale leptogenesis, especially for $\MPBHini \gtrsim 10^3$ g \cite{Perez-Gonzalez:2020vnz}. The possibility of producing both the BAU and DM relic abundance from PBHs in this mass range was studied in Ref.~\cite{JyotiDas:2021shi}, using the scotogenic model of leptogenesis \cite{Ma:2006km}. Additionally, Kerr black holes can drive non-thermal leptogenesis over a broad range of PBH masses, particularly when heavy RHNs ($\mathcal{O}(10^{12}$ GeV)) interact with a scalar field \cite{Ghoshal:2023fno}.

Heavier PBHs both form and evaporate at later times. When $\MPBHini \gtrsim 10^{5.5}$g the explosive final stages of evaporation occur after the EWPT and the background universe is in the broken phase. As demonstrated in \secref{Leptogenesis/Sphalerons}, the associated suppression of the sphaleron rate means that the BAU no longer tracks the lepton asymmetry. RHNs produced by these heavier PBHs still decay and produce lepton asymmetry but that is not translated into an increase in $\YB$. Conversely, the entropy dump from evaporation still acts to suppress the existing asymmetry. Leptogenesis being reliant entirely on the efficacy of the sphalerons, could only overproduce $\YB(\Tsphal) > \YBobs$ to mitigate the dilution from the huge numbers of photons that would be produced in a post-sphaleron PBH dominated era.\\

This section explores the consequences for leptogenesis imposed by a period of post sphaleron freeze-out PBH domination. First considering the theoretically attractive high-scale leptogenesis model and then the experimentally accessible low-scale resonant leptogenesis scenario detailed in \secref{Leptogenesis/Resonant}, the strong incompatibility between PBHs heavier than $10^6$g and leptogenesis is characterised across wide swathes of parameter space. By carefully navigating the high-dimensional parameter space of leptogenesis, the maximum achievable $\YB$ is calculated and compared to the entropy injection from PBHs. Mutual exclusion limits are accordingly derived between leptogenesis and PBH parameter spaces, allowing light to be shone on each space by experimental probes of the other.

For now it is assumed that PBHs heat the universe homogeneously as opposed to forming a hot-spot (see \secref{PBHs/Hot Spots}), an assumption that will be justified \textit{a posteriori} in Appendix \ref{C}.

\newpage

\mypapertitle{ Limits on light primordial black holes from high-scale leptogenesis}
\mypaperdate{Jun 15, 2023}
\mypaperabstract{
    The role that the evaporation of light primordial black holes may have played in the production of the baryon asymmetry of the Universe through high-scale leptogenesis is explored. In particular, for masses of primordial black holes in the range [$10^6$-$10^9$]~g, a dilution of thermally generated lepton asymmetry occurs via entropy injection in the primordial plasma after the sphaleron freeze-out. As a consequence, one can put strong constraints on the primordial black hole parameters, showing the mutual exclusion limits between primordial black holes and high-scale leptogenesis. Remarkably, an interplay between the upper bound on the initial abundance of primordial black holes and the active neutrino mass scale is demonstrated.}
\subsection[Limits on light primordial black holes from high-scale leptogenesis]{\textit{ Phys.Rev.D 109 (2024) 10, 10}\\}\label{PBHLepto/High Scale}
\makepapertitle
\newpage

This section is concerned with the consequences of a post-sphaleron freeze-out period of PBH domination for high scale leptogenesis models. First, the analysis calculates the maximum achievable yield of baryon asymmetry in high scale leptogenesis. The entropy injection associated with the evaporation of PBHs is then calculated and compared to the possible yield of baryon asymmetry to uncover where in PBH and leptogenesis parameter spaces the scenarios are incompatible.

\subsection*{CP violation}
Recall that high scale models of leptogenesis are those with RHNs $N_i$ heavier than the DI limit, $\MNi > \MDI \sim 10^{9}\GeV$. Even with only one RHN to consider, the dimensionality of the model must be reduced in order to efficiently calculate the quantities appearing in the Boltzmann equations. Since $\dmatm \gg \dmsol$, from the perspective of $\nu_3$, $m_l \approx m_m$ and therefore one may consider that the $R$ matrix (given by \equaref{vMass/Casas-Ibarra/R}) is effectively described by a single mixing angle, $R = R_{13}(\theta_{13})$. Therefore, the Yukawa matrix given by \equaref{vMass/Casas-Ibarra/Y} now depends on only four unknown parameters $\{x,y, m_h, \MNo\}$ (considering also the assignments of the Majorana and Dirac PMNS phases given in \secref{vMass/Casas-Ibarra}). $m_h$ is the mass of the heaviest active neutrino and sets the overall scale of the diagonal mass matrix $\mnu = \mathrm{diag} (m_l,m_m,m_h)$.\\

When scanning the parameter space, $\MNo$ is allowed to vary over the range $\MDI  \le \MNo \le 10^{16} \GeV$, such that in principle the situation $T \leq \Tflav \sim 10^{12}\GeV$ may be relevant suggesting that flavour effects should be considered (see \equaref{Leptogenesis/BEs/Tflav}). However the key results do not depend on dynamics which occur while flavour effects are relevant, so that it is safe to perform this analysis in the unflavoured regime. Therefore, summing over $\iflav = e,\mu,\tau$, the total CP asymmetry in the decay of $N_1$ is ~\cite{Strumia:2006qk}
\begin{equation}\label{eps}
\epsilon^1=\sum_\iflav \epsilon^1_{\iflav\iflav} = - \frac{3}{16\pi}\sum_{j=2,3} \frac{\MNo}{M_{N_j}}\frac{\mathrm{Im}(Y^\dagger Y)_{1j}}{(Y^\dagger Y)_{11}} \,.
\end{equation}
Note that in \equaref{eps} the leptonic mixing matrix, $\UPMNS$, has dropped out, thus implying no connection between the low-scale Dirac CP violation and high-scale CP asymmetry. The analytical form of $\epsilon$ is given by \cite{Bernal:2022pue}
\begin{equation}
    |\epsilon| = \frac{3\MNo}{16\pi \vEW^2}\frac{|\Delta m^2_{\rm atm}|}{m_h + m_l}\frac{|\sin(2x)\sinh(2y)|}{\cosh(2y) - f\cos(2x)}\,,
\end{equation}
where $f \equiv (m_h - m_l)/(m_h + m_l)$. Note that the generational index has been dropped, in the context of high scale leptogenesis, it is implied that $\epsilon$ refers to the unflavoured CP violation associated with the state $N_1$.

\subsection*{Boltzmann equations}
To calculate $\YB$ the Boltzmann equations describing the number density of $N_1$, and the $B-\Lnum$ asymmetry, are solved numerically. It is convenient to track the evolution of the $B-\Lnum$ asymmetry since it is conserved by EW sphalerons. Assuming a VIA of $N_1$ and $B-\Lnum$, the Boltzmann equations should account for the following processes
\begin{itemize}
    \item $1\to 2$ decays of $N_1$, $N_1 \to \lep \Higgs^\dagger$ and its CP conjugate process $N_1 \to \bar{\lep} \Higgs$. The total rates are proportional to the number density of $N_1$ and the thermally averaged decay width. Decays deplete the population of $N_1$ and source the $B-L$ asymmetry.
    \item $2 \to 1$ inverse decay modes ie $\lep \Higgs^\dagger \to N_1$. The inverse decay rate is related to the decay rate by $\Gamma^{ID}_{N_1} = \Gamma_{N_1}n_{N_1}^{\rm eq}/n^{\rm eq}_{\lep}$ where $n_{N_1}^{\rm eq},n^{\rm eq}_{\lep}$ are the equilibrium number densities of $N_1$ and the leptons given by \equaref{Early Universe/Abundance/neqrel}. These processes produce $N_1$ but only wash out the asymmetry.
    \item $2 \leftrightarrow 2$ scatterings mediated by $N_1$ exchange such as $\lep \Higgs^\dagger \to \bar{\lep} \Higgs$, for which $\Delta \Lnum = 2$. These processes contribute to the washout and do not change the number density of $N_1$. Care should be taken when combining $2\leftrightarrow 2$ scatterings in the s-channel and the inverse decay rates to avoid double counting. A subtracted form of the amplitude should be used which only accounts for off-shell $N_1$ exchange, so these processes are not Boltzmann suppressed at low energies.
\end{itemize}
The resulting Boltzmann equations for the comoving number density of $N_1$ and $B$-$L$ are
\begin{eqnarray}
\frac{{\rm d}\NNo}{{\rm d}\alpha} &=&\ln(10)  \frac{\GNTo}{H}(\NNeqo-\NNo)\,, \label{PBHLepto/HighScale/Nth} \\
\frac{{\rm d} \YBL} {{\rm d}\alpha} &=&   \left.\frac{\ln (10)}{H}\right[ \epsilon(\NNo-\NNeqo)\GNTo + \\ \nonumber
&&\qquad \left. + \left(\frac{1}{2} \frac{\NNeqo}{\Nleq}\GNTo + \gamma \frac{a^3}{\Nleq}\right)  \YBL\right]\,,\label{PBHLepto/HighScale/NBL}
\end{eqnarray}
where $\gamma$ quantifies the contribution to the washout associated with the $\Delta L = 2$ scattering processes. In Ref.\,\cite{Bernal:2022pue} it is shown that 
\begin{equation}
    \gamma = \frac{3T^6}{4\pi^5 \vEW^4} \mathrm{Tr}[m_\nu^\dagger m_\nu]\,.
\end{equation}
The yield of baryons today is then defined by
\begin{equation}\label{YB}
    |\YB| = \left(\etasphal \frac{|\YBL|}{\mathcal{S}}\right)(T=\Tsphal)\,.
\end{equation}
where $\etasphal$ is the sphaleron efficiency factor at the moment of sphaleron freeze-out, in which is implicit the approximation of instantaneous freeze-out of sphalerons. This is a good approximation for this model since $\MNo \gg \Tsphal$, meaning the lepton asymmetry stops evolving when washout processes go out of equilibrium at $T \gg \Tsphal$. In principle the Boltzmann equations should also account for $2\leftrightarrow 2$ scatterings involving the gauge bosons and top quarks, this will be done for example in the following section. However since the objective is to calculate the maximum achievable asymmetry, neglecting the washout terms is a conservative approximation. Including these diagrams would be expected to tighten slightly the constraints derived. The effects of 3-body decay processes are also numerically small, $\sim 6\%$, and can be ignored~\cite{Nardi:2007jp}.\\

The EW sphalerons go out-of-equilibrium during a matter-dominated epoch if PBHs dominate, and the enhanced Hubble rate alters the sphaleron freeze-out temperature $\Tsphal$. In the standard cosmological evolution without PBHs, $\Tsphal$ is 131~GeV, see \secref{Leptogenesis/Sphalerons}, while in the presence of PBHs the temperature $\Tsphal$ may be higher. In \figref{PBHLepto/HighScale/sphaleron} the sphaleron rate (black line) is shown as a function of the temperature compared to the Hubble rate obtained without PBHs (red line) and in the presence of PBHs with masses $10^6$ (green line) and  $10^9$ (blue line) in grams, taking their abundance to be the maximal allowed value according to the current GW energy density constraints. In the considered ranges of PBH mass and abundance, $\Tsphal$ is always smaller than $\Tewpt$. Therefore, $\etasphal(\Tsphal)$ is taken to be 12/37~\cite{Harvey:1990qw}.\\

\begin{figure}[h!]
    \centering
        \includegraphics[width =0.7\linewidth]{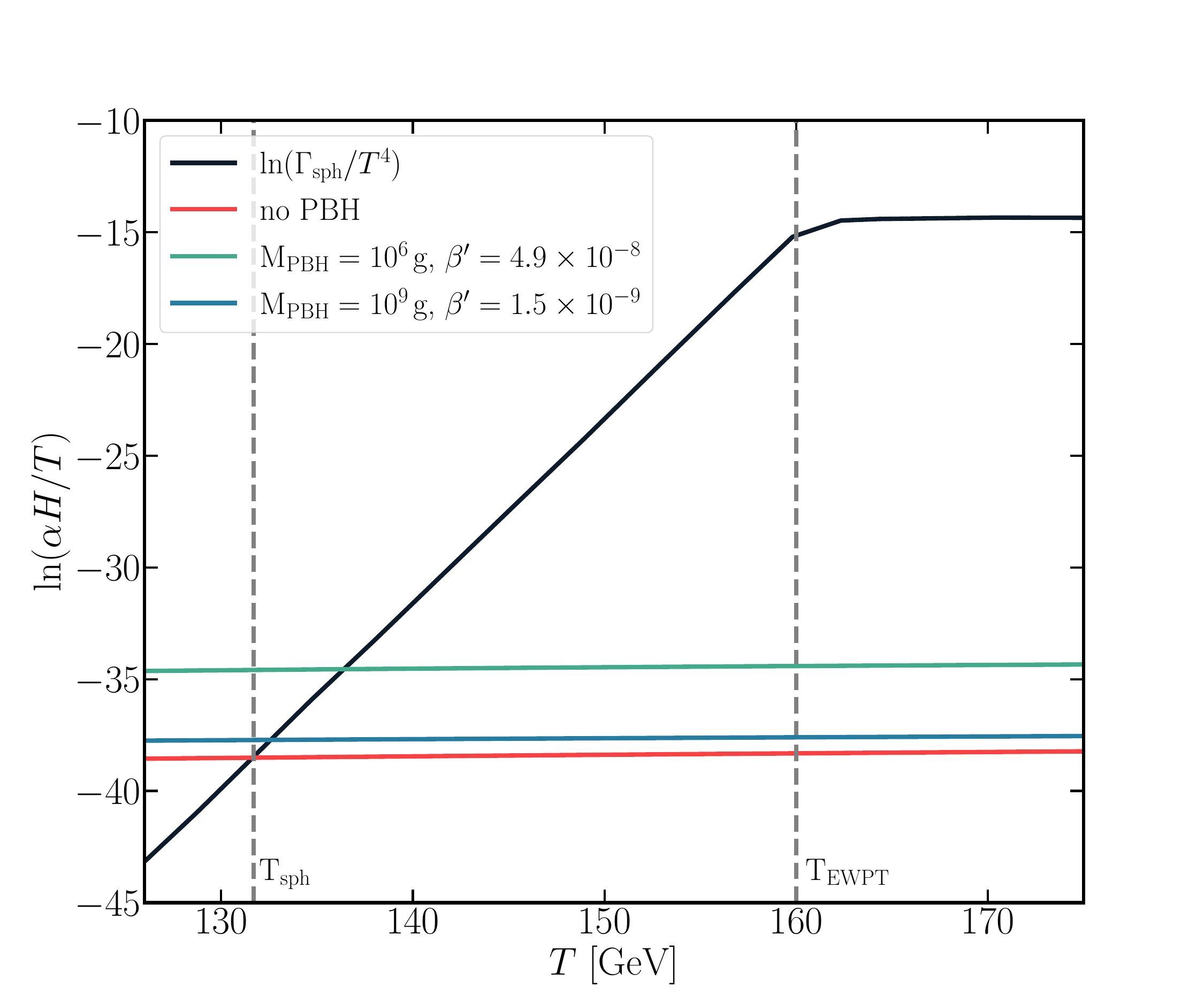}
        \caption{\label{PBHLepto/HighScale/sphaleron}The rate of sphaleron process (black line) as a function of the temperature. The coloured lines show the Hubble rate for different scenarios with and without the presence of PBHs. The crossing between $\Gamma_{\rm sphal}$ and $H$ defines the temperature $\Tsphal$ at which the sphaleron processes freeze-out. $\Tsphal < \Tewpt$ in the whole parameter space analysed.}
\end{figure}

\subsection*{The parameter space of high-scale leptogenesis}

All of the relevant quantities in \equaref{PBHLepto/HighScale/Nth} and \equaref{PBHLepto/HighScale/NBL} necessary to predict $\YB$ can be calculated in the four-dimensional parameter space $\{x,y, m_h, \MNo\}$. The following parameter ranges are scanned over:
\begin{eqnarray} \label{PBHLepto/HighScale/range}
0<x<\pi & \quad{\rm and}\quad & 0.14<y<\pi\\
0.05 < \frac{m_h}{{\rm eV}} < 0.8 & \quad{\rm and}\quad & 10^{10}< \frac{\MNo}{{\rm GeV}} < 10^{16}\nonumber
\end{eqnarray}
The strong washout regime occurs in this model for $|y| > 0.14$. The lower bound for the heaviest active neutrino mass $m_h$ comes from the neutrino oscillation data requiring $m_h > \sqrt{\dmatm^2}\approx 0.05~{\rm eV}$~\cite{Capozzi:2021fjo, Esteban:2020cvm, deSalas:2020pgw}, while the upper bound is set by the KATRIN experiment~\cite{KATRIN:2021uub}.
\begin{figure*}[h!]
    \centering
    \includegraphics[width = 0.32 \textwidth]{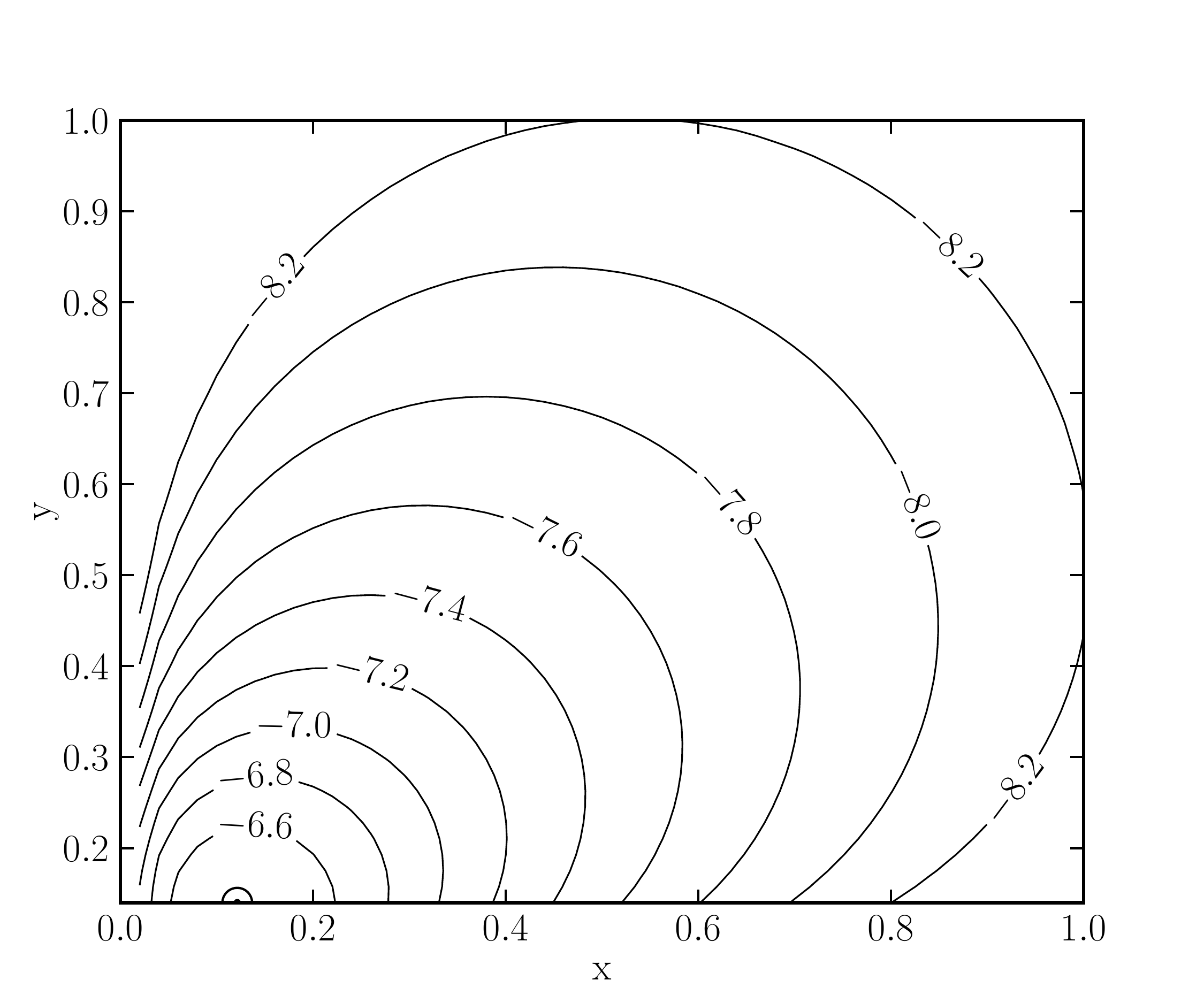}
    \includegraphics[width = 0.32 \textwidth]{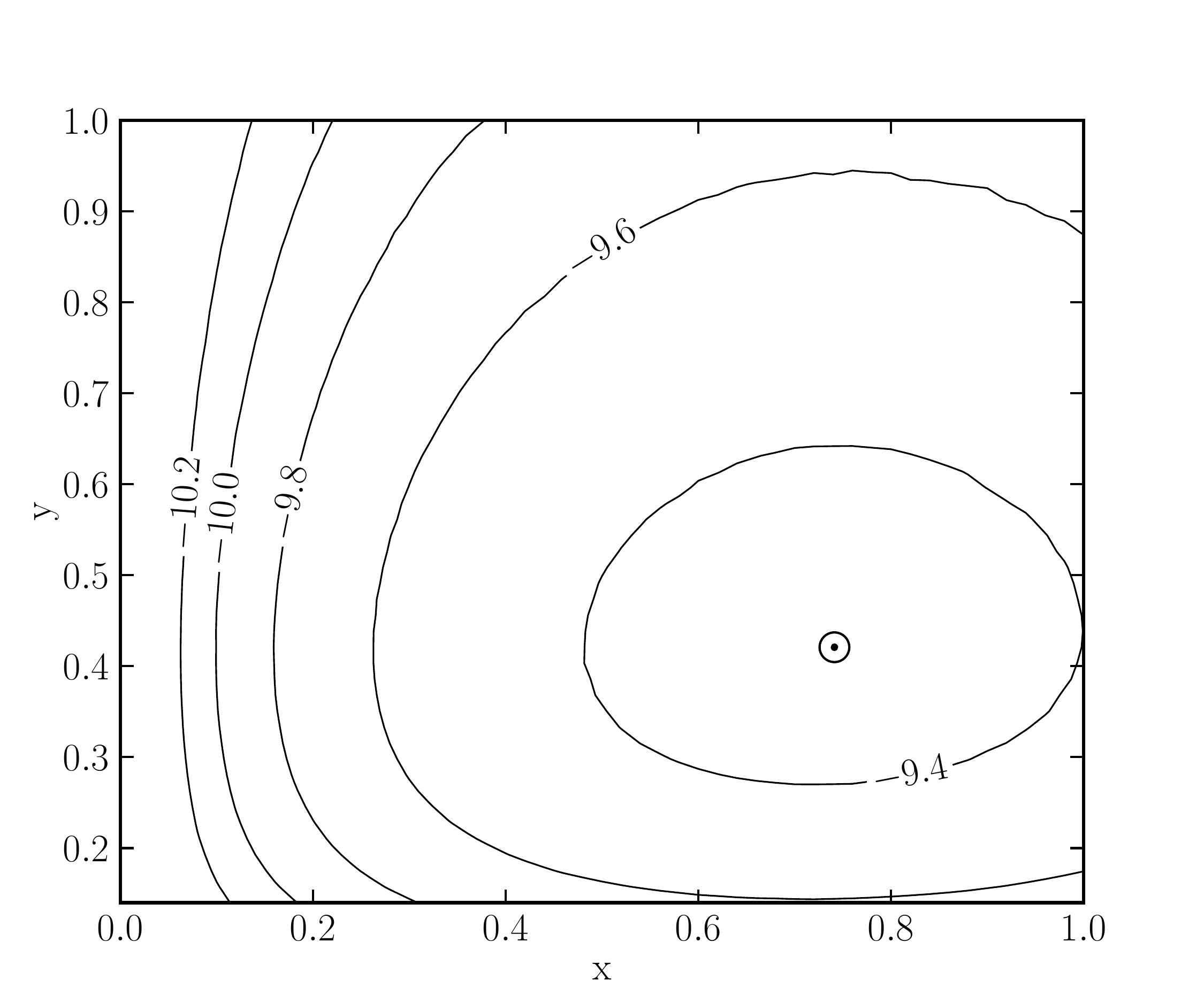}
    \includegraphics[width = 0.32 \textwidth]{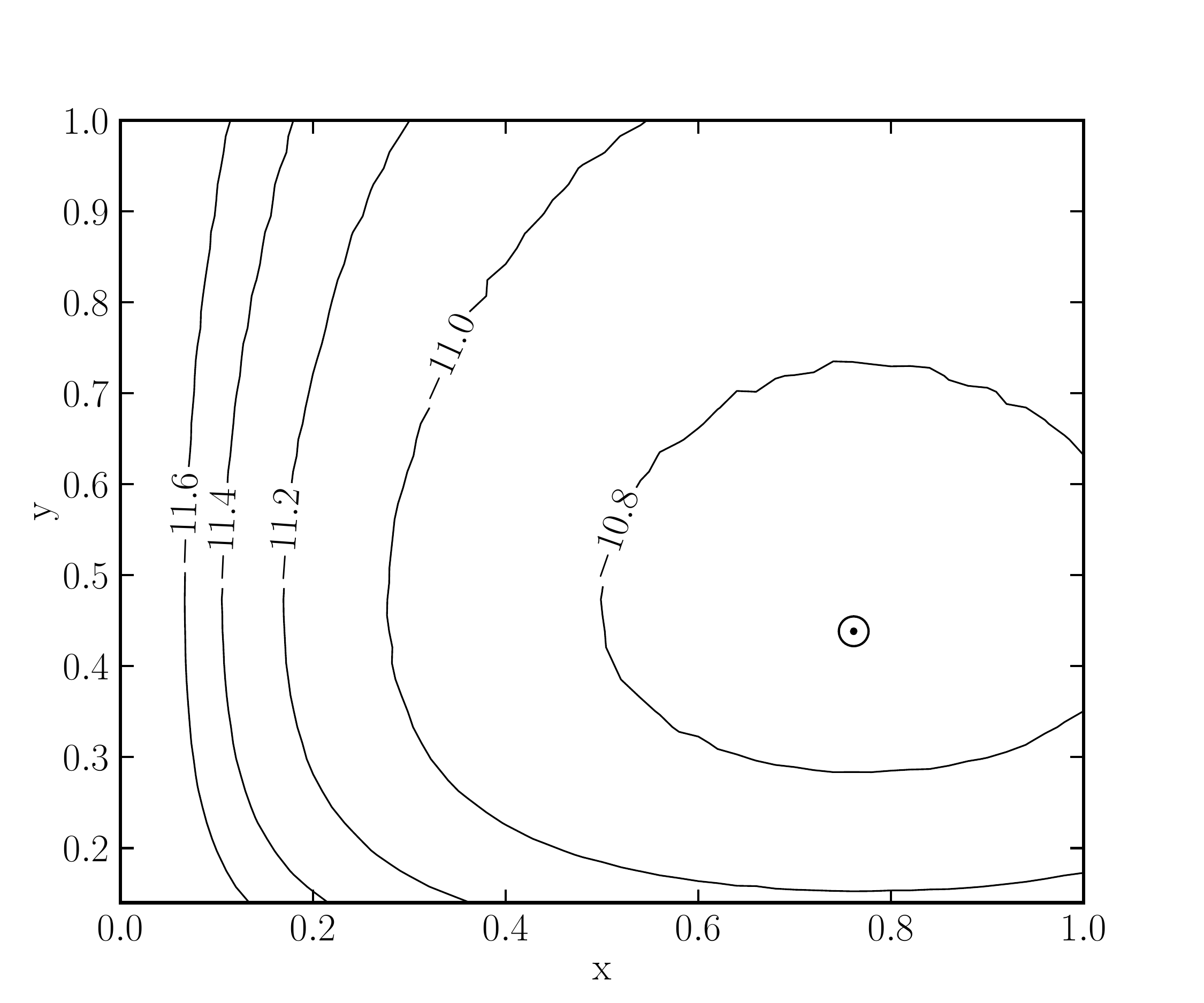}
    \caption{\label{PBHLepto/HighScale/xy}The final baryon asymmetry $\YB$ as a function of $x, \,y$ for $m_h = \sqrt{\dmatm^2} \approx 0.05$~eV (left panel), $m_h = 0.1$~eV (middle panel) and $m_h = 0.2$~eV (right panel), with $\MNo = 2.0 \times 10^{13}~{\rm GeV}$. The contours are for constant $\log_{10} \YB$ while the symbol $\odot$ indicates the point $(x,y)$ which maximises $\YB$ for the fixed values of $m_h$.}
\end{figure*}

In \figref{PBHLepto/HighScale/xy} contours of constant $\log_{10} \YB$ are shown in the plane $x, \,y$, taking $\MNo=2.0 \times 10^{13}~{\rm GeV}$, for $m_h = \sqrt{m_{\rm atm}^2} \approx 0.05$~eV (left panel), $m_h = 0.1$~eV (middle panel) and $m_h = 0.2$~eV (right panel). Moreover, the symbol $\odot$ indicates the point $(x,y)$ which maximises $\YB$. For $m_h \gg \sqrt{m_{\rm atm}^2}$, the asymmetry $|\YB|$ is maximised for $y = 0.44$ and $x = \pi/4$~\cite{Bernal:2022pue} (middle and right panels), while for smaller $m_h$ (left panel) the dependence on the angles $x,y$ is non-trivial, requiring a detailed numerical scan. Between these two regimes, the point maximising $\YB$ is continuously translated from the position on the left panel to that on the right panel. The angles $x,y$ are chosen for all points considered to maximise the final baryon asymmetry, defining the quantity
\begin{equation}\label{PBHLepto/HighScale/YBm}
    \YBm(m_h,\MNo)=\max_{x,y} \YB (x,y,m_h,\MNo)\,.
\end{equation}
\begin{figure*}[h!]
    \centering
    \includegraphics[width=\linewidth]
    {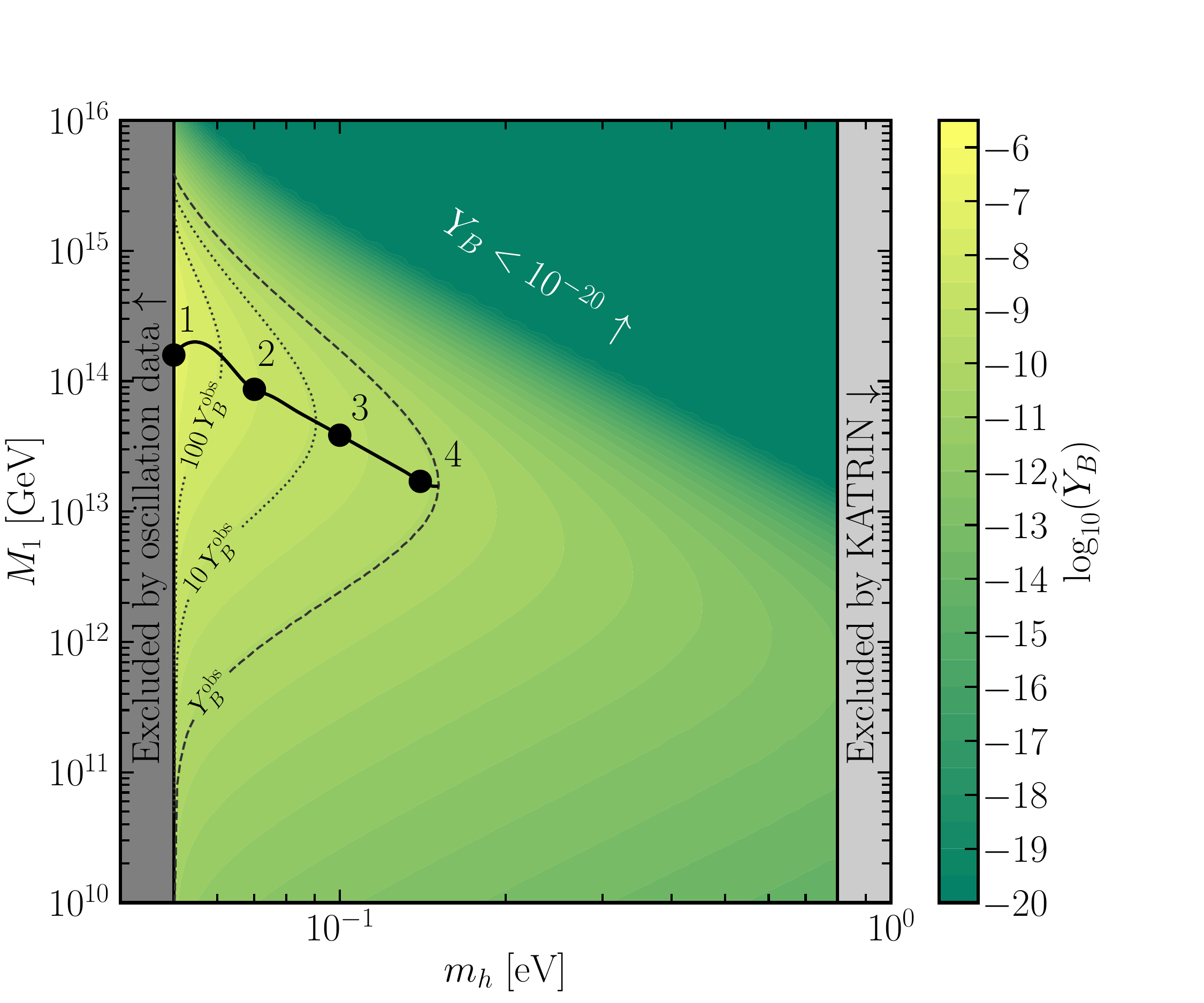}
    \caption{\label{PBHLepto/HighScale/YBmfig}The quantity $\YBm$ defined in \equaref{PBHLepto/HighScale/YBm} as a function of $m_h$ and $\MNo$. The dashed contour marks the observed baryon asymmetry of the Universe, while the dotted ones represent larger values as reported in the labels. The solid line shows the parameters that maximise the baryon asymmetry, with points from 1 to 4 highlighting the benchmark cases reported in Table \ref{Table}. The grey bands are excluded by experimental limits from oscillation data~\cite{Capozzi:2021fjo, Esteban:2020cvm, deSalas:2020pgw} (left) and KATRIN searches~\cite{KATRIN:2021uub} (right).}
\end{figure*}
\renewcommand{\arraystretch}{2}
\begin{table}[h!]
    \centering
    \begin{tabular}{c|c|c|c}
        Bench. pt & $m_h~[{\rm eV}]$ & $\MNo~[{\rm GeV}]$ & $\YBm$ \\ \hline
         {\bf 1} & $0.05$ & $1.5 \times 10^{14}$ & $1.5 \times 10^{-6} $ \\ \hline
         {\bf 2} & $0.07$ & $1.0 \times 10^{14}$ & $3.6 \times 10^{-9} $ \\ \hline
         {\bf 3} & $0.10$ & $4.0 \times 10^{13}$ & $5.5 \times 10^{-10}$ \\ \hline
         {\bf 4} & $0.14$ & $2.0 \times 10^{13}$ & $1.2 \times 10^{-10}$ \\ 
    \end{tabular}
    \caption{Benchmark points appearing in \figref{PBHLepto/HighScale/YBmfig}. \label{Table}}
\end{table}
\newpage
In \figref{PBHLepto/HighScale/YBmfig}, $\YBm$ is shown in colour in the plane $\{m_h,\MNo\}$. The dashed line encloses the region in which $\YBm$ can satisfy $\YBobs$, while the dotted lines show different increasing values for the ratio $\YBm/\YBobs$. Moreover, the solid line represents the relation between $m_h$ and $\MNo$ that maximises the baryon asymmetry $\YB$. The numbered points highlight benchmark cases for different active neutrino mass $m_h$ that will be used as reference in later plots. For each value of $\MNo$, the final baryon asymmetry increases for decreasing $m_h$. The maximal possible yield of baryon asymmetry in the high-scale thermal leptogenesis scenario is found to be
\begin{equation}\label{benchmark}
\YBmax = 1.5\times 10^{-6}
\end{equation}
achieved for $x=0.1$, $y = 0.14$, $m_h = 0.05~{\rm eV}$ and $\MNo = 1.5 \times 10^{14}\,{\rm GeV}$. With this choice of parameters, an entropy injection of about four orders of magnitude at later times (e.g. from PBHs' evaporation) would be required to obtain $\YBobs$. The amount of entropy injection from PBHs is given by
\begin{equation}
    \Sr \equiv \frac{\mathcal{S}(\alphaevap)}{\mathcal{S}(\alphaform)}
\end{equation}
such that if a population of PBHs fulfills the condition
\begin{equation}
    \Sr \geq \frac{\YBmax}{\YBobs}
\end{equation}
the PBHs would rule out high-scale thermal leptogenesis, forcing $\YB < \YB^{\rm obs}$ for any choice of parameters.

%==================================================
\subsubsection{Results}\label{sec:res}
%==================================================

In addition to the four leptogenesis parameters scanned over, $\{x,\,y,\, m_h,\,\MNo\}$, taken in the ranges in \equaref{PBHLepto/HighScale/range}, are those describing the PBH physics $\{\MPBHini,\, \beta^\prime\}$, which are scanned in the range 
\begin{equation}
    10^6 < \MPBHini/ {\rm g} < 10^9 \qquad {\rm and} \qquad 10^{-15}< \beta^\prime <0.1 \,.
    \label{PBHLepto/HighScale/PBHrange}
\end{equation}
By scanning the leptogenesis parameter space, the maximum $\YB$ attainable for fixed combinations of $m_h$ and $\MNo$ was calculated (see \figref{PBHLepto/HighScale/YBmfig}). Then, for each choice of $\MPBH$ and $\beta^\prime$, by solving \equaref{Early Universe/Cosmology/Friedmann} and \equaref{PBHs/Cosmology/S} the entropy injection and resulting dilution of asymmetry due to PBHs is calculated.\\

\begin{figure*}[h]
    \centering
        \includegraphics[width = 0.49\textwidth]{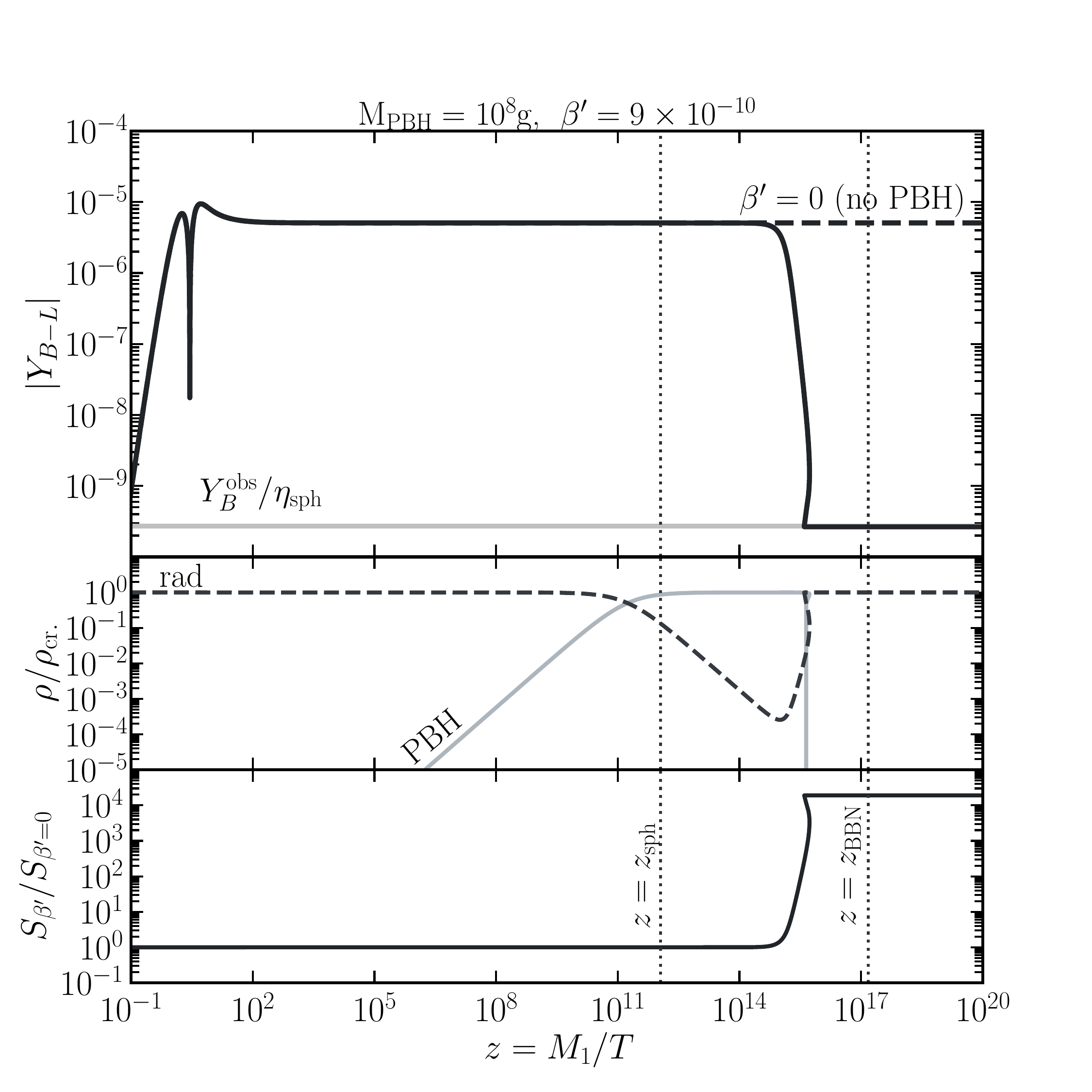}
        \includegraphics[width = 0.49\textwidth]{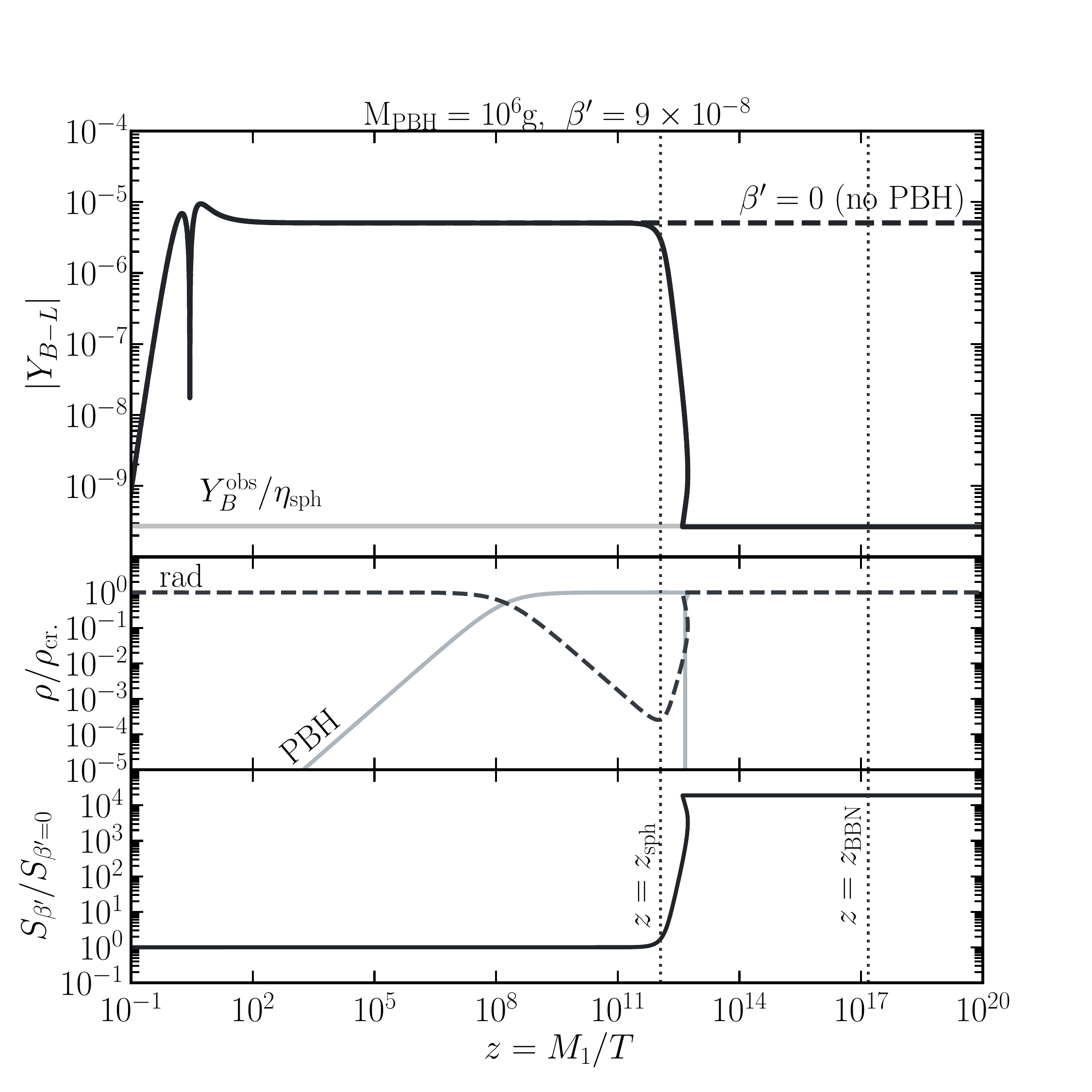}
       \caption{\label{PBHLepto/HighScale/benchmark} The yield $Y_{\rm B-L}$ as a function of $z=\MNo/T$ (top panels) with and without the presence of PBHs taking $\MPBHini = 10^6~{\rm g}$ (left plot) and $\MPBHini = 10^8~{\rm g}$ (right plot). The PBH initial abundance is fixed in order to obtain the observed baryon asymmetry (solid grey horizontal line) after PBH evaporation. The neutrino parameters are fixed according to the benchmark case 1 in \figref{PBHLepto/HighScale/YBmfig}. The middle and bottom panels show the evolution of energy densities and the entropy ratio $\Sr$.}
\end{figure*}
\begin{figure*}[t!]
    \centering
        \includegraphics[width = 0.49\textwidth]{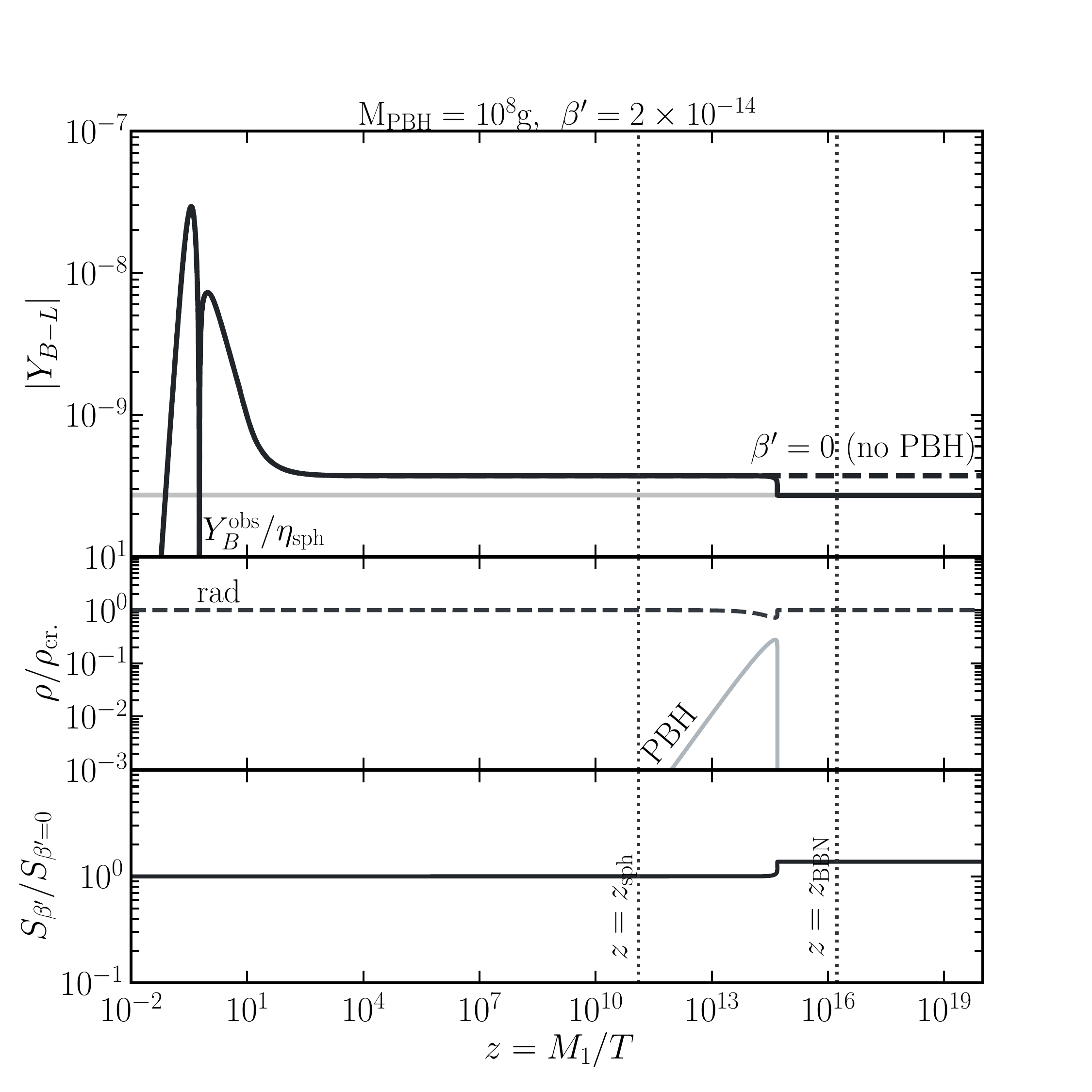}
        \includegraphics[width = 0.49\textwidth]{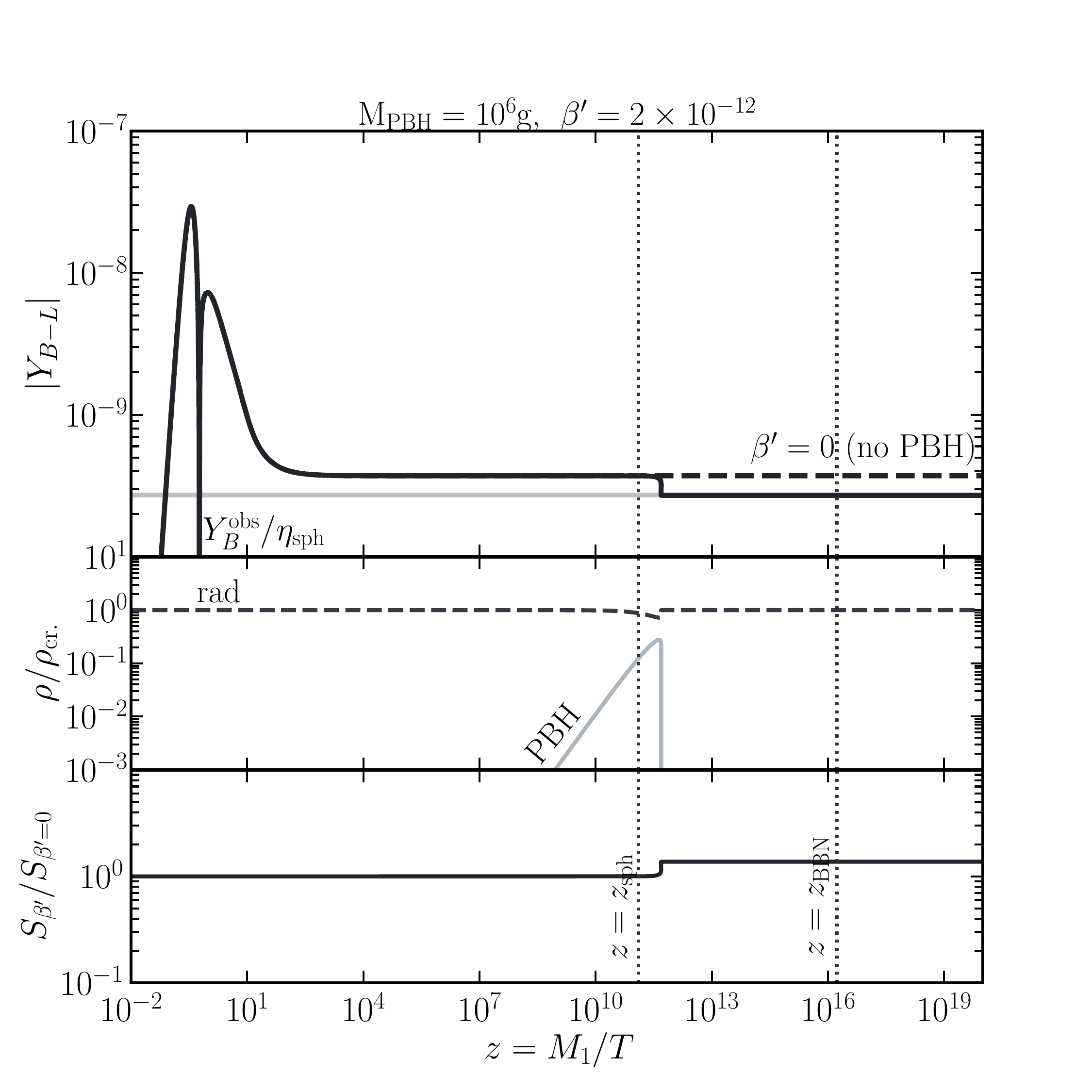}
       \caption{\label{PBHLepto/HighScale/benchmark2} Same as in \figref{PBHLepto/HighScale/benchmark} but fixing the neutrino parameters according to the benchmark case 4 reported in Table \ref{Table}.}
\end{figure*}

\figref{PBHLepto/HighScale/benchmark} and \figref{PBHLepto/HighScale/benchmark2}, illustrate how the presence of light PBHs alters the evolution of the $B-\Lnum$ yield (top panel) as a function of $z = \MNo/T$. In the former figure, the neutrino parameters are fixed according to the benchmark point 1 in \figref{PBHLepto/HighScale/YBmfig}, while in the latter they correspond to the benchmark point 4. In the left (right) plots, the initial PBH mass is taken be to $10^6~{\rm g}$ ($10^8~{\rm g}$), in both cases $\beta^\prime$ is fixed to achieve the observed baryon asymmetry after evaporation. In the top panels of the plots, the dashed (solid) lines show the evolution of $\YBL$ without (with) PBHs, for comparison the observed value is represented by the solid horizontal grey lines. Also reported are the evolution of the comoving energy densities of radiation and PBHs (middle panels) and the ratio $\Sr$ (bottom panels). In \figref{PBHLepto/HighScale/benchmark}, the $\beta^\prime$ required to dilute $\YB$ to match $\YBobs$ is large enough that PBHs dominate the radiation energy density, until they evaporate completely leading to a large entropy injection. In \figref{PBHLepto/HighScale/benchmark2} $\YB$ is much smaller so only a relatively small population of PBHs is required to dilute $\YB \to \YBobs$, which would never dominate the energy content of the Universe. A period of PBH domination is therefore not a necessary condition for PBHs to erase asymmetry produced during leptogenesis. In both of the benchmark cases discussed, a population of PBHs with the same initial mass but $\beta^\prime$ larger than the values indicated would lead to $\YB < \YBobs$ necessarily. Therefore, such a population of PBHs would be mutually exclusive with the model of high scale leptogenesis analysed here. Specifically, the PBHs would produce so much entropy that it follows that high scale leptogenesis could not reproduce $\YBobs$.

\begin{figure}[t!]
    \centering
        \includegraphics[width =\linewidth]{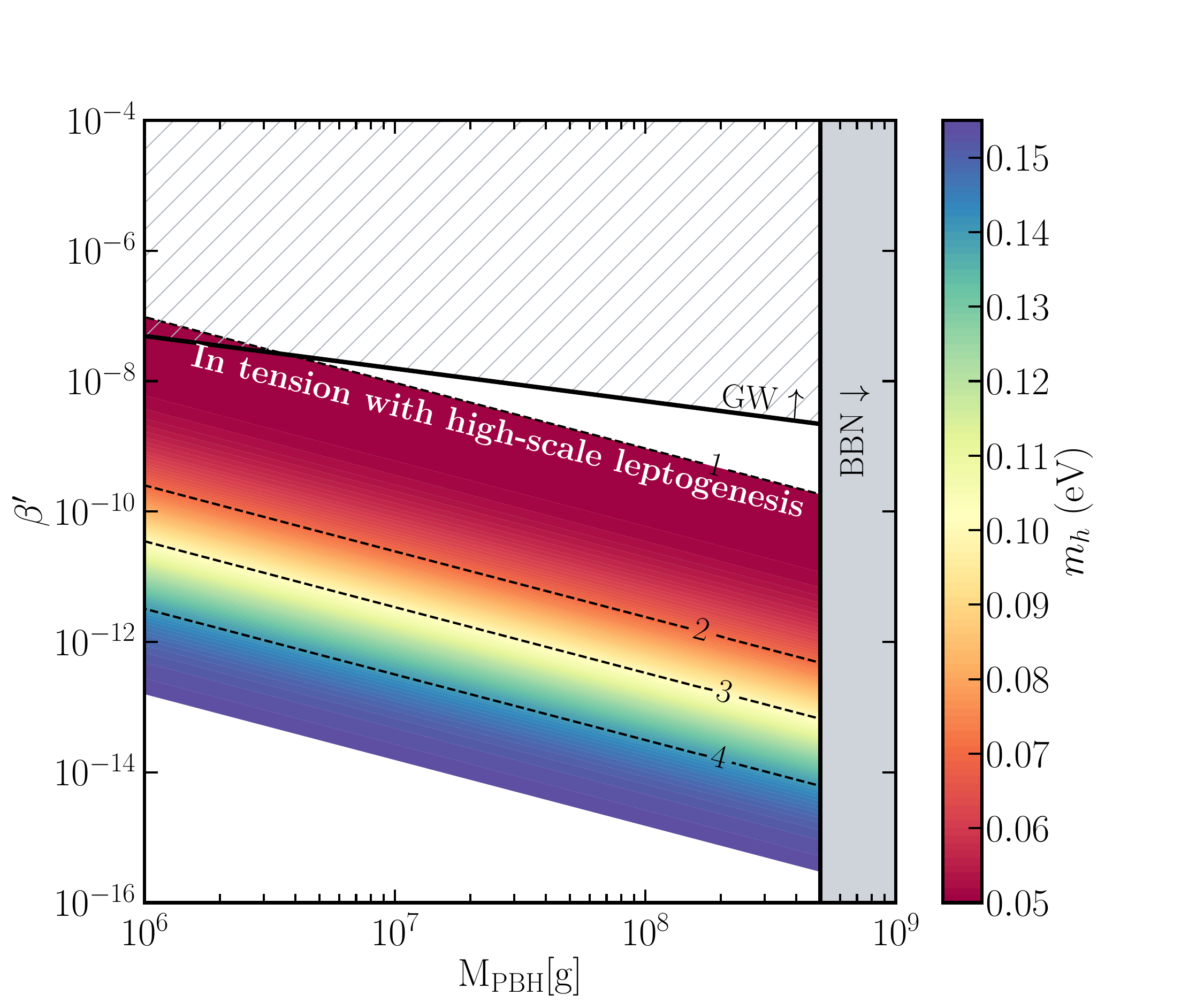}
        \caption{\label{PBHLepto/HighScale/constraints}Mutual exclusion limits between PBHs and the minimal high-scale leptogenesis in the $\{\MPBHini,\beta^\prime\}$ plane. The different colors correspond to the upper bounds obtained assuming different masses $m_h$ for the heaviest active neutrino, while the other neutrino parameters are fixed to maximise the baryon asymmetry. The numbered dashed lines refer to the benchmark cases reported in \figref{PBHLepto/HighScale/YBmfig}. The hatched region is excluded according to the constraints on the GW energy density from BBN observations~\cite{Domenech:2020ssp}. The shaded grey region highlights PBHs whose evaporation occurs during BBN.}
\end{figure}

The main results of the analysis are displayed in \figref{PBHLepto/HighScale/constraints} where the mutual exclusion limits between PBHs and high-scale leptogenesis are projected in the PBH parameter space. In particular, for each active neutrino mass $m_h$ an upper bound in the PBH parameter space is derived, further fixing the right-handed neutrino mass $\MNo$ to maximise the baryon asymmetry. In the leptogenesis parameter space, this corresponds to considering the values along the solid black line in \figref{PBHLepto/HighScale/YBmfig}. Indeed, the four numbered dashed lines refers to the four benchmark cases in \figref{PBHLepto/HighScale/YBmfig}. $\YBm$ decreases as $m_h$ increases, thus fewer PBHs are needed to dilute $\YBm$, placing stronger constraints on $\beta^\prime$. The strongest upper bound on $\beta^\prime$ (darkest violet line) corresponds to the largest $m_h$ for which thermal high scale leptogenesis is viable (benchmark point 4). Conversely, the most conservative constraint (darkest red line) corresponds to $m_h\approx 0.05~{\rm eV}$, the smallest $m_h$ allowed according to neutrino oscillation data. Even the most conservative bound is stronger than that imposed by the limit on the energy density of gravitation waves (GWs) ~\cite{Domenech:2020ssp, Papanikolaou:2020qtd} (hatched area in the plot). The shaded grey area illustrates where PBHs would evaporate during BBN. \footnote{Note that more recent projected constraints on PBH evaporation during BBN place the grey shaded region in \figref{PBHLepto/HighScale/constraints} closer to $10^8\g$ \cite{Boccia:2024nly}}

%==================================================
\subsubsection{Conclusions}\label{sec:concl}
%==================================================

This section analysed the impact of the non-standard cosmology driven by the presence and the evaporation of light PBHs on the production of the baryon asymmetry of the Universe through high-scale leptogenesis. Crucially, the evaporation of PBHs is associated with (a sudden) injection of entropy and reheating of the Universe. When this occurs after the sphaleron freeze-out, the baryon asymmetry is diluted. Firstly the four-dimensional parameter space of high-scale leptogenesis in its minimal version is explored, demonstrating the existence of a finite, maximum achievable asymmetry $\YB^{\rm max}$. It was shown that if the entropy injection from PBH evaporation is large enough, even $\YB^{\rm max}$ does not reproduce the observed asymmetry of the Universe. This effectively rules out high-scale leptogenesis as the baryogenesis mechanism. In this fashion, mutual exclusion limits in the PBH and leptogenesis parameter spaces are derived, characterising the regions which are incompatible. For an entropy increase greater than around four orders of magnitude, the minimal model for high-scale leptogenesis is insufficient to account for the baryon asymmetry of the Universe. This corresponds to a PBH population with masses from $10^6$ to $10^9$ grams with an initial abundance $\beta^\prime \gtrsim 10^{-9}$. A relation was found to exist between the heaviest active neutrino mass $m_h$ and the bound in the PBH parameter space, meaning that the impact of future experimental results on the mutual exclusion limits has been anticipated. Interestingly,any measurement of the lightest neutrino mass $m_l$ different from 0 will strengthen the constraints derived in this work.

\newpage

\mypapertitle{Impact of primordial black holes on heavy neutral leptons searches in the framework of resonant leptogenesis}
\mypaperdate{May 1, 2024}
\mypaperabstract{The effects on sub-TeV resonant leptogenesis of Primordial Black Holes with masses from $10^6$ to $\sim 10^9$g are investigated. The latter might dominate the energy content of the Universe altering its evolution and, eventually, diluting the final baryon asymmetry. Depending on the mass and abundance of Primordial Black Holes, the parameter space of sub-TeV resonant leptogenesis is shown to shrink towards higher Right Handed Neutrino masses and smaller active-sterile mixing. Remarkably, this translates into important implications for the experimental searches of heavy neutral leptons. Conversely, a possible future detection of sub-TeV heavy neutral leptons would disfavour regions of the parameter space of Primordial Black Holes currently allowed.}
\subsection[Impact of primordial black holes on heavy neutral leptons searches in the framework of resonant leptogenesis]{\textit{Phys.Rev.D 109 (2024) 10, 10}}
\makepapertitle
\newpage
Having analysed the asymmetry diluting effect of light PBHs on the theoretically attractive high scale leptogenesis mechanism in the previous section, in this section attention is turned to the experimentally accessible low scale, resonant leptogenesis model established in \secref{Leptogenesis/Resonant}. Therefore the model contains two nearly degenerate RHNs ($\Mdegen \sim \MNo \sim \MNt$) with a tiny mass splitting $\Delta M = \MNt - \MNo \ll M_{N_{1,2}}$ and a third effectively decoupled RHN ($\MNth \gg \Mdegen$). Due to radiative corrections the mass splitting is expected to be at least at the level of $10^{-16}\GeV$ (see \equaref{Leptogenesis/Resonant/dMrad}). In this work the relative mass splitting $\Delta \Mdegen/\Mdegen$ is treated as a free parameter. In this framework, the total mixing between the active neutrinos and RHNs is given by ~\cite{Chianese:2018agp}
\begin{eqnarray}
    U^2 & = & \sum_{\iflav\,i} |U_{\iflav (i+3)}|^2 \\
    & = & \frac{m_m-m_h}{2} \frac{\Delta M}{\Mdegen^2}\cos(2 x) + \frac{(m_m+m_h)}{\Mdegen} \cosh(2y) \nonumber \,,
    \label{PBHLepto/LowScale/U2}
\end{eqnarray}
since the contribution from $N_3$ is negligible. $U^2$ along with the common mass $\Mdegen$, forms the parameter space searched by current and future generation experiments~\cite{Bauer:2019vqk,MATHUSLA:2022sze, Baldini:2021hfw,Feng:2022inv,Abdullahi:2022jlv}. The requirement of successful baryogenesis via resonant leptogenesis provides an allowed region in the plane $\{\Mdegen,U^2\}$, see for instance Ref.~\cite{Klaric:2020phc,Granelli:2020ysj,Reconciling}. This region could extend to $\Mdegen \sim 0.1\GeV$, which some proposed and upcoming experiments project sensitivity curves able to probe \cite{MATHUSLA:2022sze,SHiP:2018xqw,SHiP:2021nfo,Baldini:2021hfw,Hirsch:2020klk}. The lightest active neutrino mass is assumed to be $m_l=0$, such that $m_m = \sqrt{\dmsol^2}$ and $m_h \approx \sqrt{\dmatm^2}$. In this case, which is equivalent to the one discussed in Ref.~\cite{Granelli:2020ysj}, the matrix $R$ once again has only one physically relevant angle, in this case $z_{23}=x+i\,y$, with real $x,y$.

%================================
\subsection*{Boltzmann equations}
%================================
The focus of this section is the case of TIA, see \secref{Leptogenesis/Out-of-Equilibrium}. When RHNs reach equilibrium abundance before the sphaleron freeze-out, any additional asymmetry generated via $N_i \leftrightarrow N_j$ oscillations vanishes \cite{Klaric:2021cpi, Akhmedov:1998qx, Drewes:2016gmt}, see \secref{Leptogenesis/Resonant}. Hence, the evolution of $\mathcal{N}_{\Delta \iflav}$ is not affected by the evolution of quantum correlation between the RHN states, but can be simply described by two coupled Boltzmann equations which account for the changes in the number density due to the processes outlined in \equaref{Leptogenesis/BEs/12} and \equaref{Leptogenesis/BEs/Nscatter}. The Boltzmann equations in terms of $\alpha = \log_{10} a$ read~\cite{Hambye:2016sby,Giudice:2003jh}, 
\begin{eqnarray} \label{PBHLepto/LowScale/BEN} \frac{{\rm d}\NNi}{{\rm d}\alpha} &=& \frac{a^3\ln(10)}{H}\left(1-\frac{\NNi}{\NNeq}\right)(\gamma_{D} + 2\gamma_{S_s} + 4\gamma_{S_t}) \\
\label{PBHLepto/LowScale/BEL}
\frac{{\rm d}\mathcal{N}_{B-\Lnum}}{{\rm d}\alpha} &=&\frac{a^3\ln(10)}{H} \sum_i \left[ \epsilon_{\iflav \iflav}^i \left(\frac{\NNi}{\NNeq} - 1\right)\gamma_{D} \right. \\
\nonumber &&\left. -P_{\iflav\,i}\frac{\mathcal{N}_{B-\Lnum}}{\mathcal{N}_\iflav^{\rm eq}} \left(2\gamma_{D} + 2\gamma_{S_t} + \frac{\NNi}{\NNeq} \gamma_{S_s}\right) \right] \,,
\end{eqnarray}
where compared to the notation in \secref{Leptogenesis/BEs} the $\Gamma_{NH}$ has been recast in terms of the reaction density $\gamma_D$. In the case of high scale leptogenesis all of the important dynamics occured at $T \gg \Tflav$. In the case of resonant leptogenesis the reverse is true and flavour effects are always important. The flavoured CP asymmetry parameter $\epsilon^i_{\iflav \iflav}$ (given by \equaref{Leptogenesis/Resonant/epsi}) and the flavour projection probability ($P_{\iflav i} = |Y_{\iflav i}|^2/(Y^\dagger Y)_{ii}$) of heavy neutrino mass state $N_i$ on to flavour state $\iflav$, appear in the Boltzmann equations for this reason. The reaction density $\gamma_{D}$ for the $1\leftrightarrow2$ decay and inverse decay processes is given by
\begin{eqnarray}
    \gamma_{D} &=&
    \frac{\Mdegen^3}{\pi^2 z} K_1\left(z\right)\GNiT \Theta(M - M_\lep + M_\Higgs) \nonumber \\
    &+& \frac{2 M_\Higgs^2 \Mdegen}{\pi^2 z}K_1\left(z_\Higgs\right)\Gamma_{\Higgs \to N_i \lep}^T \Theta( M_\Higgs - M_\lep + \Mdegen)
\end{eqnarray}
where $z_\Higgs = M_\Higgs / T$, and $\GNiT$ and $\Gamma_{\Higgs \to N_i \lep}^T$ are the thermally averaged decay widths corrected by the Higgs and lepton thermal masses, given by \equaref{Leptogenesis/Out-of-Equilibrium/GNifull} and \equaref{Leptogenesis/Thermal/GHiggs} respectively. Expressions for the quantities $\gamma_{S_s}$ and $\gamma_{S_t}$ of the $2 \leftrightarrow 2$ quark and gauge boson scatterings can be found in \cite{Giudice:2003jh} while the cross section formulae given in \cite{Pilaftsis:2003gt} are used in this work as explained in \secref{Leptogenesis/BEs}.

It is important to track accurately the thermal Higgs, lepton, quark, and boson masses in the Boltzmann equations. Thermal corrections to the RHN masses are suppressed by the tiny Yukawa couplings required for successful resonant leptogenesis, and so neglected. However, thermal corrections to the mass splitting $\Delta M$ can be important when the splitting is very small, and are accounted for in the expression for the CP asymmetry parameter \equaref{Leptogenesis/Resonant/epsi}.\\

Differently from high-scale leptogenesis where the lepton asymmetry is frozen long before the sphaleron decoupling at $T=\Tsphal$, in resonant leptogenesis the lepton asymmetry evolves during and even after the sphaleron decoupling, especially when $\Mdegen \lesssim \Tsphal$. Therefore, to compute the final baryon asymmetry it is important to account for the non-instantaneous freeze-out of sphalerons, as in \secref{Leptogenesis/Sphalerons}. Furthermore, while the PBHs considered here do not evaporate before the EWPT, they may still dominate the energy budget of the universe at this time and therefore enhance the Hubble parameter. This is potentially important in the sphaleron freeze-out calculation, and the PBH-enhanced Hubble parameter must be consistently accounted for. In order to calculate $\YB$, the differential equation governing the sphaleron freeze-out, \equaref{Leptogenesis/Sphalerons/dNB} is solved considering the PBH impact on Hubble and evolution of the leptonic asymmetry across the EWPT.

\subsection*{The parameter space of sub-TeV resonant leptogenesis}

The three Boltzmann equations given by \equaref{PBHLepto/LowScale/BEN}, \equaref{PBHLepto/LowScale/BEL} and \equaref{Leptogenesis/Sphalerons/dNB} are solved taking $\NNi = \NNeq$ (TIA), $\mathcal{N}_{\Delta \iflav} = 0$ and $\mathcal{N}_{B} = 0$ as initial conditions. The final yield $\YB$ depends on four unknown parameters: the RHN mass scale $\Mdegen \sim \MNo \sim \MNt$, the tiny mass splitting $\Delta M = \MNt- \MNo$, and the two angles $x$ and $y$ of the Casas-Ibarra matrix. However, its dependence on $x$ is periodic, $x=(2n+1)\pi/4$ (for integer $n$) maximises the final baryon asymmetry therefore will provide the most conservative constraints (for more see \secref{Leptogenesis/Resonant}). Hence, the dependence of $\YB$ on the remaining three parameters $\Mdegen$, $\Delta M$ and $y$ is investigated. These parameters together determine $U^2$ , and thus the regions achieving successful leptogenesis ($\YB \geq \YBobs$) in the experimentally sensitive plane $\Mdegen$-$U^2$.

\begin{figure}[h!]
    \centering
    \includegraphics[width = 0.45 \linewidth]{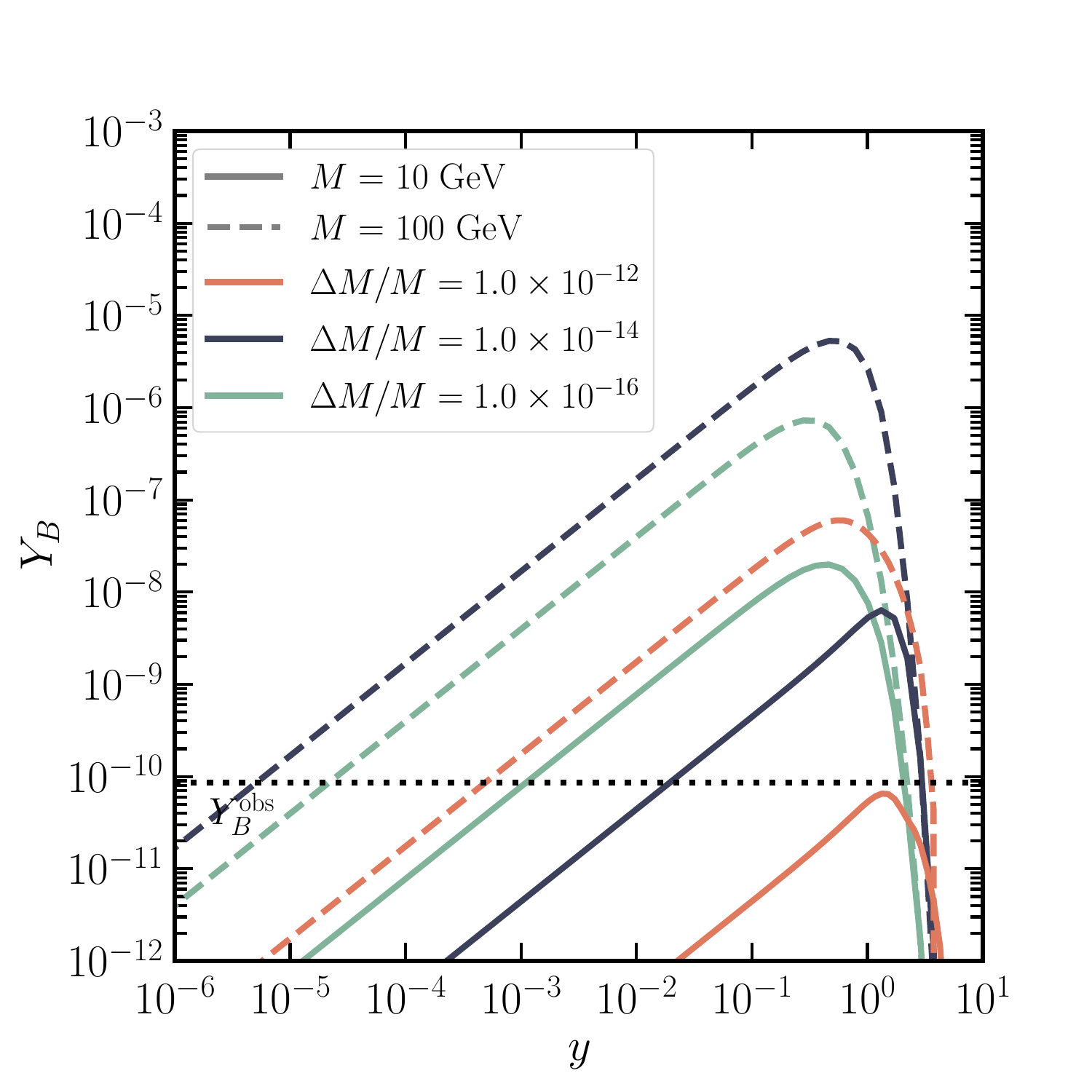}
    \includegraphics[width = 0.45 \linewidth]{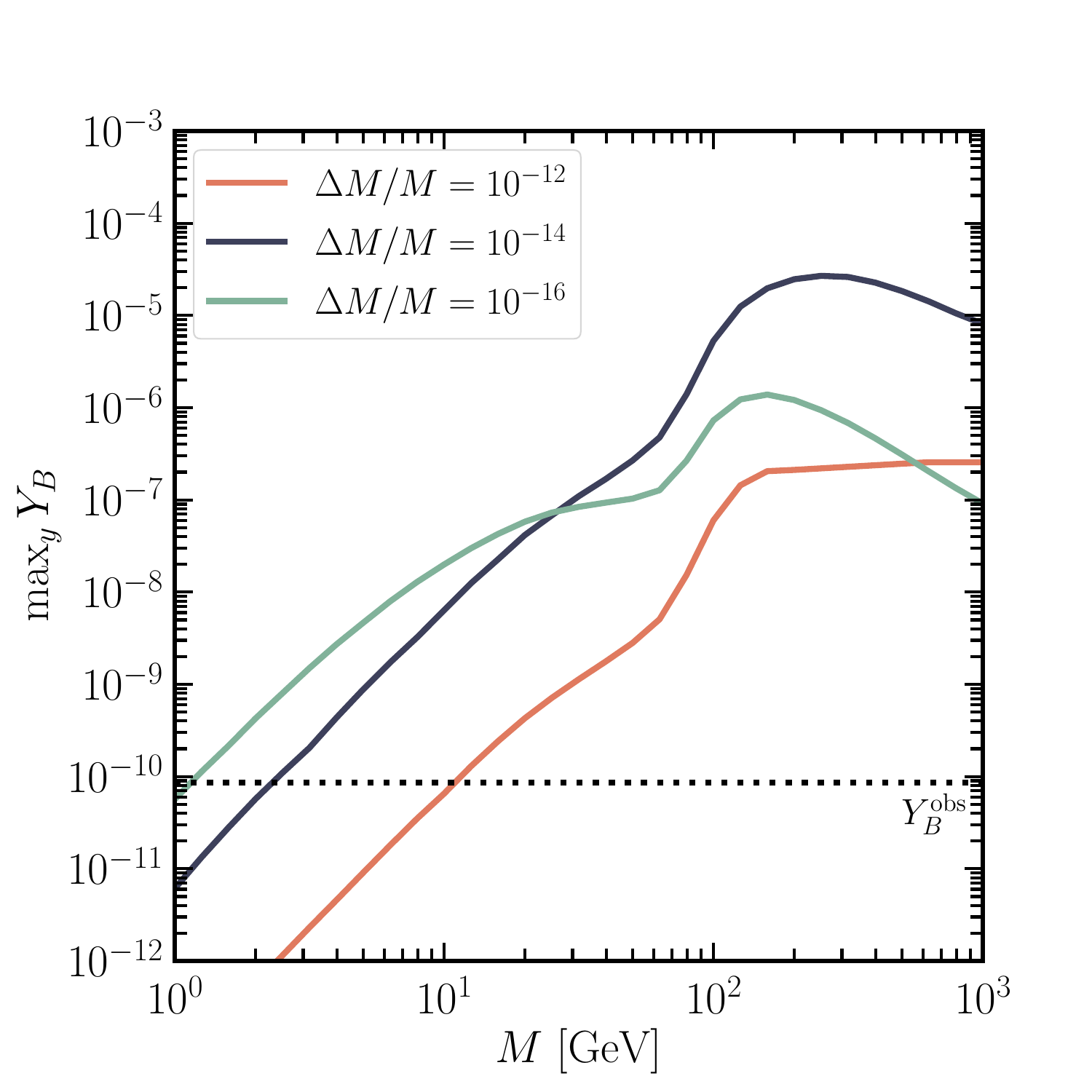}
    \caption{\label{PBHLepto/LowScale/YBy} Left panel: The behaviour of the final baryon asymmetry yield as a function of the angle $y$ for $\Mdegen = 10\GeV$ (solid lines) and $\Mdegen = 100\GeV$ (dashed lines). Different relative mass splittings $\Delta M /\Mdegen$ are indicated by different colours. The horizontal dotted line represents the observed baryon yield. Right panel: The maximum value of the baryon asymmetry yield (maximised over $y$) as a function of the RHN mass scale $\Mdegen$, for different relative mass splittings $\Delta M / \Mdegen$. The horizontal dotted line represents the observed baryon yield.}
\end{figure}

In \figref{PBHLepto/LowScale/YBy} the behaviour of $\YB$ is shown with respect to $y$ (left panel) and $\Mdegen$ (right panel) in some benchmark cases and compared with its observed value (horizontal dotted line). In particular, two benchmark values for the RHN mass scale are considered, $\Mdegen=10~{\rm GeV}$ (solid lines) and $\Mdegen=100~{\rm GeV}$ (dashed lines), for three different values for the relative mass splitting $\Delta M / \Mdegen = 10^{-12},\,10^{-14},\,10^{-16}$. In the benchmark cases shown, enough baryon asymmetry can be achieved between an upper and lower limit in $y$. These translate to upper and lower limits in $U^2$. As discussed in \secref{Leptogenesis/Resonant}, a small mass splitting on the order of the active neutrino mass splittings can be naturally generated by radiative corrections if one assumes exactly degenerate $N_{2,3}$ at some high scale. However, in this case the resulting asymmetry vanishes, some additional mass splitting is required to generate the BAU beyond the radiative splitting. In order to approximate the 3 dimensional parameter space $\{\Mdegen,U^2,\Delta M \}$,  different values of the relative mass splitting $\Delta M/\Mdegen$ are fixed and the union of the regions formed by scanning in $y$ and $\Mdegen$ is considered. \\

The relative mass splitting $\Delta M/\Mdegen$ is considered in the practically unobservable range $10^{-16}$ to $10^{-11}$. This range in $\Delta M/\Mdegen$ is chosen in order to maximise the available parameter space for leptogenesis. Considering a wider range of values does not change the results. \\

\begin{figure}[h!]
    \centering
    \includegraphics[width = 0.75 \linewidth]{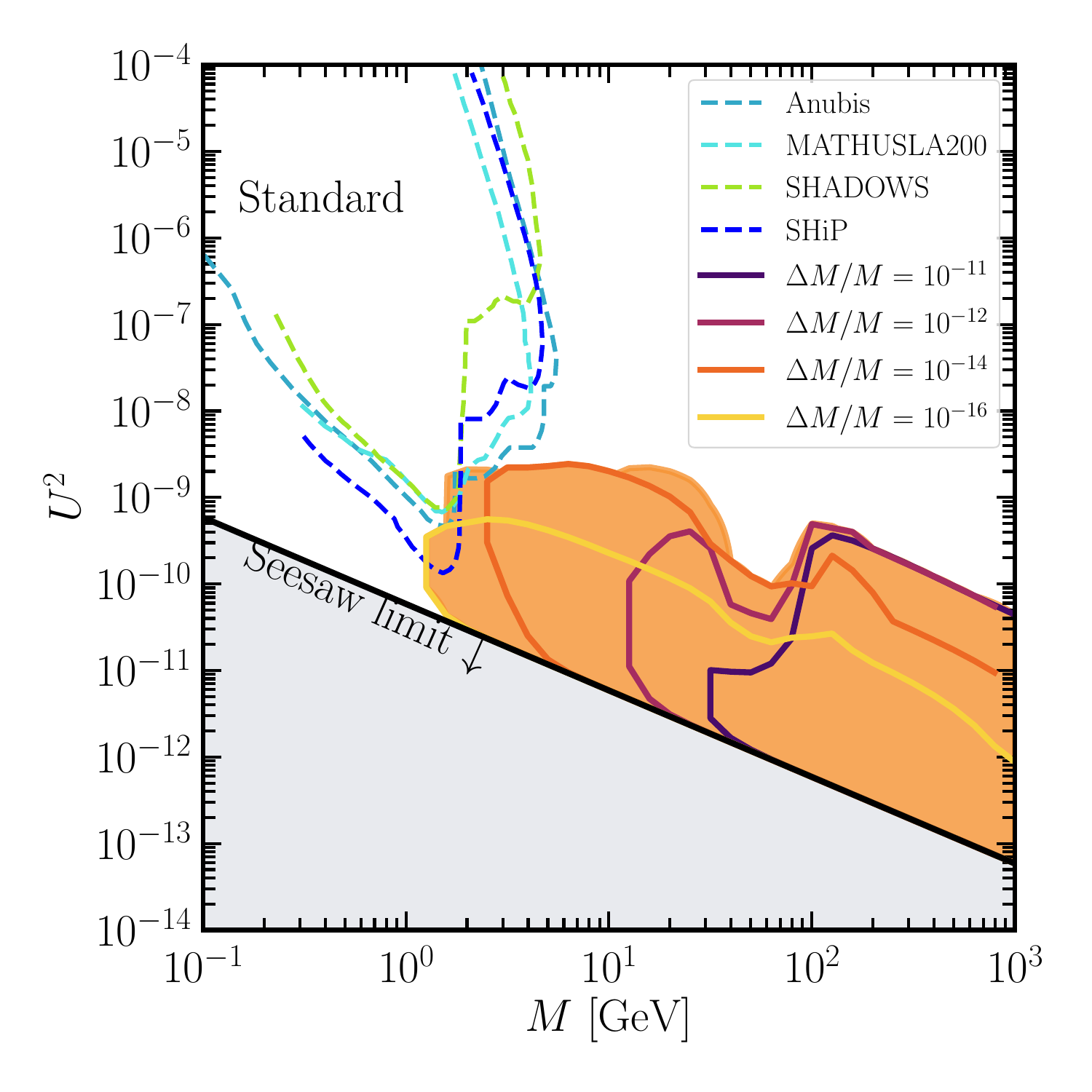}
    \caption{\label{PBHLepto/LowScale/standard} The viable leptogenesis parameter space is shown as the orange shaded region in the plane $(\Mdegen,U^2)$ , with $U^2$ defined in \equaref{PBHLepto/LowScale/U2}. The coloured outlines correspond to different values of the relative mass splitting $\Delta M / \Mdegen$. The gray area is theoretically excluded by the seesaw mechanism. The projected constraints from the SHiP \cite{SHiP:2018xqw}, SHADOWS \cite{Baldini:2021hfw}, MATHUSLA \cite{MATHUSLA:2022sze} and ANUBIS \cite{Hirsch:2020klk} experiments are reported for comparison. }
\end{figure}

The resulting viable parameter space is shown in \figref{PBHLepto/LowScale/standard} by the orange region, and is comparable to similar results presented in \cite{Granelli:2020ysj, Klaric:2020phc}. However in contrast to ~\cite{Granelli:2020ysj}, $\YB$ is calculated using the full thermally corrected decay rates as done in \cite{Hambye:2016sby}. Furthermore, the freeze-out of sphalerons is not approximated to be instantaneous as is the case in both \cite{Granelli:2020ysj, Klaric:2020phc}. Moreover, it should be noted that the RHN self-energy function $\gamma$, given by \equaref{Leptogenesis/Thermal/gamma}, is not interpolated in order to approximate the effect of soft gauge scatterings. This results in the dip feature for $\Mdegen \lesssim 100~{\rm GeV}$ which does not appear in \cite{Granelli:2020ysj} because the authors interpolate the RHN self energy in the kinematically suppressed regime $ M_\Higgs - M_\lep \leq M \leq M_\Higgs + M_\lep$.

\subsubsection{Results \label{PBHLepto/LowScale/Results}}
%========================

$\YB$ is computed by solving the set of coupled Boltzmann equations \equaref{PBHLepto/LowScale/BEN} and \equaref{PBHLepto/LowScale/BEL}, the differential equation tracking the freeze-out of sphalerons \equaref{Leptogenesis/Sphalerons/dNB}, and the Friedmann equations describing the evolution of PBHs, given by \equaref{PBHs/Cosmology/Friedmann}. Hence, the allowed region achieving the correct final baryon asymmetry is obtained in the combined parameter space of RHNs and PBHs, which consists of five parameters: the RHNs mass scale $\Mdegen$, the RHNs mass splitting $\Delta \Mdegen$, the angle $y$ in the Casas-Ibarra matrix, the PBH mass scale $\MPBHini$, and the initial abundance of PBHs $\beta^\prime$.

\begin{figure*}[h]
    \centering
        \includegraphics[width = 0.45\textwidth]{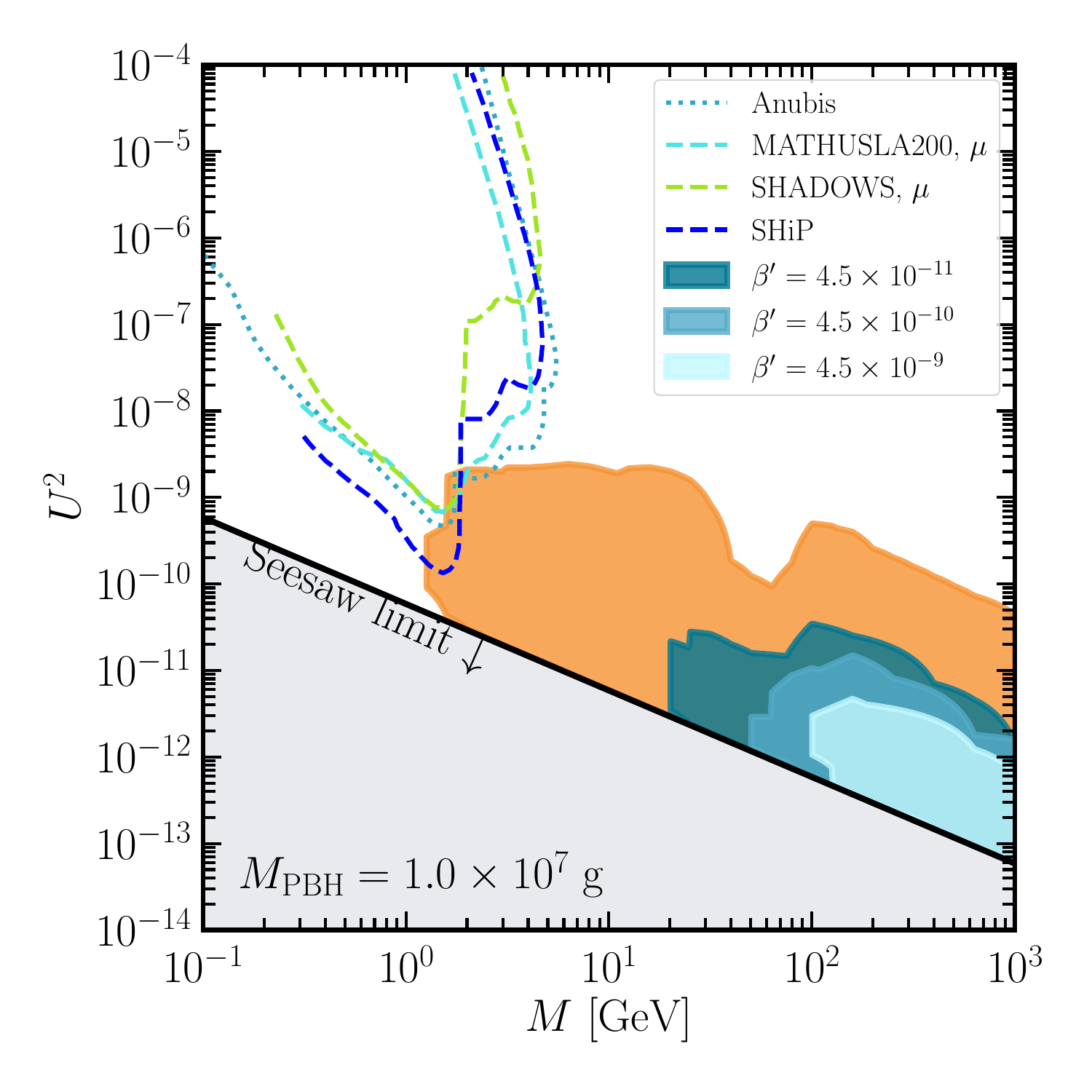}
        \hspace{0.05\textwidth}
        \includegraphics[width = 0.45\textwidth]{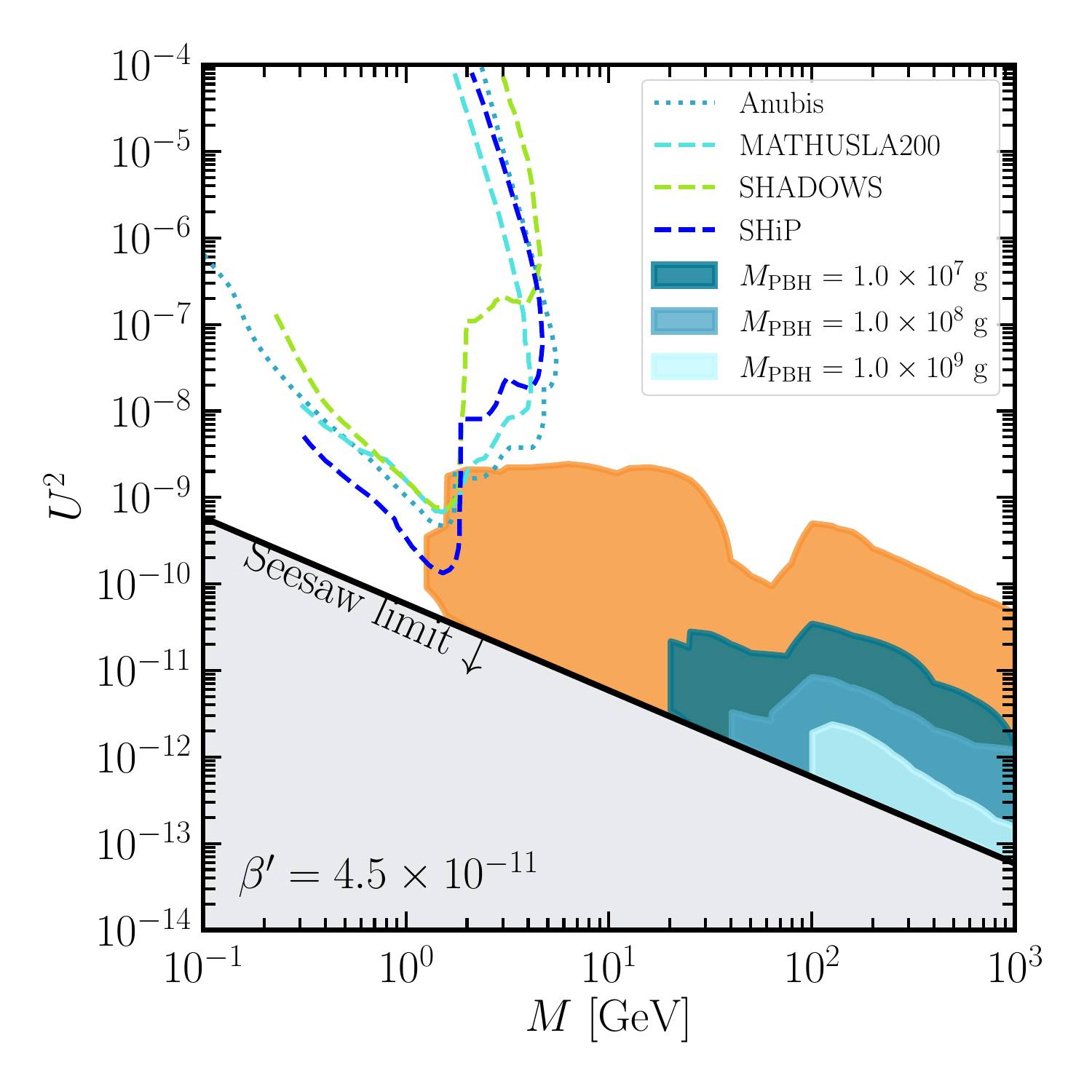}
    \caption{\label{PBHLepto/LowScale/MPBHbeta} In the left (right) panel, the effect of PBHs on the viable parameter space for resonant leptogenesis is shown fixing $M_{\rm PBH} = 1.0\times 10^7~{\rm GeV}$ ($\beta^\prime = 4.5 \times 10^{-11}$) and varying $\beta^\prime$ ($\MPBHini$). The orange region is the standard result without PBHs already discussed in \figref{PBHLepto/LowScale/standard}. For comparison the projected constraints from the SHiP \cite{SHiP:2018xqw}, SHADOWS \cite{Baldini:2021hfw}, MATHUSLA \cite{MATHUSLA:2022sze} and ANUBIS \cite{Hirsch:2020klk} experiments are shown.}
\end{figure*}

In \figref{PBHLepto/LowScale/MPBHbeta} the effects of PBHs on the viable leptogenesis parameter space are demonstrated for different initial abundances (left panel) and different initial masses (right panel) of the PBHs. Each region is obtained as the union of those resulting from considering each of the same $\Delta M$ as in \figref{PBHLepto/LowScale/standard}. Hence, the effect of PBHs is to shrink the allowed region towards higher masses $\Mdegen$ and smaller mixing parameters $U^2$. Also reported in \figref{PBHLepto/LowScale/MPBHbeta} are the projected sensitivity curves for the planned SHiP \cite{SHiP:2018xqw}, SHADOWS \cite{Baldini:2021hfw},MATHUSLA \cite{MATHUSLA:2022sze} and ANUBIS \cite{Hirsch:2020klk} experiments. While these experiments can probe the very edges of resonant leptogenesis parameter space, much larger regions would be ruled out by even the tiny populations of PBHs considered. One might imagine some future evidence for ultralight PBHs returned by GW searches (see \secref{PBHs/Constraints}), then knowing $\MPBHini$ and $\beta^\prime$ the viable parameter space for resonant leptogenesis can be found as in \figref{PBHLepto/LowScale/MPBHbeta}. From the figure it is clear that even for initial abundances as small as $10^{-11}$, PBHs would rule out a region of leptogenesis parameter space comfortably beyond the reach of experiments sensitive to RHNs. \\
\begin{figure}[h!]
    \centering
        \includegraphics[height= 0.8\linewidth]{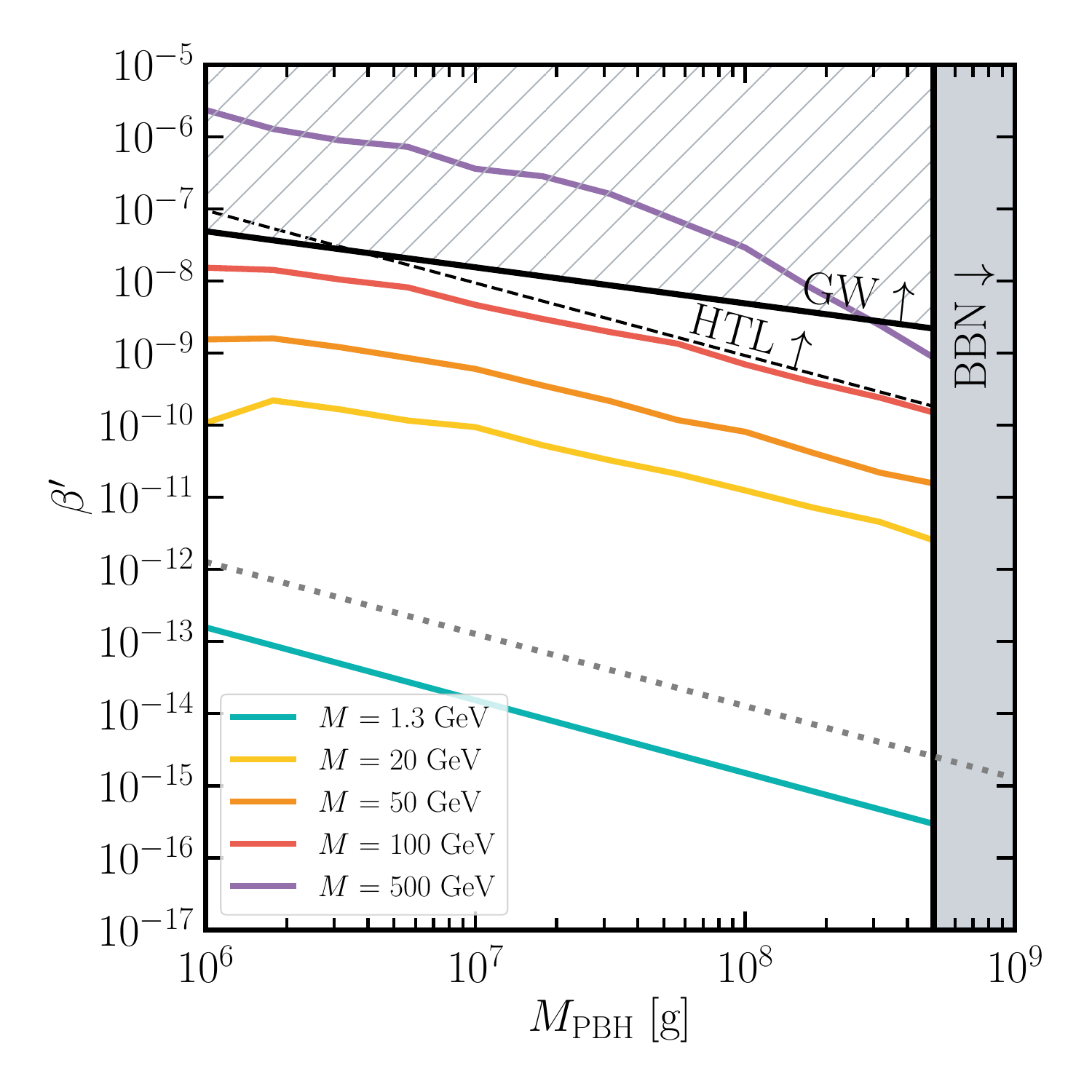}\,\,
        \caption{\label{PBHLepto/LowScale/constraints} Upper bounds on the initial abundance of PBHs $\beta^\prime$ as a function of the PBH mass $\MPBHini$ for different values of the RHN mass scale $\Mdegen$. The hatched region is excluded by the constraints from GW energy density, while the gray region is excluded by the impact of PBH evaporation on BBN. For comparison, the dashed line bounds from below the region excluded with respect to high scale thermal leptogenesis, see \secref{PBHLepto/High Scale}. The gray dotted line bounds from below the region where PBHs dominate the energy budget before evaporating.}
\end{figure}

In \figref{PBHLepto/LowScale/constraints} the PBH-resonant leptogenesis mutual exclusion limits are displayed, showing the regions of PBH parameter space which are incompatible with sub-TeV resonant leptogenesis for different RHN mass scales. The smaller the mass scale $\Mdegen$, the stronger the constraints on $\beta^\prime$. In the same plot, the limits from BBN (shaded gray region) and from the energy density of primordial GWs~\cite{Papanikolaou:2020qtd,Domenech:2020ssp,Papanikolaou:2022chm} (hatched region) are shown. The most conservative constraint placed on PBHs from resonant leptogenesis is the purple line in \figref{PBHLepto/LowScale/constraints}, such that any population of PBHs residing above this constraint is incompatible with resonant leptogenesis for all $\Mdegen \leq 500\GeV$. It is interesting that the most abundant populations of PBHs constrain resonant leptogenesis for a greater range in $\Mdegen$. \\

Near future direct detection experiments may hope to be sensitive to RHNs with $\Mdegen \sim \GeV$ (see either panel of \figref{PBHLepto/LowScale/MPBHbeta}). There is a very small region of overlap with the successful resonant leptogenesis parameter space, underscoring the extreme difficulty of detecting RHNs with such small couplings. However, any evidence which may be forthcoming in these experiments or future generation experiments could be translated directly into constraints on PBHs according to the results presented in \figref{PBHLepto/LowScale/constraints}. The constraints which would arise from the smallest, most experimentally accessible mass scales are the strongest and considerably more severe than the best current GW constraints. It should be understood that a future detection of RHNs in the successful resonant leptogenesis parameter space does not necessarily lead to the conclusion that resonant leptogensis occurred and produced the BAU. More evidence would be required to support this conclusion not least to distinguish between detecting a single state and 2 mass degenerate states.

%============================
\subsubsection{Conclusions \label{PBHLepto/LowScale/conclusions}}
%============================

This section investigated the impact of PBHs with masses from $10^6$ to $\sim10^9$~g on the predictions of sub-TeV resonant leptogenesis. Differently from case of high-scale leptogenesis discussed in the previous section \secref{PBHLepto/High Scale}, in the framework of sub-TeV resonant leptogenesis the observed BAU is generated by (at least) two almost-degenerate RHNs with masses from 1~GeV to a few TeVs. This makes resonant leptogenesis experimentally appealing since it is potentially accessible to upcoming experiments. The coupled Boltzmann equations for the number densities of the RHNs and the baryon asymmetry were solved with TIA, consistently taking into account the non-instantaneous sphaleron freeze-out and the correction due to PBH domination, thermal corrections to rates and particle masses, and flavour effects. The region in the neutrino parameter space in which resonant leptogenesis is successful was therefore derived, considering a range of possible mass splittings for the nearly degenerate RHNs.\\

The impact on resonant leptogenesis of the post-sphaleron evaporation of PBHs was then studied by solving the Friedmann equations describing their evolution (see \equaref{PBHs/Cosmology/Friedmann}). Interestingly it was shown that PBHs shrink the allowed neutrino parameter space towards heavier $\Mdegen$ and smaller active-sterile mixing, with even tiny populations of PBHs significantly restricting the parameter space for leptogenesis (see \figref{PBHLepto/LowScale/MPBHbeta}). Mutual exclusion limits were consequently derived and projected into the PBH parameter space (see \figref{PBHLepto/LowScale/constraints}), characterising the regimes in which PBHs and resonant leptogenesis are incompatible. 
Resonant leptogenesis in the experimentally accessible parameter space can be ruled out by even very small populations of PBHs. Larger populations would have the capacity to rule out resonant leptogenesis up to the $\Mdegen \sim 10^2\GeV$ scale, which is simply out of the question for RHN detection experiments. Therefore in anticipation of near future GW experiments having the capacity to constrain (or detect) ultralight PBHs, it has been shown that detecting their GW signatures would have important implications for RHN searches and leptogenesis. Conversely it follows that evidence for RHNs around $\Mdegen ~ \GeV$ (which is possible in marginal successful-resonant-leptogenesis parameter space) has the potential to rule out PBH populations. Remarkably, the lightest RHNs would confer the tightest constraints on PBHs. It is important to note the caveat that for RHNs to constrain PBHs, highly mass degenerate states would have to be resolved and separately shown to have been responsible for resonant leptogenesis in the early universe.
\label{PBHLepto/Low Scale}
\newpage

\section{Cosmological Phenomenology of Hot Spots}\label{Hot Spots}

In the previous two sections it was shown that Hawking radiation impacts leptogenesis so strongly as to potentially exclude it entirely, and the resulting incompatibility between ultralight PBHs and leptogenesis was leveraged to derive mutual exclusion limits in the combined parameter space. In particular, the injection of entropy (primarily in the form of photons) by PBHs into the background universe suppresses existing asymmetries there. This process was assumed to be homogeneous throughout the universe, with cosmological abundances, observables and temperatures described by a single value everywhere simultaneously. Such is the standard treatment of PBH physics in the early universe. 

However, recent studies have shown that Hawking radiation heats the primordial plasma around PBHs in a highly non-uniform manner, producing localised hot spots with dramatic temperature gradients that can exceed the average plasma temperature by some orders of magnitude (see \secref{PBHs/Hot Spots}). This section delves into the impact of PBH hot spots on out-of-equilibrium dynamics, establishing a formalism for the propagation of Hawking radiation through hot-spots, then revisiting the PBH-leptogenesis scenario discussed in the previous sections and further considering the WIMP DM scenario when DM is produced by PBHs.

For DM, if the hot spot's temperature exceeds the freeze-out temperature, DM particles may scatter off the heated plasma within the hot spot and thermalise, thereby erasing their contribution to the relic abundance. Only DM which escapes the hot spot contributes to the relic abundance. In contrast, if the hot spot's temperature exceeds the sphaleron freeze-out temperature for leptogenesis, only right-handed neutrinos that decay within the hot spot would contribute to the observed BAU. Those RHNs that escape the hot spots seed a leptonic asymmetry that is never converted to BAU.
\newpage

\mypapertitle{Primordial Black Hole Hot Spots
and Out-of-Equilibrium Dynamics}
\mypaperdate{Sep 16, 2024}
\mypaperabstract{When light primordial black holes (PBHs) evaporate in the early Universe, they locally reheat the surrounding plasma, creating hot spots with temperatures that can be significantly higher than the average plasma temperature. In this section, a general framework for calculating the probability that a particle interacting with the Standard Model can escape the hot spot is provided. More specifically, it is considered how these hot spots influence the generation of the baryon asymmetry of the Universe (BAU) in leptogenesis scenarios, as well as the production of dark matter (DM). For leptogenesis, PBH-produced right-handed neutrinos can contribute to the BAU even if the temperature of the Universe is below the electroweak phase transition temperature, since sphaleron processes may still be active within the hot spot. For DM, particles emitted by PBHs may thermalise with the heated plasma within the hot spot, effectively preventing them from contributing to the observed relic abundance. These effefts highlight the importance of including hot spots in the interplay of PBHs and early Universe observables.}
\subsection*{\textit{arXiv/2409.02173 - accepted for publication at JCAP}}
\makepapertitle 
\newpage

Here the impact of PBH hot spots on non-equilibrium processes is examined, specifically leptogenesis and DM generation via freeze-out, as illustrated schematically in \figref{Hot Spots/Intro/schematic}. Interestingly, the presence of the hot spot influences these scenarios differently. This work is organized as follows. First the evolution of hot spots in an expanding universe is studied in \secref{Hot Spots/Evolution}. A general formalism used to calculate the probability for a PBH-produced gauge singlet particle to escape a hot spot is established in \secref{Hot Spots/Hawking}. In \secref{Hot Spots/Lepto} the general formalism is applied to low scale leptogenesis, demonstrating that the BAU can be produced by RHNs inside a hot spot even after sphalerons have frozen out in the background universe. \secref{Hot Spots/DM} extends the considerations to a simple WIMP DM model, where the scattering of PBH-produced DM by the hot spot thermalises the DM and prevents the overclosing of the universe. Conclusions are drawn in \secref{Hot Spots/Conclusions}.

\begin{figure}[h]
 \centering
 \includegraphics[width=\linewidth]{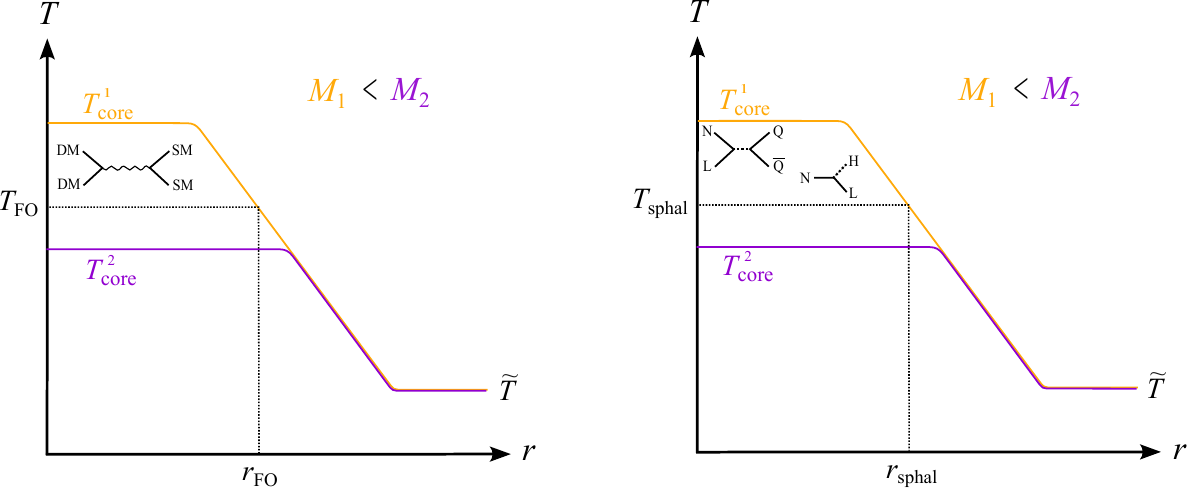}
 \caption{\label{Hot Spots/Intro/schematic} Schematic illustration of the effects of hot spots around PBHs on DM production and leptogenesis for two different PBH masses, $\MPBH^1$ (orange) and $\MPBH^2$ (purple), with $\MPBH^1<\MPBH^2$. Left: DM production. The radius $\rFO$ represents the distance within the hot spot where the DM freeze-out temperature $\TFO$ occurs. DM particles will thermalise for distances $r < \rFO$. For the heavier PBH mass $\MPBH^2$, DM can escape the hot spot and contribute to the relic abundance since the plasma temperature is below $\TFO$ everywhere. Right: Leptogenesis. The radius $\rsphal$ indicates the region where sphalerons remain active, with $T > \Tsphal$. Right-handed neutrinos decaying within $r < \rsphal$ will seed additional baryon asymmetry.}
\end{figure}

\subsection{Hot Spots in a Expanding Universe}\label{Hot Spots/Evolution}\label{Hot Spots/Evolution}
This section extends the work of \cite{He:2022wwy,He:2024wvt}, reviewed in \secref{PBHs/Hot Spots}, to account for the expansion of the background universe around a PBH and build a picture of how hot spots form and evolve throughout the PBH lifetime.\\

The temperature of the homogeneous core of the hot spot is given by \equaref{PBHs/Hot Spots/Tcore} and is always about 4 orders of magnitude below the Hawking temperature $\TBH$.
Therefore after the PBH starts to evaporate (when $T = \TBH$) the entire Universe expands as normal (homogeneously) for another four decades in the scale factor. During this time, the energy radiated by the PBH is insufficient to heat the region $r \leq \rcore$, because the ambient plasma temperature is hotter than $\Tcore$. The hot spot can be said to form at the moment 
\begin{equation}\label{Hot Spots/Evolution/Tform}
    \Tplasma(\alpha_{HS}) = \Tcore(\MPBHini,\alpha_{HS})\,,
\end{equation}
since from this moment the PBH is radiating enough energy to heat its surroundings. It is clear then, that the initial temperature of the hot spot core is exactly the background temperature at the moment of formation, $\Tplasma(\alpha_{HS})$. Since in the early phase of the PBH life, its mass is quasi-static, the temperature of the core formed by evaporation is almost constant too, $\Tcore(\alpha) \approx \Tcore(\alpha_{HS})$ and $\rcore(\alpha) \approx \rcore(\alpha_{HS})$. In the background universe however, the expansion of the universe is cooling the radiation bath and $\Tplasma$ continues to drop with the scale factor. The hot spot profile outside the core therefore extends further and further, the local temperature falling as $r^{-1/3}$ until $\Tplasma$ is reached.
\begin{figure}[h]
 \centering
 \includegraphics[width = 0.7\textwidth]{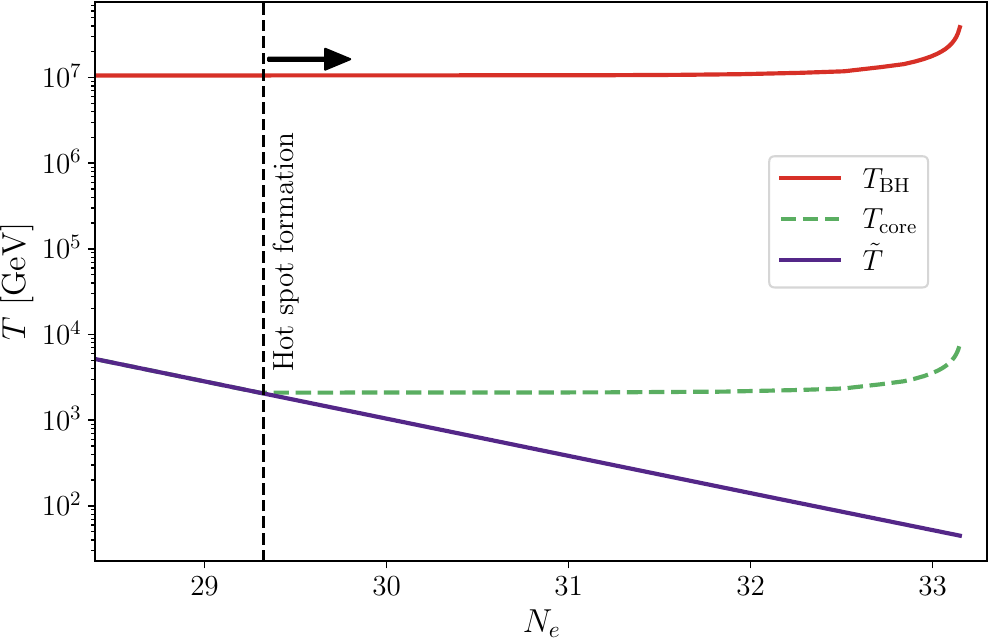}\label{PBHs/Hot Spots/Tcomp}
 \caption{The relevant temperatures involved in the formation and evolution of a PBH hot spot around a PBH with mass $\MPBH = 10^6$g are compared. The purple solid line indicates the background plasma temperature, the red solid line indicates the Hawking temperature of the BH and the green dashed line shows the temperature of the hot spot, diverging in temperature from the background at a well-defined moment. In this plot, $N_e \equiv \ln(a)$ is the natural logarithm of the scale factor.}
\end{figure}

The plasma, hot spot and Hawking temperatures are shown in \figref{PBHs/Hot Spots/Tcomp} in the case where $\MPBHini = 10^{6}$ g. The solid purple line shows the evolution of $\Tplasma$ which is the temperature far enough from the PBH to be unaffected. The Hawking temperature, given by \equaref{PBHs/Evaporation/TBH}, is shown by the solid red line and the dashed green line shows the hot spot temperature. Initially the plasma temperature is above the core temperature and no hot spot exists. The hot spot develops when $T(\alpha_{HS}) = \Tcore(\MPBHini,\alpha_{HS})$, and the region $r \leq \rcore$ remains at an approximately constant temperature for as long as the PBH mass is quasi-static also. The core of the hot spot is maintained at roughly the temperature of the universe when the hot spot formed, while the background universe cools around it. The radius of the hot spot can be defined through
\begin{equation}
T(\rHS) \equiv \Tplasma
\end{equation}
such that $\rHS$ is the maximum extent of the hot spot temperature profile before the background temperature is reached. Initially it is always the case that $\rHS \ll \rdec$, the entire hot spot profile is in thermal contact with the PBH and so scales like $r^{-1/3}$ outside the core. Towards the end of the PBH lifetime, the rate of mass loss sharply increases and consequently the hot spot temperature also rises. An increasing rate of evaporation means diffusion may no longer be efficient on large length scales (see \equaref{PBHs/Hot Spots/rdec}) and at some late time $\rdec \leq \rHS$. Following this time, radii $\rdec \leq r \leq \rHS$ are frozen out of thermal contact with the PBH and a new scaling emerges of $T(r) \propto r^{-7/11}$. At the very end of the PBH life, $\Tcore \to \Tmax$ (see \equaref{PBHs/Hot Spots/Tmax}), $\rdec \to \rcore$ and the entire profile will scale as $r^{-7/11}$ outside the core. \\

\begin{figure}[h]
    \centering
    \includegraphics[width=0.8\linewidth]{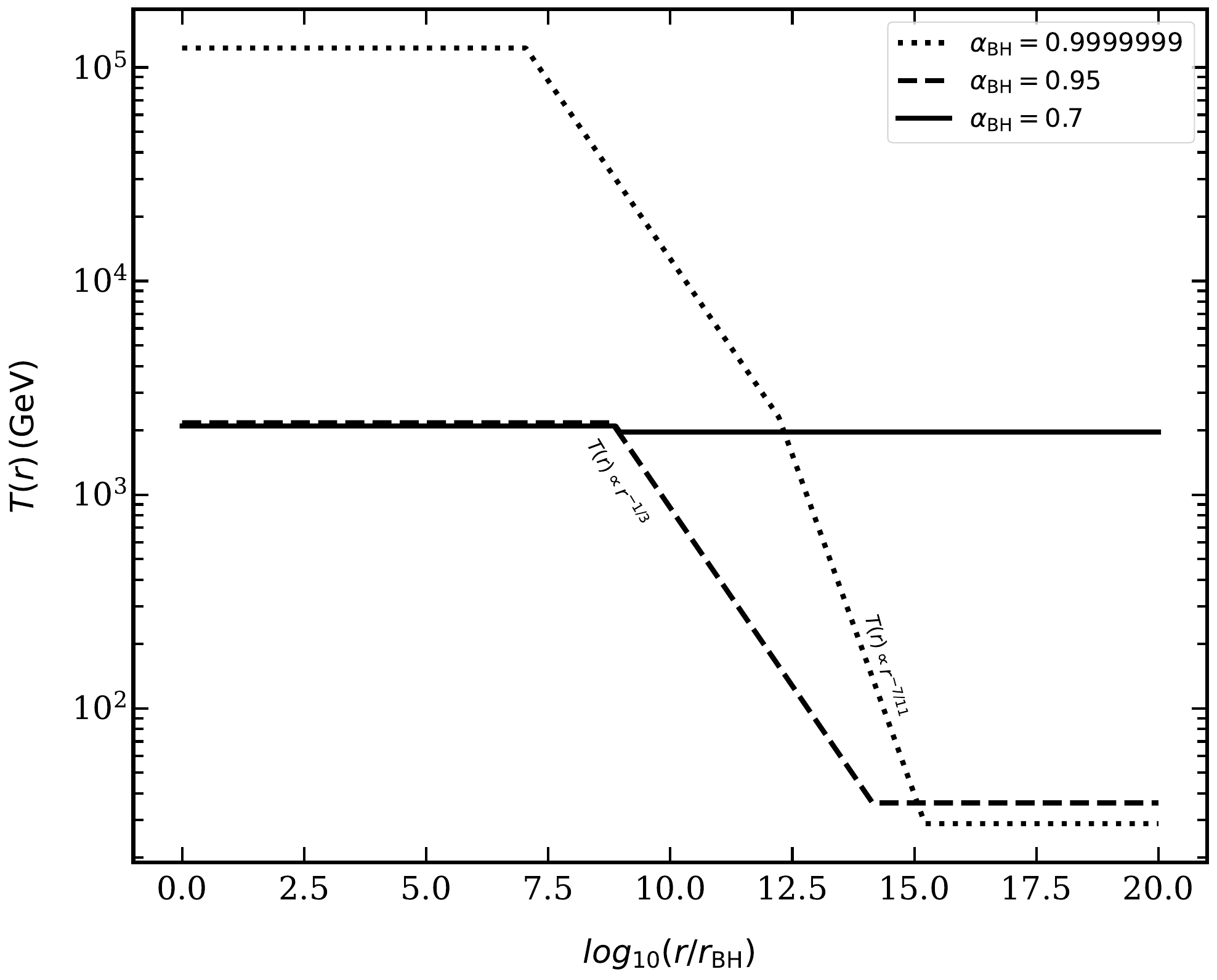}
    \caption{The hot spot profile around a PBH with $\MPBHini = 10^6\g$ is shown at three different moments in the PBH life. Here, $\alpha_{\rm BH} = \alpha/\alpha_{\rm evap}$. The solid line shows the PBH very soon after hot spot formation. The dashed line shows the hot spot profile at a later time, but early enough in the PBH life that $\MPBH \approx \MPBHini$. The dotted line shows the PBH at a late stage in its life, where now $\MPBH \gg \MPBHini$. }
    \label{Hot Spots/Evolution/profile}
\end{figure}
The evolution of a PBH hot spot is depicted in \figref{Hot Spots/Evolution/profile}, where the solid line shows the PBH very soon after hot spot formation, the plasma temperature has barely cooled at all following formation so there is barely any temperature gradient present. The dashed line shows the hot spot profile at a later time, but early enough in the PBH life that $\MPBH \approx \MPBHini$ and so the core temperature has barely changed. Between the early time and this intermediate time, the background universe has cooled by around two orders of magnitude revealing an extended profile where $T(r) \propto r^{-1/3}$. At this intermediate time it is still the case that $\rdec \gg \rHS$, so only the $r^{-1/3}$ scaling is present. The dotted line shows the PBH at a late stage in its life, where now $\MPBH \ll \MPBHini$. Note that since the time $\alpha_{\rm BH} = 0.95$ (dashed line), the background universe has barely cooled at all but the core temperature of the hot spot has heated significantly. At this late time, $\rdec \leq \rHS$ and so a region where $T(r) \propto r^{-7/11}$ appears at the outer reaches of the profile.
\newpage

\subsection{Can Hawking Radiation Escape Hot Spots?}\label{Hot Spots/Hawking}
\noindent In this section, the propagation of a generic Hawking radiation particle, $X$, travelling through the non-homogeneous local hot spot which has a temperature profile given by \equaref{eq:tprof} is considered. The mean free path for particle $X$ to deposit energy into the plasma via $1\to 2$ and $2 \to 2$ processes is derived and the likelihood of $X$ to escape the hot spot calculated. Hawking radiation travels radially outward with average momentum $\langle \vec{p} \rangle \sim \TBH \gg T(r)$ once the hot spot has formed. Eventually interactions with the surrounding plasma cause the boosted Hawking radiation to thermalise. As discussed in \secref{Early Universe/TFT} and \secref{PBHs/Hot Spots}, the dominant mechanism of energy deposition is the LPM-suppressed collinear splittings for any particles with gauge couplings. However, PBHs radiate into all available degrees of freedom. One may anticipate that particles with no or very small gauge couplings may be expected to free-stream out of the hot spot while relatively strongly interacting particles may be trapped within the hot spot. The probability for such a gauge singlet particle to decay or scatter within the hot spot is considered in the following. \\

In a homogeneous environment, the mean free path of a particle $X$ is given by
\begin{equation}
\lambda = \frac{1}{\Gamma_X}\,, \end{equation}
where $\Gamma_X$ is the total interaction rate of $X$ with the thermal bath, summing all allowed diagrams. One can see immediately that the above definition of $\lambda$ is insufficient in the case of Hawking radiation travelling through a hot spot since the interaction rate varies with the local temperature and density of the surrounding plasma
\begin{equation}
\lambda\left(T(r)\right) = \frac{1}{\Gamma_X\left(T(r)\right)}\,.
\end{equation}
It is clear that a physically meaningful definition of the mean free path should not depend on $r$. To address this, the probability for $X$ to reach a radius $r$ before interacting can be expressed
\begin{equation}\label{Hot Spots/P}
P(r) = e^{-\int^r_0\Gamma_X(r^\prime) \mathrm{d}r^\prime }\,,
\end{equation}
such that by setting
\begin{equation}
 P\left( \lambda \right) \equiv \frac{1}{2}\,,
\end{equation}
the mean free path in a hot spot can be understood as the radius beyond which it is more likely than not that $X$ will have undergone some interaction. This definition of $\lambda$ permits quantitative estimates of the average distance travelled by Hawking radiation for a specific hot spot profile $T(r)$, which is calculated as in \ref{PBHs/Hot Spots/profile}. However, since the profile of any hot spot evolves with $\alpha$, this measure does not provide a complete picture at all times. In particular, it will be of interest to calculate the comoving number density of $X$ produced by a PBH which escapes ($\mathcal{N}_X^{\rm escape}(r)$) to or is trapped within ($\mathcal{N}_X^{\rm trap}(r)$) radius $r$ across the entire hot spot lifetime. These quantities are found by solving the following set of equations
\begin{equation}\label{Hot Spots/Hawking/Nescape}
\frac{d \mathcal{N}_X^{\rm escape}(r)}{d\alpha} = \ln(10)\frac{\Gamma_{\rm{PBH} \to X}}{H}\, P\left( r \right)\,\,\,,
\end{equation}
\begin{equation}\label{Hot Spots/Hawking/Ntrapped}
\frac{d \mathcal{N}_X^{\rm trap}(r)}{d\alpha } = \ln(10)\frac{\Gamma_{\rm{PBH} \to X}}{H}\,(1 - P\left( r \right) )\,\,\,,
\end{equation}
where $\Gamma_{\rm{PBH} \to X}$  is the instantaneous rate of production of particle $X$ by a PBH, found by integrating \equaref{PBHs/Evaporation/dn} with respect to the particle energy. Integrating \equaref{Hot Spots/Hawking/Nescape} (\equaref{Hot Spots/Hawking/Ntrapped}) between the formation time of the hot spot, $\alphaform$, and evaporation time of the black hole, $\alphaevap$, captures the total number density of $X$ that escapes to (is trapped within) $r$ over the hot spot lifetime \footnote{The tiny amount of $X$ produced before the hot spot forms is neglected.}. \\

The total interaction rate appearing in $P(r)$, $\Gamma_X(r)$, is a complicated function of $r$ because $X$ is moving through inhomogeneous plasma in which the temperature $T(r)$, particle comoving densities $\mathcal{N}(r)$, cross sections $\sigma(r)$, and degrees of freedom $g_*(r)$, all depend on the radial coordinate, $r$. $\Gamma_X(r)$ is given by
\begin{equation}
\Gamma_X(r) = \Gamma_S(r) + \Gamma_D(r)\,, \end{equation}
where $\Gamma_S$ sums all scattering diagrams of $X$ with particles in the plasma while $\Gamma_D$ describes all available decay modes of $X$.

\subsubsection*{Decays in hot spots}
\noindent Consider that $X$ is a generic particle with mass $m_X$, which can undergo a two-body decay into particles $Y$ and $Z$. While \secref{Hot Spots/Lepto} will focus on the specific case of the CP asymmetric decay of RHNs, the discussion here remains general. The particle $X$ is produced via Hawking radiation, assuming $\TBH \gtrsim m_X$, following an approximate blackbody spectrum. If $\TBH \gg m_{X}$, the $X$ particles can be highly boosted. Consequently the vacuum decay rate $\Gamma_{X \to YZ}$, must be adjusted for time dilation, thermally averaged over the BH’s instantaneous spectrum, and it is given by (see \secref{Leptogenesis/Out-of-Equilibrium} for a derivation)
\begin{equation}\label{Hot Spots/Hawking/GXtd}
\Gamma_{X \to YZ} \left< \frac{m_{X}}{\TBH} \right>_{\rm BH} \approx \Gamma_{X \to YZ}\frac{K_1(z_{\rm BH})}{K_2(z_{\rm BH})}\,,
\end{equation}
where $z_{\rm BH} = m_{X}/\TBH$ and Maxwell-Boltzmann statistics have been assumed, taking $f(E) = e^{-E_X/T_X}$. In the results which follow, the analytical approximation \equaref{Hot Spots/Hawking/GXtd} is not used; instead, a full numerical solution that averages over the Hawking spectrum and accounts for the quantum statistics of the decaying particles is used.\\

The vacuum decay rate $\Gamma_{X \to YZ}$ is kinematically unsuppressed if the mass of $X$ is greater than or equal to the sum of the masses of the final states. If the bare masses of any of these particles are small relative to the temperature of the thermal bath, thermal corrections on the order of $M^2 \sim \fiducial T^2$ will be significant (see \secref{Early Universe/TFT}). Since the temperature in the hot spot varies radially, the thermal masses of the particles will also vary accordingly. Therefore at radius $r$ the decay is kinematically accessible if the condition $M_X(r) \geq M_Y(r) + M_Z(r)$ is satisfied, where capital $M$ indicates the thermally corrected mass. Where the kinematics vary radially, the full decay rate for $X$ in a hot spot should be given by
\begin{equation}\label{GammaD}
\Gamma_D(r) = \left< \frac{m_{X}}{\TBH} \right>_{\rm BH} \Gamma_X^T(r) \,,
\end{equation}
where $\Gamma_X^T(r)$ is the rest frame decay width of $X$ including any radially-varying thermal corrections.

\subsubsection{Scattering in hot spots}
\noindent Hawking radiation may also scatter off the plasma as it travels through the hot spot. Considering only $2\to 2$ processes, the total rate of scattering of a PBH-emitted particle, $X$, with temperature $\TBH$ and a particle, $j$, drawn from the hot spot thermal bath with temperature $T(r)$ and mass $m_{j}$ is given by
\begin{equation}\label{Hot Spots/Hawking/GammaS}
\Gamma_S = \sum_\sigma n_j\langle \sigma \cdot v_{\text{Mol}} \rangle_{\TBH T(r)} \,,
\end{equation}
where the summation is over all allowed scattering diagrams, $\langle \sigma \cdot v_{Mol} \rangle_{\TBH T(r)}$ indicates the thermally averaged cross-section for the process $X,j \to k,l$, and $n_j(r)$ is the local number density of the particle species off which $X$ scatters. Further, it is assumed that particle $j$ is in thermal equilibrium within the hot spot. Note that the local density of particle $j$ in the hot spot, $n_j(r)$, will therefore vary as a function of radial distance since the local temperature varies. For instance, the hot spot temperature is largest within the ``core'', so the number density of SM particles in the core will be larger than in the cooler regions, $r> \rcore$. In general, $M_{X}\neq M_{j}$ and $\TBH \neq T(r)$ so that the cross-section must therefore be thermally averaged with respect to the two (in general different) temperatures and masses of the incoming states. In Appendix \ref{A}, the exact form of this thermal averaging factor is derived. The result is
\begin{equation}\label{Hot Spots/Hawking/thermalAverage}
    \langle \sigma \cdot v_{\text{Mol}} \rangle_{T_{\rm BH}T(r)} = B \int^{\infty}_{s_\text{low}} \sigma(s) F \left ( \frac{C_1}{D}e^{-x_+^{\rm min}}(1 + x_+^{\rm min}) + \frac{\sqrt{C_2}}{\sqrt{D}}K_1 \left( \frac{\sqrt{D}}{T_{\rm BH}T(r)} \right) \right) \, ds\,.
\end{equation}
Since the specific hot spot profile, as well as the PBH temperature are functions of the scale factor, the thermal averaging of the cross sections evolves with $\alpha$ in \equaref{Hot Spots/Hawking/Nescape}. Note that in the limit $m_X = m_j$, this result reduces to that found in \cite{Cheek:2021cfe}, which in turn is equal to the results in \cite{Gondolo:1990dk} if one sets $T(r) = \TBH$.

\newpage

\subsection{Leptogenesis after Sphaleron Freeze-Out}
In this section, the considerations of the hot spot effect on Hawking radiation made in the previous section are applied to the low scale model of thermal resonant leptogenesis introduced in \secref{Leptogenesis/Resonant}.
In \secref{PBHLepto/Low Scale} when calculating the impact of PBHs on resonant leptogenesis, the contribution of RHNs produced by PBHs to $\YB$ was ignored since for $\MPBHini \gtrsim 10^{5.5}$g, the temperature of the background Universe is below $\Tsphal$ by the end of the PBH lifetime. Therefore, in principle, PBHs heavier than this value cannot produce any baryon asymmetry in the background through the production of RHNs because the sphaleron processes are exponentially suppressed. The most important effect of such heavy PBHs was shown to be to washout existing asymmetry through the production of a large number of photons. However inside a hot spot, sphalerons remain active where the local temperature $T(r)$ is heated above $\Tsphal$ by the thermalisation of the evaporation products of the PBH\footnote{Note that BHs act as seeds of baryon number violation, altering the sphaleron rate, when the Schwarzschild radius is of the same order as the electroweak scale, i.e., $\MPBHini\sim 10^{11}~{\rm g}$~\cite{DeLuca:2021oer}. However, for the BH masses that we are interested in, this effect is negligible.}. In this region, a leptonic asymmetry would continue to be converted into baryon asymmetry. The treatment in \secref{PBHLepto/Low Scale} therefore is equivalent to assuming that all PBH produced RHNs easily escape the hot spot and free stream into the background universe where sphalerons are inactive.\\

This assumption is improved upon in this section by carefully accounting for the effect of hot spots on the RHNs produced by PBHs. The conversion of lepton asymmetry into baryon asymmetry by sphalerons is assumed to be instantaneous. Further, it has been numerically verified that the rate of washout processes that would erase the lepton asymmetry is always much smaller than the rate of outward diffusion within the hot spot. This follows from the relatively small Yukawa couplings of leptons to the RHNs which control the washout, while the outward diffusion proceeds by the summation of all other possible diagrams. Therefore washout occurring in the hot spot can be neglected. Excess (anti)leptons produced as RHN decay products in the hot spot are almost certain to diffuse out of the hot spot via $\Lnum$ conserving diffusion before they scatter in a way as to erase the asymmetry. In this way, baryonic asymmetry produced in the hot-spot is protected since there exists no baryonic washout processes anywhere. Therefore, when RHNs produced by a PBH travel radially through its hot spot, their decays produce a lepton asymmetry which is converted into baryon asymmetry only if the decay occurs at $r \leq \rsphal$. The length scale $\rsphal$ is defined such that
\begin{equation}
 T(\rsphal) = \Tsphal\,,
\end{equation}
so that the number density of RHNs which decay within $r \leq \rsphal$ is $\mathcal{N}_{N}^{\rm trap}(\rsphal)$. 
In \secref{PBHLepto/High Scale} it was shown that a period of PBH domination can alter $\Tsphal$ in the background, in this analysis the freeze-out of sphalerons is local to each hot spot, so $\Tsphal \approx 130$GeV is fixed \cite{DOnofrio:2014rug}. By integrating \equaref{Hot Spots/Hawking/Nescape} over the PBH lifetime, with $\Gamma(r)$ the total interaction rate for the RHNs, one can calculate $\mathcal{N}_{N}^{\rm trap}(\rsphal)$ and predict the corresponding contribution to $\YB$.

$\Gamma_S(r)$ is the total scattering rate for $N_i$, given by \equaref{Hot Spots/Hawking/GammaS} where the sum over $\sigma$ includes all possible $2\to 2$ diagrams with $N_i$ in the initial state. Using the expressions given in \cite{Pilaftsis:2003gt} which include thermal corrections, it turns out $\Gamma_S \ll \Gamma_D$ inside the hot spot and therefore it is safe to neglect scattering such that $\Gamma(r) \approx \Gamma_D(r)$. $\Gamma_D(r)$ is the decay rate of the RHNs as a function of the radial distance from the PBH which is given by (see \secref{Leptogenesis/Thermal})
\begin{equation} 
 \Gamma_D(r) = \sum_{i = 2,3} \left< \frac{M_{N_i}}{\TBH} \right>_{\rm BH} \GNiT(r)
\end{equation}
where $\GNiT(r)$ is the thermally corrected rest frame decay rate of $N_i$, given by \equaref{Leptogenesis/Out-of-Equilibrium/GNifull}. Depending on the local plasma temperature, $T(r)$, the lepton and Higgs thermal masses may suppress the decay rate of $N_i\to \lep \Higgs$. For the decay $N_i \to \lep \Higgs$ to be kinematically accessible at $r \leq \rsphal$, the condition $\MNi \geq \mHiggsT + \mLepT \approx 0.1 \Tsphal$ must be fulfilled, where $0.1$ is taken as a fiducial value representing the SM gauge couplings.

The following points in neutrino parameter space, which are testable at the ILC and CEPC experiments \cite{Yang:2023ice}, 
\begin{equation}\label{Lepto:BM}
\begin{aligned}
 U^2 &= 10^{-5} \, , \, \Mdegen = 50\,\rm{GeV}\,, \\
 U^2 &= 10^{-7} \, , \, \Mdegen = 100\,\rm{GeV}\,,\\
 U^2 &= 10^{-4} \, , \, \Mdegen = 150\,\rm{GeV}\,.
\end{aligned}
\end{equation}
are chosen to predict the region in PBH parameter space for which successful leptogenesis can occur in hot spots. Due to strong washout effects, these benchmark points are also incompatible with successful thermal leptogenesis in the two right-handed neutrino scenario. The resulting yield of baryon asymmetry can be estimated by solving the differential equation 
\begin{eqnarray}
\frac{d|\YB|}{d \alpha} \approx \eta_{\rm sphal}\frac{|\epsilon|}{\mathcal{S}} \frac{d\mathcal{N}_{N}^{\rm trap}(\rsphal)}{d \alpha}\NPBH\,,
\end{eqnarray}
where $d\mathcal{N}_{N}^{\rm trap}(\rsphal)/d \alpha$ is given by \equaref{Hot Spots/Hawking/Ntrapped}, $\epsilon = \epsilon_2 + \epsilon_3$, with $\epsilon_i$ the CP violation parameter for $N_i$ summed over all flavours, given by \equaref{Leptogenesis/Resonant/epsi}. As discussed earlier, the washout processes within the hot spot are inefficient compared to the outward diffusion and do not erase $\YB$. This conclusion is not contradicted by the fact that for the points in neutrino parameter space considered, washout is too strong for purely thermal leptogenesis to be successful. Leptons and antileptons produced as RHN decay products at $r \leq \rsphal$ scatter elastically much more efficiently than is provided by washout processes and are quickly moved outwards towards $r \geq \rsphal$. Once the leptons pass $r=\rsphal$ it is irrelevant whether they are washed out or not since the baryon asymmetry they generated via sphalerons is protected by the $B$ conserving universe at $T \leq \Tsphal$.

%%%%%%%%%%%%%%
\begin{figure}[h] 
 \includegraphics[width = 0.49\textwidth]{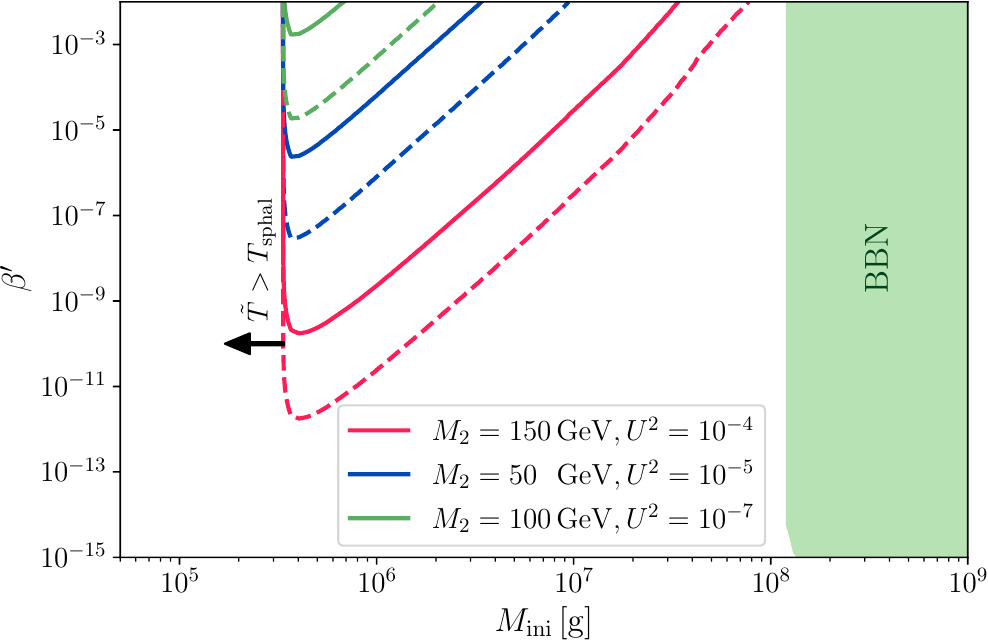}
 \includegraphics[width = 0.49\textwidth]{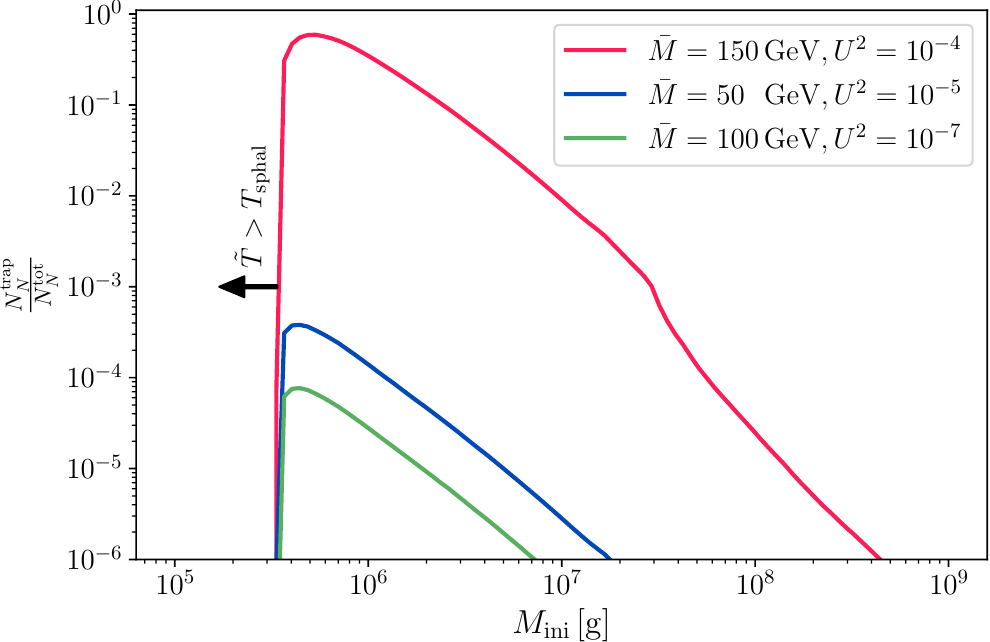}
 \caption{Left panel: Contours of $\YB = \YBobs$ for each benchmark point in \equaref{Lepto:BM}. The solid and dashed lines correspond respectively to $\Delta M/\Mdegen = 10^{-1}$ and $\Delta M/\Mdegen = 10^{-3}$. The most recent constraints on PBHs from BBN are reported as a green shaded region \cite{Boccia:2024nly}. Right panel: The fraction of RHNs produced by a PBH which are trapped within $\rsphal$ and therefore may contribute to $\YB$ is shown as a function of $\MPBHini$ for each benchmark point shown in \equaref{Lepto:BM}. In both panels, the lower limit in $\MPBHini$ occurs for $\Tevap = \Tsphal$, such that lighter PBHs never produce RHNs after the sphaleron freeze-out in the background Universe. \label{Hot Spots/Lepto/YBplot}}
\end{figure}
%%%%%%%%%%%%%%

In the left panel of \figref{Hot Spots/Lepto/YBplot} contours of $\YB = \YBobs$ are projected in the PBH parameter space ($\MPBHini,\beta^\prime)$, for the three benchmark points shown in \equaref{Lepto:BM} with two different relative mass splittings. The green shaded region is the parameter space excluded by BBN \cite{Boccia:2024nly}. In the regions enclosed by the contours, $\YBobs$ can be reproduced entirely from decays of RHNs inside the sphaleron radius of hot spots, while no contribution to $\YB$ would be possible in the absence of hot spots. Remarkably, a baryon asymmetry is produced despite EW sphalerons having frozen out in the background Universe and only remaining active in isolated hot spots. As expected, allowing for a smaller $\Delta M$ (dashed lines in \figref{Hot Spots/Lepto/YBplot}) can produce the observed BAU for a smaller initial number density of PBHs ($i.e.$ smaller $\beta^\prime$) due to the resonant enhancement in the decay asymmetry parameter, see \secref{Leptogenesis/Resonant}. It is also interesting to note that only a mild ($\Delta M / \Mdegen \sim \mathcal{O}(10^{-1})$) mass degeneracy is required, which could be detected given the anticipated level of experimental resolution \cite{SHiP:2021nfo}. This scenario may therefore be testable at the ILC and CEPC experiments \cite{Yang:2023ice}. 

The right-hand plot of \figref{Hot Spots/Lepto/YBplot} shows the fraction of RHNs produced by PBHs which are trapped within $\rsphal$ as a function of $\MPBHini$. For smaller PBH masses, $\MPBHini \lesssim 10^{5.5}$ g, the whole universe remains hotter than $\Tsphal$, therefore $\mathcal{N}^{\rm trap}_N$ is undefined. Since the RHNs or their decay products cannot escape to a region where $\Tplasma < \Tsphal$, the resulting baryon asymmetry is expected to be washed out in this region. Lighter PBHs have hotter Hawking temperatures and therefore, hotter hot spots. Therefore when it is defined, $\rsphal$ is larger for a lighter PBH (see for example \figref{Hot Spots/Evolution/profile}). As $\MPBHini$ increases from $\sim10^{5.5}$g, $\rsphal$ becomes shorter (considering the same stage of the PBH lifetime) meaning it is easier for RHNs to escape.
Furthermore, as $\MPBHini$ increases, the initial temperature of the hot spot decreases (see \secref{Hot Spots/Evolution}) so that for $\MPBHini \gtrsim 2\times10^7$g the hot spot forms at a temperature below $\Tsphal$. Only RHNs produced at late times when the core temperature of the hot spot increases can contribute to $\YB$ (see \figref{PBHs/Hot Spots/Tcomp}) in this case. Thus for larger $\MPBHini$ the amount of time the hot spots are hotter than $\Tsphal$ decreases, trapping fewer RHNs within $r < \rsphal$. Further, the number of trapped RHNs increases with $U^2$ as increasing the mixing enhances the decay rate.\label{Hot Spots/Lepto}
\newpage

\subsection{Hot Spots Absorb Dark Matter}
The production of a DM candidate in the early Universe can be particularly sensitive to the abundance of light evaporating PBHs ~\cite{Cheek:2021cfe, Gondolo:2020uqv,Masina:2020xhk,Baldes:2020nuv}. In the absence of any non-gravitational interaction between DM and the SM, the evaporation of PBHs simply constitutes an additional contribution to the DM relic abundance today~\cite{Gondolo:2020uqv,Baldes:2020nuv}. However, in the presence of interactions between dark-sector and SM particles (e.g. via the exchange of a dark mediator), the size of this contribution can vary widely across the parameter space~\cite{Cheek:2021cfe}. In particular, the variety of energy scales present in the problem --- the Hawking temperature of the evaporating PBH, the plasma temperature around the black holes, and the dark-sector particle masses --- can dramatically affect the capacity of DM particles produced via Hawking evaporation to propagate freely through the cosmos when they are produced. In Ref.~\cite{Cheek:2021cfe}, such interactions were considered and demonstrated to be relevant in the context of {\em Freeze-Out} and {\em Freeze-In} scenarios of DM production.

In this section, the influence of the hot spot around a PBH on DM production in the context of {\em Freeze-Out} and {\em Freeze-In} scenarios is examined. Using a simple toy-model the extent to which the contribution to the relic abundance is affected is calculated.

\noindent In what follows, the so-called $Z^\prime$ portal DM model~\cite{Arcadi:2013qia,Chu:2013jja,Lebedev:2014bba,Bhattacharyya:2018evo,Mambrini:2010dq,Dudas:2012pb,Dudas:2013sia,Dudas:2009uq} is employed. In particular, a massive Dirac fermion $\DM$ is taken to be the DM candidate, singlet under the SM gauge group but charged under a dark Abelian symmetry $U(1)_D$. The gauge boson associated with this new symmetry is denoted as $Z'_\mu$. The mass of $Z'_\mu$ is assumed to originate from a St\"uckelberg mechanism \cite{Stueckelberg:1938hvi} so that the existence of any other BSM particles in the spectrum of particles evaporated by the black hole can be safely ignored. The Lagrangian is
\begin{eqnarray}\label{eq:lag}
\Lagr &=& \Lagr_{\rm SM}+\bar\DM(i\cancel{\partial}-\mDM)\DM
+\frac{1}{4}Z'_{\mu\nu}{Z'}^{\mu\nu}-\frac{1}{2}\mZp^2 {Z'}_\mu {Z'}^\mu- g_{\rm D} {Z'}_\mu\bar\DM\gamma^\mu\DM- g_{\rm V} {Z'}_\mu\bar f\gamma^\mu f\,,\nonumber\\
\end{eqnarray}
where $f$ denotes any fermion in equilibrium with the SM bath, $\mDM$ ($\mZp$) is the mass of the DM particle $\DM$ ($Z^\prime$), $Z_{\mu\nu}$ is the field strength tensor of $U(1)_D$ and $g_{\rm D}$ ($g_{\rm V}$) denotes the coupling of the $Z^\prime$ to the dark (visible) sector.

The annihilation and scattering of $\DM$ DM produced by a PBH and coupled to the visible sector via a $Z^\prime$ mediator proceeds via the processes
\begin{eqnarray}
\DM\bar\DM\to &Z'& \to \bar f f\\
\DM\bar\DM &\to& Z' Z'\\
\DM f &\to& \DM f\\
\end{eqnarray}
where the $\bar\DM$ and $f$ particles off which the Hawking radiation $\DM$ scatters come from the hot spot and so have radially varying temperature, number density and mass. Expressions for the cross sections of each of the above rates are given in Appendix \ref{B}.
\newpage
%%%%%%%%%%%%%%%%%%%%%%%%%%%%%%%%%%%%%%%%%%%%
\subsubsection{Thermal Production}
%%%%%%%%%%%%%%%%%%%%%%%%%%%%%%%%%%%%%%%%%%%%
\noindent The capacity of DM to thermalise with the SM plasma depends on the values of the dark and visible couplings, $g_{\rm D}$ and $g_{\rm V}$. If these couplings are large enough, such that DM particles thermalise with the plasma before they become non-relativistic, the DM relic abundance will arise from DM particles {\em freezing out} of equilibrium at $T\lesssim \mDM$ (the FO scenario). In the opposite case where these couplings are too small to establish a thermal equilibrium within the dark sector, then DM particles are slowly produced out-of-equilibrium from the thermal annihilation of SM particles and the DM relic abundance is said to {\em freeze in} (the FI scenario) (see also \secref{Early Universe/Out-of-Equilibrium} for discussion of FO and FI scenarios). In the absence of PBHs, the evolution of the relic density is described by the Boltzmann equation \equaref{Early Universe/Abundance/BEalpha} specialised to the Z-portal DM model under study
\begin{eqnarray}
    \frac{\drm \mathcal{N}_{\rm DM}}{\drm \alpha} &=& -\frac{\ln(10)}{H}\langle\sigma v_{\rm Mol}\rangle_{\rm DM\to SM} (\mathcal{N}_{\rm DM}
- \mathcal{N}_{\rm DM}^{\rm eq})n_{\rm DM}^{\rm eq}\
\end{eqnarray}
where $\mathcal{N}_{\rm DM}$ denotes the comoving number density of DM particles, $n_{\rm DM,\, eq}$ is the temperature-dependent number density of DM particles in thermal equilibrium, and $\langle \sigma v_{\rm Mol}\rangle_{\rm DM\to SM}$ is the thermally averaged cross-section of DM annihilation.

In the presence of PBHs, this equation is modified by the inclusion of a source term corresponding to the injection of DM particles in the plasma from the Hawking evaporation of PBHs. In Ref.~\cite{Cheek:2021cfe}, the complete Boltzmann equations involving DM and SM particles, including PBH evaporation, were studied in detail. Whereas it is not the aim of this section to reproduce the results of Ref.~\cite{Cheek:2021cfe}, it is of interest to learn to what extent the conclusions of this earlier work are affected by the presence of hot spots. In the two scenarios mentioned above, the following strategy can be employed:
\begin{enumerate}
 \item The parameters of the model are adjusted such that DM is produced with the correct relic abundance in the absence of PBHs;
 \item Using these parameters, the annihilation and scattering cross sections off the SM and DM particles in the plasma are calculated as above;
 \item Following the methodology detailed in \secref{Hot Spots/Hawking} and given these cross sections, the fraction of DM which is trapped by the hot spot is calculated;
 \item Finally the region that would accordingly contribute to the relic density of DM at the $\mathcal O(1)$ level is found by scanning in the PBH parameter space $(\MPBHini,\beta^\prime)$.

\end{enumerate}

\paragraph{\bf Freeze-In Production --}
In the case of FI production, the couplings $g_{\rm V}$ and $g_{\rm D}$ are typically very small, ensuring both that DM particles do not thermalise with the plasma and that the production of DM does not overclose the Universe. In this case, the effect of the hot spot on the direct production of DM from PBH evaporation is always negligible. The escape probability is always close to one, the DM particles produced through Hawking radiation manage to escape the hot spot and propagate freely in the Universe without interacting with the plasma. The phenomenology described in Ref.~\cite{Cheek:2021cfe} regarding the FI case is thus robust and unaffected by the presence of PBH hot spots.

\paragraph{\bf Freeze-Out Production --}
In the FO scenario, in which couplings can be $\mathcal O(1)$, ultra-relativistic DM particles produced by Hawking evaporation of PBHs easily thermalise with the plasma outside the black hole. In the absence of PBHs, the FO of DM particles happens when the plasma temperature approaches the FO temperature, at which DM particles become non-relativistic and decouple from the thermal bath. In the presence of PBHs, this homogeneous decoupling is affected in two ways: $(i)$ PBHs produce DM particles via Hawking radiation and $(ii)$ these PBHs reheat the Universe, forming hot spots around them, which affects the time-evolution of the plasma temperature locally, hence the dynamics of the DM decoupling within these hot spots. The first aspect was already studied in previous works~\cite{Bernal:2020bjf, Bernal:2020ili, Bernal:2020kse, Bernal:2021bbv, Bernal:2021yyb, Bernal:2022oha, Bernal:2022pue, Masina:2020xhk, Masina:2021zpu, Gondolo:2020uqv, Baldes:2020nuv, Hooper:2019gtx, Cheek:2022dbx, Cheek:2022mmy}, in which the DM particles produced via Hawking radiation were assumed to either free-stream through the Universe, or to instantaneously thermalise with the plasma. However, throughout these references, the plasma surrounding PBHs was assumed to remain homogeneous even during reheating by PBHs. In the highly non-homogeneous environment of a hot spot, DM produced as Hawking radiation may be trapped close to the PBH where the local temperature exceeds the freeze-out temperature $\TFO$. This partly erases the contribution of PBH-produced DM, which one may naively think would free-stream and therefore contribute to the DM relic density if hot spots were not considered.

\begin{figure}[h!]
\centering
 \includegraphics[width = 0.65\textwidth]{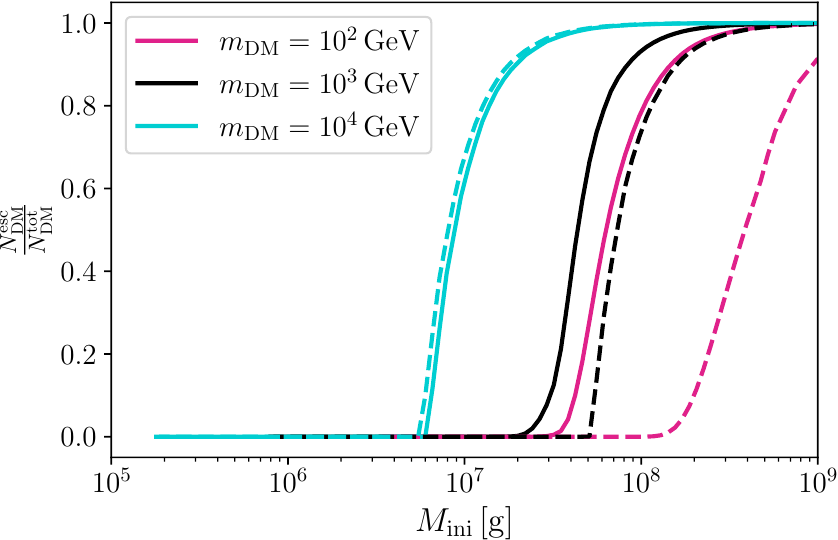}
 \caption{\label{Hot Spots/DM/DMP} Characteristic curves showing the fraction of DM produced by a PBH which escapes to the freeze-out radius. The solid lines indicate  $\mZp/\mDM = 10$ while the dashed lines indicate $\mZp/\mDM = 10^{-3}$.}
\end{figure}

\figref{Hot Spots/DM/DMP} shows the ratio between the number of DM particles that manage to escape the hot spot and the total number of DM particles emitted by PBHs, as a function of the PBH mass {$\MPBHini$} and for a few benchmark values of the DM mass. The solid (dashed) lines correspond to the case of a heavy (light) $Z'$ mediator with $\mZp = 10 \mDM$ ($\mZp = 10^{-3}\mDM$). The DM that escapes the hot spot will contribute to the relic density, and therefore the effect of the hot spot is that for $\MPBHini \lesssim 10^{6.5}$ g the PBH-produced DM cannot contribute at all to the relic density.\\

The difference between the heavy and light mediator cases is most pronounced for lighter DM masses because the t-channel diagram $\bar{\psi}\psi \to Z^\prime Z^\prime$ opens up for light DM and provides an additional scattering channel which further increases the absorption due to the hot spot. Given the ratios displayed in \figref{Hot Spots/DM/DMP}, one can infer the PBH capacity to overclose the Universe, hence providing an upper limit on their abundance at formation, which is represented in \figref{Hot Spots/DM/DM} for the same benchmark parameters, the left (right) panel corresponding to the case of a heavy (light) mediator. In both \figref{Hot Spots/DM/DMP} and \figref{Hot Spots/DM/DM}, the couplings are set to $g_{\rm V}=g_{\rm D}$ and such that DM is produced with the correct relic abundance in the plasma, far away from the hot spot. In \figref{Hot Spots/DM/DM}, indicated with dotted lines are the results when ignoring the presence of the hot spot. 
\newpage

\begin{figure}[h!]
\centering
  \includegraphics[width = 0.45\textwidth]{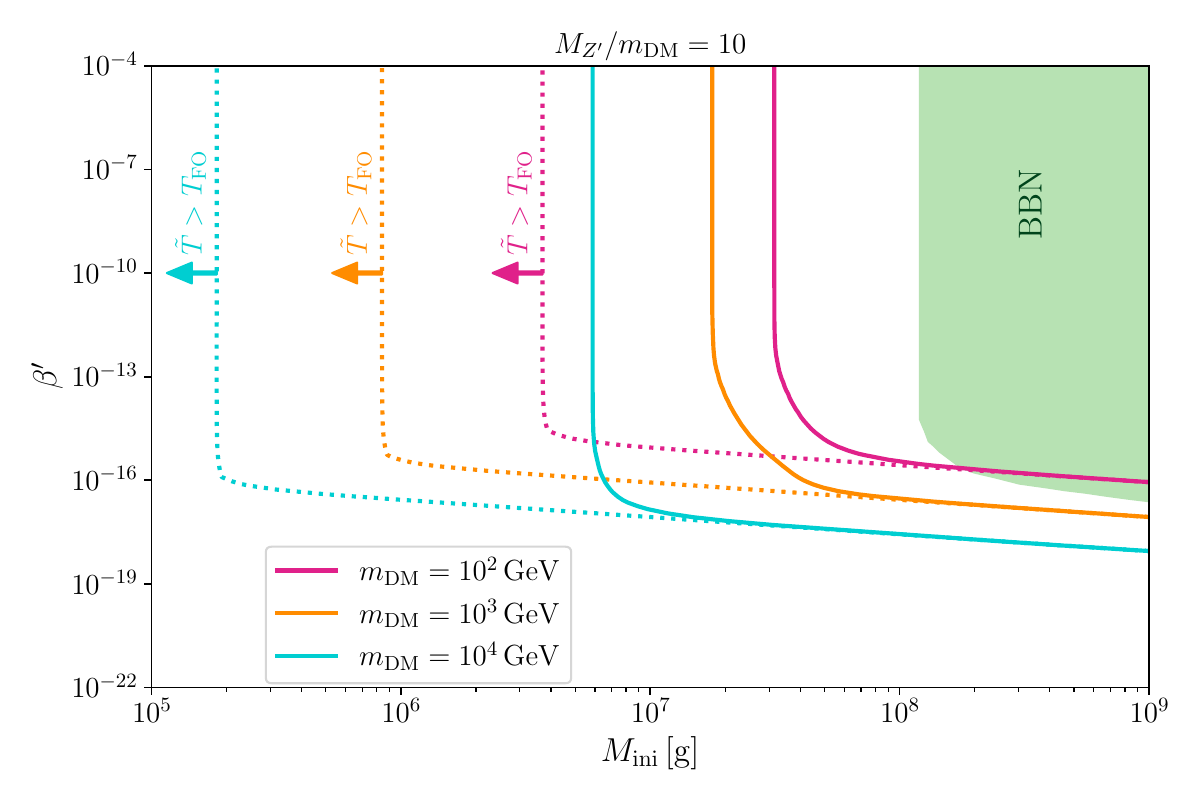}
 \includegraphics[width = 0.45\textwidth]{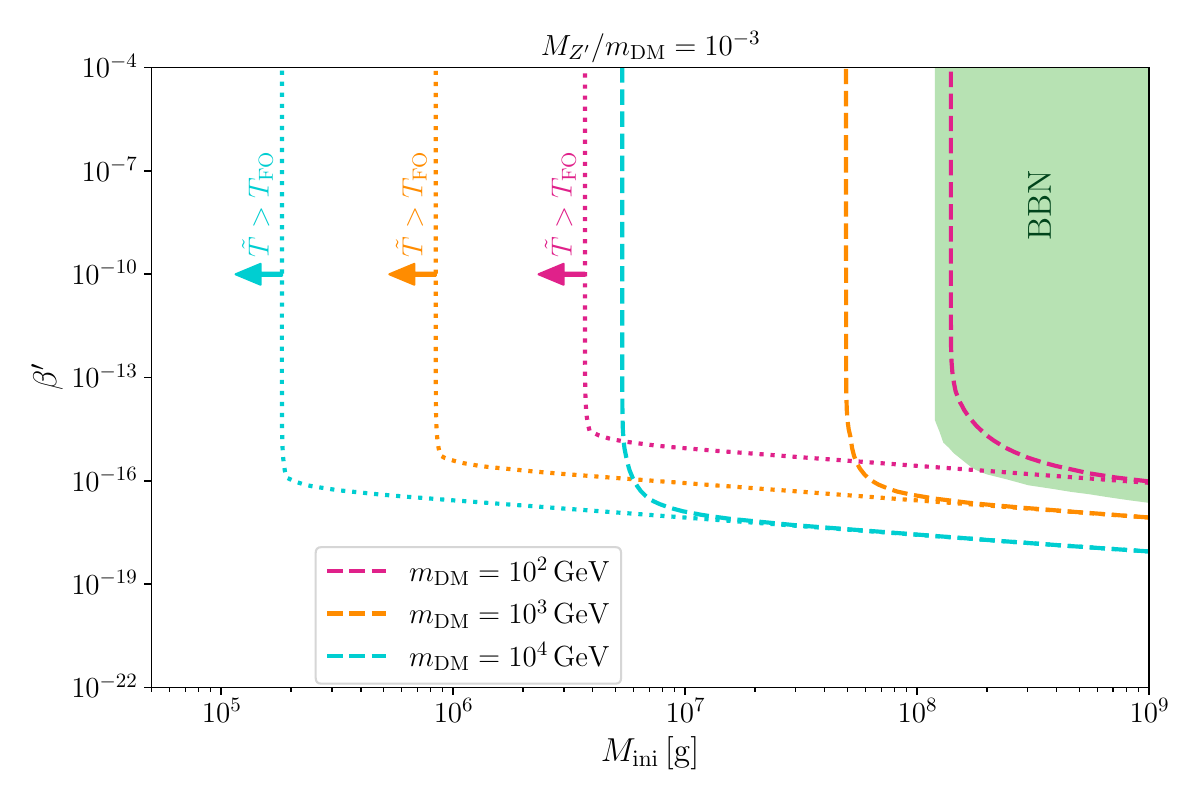}
 \caption{\label{Hot Spots/DM/DM}Contours of $\Omega_{\rm DM}h^2= 0.12$ for three different masses of DM. In the left (right) panel, $\mZp/\mDM = 10$ ($\mZp/\mDM = 10^{-3}$). The dotted lines show the na\"ive calculation, that is without treating the hot spot properly whereas the solid (dashed) lines show the full calculation including the DM absorbing effect of the hot spots. In both panels, $g_{\rm D} = g_{\rm V} = g$ where $g$ is fixed such that the relic density of DM is reproduced via thermal freeze-out in the absence of PBHs. The most recent constraints on PBHs from BBN are shown as a green shaded region \cite{Boccia:2024nly}.}
\end{figure}

As one can see from \figref{Hot Spots/DM/DM},  for heavy enough PBH masses the effect of the hot spots is negligible, as heavier PBHs support colder hot spots that fail to trap DM particles and allow the PBH evaporation to contribute to the DM relic density freely. In the opposite case of lighter PBHs, the hot spot is hotter, and the scattering of DM particles is sufficient to trap these particles inside the hot spot and force them to thermalise instead of free streaming. The contribution of the PBHs to the DM relic abundance is thus reduced only to the thermal population of DM particles freezing out inside the hot spot when the latter cools down, which is comparable to the contribution from the DM freeze-out outside the hot spots. This effect is somewhat the opposite of the leptogenesis case, where RHNs were needed to decay within the sphaleron radius $\rsphal$ to be able to contribute to BAU, favouring light PBH masses. Here, the absorption of DM by the hot spot is significant for each DM mass considered, shifting the parameter space which overcloses the universe towards heavier $\MPBHini$. \\

The hot spots of lighter PBHs less efficiently absorb DM with larger masses, the PBH mass for which the hot spot effect becomes negligible is lighter for heavier DM. This is not intuitive, as heavy DM particles usually require larger couplings with SM particles to annihilate more efficiently during the FO process. However heavier DM has a hotter freeze-out temperature such that for a fixed PBH mass, the region $r \leq \rFO$ is smaller and the DM does not have to travel as far in order to escape. For each choice of the DM mass, there exists a minimum PBH mass below which the temperature of the universe at the time of evaporation $\Tplasma$ exceeds the FO temperature. In that case, DM particles emitted by the black hole would always thermalise with the surrounding plasma, and the PBH cannot contribute to any excess in the DM relic density. This region is indicated in \figref{Hot Spots/DM} by the coloured arrows, explaining why the relic density parameter space closes, even in the absence of a hot spot.

A qualitative difference between the light and heavy mediator cases is that the DM annihilation (which is the main process controlling its freezing out from the plasma) is entirely given by the $s$-channel annihilation process $\bar \DM\DM \to Z'\to \bar f f$ in the heavy mediator case, but also by the $t$-channel annihilation process $\bar \DM \DM\to Z'Z'$ in the light mediator case. As can be seen from the expressions \equaref{Hot Spots/DM/schannel} and \equaref{Hot Spots/DM/tchannel}, the $s$-channel annihilation cross-section scales like $g_{\rm V}^2g_{\rm D}^2$ whereas the $t$-channel annihilation cross-section scales like $g_{\rm D}^4$. Therefore, in the light mediator case, the annihilation cross section can be mainly controlled by the dark coupling, while its interaction with the hot spot plasma can be suppressed in the case of a small visible coupling. \\
\newpage
\begin{figure}[h!]
    \centering
    \includegraphics[width=0.6\linewidth]{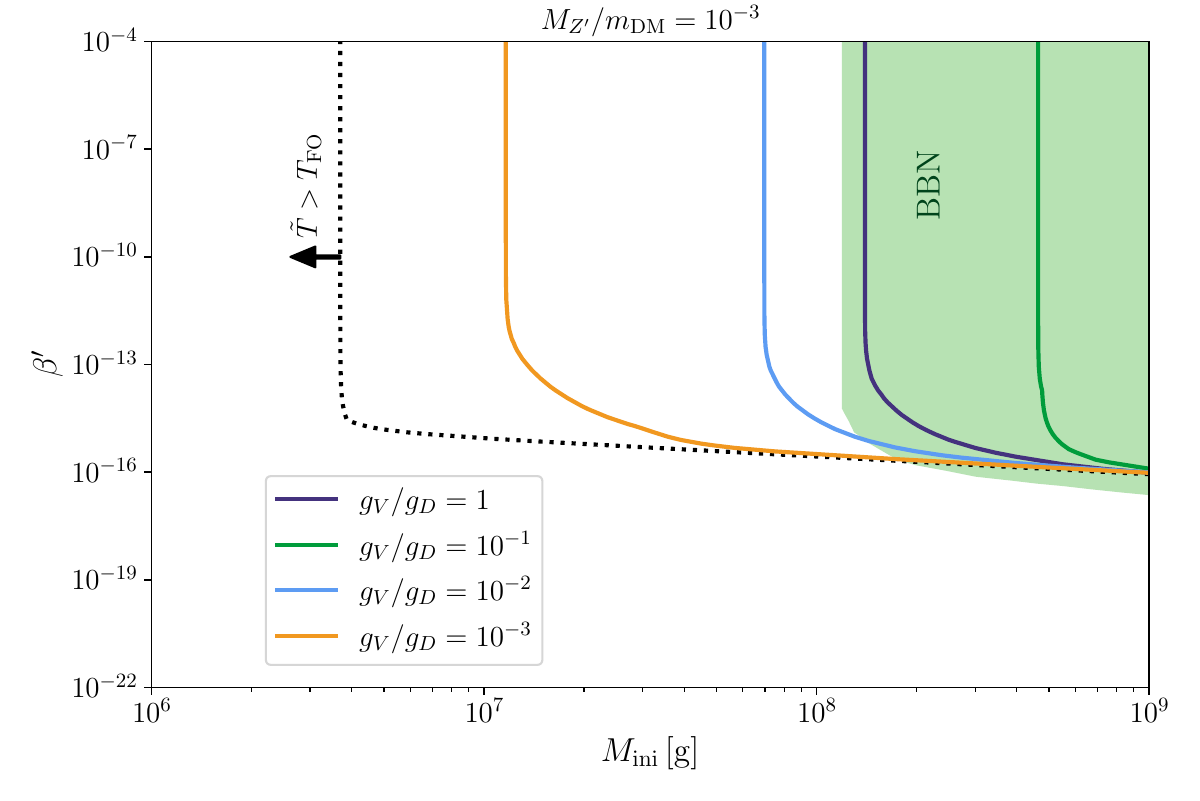}
    \caption{\label{Hot Spots/DM/ratios}Same results as \figref{Hot Spots/DM/DM}, but for $\mDM= 10^3 \mZp=100\, \mathrm{GeV}$ and various values of the ratio $g_{\rm V}/g_{\rm D} = 1$ (purple), $10^{-1}$(green), $10^{-2}$ (light-blue), and $10^{-3}$ (orange). The dotted line shows the calculation without considering the effect of hot spots.}
\end{figure}
In \figref{Hot Spots/DM/ratios}, the ratio between the visible and dark couplings is varied in the light mediator case, showing that the effect of the hot spot is reduced in this limit. Nonetheless, such a situation cannot be pushed to the limit where the visible coupling is exactly zero, as this would imply that the mediator is not maintained in equilibrium with the SM plasma and the dark sector would have its own equilibrium, leading to a different phenomenology~\cite{Heurtier:2019eou,Berlin:2016gtr,Berlin:2016vnh}. Therefore, it is expected that the effect of the hot spot is always significant in the case of a WIMP candidate since all the cases we surveyed show that hot spots tend to suppress the escape probability of DM for small PBH masses.
\label{Hot Spots/DM}
\newpage

\subsection*{Conclusions}
% \lh{Note that in principle, the mediator particle $Z'$ is also produced from the evaporation of PBHs and present in the thermal plasma so long as $T\gtrsim M_{Z'}$. In the heavy mediator case, producing $Z'$ particles from PBH evaporation may lead to an additional source of out-of-equilibrium DM particles after the $Z'$ particle eventually decays. However, the $Z'$ interacts directly with SM particles, and because the visible coupling $g_{\rm V}$ is expected to be relatively large in the heavy mediator case, we could check that the mean free path of the $Z'$ in the hot spot is much smaller than that of the DM particles in the hot spot in the examples we consider. In the case of the light mediator, this conclusion may not hold anymore, as the ratio $g_{\rm V}/g_{\rm D}$ could be relatively small and the $Z'$ could propagate further as dark matter has decoupled from the thermal bath. However, in that case, the $Z'$ cannot decay into DM particles for kinematic reasons and can only decay into SM particles. Such a late energy deposition may, in principle, affect slightly the dynamics of the hot spot temperature profile. However, we expect this effect to be negligible as the fraction of the energy released by the black hole under the form of $Z'$ particles during the evaporation is negligible compared to the fraction released into SM particles.}

\noindent This section investigated the role that PBH hot spots play in addressing two of the most pressing questions in cosmology: the origin of the BAU and the nature of DM. Properly accounting for the localised hot spots formed around PBHs due to Hawking radiation \cite{He:2022wwy,He:2024wvt} reveals deviations from previous studies that assumed homogeneous heating of the primordial plasma. For low-scale leptogenesis, the presence of hot spots can sustain electroweak sphaleron processes locally, even after they have frozen out in the background Universe. This allows the lepton asymmetry produced by the decay of right-handed neutrinos within these hot spots to be converted into baryon asymmetry. The mechanism enables the generation of the BAU in regions of leptogenesis parameter space where thermal leptogenesis would not be viable in standard cosmology.  Additionally it allows PBHs which could not produce baryon asymmetry under the homogeneous heating assumptions commonly used in the literature, to do so within the hot spot framework. Importantly, these regions of the parameter space tend to favour larger mixing of the RHNs with the active neutrinos and so are in testable regions of the RHN parameter space. This underscores the importance of including the effects of hot spots in cosmological models.\\

In the context of thermal DM production, the FO and FI scenarios were analysed in the presence of PBH hot spots. In FI scenarios, the impact of hot spots is minimal, with DM particles produced via Hawking evaporation largely escaping and contributing to the relic density as previously established in the literature. 
For FO scenarios, since the temperature inside the hot spot can exceed the DM FO temperature, PBH-produced DM can be trapped in the hot inner region, and thus it can be efficiently thermalised with the plasma, erasing its contribution to the relic abundance.
Only the DM particles which travel far from the PBH and free stream in the Universe contribute to the observed abundance of DM today. This effect leads to large corrections in the regions of PBH parameter space which would overclose the Universe in DM scenarios.\\

Before concluding, it is worth commenting on a few aspects that would deserve further exploration in the future. First, it was assumed throughout this section that the production and scattering of BSM particles do not affect the time evolution of the hot spot temperature profile. In reality, the fact that particles may be emitted by the black hole and deposit their energy at distances that are larger than what is expected from SM particles may affect this evolution and the overall shape of the temperature profile. This would be particularly interesting in the case PBHs which would emit more than one dark sector particle~\cite{Baker:2022rkn, Calza:2021czr, Perez-Gonzalez:2023uoi, Calza:2023iqa, Calza:2023rjt}, in which case the morphology of the energy deposition throughout the hot spot could be significantly altered by the presence of BSM physics. Second, it was assumed that PBHs contribute to the BAU or the DM relic abundance in a homogeneous fashion following the equilibration of the plasma after the PBH lifetime. In reality, the thermal history of the Universe is locally affected by the presence of PBH hot spots. Whereas the distance that separates PBHs from one another when they evaporate may be much smaller than the Hubble radius at that time, --- which is commonly used to argue that PBHs inject particles in a homogeneous way in cosmology --- this modified evolution of the thermal bath on small scales could lead to spatial variation in the DM abundance, or the baryon/lepton asymmetry of the Universe, leading to potentially observable signals. \\

The findings of this section underscore the necessity of accounting for the non-uniform temperature profiles around PBHs when considering their role in early universe cosmology. It was demonstrated that the interplay between PBHs, hot spots, and non-equilibrium processes like leptogenesis and DM production could provide new insights into conditions in the early universe, potentially altering these scenarios' testability in future experiments. More generally, hot spots may {\em screen} the emission of Hawking radiation on scales much smaller than the average distance between PBHs. This may significantly impact the way PBHs affect cosmology and particle physics beyond the sole cases of DM and baryon asymmetry production. \label{Hot Spots/Conclusions}
\newpage

\chapter{Summary and Outlook}\label{Conclusions}

This thesis set out to explore the impacts of PBHs on early universe particle physics processes, seeking to anticipate results from upcoming GW experiments in order to shed light on crucial cosmological phenomena. A key goal was to demonstrate that cosmological populations of PBHs would exclude leptogenesis scenarios which are far beyond the reach of direct experimental probes. Another was to discover how hot spots influence Hawking radiation and the phenomenology of PBH evaporation. In a broad sense the purpose of this thesis was to progress understanding of how PBHs interact with the early universe to alter cosmological observables.\\

In \chapref{PBHLepto} the consequences of a period of post-sphaleron PBH domination were analysed in the context of high scale and low scale leptogenesis scenarios.  The results of found demonstrate the severe incompatibility of cosmological PBH populations and leptogenesis scenarios. Of particular importance is the discovery and precise computation of mutual exclusion limits between the parameter spaces of leptogenesis and PBHs. This means that in the event of a GW experiment returning evidence for ultralight PBHs, that information can be translated into information on leptogenesis. In fact via ultralight PBHs, GW experiments may be able to exclude the entire high scale leptogenesis scenario, something which is far beyond the capabilities of any current or future generation RHN detection experiment. In the case of resonant leptogenesis the reverse argument also works, in case an experiment finds light RHNs at the $\sim\GeV$ scale one can rule out PBH parameter space above the corresponding exclusion limit. An important caveat in this case is that the detected RHN must be assumed or otherwise shown to be responsible for resonant leptogenesis in the early universe. Therefore, \chapref{PBHLepto} advances the quest for knowledge of the early universe by proving that GW experiments can constrain and even rule out leptogenesis entirely, by detecting the GW signatures of ultralight PBHs.\\

\secref{PBHLepto/High Scale} considered the high scale case, characterised by RHNs heavier than the DI limit. It was shown that PBH domination affects the sphaleron freeze-out temperature, though not by enough to change the sphaleron efficiency coefficient. By scanning over leptogenesis parameter space for the maximum achievable asymmetry, mutual exclusion limits were projected into the PBH parameter space, delimiting the regions where PBHs would be in tension with high scale leptogenesis. It was demonstrated that the entropy injection due to PBHs can be so severe as to completely exclude high scale leptogenesis even for very small populations. The dependency of the constraints on the active neutrino mass scale $m_h$ is such that lighter neutrinos confer tighter bounds on the PBH parameter space. \secref{PBHLepto/Low Scale} then extends this argument to the case of low scale, resonant leptogenesis. Since light sterile neutrinos continue to produce lepton asymmetry across the EWPT, the effect on the sphaleron decoupling demonstrated in \secref{PBHLepto/High Scale} is relevant in the low scale case. PBHs were shown to shrink the viable resonant leptogenesis parameter space towards heavier RHN masses and smaller couplings, away from the reach of experiments. Mutual exclusion limits were again projected onto the PBH parameter space, in contrast to the high scale case these limits are dependent on the sterile neutrino mass scale. Resonant leptogenesis with lighter RHNs is excluded more easily by PBHs, such that for $\sim \GeV$ scale RHNs, constraints are around a million times more severe than for $500\GeV$ scale RHNs. \\

\chapref{Hot Spots} tackled the propagation of Hawking radiation through hot spots by developing a formalism for calculating the probability for a Hawking particle to escape to a particular distance. The evolution of a hot spot in an expanding universe was considered for the first time, revealing how hot spots emerge and grow as the background expands and cools, before heating dramatically at the end of the PBH life. Applying the probabilistic formalism developed for a static profile to an evolving hot spot, the rich phenomenology of hot spots was revealed. It was demonstrated that PBH hot spots are able to sustain leptogenesis even after the freeze-out of sphalerons, because RHNs produced as Hawking radiation may be trapped in the hot regions where sphalerons are still active. Remarkably, this mechanism produces enough baryon asymmetry considering testable neutrino parameter space points with mild mass degeneracies ($\Delta \Mdegen \sim 10^{-1}$). DM production by a PBH was also shown to be strongly affected; hot spots which remain above the DM freeze-out temperature absorb DM produced as Hawking radiation. Constraints on PBHs from the overproduction of DM were shown to be significantly modified when accounting for DM absorption in hot spots.\\

Historically missed in the literature, the effects of PBH hot spots on Hawking radiation revealed in \chapref{Hot Spots} set the stage for a new approach to PBH physics, in which the crucial interaction between Hawking radiation and hot spots is understood and can be applied to any scenario. The formalism developed in this chapter, including the full thermal averaging factor derived in Appendix \ref{A}, permits calculation of the escape probability for Hawking radiation. In general, the escape probability is lower for heavier and more strongly interacting particles but precisely where the line lies requires careful calculation in all cases. Hot spots have a broad capacity to affect Hawking radiation which in future must be accounted for in any study of PBH evaporation. This significant advancement progresses understanding of PBH physics and reveals an intriguing phenomenology, exemplified by the surprising effects on leptogenesis and WIMP DM which were demonstrated. The neutrino parameters considered were chosen to be accessible at upcoming experiments so this scenario will soon be subject to experimental testing. Moreover the considerations of how hot spots evolve in an expanding universe given in \secref{Hot Spots/Evolution} tackle this problem for the first time, painting an intuitive picture of how hot spots form and evolve where previous considerations had imagined hot spots in static backgrounds. The field of hot spot phenomenology is truly in its infancy and the contributions found in \chapref{Hot Spots} help lay the foundations for future work. \\

Many questions remain open. Regarding the interplay between leptogenesis and PBHs, more work is needed to understand exactly how different leptogenesis scenarios would be constrained by PBHs. The analysis in \chapref{PBHLepto} employs models of leptogenesis where one or more of the RHNs are decoupled and do not contribute to the baryon asymmetry. Allowing additional RHNs to contribute is expected to ameliorate the mutual exclusion limits derived, but exactly how so remains unknown. The parameter space of models involving multiple active RHNs are higher dimensional necessarily and more computationally expensive to scan over. Monochromatic mass distributions were considered throughout this work, therefore the constraints drawn apply to extended mass distributions only so far as the PBHs with different masses can be treated independently. Further study is required to understand if extended PBH mass distributions affect leptogenesis differently to monochromatic ones. \\

Regarding hot spots, perhaps the most pressing question to answer going forwards is how hot spots would evolve following the cessation of PBH evaporation. Analytical estimates made in \cite{Das:2021wei,He:2022wwy} suggest that the core temperature cools as $t^{-7/15}$, slightly slower than the rate of cooling in the radiation dominated background. This suggests that hot spots may continue to expand even after PBH evaporation ceases, thus it is not clear whether or not the hot spots would ever fully equilibriate with the rest of the universe. This may leave observable signatures. The propensity for hot spots to absorb Hawking radiation, proved in \secref{Hot Spots/DM}, also invites revisiting existing constraints on PBHs from the non-observation of gamma rays, since naively one may imagine that the primary signal from a PBH would be attenuated by a hot spot, and the hot spot could produce a secondary signal. To approach this question one would have to further study how hot spots form and evolve in the late time universe, far from radiation domination and in complex astrophysical environments where accretion and other local phenomena may strongly impact the hot spot profile. Another interesting and unexplored question is the role of hot spots during strong first order phase transitions. Depending on the initial mass of the PBH, and the symmetry breaking scale of interest, one can imagine abundant scenarios deserving of investigation; hot spot cores above the symmetry breaking scale may exist as isolated symmetric regions impinged upon by bubbles of broken phase once the background undergoes the phase transition, or hot spot cores which suddenly heat back to temperatures above the symmetry breaking scale could bring about symmetry restoration. Both of these scenarios would likely be associated with a smoking gun GW signature.\\

Before concluding it is appropriate to make a few comments on the emerging picture of the memory burden effect. The expectation for PBHs to experience memory burden depends on the analogy to field theoretic configurations known as saturons which share other properties with black holes. Burdened PBHs would evaporate at a suppressed rate after the Page time, when they have lost around half of their mass. Exactly how a PBH should evaporate differently from the semiclassical rate demonstrated by Hawking is a matter of pure speculation at the time of writing. What is clear is that any significant suppression of the evaporation rate would reduce the impacts of Hawking radiation in the early universe and in particular suppress the explosive final stages of evaporation. Memory burden, if it really applies to black holes, is likely to numerically affect the constraints derived in \chapref{PBHLepto} and the contours of the leptogenesis and DM production scenarios in \chapref{Hot Spots}. By contrast, the hot spot effects on Hawking radiation discussed \chapref{Hot Spots} appear before memory burden becomes active. PBHs form hot spots significantly earlier than the Page time, meaning studies of PBH evaporation cannot simply neglect the impact of hot spots by appealing to the memory burden effect. To reiterate, the precise impact that the memory burden effect would have on the constraints drawn in \chapref{PBHLepto} is impossible to realistically analyse, not only because it is still unclear if black holes become burdened, but because the nature of burdened evaporation is also unknown.

To conclude, the work presented in this thesis draws back the curtain on the physics of the primordial universe by leveraging PBHs as cosmological probes. The violent deaths of ultralight PBHs were shown to severely inhibit leptogenesis, alter the sphaleron freeze-out temperature and provide an opportunity to rule out (through GW signals) models of leptogenesis which would otherwise evade direct detection experiments for effective eternity. Turning to the extreme temperature gradients of hot spots, the evolution and formation of hot spots was considered in an expanding background for the first time, and a formalism treating the propagation of Hawking radiation was introduced. Upon addressing the previously unexplored effects of hot spots on Hawking radiation, it was found that hot spots can support successful leptogenesis after sphaleron freeze-out, and efficiently absorb radiated DM. These examples hint at an intricate and rich tapestry of BSM physics in hot spots. Though shrouded by the fog of 14 billion years, understanding the origins of our universe is a tractable and crucial project for modern physics, it is hoped that this thesis has helped to illuminate the long road there.

\newpage
\begin{appendices}
    \chapter{Thermal averaging of Hawking radiation scattering cross section}\label{A}
    
The velocity averaged cross section for a $2 \to 2$ process where the incoming particles have temperatures and momenta $T_{1,2}$\,,\,$p_{1,2}$ is defined as
\begin{equation} \label{Hot Spots/Hawking/int}
 \langle \sigma \cdot v_{Mol} \rangle_{T_{1} T_{2}} \equiv \frac{\int
 \sigma \cdot v_{Mol} f_1 f_2\, d^3p_1 \, d^3p_2}{\int d^3p_1 f_1 \, \cdot \int d^3p_2f_2}
\end{equation}
where $v_{Mol}$ is the Moller velocity, and both interacting particles are taken to be Maxwell-Boltzmann distributed, i.e., $f_i = e^{-E_i/T_i}$. The volume element can be recast as 
\begin{equation}
 d^3p_1\,d^3p_2 = 2\pi E_1 E_2 \, dE_1 \,dE_2\, d \cos \theta
\end{equation}
where $\theta$ is the angle between the momentum vectors of the two incoming particles which have energy $E_{1,2}$. Changing variables, 
\begin{eqnarray}
x_+ \equiv \frac{E_1}{T_1} + \frac{E_2}{T_2} \\
x_- \equiv \frac{E_1}{T_1} - \frac{E_2}{T_2} \\
s = (p_1^\mu + p_2^\mu)^2
\end{eqnarray}
where $p^\mu_i$ is the four-momentum of particle $i$. Calculating the Jacobian leads to 
\begin{equation}
d^3p_1\,d^3p_2 = 2\pi^2 T_1 T_2 E_1 E_2 dx_-\, dx_+ \, ds
\end{equation}
so that the numerator of the integral \equaref{Hot Spots/Hawking/int} becomes
\begin{equation}
 2\pi^2 T_1 T_2 \int \sigma \cdot v_{Mol} f_1 f_2 E_1 E_2 dx_- \, dx_+ \, ds = 2\pi^2 T_1 T_2 \int \sigma F e^{-x_+} dx_- \, dx_+ \, ds 
\end{equation}
where $F = v_{Mol}E_1 E_2$. The limits of integration transform as
\begin{eqnarray}
 E_1 \geq m_1 \\
 E_2 \geq m_2 \\
 -1 \leq \cos \theta \leq 1
\end{eqnarray}
and since in general $m_1 \neq m_2$, it follows
\begin{equation}
 s = m_1^2 + m_2^2 + 2E_1E_2 - 2|p_1||p_2|\cos \theta
\end{equation}
so that 
\begin{equation}
 s \geq m_1^2 + m_2^2 + 2E_1E_2 - 2|p_1||p_2|
\end{equation}
This inequality can be solved for $x_-^{\rm max}$, the result reads
\begin{equation}
 x_-^{\rm max} = \frac{ x_+ C_1 + \sqrt{C_2(T_1^2T_2^2x_+^2 - D})}{D}
\end{equation}
making the identifications $C_1 = m_1^2T_2^2 - m_2^2T_1^2$, $C_2 = m_1^4 + \,(\,m_2^2 -s\,)\,^2 -2m_1^2\,(\,m_2^2 + s\,)$, and $D = sT_1T_2 + \,(\,T_2 - T_1\,)\,\,(\,m_1^2T_2 - m_2^2 T_1)$.
Clearly $x_-^{\rm min} = 0$. For $x_+$ the upper limit is of course $\infty$, while the lower limit can be obtained straightforwardly by requiring that $x_-^{\rm max}$ be real, so that
\begin{eqnarray}
 x_+ \geq \frac{\sqrt{D}}{T_1T_2}
\end{eqnarray}
and finally $s_{\rm min} = (m_1 + m_2)^2$.

The numerator can now be calculated as
\begin{equation}
2\pi^2 T_1 T_2 \int^{\infty}_{s_{\rm min}} \sigma F \, ds \int^{\infty}_{x_+^{\rm min}} e^{x_+} \, dx_+ \int^{x_-^{\rm max}}_0 dx_- = 2\pi^2 T_1 T_2 \int^{\infty}_{s_{\rm min}} \sigma F \, ds \int^{\infty}_{x_+^{\rm min}} e^{x_+}x_-^{\rm max} \, dx_+ 
\end{equation}
First considering the integration over $x_+$, the integral splits into
\begin{equation}
 \int e^{-x_+}x_-^{\rm max} \, dx_+ = \frac{C_1}{D}\int e^{-x_+}x_+ dx_+ + \frac{\sqrt{C_2}}{D}\int e^{-x_+}\sqrt{ T_1^2T_2^2x_+^2 - D}dx_+
\end{equation}
where the first term gives
\begin{equation}
\frac{C_1}{D}\int e^{-x_+}x_+ dx_+ = \frac{C_1}{D}[-e^{-x_+}(1+x_+)]^{\infty}_{x_+^{\rm min}} = \frac{C_1}{D}e^{-x_+^{\rm min}}(1 + x_+^{\rm min})
\end{equation}
while for the second term, one can define
\begin{align}
 u &\equiv \frac{T_1T_2}{\sqrt{D}}x_+\\
 du &= \frac{T_1T_2}{\sqrt{D}}dx_+\\
 u_{\rm min} &= 1
\end{align}
so that the integral becomes
\begin{equation}
 \frac{\sqrt{C_2}}{T_1T_2}\int^{\infty}_1 e^{-u \sqrt{D}/(T_1T_2)}\sqrt{u^2 - 1}du
\end{equation}
This integral can be evaluated by using the integral form of the modified Bessel functions \equaref{Leptogenesis/Out-of-equilibrium/Bessel}
so that by setting $n = 1$ and $z = \sqrt{D}/(T_1T_2)$ it follows
\begin{equation}
 \frac{\sqrt{C_2}}{T_1T_2}\int^{\infty}_1 e^{-u \sqrt{D}/(T_1T_2)}\sqrt{u^2 - 1}du = \frac{2\Gamma\left(\frac{3}{2} \right)\sqrt{C_2}}{\sqrt{\pi D}}K_1 \left( \frac{\sqrt{D}}{T_1T_2} \right) = \frac{\sqrt{C_2}}{\sqrt{D}}K_1 \left( \frac{\sqrt{D}}{T_1T_2} \right)
\end{equation}
Therefore, the numerator of the thermally averaged cross section can be written as
\begin{equation}
 \int \sigma \cdot v_{mol} f_1f_2\, d^3p_1 \, d^3p_2 = 2\pi^2 T_1 T_2 \int^{\infty}_{(m_1 + m_2)^2} \sigma F \left ( \frac{C_1}{D}e^{-x_+^{\rm min}}(1 + x_+^{\rm min}) + \frac{\sqrt{C_2}}{\sqrt{D}}K_1 \left( \frac{\sqrt{D}}{T_1T_2} \right) \right) \, ds
\end{equation}
For $m_1 \neq m_2$, the correct expression for $F$ is $F = \frac{1}{2}\sqrt{(s-m_1^2-m_2^2)^2 - m_1^2m_2^2}$ \cite{Gondolo:1990dk}.
For the denominator, the integral becomes
\begin{equation}
 16\pi^2\int^\infty_{m_1} e^{-E_1/T_1} \sqrt{E_1^2-m_1^2} E_1 dE_1\int^\infty_{m_2} e^{-E_2/T_2}\sqrt{E_2^2-m_2^2} E_2 dE_2
\end{equation}
where the $E_1$ integral gives
\begin{equation}
 m_1^2 T_1 K_2\left(\frac{m_1}{T_1}\right)
\end{equation}
so that the full result for the denominator is
\begin{equation}
 16\pi^2 T_1 T_2 m_1^2 m_2^2 K_2 \left( \frac{m_1}{T_1}\right) K_2 \left( \frac{m_2}{T_2} \right)
\end{equation}
Finally, the thermally averaged cross section can be expressed as
\begin{equation} \label{Scatter:TA}
 \langle \sigma \cdot v_{Mol} \rangle_{T_1T_2} = B \int^{\infty}_{(m_1 + m_2)^2} \sigma(s) F \left ( \frac{C_1}{D}e^{-x_+^{\rm min}}(1 + x_+^{\rm min}) + \frac{\sqrt{C_2}}{\sqrt{D}}K_1 \left( \frac{\sqrt{D}}{T_1T_2} \right) \right) \, ds
\end{equation}
where 
\begin{equation}
 B = \frac{1}{8 m_1^2 m_2^2 K_2 \left( \frac{m_1}{T_1}\right) K_2 \left( \frac{m_2}{T_2} \right)}
\end{equation}
Note that in the limit $m_1 = m_2$, this result reduces to that found in \cite{Cheek:2021cfe}, which in turn is equal to the results in \cite{Gondolo:1990dk} if one sets $T_1 = T_2$. 

Returning to the case of Hawking radiation scattering in a hot spot, the identifications $T_1 = \TBH$, $T_2 = T(r)$, $m_1 = M_X$, $m_2 = m_j$ lead to 
\begin{equation}\label{Hot Spots/Hawking/thermalAverage}
    \langle \sigma \cdot v_{\text{Mol}} \rangle_{T_{\rm BH}T(r)} = B \int^{\infty}_{s_\text{low}} \sigma(s) F \left ( \frac{C_1}{D}e^{-x_+^{\rm min}}(1 + x_+^{\rm min}) + \frac{\sqrt{C_2}}{\sqrt{D}}K_1 \left( \frac{\sqrt{D}}{T_{\rm BH}T(r)} \right) \right) \, ds\,,
\end{equation}
where the masses appearing should include any radially dependent thermal corrections. Since the specific hot spot profile, as well as the PBH temperature are functions of the scale factor, the thermal averaging of the cross sections evolves with $\alpha$ in \equaref{Hot Spots/Hawking/Nescape}.

    \chapter{DM scattering cross sections}\label{B}
    The cross section for $s$-channel DM annihilation is given by
  \begin{equation}\label{Hot Spots/DM/schannel}
  \sigma_{\bar\DM\DM\to Z' \to \bar f f}(s)=\frac{g_{\rm D}^2 g_{\rm V}^2 }{12 \pi s}\left(\frac{\left(2 \mDM^2+s\right) \left(2 m_{ f}^2+s\right) }{ \Gamma_{Z'}^2+\left(M_{Z'}^2-s\right)^2}\right)\sqrt{\frac{s-4 m_{ f}^2}{s-4 \mDM^2}}\,.
  \end{equation}
whereas annihilation in the $t$-channel has cross section
\begin{eqnarray}\label{Hot Spots/DM/tchannel}
  \sigma_{\bar\DM\DM\to Z' Z'}(s)&=&-\frac{g_{\rm D}^4 }{8 \pi s (\mDM^2 (s-4 M_{Z'}^2)+M_{Z'}^4)} \sqrt{\frac{s-4 M_{Z'}^2}{s-4 \mDM^2}}\left[4 \mDM^4+\mDM^2 s +2 M_{Z'}^4\right.\nonumber\\
  &-&\frac{2 (\mDM^2 \left(s-4 M_{Z'}^2\right)+M_{Z'}^4) (-8 \mDM^4+4 \mDM^2 (s-2 M_{Z'}^2)+4 M_{Z'}^4+s^2)
  }{\sqrt{s-4 \mDM^2} \sqrt{s-4 M_{Z'}^2} (s-2 M_{Z'}^2)}\times\nonumber\\
  &&\left.\coth ^{-1}\left(\frac{s-2 M_{Z'}^2}{\sqrt{s-4 \mDM^2} \sqrt{s-4 M_{Z'}^2}}\right)\right]  
  \end{eqnarray}
and finally the cross section for $t$-channel DM scattering off SM particles is given by
  \begin{eqnarray}\label{Hot Spots/DM/tchannelZ}
  \sigma_{\DM f\to \DM f}(s)&=&\frac{g_{\rm D}^2g_{\rm V}^2}{16 \pi s}\left[2+\frac{4 s (M_{Z'}^2+s)}{\mDM^4-2 \mDM^2 (m_f^2+s)+(m_f^2-s)^2} \right.\times\nonumber\\
  &&\log \left(\frac{M_{Z'}^2 s}{\mDM^4-2 \mDM^2 (m_f^2+s)+m_f^4-2 m_f^2 s+s (M_{Z'}^2+s)}\right)\nonumber\\
  &+&\left.\frac{2 s}{M_{Z'}^2} \left(\frac{8 \mDM^2 m_f^2+M_{Z'}^4}{\mDM^4-2 \mDM^2 (m_f^2+s)+m_f^4-2 m_f^2 s+s (M_{Z'}^2+s)}+2\right)\right]\nonumber\\
  \end{eqnarray}
In these expressions, the width of the $Z'$ is given by
\begin{eqnarray}
\Gamma_{Z'}&=&\frac{g_{\rm D}^2 \sqrt{M_{Z'}^2-4 \mDM^2} \left(2 \mDM^2+M_{Z'}^2\right)}{12 \pi M_{Z'}^2}\theta (M_{Z'}^2-4 \mDM^2)\nonumber\\
&+&\frac{g_{\rm V}^2 \sqrt{M_{Z'}^2-4 m_{\rm SM}^2} \left(2 m_{\rm SM}^2+M_{Z'}^2\right)}{12 \pi M_{Z'}^2}\theta (M_{Z'}^2-4 m_{\rm SM}^2)\,.
\end{eqnarray}
In general the $\DM$ radiated from PBHs may have different mass and different temperature than the particles off which it scatters, so that the above cross sections must be thermally averaged using \equaref{Hot Spots/Hawking/thermalAverage}.
    \chapter{Hot spot effects on thermal leptogenesis}\label{C}
    In light of the findings of \secref{Hot Spots/Lepto}, it is appropriate to revisit the assumption of homogeneous heating made in \chapref{PBHLepto}. Homogeneous heating is equivalent to assuming that all RHNs produced by PBHs easily escape the PBH hot spots (and that diffusion re-equilibrates the universe before BBN, this assumption is retained here). With the machinery developed in \secref{Hot Spots/Hawking}, it is now possible to assess the validity of this assumption in the case of high scale leptogenesis, and resonant LowScale leptogenesis.

\subsection*{High scale leptogenesis}
Taking the benchmark point 1 from Table \ref{Table}, which produces the maximum possible $\YB$ in high scale leptogenesis, the probability for RHNs to decay within the sphaleron radius is $\approx 1$ for all relevant $\MPBHini$. However since the RHNs are so heavy, the number density produced by PBHs is very small relative to the number of photons produced. Therefore, the contribution to $\YB$ is extremely small compared to $\YBobs$ for all regions of PBH parameter space. \\

It is concluded that hot spots do not alleviate the constraints shown in \figref{PBHLepto/HighScale/constraints}. 

\subsection*{Resonant leptogenesis}
Since the most likely RHNs to decay inside $r\leq \rsphal$ are heavy and strongly coupled, choosing $\Mdegen = 500\GeV$ and the largest allowed value of $U^2$ according to the parameter space calculated in \figref{PBHLepto/LowScale/standard}, which is $U^2 \approx 10^{-10}$, gives a conservative picture of the effect of hot spots on the constraints shown in \figref{PBHLepto/LowScale/constraints}. $\Delta M/M$ is fixed to maximise the CP asymmetry parameter. With these choices of parameters, there is a region of PBH parameter space in which a contribution to the baryon asymmetry $\sim \YBobs$ can be generated inside hot spots. Therefore, the constraints in \figref{PBHLepto/LowScale/constraints} should be corrected to account for this.\\

Lighter RHNs are expected to be less likely to decay inside $r\leq \rsphal$. Choosing $\Mdegen = 100\GeV$, there is no significant contribution to $\YB$ from PBH hot spots. Therefore the only constraint affected by the presence of hot spots is that for $\Mdegen = 500\GeV$. \\
\newpage
\begin{figure}[h]
    \centering
    \includegraphics[width=\linewidth]{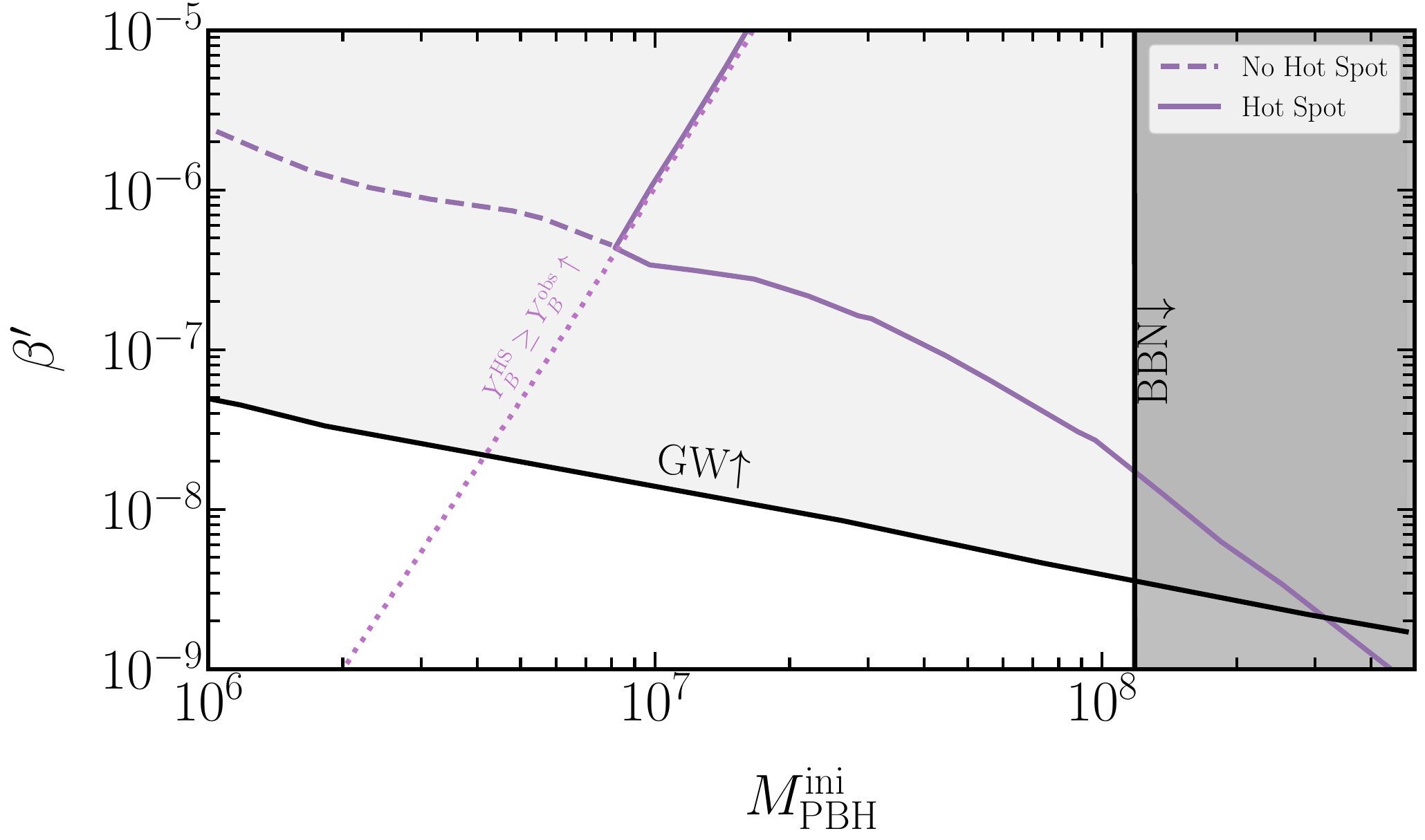}\caption{\label{Appendices/HSLepto/constraints} The bound calculated in \secref{PBHLepto/Low Scale} for $\Mdegen = 500\GeV$ is shown as the purple dashed line, which assumes homogeneous heating. The purple dotted line bounds the region where the contribution to $\YB$ from hot spots is greater than $\YBobs$. The constraint for $\Mdegen = 500\GeV$ is therefore modified due to the presence of hot spots, with the modified constraint shown as the solid purple line, which tracks the dotted line because in the region above the dotted line the contribution to $\YB$ from the hot spots ameliorates the entropy dilution, an effect which was not accounted for in \secref{PBHLepto/Low Scale}. The light grey shaded region is excluded by the constraints from GW energy density \cite{Papanikolaou:2020qtd,Papanikolaou:2022chm}, while the dark grey region is excluded by the impact of PBH evaporation on BBN \cite{Boccia:2024nly}.}
\end{figure}

In \figref{Appendices/HSLepto/constraints} the mutual exclusion limit appearing in \figref{PBHLepto/LowScale/constraints} for $\Mdegen = 500\GeV$ is shown as the dashed purple line, while the modified constraint accounting for the hot spot contribution to $\YB$ as in \secref{Hot Spots/Lepto} is shown as the solid purple line. The constraint is modified because in the region above the purple dotted line, $\YB \geq \YBobs$ is produced in the PBH hot spots. None of the other constraints in \figref{PBHLepto/LowScale/constraints} are modified by hot spots, the lighter RHNs are simply so weakly coupled as to easily escape beyond the sphaleron radius.  The BBN constraints have also been updated to account for the results reported in \cite{Boccia:2024nly}. However, the modification only occurs in the region where the GW energy density from PBHs exceeds the background energy density, leading to a back reaction problem. It is reasonable to conclude that the treatment in \secref{PBHLepto} was appropriate, since having improved upon it in this appendix, the conclusions do not change appreciably in either the high scale or resonant models.
\end{appendices}
\newpage

\bibliographystyle{unsrt}
\bibliography{Bibliography}

\newpage

\end{document}